\documentclass[aps,prd,onecolumn,superscriptaddress,preprintnumbers,floatfix,nofootinbib,longbibliography,eqsecnum]{revtex4-2}

\usepackage[german,english]{babel}
\usepackage{amsmath,amssymb}
\usepackage{hyperref}
\usepackage{titlesec}
\usepackage{bm}
\usepackage{xcolor}
\usepackage{bbm}
\usepackage{bbold}
\usepackage{ulem}
\usepackage{braket}
\usepackage{tablefootnote}
\usepackage{slashed}
\usepackage{multirow}
\usepackage{epsfig}
\usepackage{epstopdf}
\usepackage{diagbox}
\usepackage[mathlines]{lineno}
\usepackage[utf8]{inputenc}
\usepackage{comment}
\usepackage{graphicx}
\usepackage{mathrsfs}
\usepackage{tipa}
\usepackage{mathtools}
\usepackage{graphicx}
\usepackage{enumitem}
\usepackage{xcolor}
\usepackage{soul} 
\usepackage{soul}  

     \def\dag{\dagger}

\newcommand{\RN}[1]{%
  \textup{\uppercase\expandafter{\romannumeral#1}}%
}

\newcommand{\beq}{\begin{equation}}
\newcommand{\eeq}{\end{equation}}

\newcommand{\be}{\begin{equation}}
\newcommand{\ee}{\end{equation}}
\newcommand{\bea}{\begin{eqnarray}}
\newcommand{\eea}{\end{eqnarray}}

\makeatletter
\let\LN@align\align
\let\LN@endalign\endalign
\renewcommand{\align}{\linenomath\LN@align}
\renewcommand{\endalign}{\LN@endalign\endlinenomath}
\let\LN@gather\gather
\let\LN@endgather\endgather
\renewcommand{\gather}{\linenomath\LN@gather}
\renewcommand{\endgather}{\LN@endgather\endlinenomath}
\makeatother
\allowdisplaybreaks

\begin{document}

\preprint{TUM-EFT 199/25}
\title{Open-flavor threshold effects on quarkonium spectrum in the BOEFT}
\author{Nora Brambilla}
\email{nora.brambilla@tum.de}
\affiliation{Technical University of Munich,\\
TUM School of Natural Sciences, Physics
Department,\\   
James-Franck-Strasse~1, 85748 Garching, Germany.}
\affiliation{Technical University of Munich, Institute for Advanced Study, \\ 
Lichtenbergstrasse 2 a, 85748 Garching, Germany.}
\affiliation{Technical University of Munich, Munich Data Science Institute, \\ 
Walther-von-Dyck-Strasse 10, 85748 Garching, Germany.}
\author{Abhishek Mohapatra}
\email{abhishek.mohapatra@tum.de}
\affiliation{Technical University of Munich,\\
TUM School of Natural Sciences, Physics Department,\\   
James-Franck-Str.~1, 85748 Garching, Germany.}
\affiliation{
Theoretical Physics Division, Physical Research Laboratory, \\ 
Navrangpura, Ahmedabad 380009, India.}
\author{Tommaso Scirpa}
\email{tommaso.scirpa@tum.de}
\affiliation{Technical University of Munich,\\
TUM School of Natural Sciences, Physics Department,\\   
James-Franck-Str.~1, 85748 Garching, Germany.}
\author{Antonio Vairo}
\email{antonio.vairo@tum.de}
\affiliation{Technical University of Munich,\\
TUM School of Natural Sciences, Physics Department,\\   
James-Franck-Str.~1, 85748 Garching, Germany.}

\begin{abstract}
The impact of open-flavor thresholds on the quarkonium spectrum has been a subject of study since the introduction of the Cornell potential and has been quantified through various phenomenological approaches, most notably the $^3P_0$ model. 
We revisit this problem using the Born--Oppenheimer effective field theory (BOEFT), an effective field theory systematically derived from QCD by exploiting hierarchies of energy scales and symmetries. 
Within the BOEFT, open-flavor threshold effects emerge from the mixing between quarkonium and tetraquark static potentials sharing the same Born--Oppenheimer quantum numbers. 
The shapes of the static potentials are constrained by lattice QCD calculations.
Furthermore, we account for the distinctive behavior of the BOEFT tetraquark static potentials at short and large distances: 
at short distances they are repulsive, reflecting the color-octet configuration of the heavy quark-antiquark pair, 
while at large distances they asymptotically approach heavy-light meson-antimeson thresholds. 
To quantify threshold effects on the quarkonium spectrum below threshold, we solve a set of coupled Schr\"{o}dinger equations dictated by the BOEFT, 
whose only free parameter, the adjoint meson mass, is fixed to the mass of the $\chi_{c1}(3872)$ state. 
These coupled equations are solved both in the spin-isospin averaged threshold limit and, for the first time, including  the spin splittings of the physical thresholds. 
We validate our results by computing the same threshold effects as self-energy corrections to the quarkonium propagator.
We compare our predictions with existing experimental data and previous literature.
Finally, we provide a field-theoretical interpretation of the pair-creation constant $\gamma$ appearing in the $^3P_0$ model.
\end{abstract}


\maketitle 
\clearpage

\section{Introduction}\label{sec:intro}
The November revolution of 1974 marked a turning point in particle physics with the discovery of the $J/\psi$ and $\psi(2S)$ states~\cite{E598:1974sol,SLAC-SP-017:1974ind,Bacci:1974za,Abrams:1974yy}. 
This breakthrough triggered substantial theoretical efforts to develop a QCD based framework, incorporating both confinement and asymptotic freedom, to explain the masses and decay properties of these particles. 
The states were successfully interpreted as non-relativistic charm-anticharm, $c\bar{c}$, bound states called charmonia~\cite{PhysRevLett.34.43,PhysRevLett.34.369}.
The subsequent discovery of the $\Upsilon(1S)$, $\Upsilon(2S)$, and $\Upsilon(3S)$ states in 1977~\cite{E288:1977efs,CLEO:1980qvy,PhysRevLett.44.1111}, interpreted as bottom-antibottom, $b\bar{b}$, bound states called bottomonia, 
further stimulated the development of non-relativistic potential models for quarkonium systems~\cite{Bhanot:1978mj,Richardson:1978bt,PhysRevLett.45.103,Martin:1980jx,Buchmuller:1980su} (see~\cite{Quigg} for a comprehensive review).
Moreover, the observation of new quarkonium states lying near heavy-light meson-antimeson thresholds, as the $\psi(3770)$ and $\Upsilon(4S)$ resonances~\cite{Rapidis:1977cv,Abrams:1979cx,CLEO:1980tem}, generated significant interest in the study of threshold effects on quarkonium states.

To explain strong hadronic decays, the vacuum pair creation model was initially proposed by Micu~\cite{Micu:1968mk} and subsequently refined by Carlitz and Kislinger~\cite{Carlitz:1970xb}, and by Le Yaouanc et al. specifically for charmonium strong decays~\cite{LeYaouanc:1972vsx,LeYaouanc:1977fsz}. 
This model, commonly referred to as the $^3P_0$ model, from the quantum numbers of the light quark-antiquark pair generated from the vacuum during the decay process, rapidly gained widespread adoption due to its simplicity and adaptability~\cite{Ono:1980js,Heikkila:1983wd,Ono:1983rd,Tornqvist:1984xz,Roberts:1992esl}, and remains still popular to describe the quarkonium spectrum including meson-antimeson threshold effects~\cite{Barnes:2005pb,Pennington:2007xr,Barnes:2007xu,Li:2009ad,Danilkin:2009hr,Ferretti:2012zz,Ferretti:2013faa,Ferretti:2013vua,Ferretti:2014xqa,Ferretti:2018tco,Man:2024mvl,Blossier:2024dhm,Bruschini:2025paj,Hao:2025vmw,Ni:2025gvx,M:2025gnf,Sultan:2025dfe,Ahmad:2025mue,Gao:2025tob,Man:2025vmm,Ahmad:2025hcr,Farina:2020slb,Zhao:2025kno,Chen:2025pvk,Rui:2025olj,daSilva:2008rp, Barnes:2002mu}.
However, it retains a phenomenological character due to the pair-creation constant $\gamma$, which regulates the mixing between quarkonium and threshold states, that must be fitted to experimental data along with other parameters of the potential model. 
Moreover, despite several generalisations that have been attempted, like making $\gamma$ dependent on the space coordinates to account for the flux-tube breaking~\cite{Kokoski:1985is, Godfrey:1986wj}, postulating the one-gluon exchange or confining potentials responsible for the light quark-antiquark pair production~\cite{Ackleh:1996yt}, making $\gamma$ scale dependent~\cite{Segovia:2012cd}, considering the case of $^3S_1$ quantum numbers for the light quark-antiquark pair~\cite{Kumano:1988ga}, its direct connection with QCD remains elusive.

Concurrently, Eichten et al. developed the Cornell coupled-channel model (CCCM), where,  to limit the number of free parameters, a common Cornell potential is used to describe both the non-relativistic quarkonium dynamics and the string-breaking dynamics, thereby accounting for the mixing between quarkonium and heavy-light meson-antimeson states.
Also in this case, the potential parameters are fixed to the spectrum in both the original studies~\cite{Eichten:1978tg,Eichten:1979ms} and subsequent ones~\cite{Eichten:2004uh,Eichten:2005ga}.  

In these approaches the presence of the light quark contributions, either as a part of the binding dynamics or through virtual loops, affects both the potential parameters and the transition amplitudes and decay processes. 
The treatment is model dependent by construction. 
The main open and interconnected questions that these models try to address are: 
How to account for the threshold effects and couplings of quarkonium states to heavy-light meson decay channels? 
How to incorporate these effects at the level of a nonrelativistic description? 
What is the role and the impact of light quark loops at the different energy scales? 
How to treat the light degrees of freedom when appearing as constituents  in QCD states with two heavy quarks?

Since more than two decades, it is experimentally known that heavy-light pairs may exist as constituents in states made by a heavy quark-antiquark (or two heavy quark) pair. 
These states, which theory predicted long before their discovery~\cite{Gell-Mann:1964ewy,Jaffe:1975fd,Jaffe:1976ig,Jaffe:1976ih}, are known as tetraquarks.
The first of such states to be observed was the $\chi_{c1}(3872)$ (originally known as $X(3872)$)~\cite{Belle:2003nnu}, 
which was followed by numerous other tetraquark candidates like the $Z_c$, $Z_b$, and $T_{cc}(3875)^+$~\cite{BESIII:2013ouc,Belle:2013yex,Belle:2011aa,Belle:2013urd,BESIII:2015cld,Belle:2008qeq,Belle:2014nuw,BESIII:2020qkh,LHCb:2021uow,PhysRevLett.131.131901,LHCb:2021vvq}, as well as pentaquark candidates like $P_{c\bar{c}}(4312)^+$, $P_{c\bar{c}}(4380)^+$, $P_{c\bar{c}}(4440)^+$, $P_{c\bar{c}}(4457)^+$~\cite{LHCb:2015yax,LHCb:2019kea}.
Besides tetraquark and pentaquark candidates, dozens of other heavy hadrons have been discovered at the $B$-factories, $\tau$-charm factories, and hadron colliders in the quarkonium spectrum that cannot be identified with conventional quarkonia or heavy baryons, be it for their quantum numbers, masses, or decay patterns.
For some reviews, we refer to~\cite{Guo:2017jvc,Ali:2017jda,Olsen:2017bmm,Ali:2019roi,Brambilla:2019esw,Liu:2019zoy,Chen:2022asf}.

At the same time, in the last decades, significant strides have been made in improving both the reach and the precision of lattice QCD calculations, as well as in describing quarkonium systems within QCD's non-relativistic effective field theories, 
in this way moving beyond potential models~\cite{Brambilla:2004jw}. 
In the framework of non-relativistic effective field theories originally developed for quarkonium~\cite{Bodwin:1994jh,Brambilla:1999xf} and then extended to the description of any exotic system containing two heavy quarks, potentials show up as matching coefficients in an ultimate effective field theory that lives at the scale of the binding energy. 
Potentials can be computed non-perturbatively in (lattice) QCD. 
They are distinguished by specific quantum numbers and expressed in terms of generalized static Wilson loops.
This ultimate effective field theory is called the Born--Oppenheimer effective field theory (BOEFT)~\cite{Berwein:2015vca,Brambilla:2017uyf,Soto:2020xpm,Berwein:2024ztx}.  
The BOEFT predicts the form of the coupled Schr\"odinger equations describing both quarkonium and exotic states, 
and the mixing between them.

In the case of tetraquarks, the BOEFT exploits the energy scale separations and symmetry properties of systems containing two heavy quarks and light degrees of freedom. 
It follows from integrating out modes associated with energy scales larger than the binding energy from the QCD description of the system, 
and encoding systematically their dynamical information into suitable potentials.
These modes include light quarks of energy and momentum of the order of or larger than the typical momentum transfer in the bound state.
Valence light quarks contribute to defining the Born--Oppenheimer (BO) quantum numbers of the fields describing the tetraquarks in the effective field theory Lagrangian. 
In this way, all contributions coming from light quarks are throughly considered, without the possibility of double counting. 
Open-flavor threshold effects in the quarkonium spectrum appear through the mixing of the quarkonium potential with the first tetraquark static potential of the same BO quantum numbers.
It is the long-range tail of the tetraquark potential that carries the information of the string breaking. 
So, leveraging the BOEFT with appropriate lattice QCD inputs, we are able to treat 
the impact of light quarks on the quarkonium spectrum and transitions systematically. 

Beyond the aforementioned studies based on the $^{3}P_0$ model, 
threshold effects have been partially addressed by different groups adopting the BO approximation~\cite{Bicudo:2019ymo, Bruschini:2020voj, Bicudo:2020qhp,Bruschini:2021sjh, Bicudo:2022ihz,Zhang:2025bex,Bruschini:2021cty, Bruschini:2021sjh} and even within the BOEFT formalism~\cite{TarrusCastella:2022rxb}. 
In these studies, the static tetraquark potentials are taken as a constant for all heavy quark-antiquark distances $r$, 
the value of the constant being that one of the physical meson-antimeson threshold energy.
However, as recently highlighted in~\cite{Berwein:2024ztx}, a tetraquark potential evolves to the corresponding heavy-light meson-antimeson threshold only at large $r$,  
while at small $r$ it exhibits a repulsive octet behavior due to the color configuration of the heavy quark-antiquark pair. 
Hence, in this paper we re-examine open-threshold effects in the light of the recent progress made in the BOEFT framework 
and the improved lattice determinations of the non-relativistic potentials.
We restrict our analyses to threshold effects affecting quarkonium states below the lowest meson-antimeson thresholds. 
This amounts to solving a system of coupled Schr\"odinger equations between the quarkonium and the lowest isospin-0 tetraquark states.
The same system of equations also predicts the $\chi_{c1}(3872)$ state on which we fix the only free parameter of our calculation, which is the adjoint meson mass. 
All the other predictions follow from this choice.

The structure of the paper is as follows. 
In Section~\ref{sec:BOEFT}, we summarize the BOEFT description of spin-averaged quarkonium and tetraquark states, and provide the leading order BOEFT Lagrangian describing the mixing between them, as well as the static potential parametrizations. 
In Section~\ref{sec:coupled}, we solve the coupled Schr\"{o}dinger equations and quantify the open-flavor threshold effects on the spectrum.
In Section~\ref{sec:thr-spl}, we include heavy-quark spin effects via $\mathcal{O}(1/m_Q)$ tetraquark spin-dependent potentials, $m_Q$ being the heavy quark mass, 
which induce meson–antimeson threshold splittings. 
The resulting system of coupled Schr\"{o}dinger equations is solved to compute effects on the quarkonium spectrum.
In Section~\ref{sec:self-en}, we calculate the same open-flavor threshold effects as in Section~\ref{sec:coupled}, however from the self-energy of the quarkonium propagator.
Moreover, we compare with the $^{3}P_0$ model, identifying the mixing potential required by the $^3P_0$ model 
and comparing it with the lattice QCD (LQCD) based BOEFT mixing that we adopt.
In Section~\ref{sec:literature}, we compare our results with phenomenological and BO based models, as well as with experimental data. 
Finally, in Section~\ref{sec:conclusions} we draw our conclusions and provide future outlooks.
This work is also complemented by four appendices.
Appendix~\ref{app:comparison_3P0_BOEFT} and Appendix~\ref{app:comparison_S_P} give additional details on the self-energy calculations within the $^{3}P_0$ model framework, providing the most general expression of the self-energy and analyzing the specific case of $P$-wave quarkonia.
In Appendix~\ref{app:meson-meson basis}, we present the full set of coupled Schr\"{o}dinger equations that incorporate the $\mathcal{O}(1/m_Q)$ tetraquark spin-dependent potentials.
The corresponding bound-state predictions, along with summary plots comparing our findings with experimental data, are given in Appendix~\ref{app:tetraquark spin splitting}.

\section{The BOEFT description of the system}\label{sec:BOEFT}
The BOEFT formalism has been established in several papers~\cite{Berwein:2015vca,Oncala:2017hop,Soto:2020xpm,Berwein:2024ztx,Brambilla:2018pyn,Brambilla:2019jfi,Brambilla:2024imu, Brambilla:2025xma}.
Here, we investigate how it describes the mixing 
between quarkonium, $Q\bar{Q}$, and the lowest isospin-zero tetraquark states, $Q\bar{Q}q\bar{q}$, which is responsible for open-flavor threshold effects; 
$Q$ denotes a heavy quark, $Q = c, b$, while $q$ denotes a light quark, $q = u, d$.
The notation $Q\bar{Q}$ encompasses heavy quark pairs of the same or different flavors, $Q \bar{Q} \in\{ b \bar{b}, c \bar{c}, b \bar{c}, c \bar{b}\}$, and similarly for $q\bar{q}$, unless otherwise specified.
The heavy quarks constitute the heavy degrees of freedom (HDF) of the system, while the light quarks, if present, are the light degrees of freedom (LDF). 
We introduce the BO quantum numbers in Section~\ref{subsec:quantum_numbers}, the leading-order BOEFT Lagrangian relevant to our problem in Section~\ref{subsec:BOEFT_Lagrangian}, and the parametrizations of the static BOEFT potentials entering it in Section~\ref{subsec:parametrization}.

\subsection{Born--Oppenheimer quantum numbers}\label{subsec:quantum_numbers}
Let us consider a QCD bound state consisting of a heavy quark-antiquark pair $Q \bar{Q}$ separated by a distance $r$ and some LDF, i.e. light quarks or gluons.
In the static limit $m_Q \rightarrow \infty$, the heavy quark positions are fixed and  locate the static color sources. 
The system's states and quark-antiquark static potentials are classified according to representations of the cylindrical group $D_{\infty h}$. 
These are labeled by the BO quantum numbers $\Lambda_\eta^\sigma$:
\begin{itemize}
 \item[{\it (a)}] $\Lambda \equiv \lvert \mathbf{K} \cdot \hat{\bm r} \rvert$ is the absolute value of the projection of the total LDF angular momentum $\mathbf{K}$ onto the $\hat{\bm r}$ axis joining the static $Q$ and $\bar{Q}$. Integer values $\Lambda = 0, 1, 2, \dots$ are denoted $\Sigma$, $\Pi$, $\Delta$, etc;
\item[{\it (b)}] $\eta = u, g$ is the PC eigenvalue of the LDF;   
\item[{\it (c)}] $\sigma = \pm 1$ is the eigenvalue under reflection through a plane containing $\hat{\bm r}$. This is explicitly indicated only for $\Lambda = \Sigma$, where static states are non-degenerate.
\end{itemize}
We denote a static potential for a given LDF configuration as $V_{\Lambda_\eta^\sigma}$. 
In addition to the BO quantum numbers, static potentials carry LDF flavor quantum numbers $f$ (e.g., isospin $I$ or baryon number $b$).
For a $Q\bar{Q}q\bar{q}$ tetraquark, the two heavy quarks form a color-octet $(Q\bar{Q})_8$ core in the $r \rightarrow 0$ limit. 
This binds to the LDF in the adjoint representation of $SU(3)$ to produce a color-singlet hadron~\cite{Berwein:2024ztx}.\footnote{
Color-singlet $Q\bar{Q}$ combinations yield quarkonium plus light hadrons (e.g. pions), 
which lattice QCD shows do not form tightly bound multiquark states~\cite{Alberti:2016dru,Prelovsek:2019ywc}.} 
We refer to the LDF states with quantum numbers $\kappa = \{K^{PC}, f\}$ as \textit{adjoint mesons}.

The static potentials introduced above originate from the non-perturbative matching between QCD and the BOEFT in the static limit~\cite{Berwein:2024ztx}.
Different static potentials can mix.
This happens either in the short-distance limit $r \rightarrow 0$, due to spherical symmetry restoration,\footnote{
Different static potentials having the same $K^{PC}$ quantum numbers become degenerate; $K(K+1)$ are the eigenvalues of $\mathbf{K}^2$.} 
or at finite distances if two static potentials share the same BO quantum numbers. 
In the latter case, when the two potentials in the adiabatic representation, introduced in Sections~\ref{subsec:parametrization} and ~\ref{sec:coupled},
become close to one another in some range of \( r \), they give rise to the avoided level crossing phenomenon~\cite{Berwein:2024ztx}.

\subsection{BOEFT Lagrangian}\label{subsec:BOEFT_Lagrangian}
We consider the BOEFT Lagrangian that describes the mixing between conventional quarkonium and the lowest-lying isospin-0 tetraquark encoding string breaking. 
Since both systems have isospin $I=0$, we omit this quantum number in the following discussion and denote the meson's adjoint quantum numbers simply as $\kappa = K^{PC}$.
As detailed in~\cite{Berwein:2024ztx}, BOEFT Lagrangians are non-relativistic by construction, with all the potentials expanded in powers of $1/m_Q$.
The static potentials $V_{\Lambda_\eta^\sigma}$ introduced in Section~\ref{subsec:quantum_numbers} constitute the leading order term in this expansion. 
We refer to the relative Lagrangians, retaining only the static potentials, as the \textit{leading-order BOEFT Lagrangians}.

\begin{figure}[ht]
\centering 
\includegraphics*[width=10.5cm,clip=true]{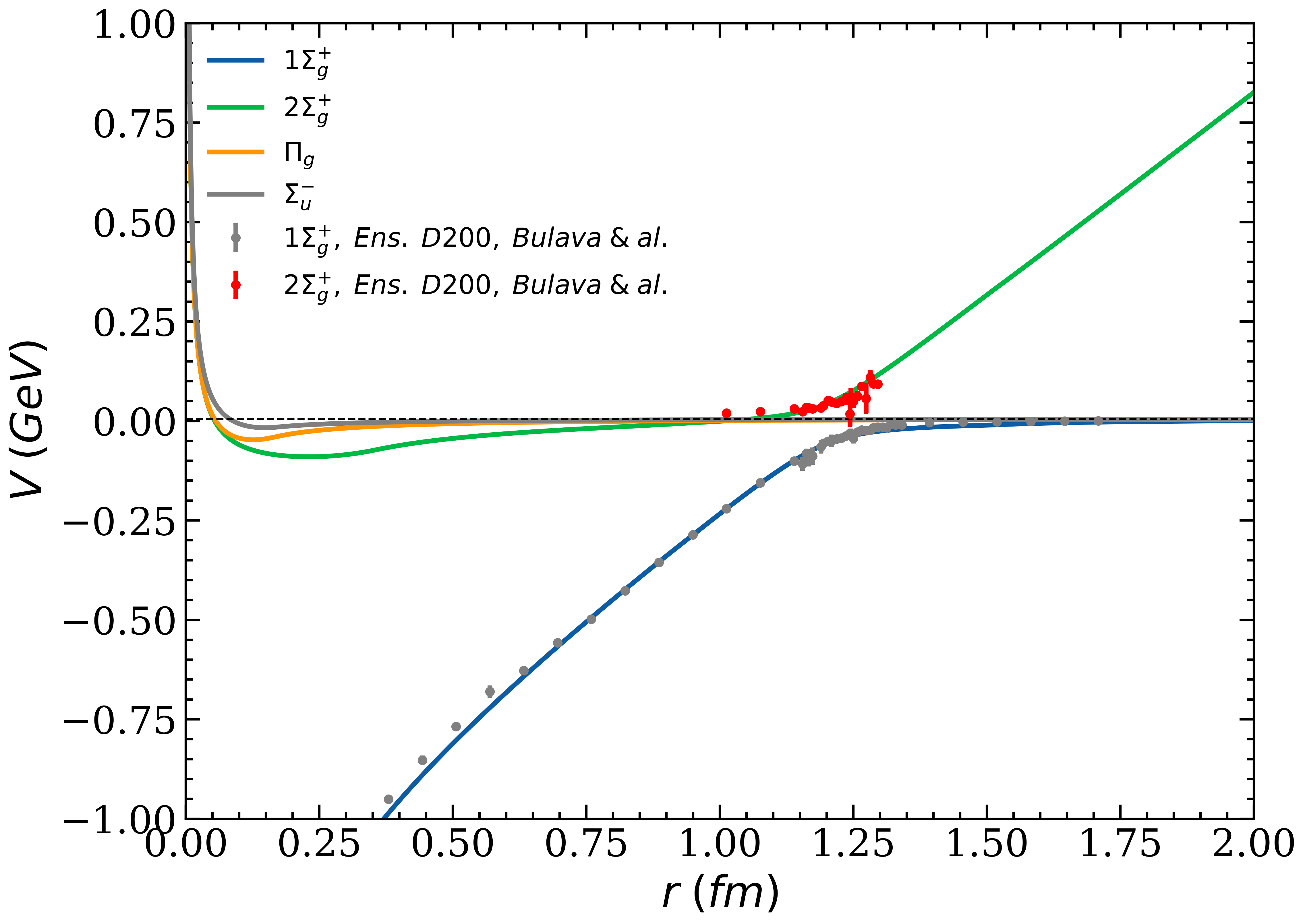}
\caption{The static BO adiabatic potentials appearing in the leading order BOEFT Lagrangian \eqref{eq:BOEFTlagrangian}. 
The $1\, \Sigma_g^+$, $2 \, \Sigma_g^+$ adiabatic levels are obtained by diagonalizing the $2 \times 2$ potential matrix having $V_{\Sigma_g^+}$, $V_{\Sigma_g^{+ \prime}}$ along the main diagonal and $V_{\Sigma_g^+ - \Sigma_g^{+ \prime}}$ in the off-diagonal entries.
Both $V_{2\Sigma_g^{+}}$ and $V_{\Pi_g}$ tend to the same $\Lambda_{1^{--}}$ adjoint meson mass at short distances, while $V_{{\Sigma_u^-}}$ tends to the $\Lambda_{0^{-+}}$ one. 
These three potentials share the same color-octet repulsive behavior at small distances and approach the meson-antimeson static threshold from below, set at $0.005\,\mathrm{GeV}$ according to the LQCD data. 
The lattice data are from the D200 ensemble of~\cite{Bulava:2024jpj}, whose $2\Sigma_g^{+}$ data is limited to the avoided level crossing region only.}
\label{fig:isospin0}
\end{figure}

The lowest static potential has BO quantum numbers $\Lambda_\eta^\sigma = \Sigma_g^+$, which for the adjoint meson translates to $K^{PC} = 0^{++}$, and is identified with the quarkonium potential.
For the lowest-lying isospin-0 tetraquarks \((Q \bar{Q} q \bar{q})\) with \( q = u, d \), the adjoint meson quantum numbers are $K^{PC} = 0^{-+}$ or $1^{--}$. 
These correspond, respectively, to the BO configurations $\Lambda_\eta^\sigma = \Sigma_u^-$ and $\Lambda_\eta^\sigma = \{ \Sigma_g^{+'}, \Pi_g \}$.
The static heavy-light $S$-wave meson-antimeson thresholds possess LDF quantum numbers $K^P = (1/2)^-$ for the light quark $q$ 
and $K^P = (1/2)^+$ for the light antiquark $\bar{q}$. 
Their total angular momentum combinations yield the same quantum numbers, $K^{PC} = 0^{-+}$ and $1^{--}$ as the adjoint meson and share the same BO quantum numbers — $\Sigma_u^-$ and $\{ \Sigma_g^{+'}, \Pi_g \}$~\cite{TarrusCastella:2022rxb, TarrusCastella:2024zps} — with  the lowest isospin-0 tetraquarks.
Due to the conservation of the BO quantum numbers $\Lambda_\eta^\sigma$, the tetraquark static potentials must connect smoothly at large distances to the corresponding static heavy-light meson-antimeson thresholds with the same $\Lambda_\eta^\sigma$~\cite{Berwein:2024ztx, Braaten:2024tbm, Brambilla:2024imu}.
Since the quarkonium potential $V_{\Sigma_g^+}$ and one of the tetraquark potentials, $V_{\Sigma_g^{+\prime}}$, share the same BO quantum numbers and cross, 
their adiabatic potentials undergo avoided level crossing, which occurs at approximately $1.2\, \mathrm{fm}$ (see Figure~\ref{fig:isospin0}). 
Moreover, the tetraquark potentials $V_{\Sigma_g^{+\prime}}$ and $V_{\Pi_g}$ are degenerate at short distances with the adjoint meson of quantum numbers
$K^{PC} = 1^{--}$.
In addition to the static potentials $V_{\Sigma_g^{+}}$, $V_{\Sigma_g^{+\prime}}$,  $V_{\Pi_g}$ and $V_{\Sigma_u^-}$, the full BOEFT Lagrangian includes the heavy quark-antiquark kinetic term, 
which enters at $\mathcal{O}(1/m_Q)$. 
According to the virial theorem, its expectation value is comparable in magnitude to that of the leading-order BO static potential. 

The leading order BOEFT Lagrangian density that accounts for the mixing between quarkonium and the lowest isospin-0 tetraquark states is then given by:\footnote{
In the more general case of non-equal heavy quark masses, as for the $b \bar{c}$ and $c \bar{b}$ cases, $m_Q$ needs to be substituted with two times the reduced mass of the system.
Consistently with the LQCD calculations discussed in Section~\ref{subsec:parametrization} for the static tetraquark potentials, the $u$, $d$ quarks are taken to be mass degenerate (isospin limit).}
\begin{equation}\label{eq:BOEFTlagrangian}
{\cal L}_{\mathrm{BOEFT}} = {\cal L}_{{\mathrm{quarkonium}}}+{\cal L}_{{\mathrm{tetraquark}}}+{\cal L}_{\mathrm{mixing}},\\
\end{equation}
with: 
\begin{align}
&{\cal L}_{{\mathrm{quarkonium}}}=
\int d^3{\bm r}\, 
 \mathrm{Tr}\Bigg[\bm{\psi}_{\Sigma_g^+}^{\dagger}(\mathbf{r},\,\mathbf{R},\,t)\left(i\partial_t+\frac{\bm{\nabla}^2_r}{m_Q}-V_{\Sigma_g^+}(r)\right)\bm{\psi}_{\Sigma_g^+}(\mathbf{r},\,\mathbf{R},\,t)\Bigg],
\label{eq:LQQ}\\
&{\cal L}_{{\mathrm{tetraquark}}} = \int d^3{\bm r} \, \sum_{\kappa=0^{-+},1^{--}} \sum_{\lambda,\lambda^{\prime} = -K}^K \mathrm{Tr}\left\{{\bm{\psi}}^{\dagger}_{\kappa\lambda}(\mathbf{r},\,\mathbf{R},\,t) \left[i\partial_t- V_{\kappa\lambda}(r) \delta_{\lambda,\lambda^{\prime}}+P^{i\dag}_{\kappa\lambda}\frac{\bm{\nabla}^2_r}{m_Q}P^i_{\kappa\lambda^{\prime}}
                   \right]{\bm{\psi}}_{\kappa\lambda^{\prime}}(\mathbf{r},\,\mathbf{R},\,t)\right\}, 
\label{eq:LQQqq}\\
&{\cal L}_{\mathrm{mixing}} = -\int d^3{\bm r}\,\mathrm{Tr}\left[\bm{\psi}_{\Sigma_g^+}^{\dagger}(\mathbf{r},\,\mathbf{R},\,t)\,V_{1^{--}\, 0}^{\mathrm{mix}}(r)\,{\bm{\psi}}_{1^{--}\, 0}(\mathbf{r},\,\mathbf{R},\,t) + \mathrm{H.c.}\right],
\label{eq:L_mixing}
\end{align}
where the trace is over the spin indices.
$\bm{\psi}_{\Sigma_g^+}$ and ${\bm{\psi}}_{\kappa\lambda}$ denote the quarkonium and the tetraquark fields, respectively.
They are functions of the relative coordinate $\bm{r}\equiv {\bm x}_1-{\bm x}_2$, and the center of mass coordinate $\bm{R}\equiv \left({\bm x}_1+{\bm x}_2\right)/2$ of the $Q\bar{Q}$ pair, where ${\bm x}_1$ and ${\bm x}_2$ are the space locations of the heavy quark and antiquark. 
We set $\bm{R} = 0$ in the following calculations. 
The tetraquark static potentials in Eq.~\eqref{eq:LQQqq} 
are understood as
\begin{equation}\label{eq:potentials}
    V_{1^{--}\,0}(r) \equiv V_{\Sigma_g^{+ \prime}}(r),\qquad
    V_{1^{--}\,\pm1}(r) \equiv V_{\Pi_g}(r),\qquad
    V_{0^{-+}\, 0} (r) \equiv V_{\Sigma_u^-}, \qquad
\end{equation}
moreover we rename $V^{\mathrm{mix}}_{1^{--}\,0}(r) \equiv V_{\Sigma_{g}^{+}-\Sigma_{g}^{+\prime}}(r)$ to emphasize that the mixing occurs between the potentials with BO quantum numbers $\Sigma_{g}^{+}$ and $\Sigma_{g}^{+\prime}$.

The projection vector $P^i_{\kappa \lambda}$ with the index $i$ denoting a vector or spin component projects the field 
onto an eigenstate of the operator $\bm{K} \cdot \hat{\bm{r}}$ with eigenvalue $\lambda = -K, \dots, 0, \dots, K$. 
For the quarkonium channel with $\Lambda_\eta^\sigma = \Sigma_g^+$ ($K^{PC}=0^{++}$) and the tetraquark channel with $\Lambda_\eta^\sigma = \Sigma_u^-$ ($K^{PC}=0^{-+}$), the projection is trivial, yielding $P_{00} = \mathbb{1}$.
For the tetraquark components $\{\Sigma_g^{+'}, \Pi_g\}$, which share the quantum numbers $K^{PC}=1^{--}$, the projection vectors $P^{i}_{1\lambda}$ are given by 
\begin{equation}
    P^i_{10} = \hat{r}^i,\qquad
    P^i_{1\pm 1}=\hat{r}_{\pm}^{i}\equiv \mp\left(\hat{\theta}^i\pm i\hat{\phi}^i\right)/\sqrt{2},\label{eq:P_1}
\end{equation}
where $\hat{r}$, $\hat{\theta}$, and $\hat{\phi}$ are the spherical unit vectors.
In the Lagrangian of Eq.~\eqref{eq:LQQqq}, the off-diagonal kinetic matrix terms $P^{i\dag}_{\kappa\lambda}\displaystyle\frac{\bm{\nabla}^2_r}{m_Q}P^i_{\kappa\lambda^{\prime}}$ with $\lambda \neq \lambda'$
follow from the kinetic energy no being $D_{\infty h}$ symmetric.

\subsection{Static potentials parameterization and BOEFT states}\label{subsec:parametrization}
In this section, we discuss the parameterizations of the BO static potentials appearing in Eq.~\eqref{eq:BOEFTlagrangian}. 
The first LQCD calculations focusing on the mixing between quarkonium and the lowest isospin-0 $Q \bar{Q} q \bar{q}$ tetraquark static potentials 
were performed in~\cite{Bali:2005fu}.\footnote{
In this and subsequent lattice studies, 
mixing with quarkonium was investigated by considering
tetraquarks in their asymptotic meson-antimeson configurations only.} 
More recent studies~\cite{Bulava:2019iut, Bulava:2024jpj} also included mixing with the $Q \bar{Q} s \bar{s}$ tetraquark, not considered here.\footnote{
Including $Q \bar{Q} s \bar{s}$ tetraquark static potentials would introduce two additional channels to the coupled Schr\"{o}dinger equations~\eqref{eq:coupledSchr} — mixing at leading order with quarkonium — and two extra free parameters (the adjoint meson masses $\Lambda_{0^{-+}}^{s \bar{s}}$ and $\Lambda_{1^{--}}^{s \bar{s}}$), which would need to be fixed on the spectrum.}
For our analysis, we employ the lattice data from~\cite{Bulava:2024jpj}, which is restricted to the string-breaking region 1~fm$\lesssim r \lesssim$~1.5~fm. 
This data constrains the behavior of the quarkonium potential $V_{\Sigma_{g}^{+}}$, the tetraquark potential $V_{\Sigma_{g}^{+\prime}}$, and the mixing potential $V_{\Sigma_{g}^{+}-\Sigma_{g}^{+\prime}}$ in that region. 
The quarkonium potential itself is reliably known also beyond the string-breaking region.
In contrast, the tetraquark potentials $V_{\Pi_g}$ and $V_{\Sigma_u^-}$ have not yet been computed in LQCD. 
Although their functional forms are constrained by symmetries, one key parameter — the adjoint meson mass — must be fixed phenomenologically, as discussed below.

We model the tetraquark static potentials $V_{\Sigma_g^{+\prime}}$, $V_{\Pi_g}$, and $V_{\Sigma_u^-}$ distinguishing between short and long distance behaviors.
At short distances, the $Q\bar{Q}$ pair is in a color-octet configuration.
Hence, the leading term of the tetraquark potentials is a repulsive Coulomb potential, whose coefficient is fixed by the one-gluon exchange. 
The next term in the $r$ expansion is the adjoint meson mass, which accounts for the $q\bar{q}$ pair mass, followed by higher-order corrections. 
The adjoint meson mass, as well as the coefficients of the $\mathcal{O}(r^2)$ and $\mathcal{O}(r^4)$ terms, encode the non-perturbative LDF dynamics and can be expressed via appropriate correlators~\cite{Berwein:2024ztx}. 
Due to the current lack of lattice data, we model the short-distance part using the parameterization of the \textit{quenched} BO potentials from pure $SU(3)$ lattice gauge theory as reported for hybrid static potentials in~\cite{Alasiri:2024nue}.
At large distances, the potentials are constrained by the conservation of the BO quantum numbers $\Lambda_\eta^\sigma$ and asymptotically approach the spin-isospin averaged $S$-wave heavy-light meson-antimeson threshold.\footnote{
The isospin average follows from taking the $u,d$ quarks degenerate, as in the lattice study~\cite{Bulava:2024jpj}. 
The spin average follows from the fact that the heavy-quark spin decouples from the LDF dynamics at leading order in the BOEFT Lagrangian, 
making the thresholds $M_1 \bar{M}_2$, $M_1^* \bar{M}_2$ (or $M_1 \bar{M}_2^*$), and $M_1^* \bar{M}_2^*$ — with $M_{1,2}^{(*)} = D^{(*)}, B^{(*)}$ — degenerate at this order.} 
We refer to this threshold as the \textit{spin-averaged threshold} or equivalently the \textit{spin-averaged meson-antimeson threshold}.
The transition to this asymptotic configuration is modeled by including residual strong interactions between the heavy-light meson and antimeson, specifically through the pion-exchange dynamics.\footnote{Unlike in our previous work~\cite{Brambilla:2024imu}, where we adopted a two-pions exchange parametrization based on Refs.~\cite{Lyu:2023xro, Aoki:2025jvi}, here we use a one-pion exchange parametrization for the tetraquark static potentials. 
Indeed, the spin-averaged one-pion contribution to the $\Sigma_g^{+ \prime}$, $\Pi_g$, and $\Sigma_u^-$ thresholds computed in heavy hadron chiral perturbation theory is non-vanishing in all these cases. 
This induces a shift in the value of the adjoint meson mass of about $90$~MeV, but leaves all the properties of the $\chi_{c1}(3872)$ state (composition, mass, radius, etc.), on which the $\Lambda_{1^{--}}$ has been tuned in both cases, nearly identical. 
Furthermore, as we discuss later, the masses of and the string-breaking corrections to the quarkonium states, which are central to this study (Secs.~\ref{sec:coupled} and~\ref{sec:thr-spl}), are affected at most at the level of a few $\mathrm{MeV}$ by the choice of one parametrization or the other.}

Based on the above considerations, we parameterize the quarkonium and tetraquark static potentials in Eq.~\eqref{eq:BOEFTlagrangian} as:
\begin{align}
&V_{\Sigma_g^+}(r) = V_0 + \frac{\gamma}{r} + \sigma r,\label{eq:Vcornell}\\
&V_{\Lambda^{\sigma}_{\eta}}(r) = 
\begin{cases}
\kappa_8/r +  \bar{\Lambda}_{K^{PC}} + A_{\Lambda_\eta^\sigma}\, r^2+ B_{\Lambda_\eta^\sigma} \, r^4 & r < R_{\Lambda_\eta^\sigma} \\
F_{\Lambda_\eta^\sigma}\,e^{-r/d}/r + E_1 & r > R_{\Lambda_\eta^\sigma}
\end{cases},
\label{eq:QQbar_V}
\end{align}
where $\Lambda^{\sigma}_{\eta}\in\{\Sigma_g^{+\prime}, \Pi_g, \Sigma_u^- \}$, $V_0 = -1.142~\mathrm{GeV}$, $\gamma = -0.434$, $\sigma = 0.198\,\mathrm{GeV}^2$,
$\kappa_8=0.037$, $A_{\Sigma_g^{+\prime}} = 0.0065 \,\mathrm{GeV}^3$, $B_{\Sigma_g^{+\prime}} = 0.0018\,\mathrm{GeV}^5$, $A_{\Pi_g} = 0.0726\,\mathrm{GeV}^3$, $B_{\Pi_g}=-0.0051\,\mathrm{GeV}^5$, $A_{\Sigma_u^-} = 0.0420 \, \mathrm{GeV}^3$, $B_{\Sigma_u^-} = 0.0001 \, \mathrm{GeV}^5$, $d =  1/m_\pi = 1/0.15\,\mathrm{GeV}^{-1} = 1.31\,\mathrm{fm}$, $E_1=0.005$~GeV. 
The parameters $F_{\Lambda_\eta^\sigma}$ and $R_{\Lambda_\eta^\sigma}$ are determined by imposing continuity up to the first derivatives for the potentials.
The terms $\Lambda_{K^{PC}} \equiv \bar{\Lambda}_{K^{PC}} + V_0$ are the adjoint meson masses $\Lambda_{0^{-+}}$ and $\Lambda_{1^{--}}$ of the $V_{\Sigma_u^-}$ and $V_{\Sigma_g^{+\prime}}$, $V_{\Pi_g}$ tetraquark static potentials, respectively. 
The isospin-0 $1^{--}$ adjoint meson mass $\Lambda_{1^{--}} 
= 1.008$~GeV,\footnote{
The given adjoint meson mass is in the Cornell mass scheme~\cite{Mateu:2018zym}.
The Cornell mass scheme can be taken as a proxy of the pole mass scheme, the only difference between the two schemes being that the octet potential in the pole mass scheme is given by the perturbative expression of the adjoint potential.}
not having been computed yet in LQCD, has been tuned to reproduce the mass of the $\chi_{c1}(3872)$ state. 
Moreover, we use the energy difference $\Lambda_{0^{-+}} - \Lambda_{1^{--}} = 0.044$~GeV obtained in~\cite{Foster:1998wu} to fix the isospin-0 $0^{-+}$ adjoint meson mass: $\Lambda_{0^{-+}} = 1.052$~GeV.
The adiabatic potentials $1\Sigma_g^+$ and $2\Sigma_g^+$ are obtained by diagonalizing the $2 \times 2$ potential matrix having  $V_{\Sigma_g^+}$, $V_{\Sigma_g^{+ \prime}}$ along the main diagonal and $V_{\Sigma_g^+ - \Sigma_g^{+ \prime}}$ in the off-diagonal entries. 
They reproduce the lattice data relative to the ensemble D200 of~\cite{Bulava:2024jpj} around the string-breaking region (see Fig.~\ref{fig:isospin0}). 
Being both $\bar{\Lambda}_{1^{--}} = -0.134$~GeV and $\bar{\Lambda}_{0^{-+}}= -0.090$~GeV below the spin-averaged threshold set by the lattice simulation at $E_1=0.005$~GeV, the corresponding tetraquark potentials approach the threshold from below, allowing for the existence of tetraquark bound states below threshold.
The Cornell-like form of $V_{\Sigma_g^{+}}$ provides a field-theoretical justification for the success of the phenomenological models developed in the 1970's and 1980's to describe the quarkonium spectrum, which assumed the same functional form for this potential~\cite{PhysRevLett.34.369, Bhanot:1978mj, Eichten:1978tg, Eichten:1979ms, Ono:1980js}.  
\newpage

The mixing potential $V_{\Sigma_g^+ -\Sigma_g^{+\prime}}$ has to vanish linearly for $r\rightarrow 0$ based on pNRQCD~\cite{TarrusCastella:2022rxb} and approaches zero asymptotically for $r \to \infty$ where $V_{\Sigma_g^+}$ and $V_{\Sigma_g^{+'}}$ get far away from each other, peaking at the string-breaking region around 1.2~fm where mixing is significant. 
We parametrize it as:
\begin{align}
V_{\Sigma_g^+ -\Sigma_g^{+\prime}}(r) = 
\begin{cases}
g\,r/r_1 & r < r_1 \\
g & r_1 \le r \le r_2 \\
g \, e^{-(r-r_2)/r_0} & r > r_2
\end{cases}, 
\label{eq:Vmix}
\end{align}
where $r_0=0.5\, \mathrm{fm}$ is the Sommer scale, and $g=0.05\, \mathrm{GeV}$, $r_1=0.95\, \mathrm{fm}$, and $r_2=1.51\, \mathrm{fm}$ are fixed to lattice data and continuity requirements. The flat parametrization in the mixing region was proposed in~\cite{Bulava:2024jpj} to fit lattice results. 

As for the values of $m_Q$, consistently with~\cite{Brambilla:2024imu}, we fix them to the spin-averaged $D$ meson and $B$ meson masses: 
$m_c = m_{D^{\rm spin \, avg.}} =  1.973\, \mathrm{GeV}$ and $m_b = m_{B^{\rm spin \, avg.}} = 5.313\, \mathrm{GeV}$, this being a natural choice in calculations involving threshold interactions.
The masses of the spin-averaged meson-antimeson thresholds are: $m_{D\bar{D}^{\rm spin\, avg.}} = 3.946 \, \mathrm{GeV}$, $m_{D \bar{B}^{\rm spin\, avg.}} = m_{\bar{D} B^{\rm spin \, avg.}} = 7.287\, \mathrm{GeV}$, $m_{B \bar{B}^{\rm spin \, avg.}} = 10.627 \, \mathrm{GeV}$.

\begin{table}[ht]
\centering
\renewcommand{\arraystretch}{1.6}
\begin{tabular}{| c || c c c c c |}  
\hline
 & & & & $J^{P(C)}$ & $J^{P(C)}$ \\
\hline
$K^{PC}$ & & $D_{\infty h}$ & $l$ & $s_{Q \bar{Q}} = 0$ & $s_{Q \bar{Q}} = 1$ \\
\hline\hline
$(Q \bar{Q})$ & & & & & \\
$0^{++}$ & & $\Sigma_g^+$ & $0$ & $0^{-+}$ & $1^{--}$ \\
$0^{++}$ & & $\Sigma_g^+$ & $1$ & $1^{+-}$ & $(0,1,2)^{++}$ \\
$0^{++}$ & & $\Sigma_g^+$ & $2$ & $2^{-+}$ & $(1,2,3)^{--}$ \\
\hline\hline
$(Q \bar{Q} q \bar{q})$ & & & & & \\
$1^{--}$ & & $\Sigma_g^{+'}$, $\Pi_g$ & $1$ & $1^{+-}$ & $(0,1,2)^{++}$ \\
$1^{--}$ & & $\Sigma_g^{+'}$ & $0$ & $0^{-+}$ & $1^{--}$ \\
$1^{--}$ & & $\Sigma_g^{+'}$, $\Pi_g$ & $2$ & $2^{-+}$ & $(1,2,3)^{--}$ \\
\hline
\end{tabular}
\caption{The Table shows the $J^{P(C)}$ multiplets for quarkonium ($K^{PC} = 0^{++}$) and the isospin-0  tetraquarks ($K^{PC} = 1^{--}$), whose static potentials mix at leading order due to avoided level crossing, for $l = 0, 1, 2$. 
The multiplets are ordered by increasing energies of the relative states. $J^{P(C)} = J^{PC}$ if the heavy quark-antiquark pair is flavor hidden $Q \bar{Q} = b \bar{b}, c \bar{c}$, and $J^{P(C)} = J^P$ if $Q \bar{Q} = b \bar{c}, c \bar{b}$.
}
\label{tab:multiplets}
\end{table}

For a flavor-hidden heavy quark-antiquark pair with $Q\bar{Q} = b\bar{b}, c\bar{c}$, states are labeled by $J^{PC}$, 
while if $Q\bar{Q} = b\bar{c}, c\bar{b}$, states are labeled by $J^P$.
$J$ is the total angular momentum quantum number, while $P$ and $C$ denote respectively parity and charge conjugation. 
For both quarkonium and tetraquark systems, $\bm{J} = \bm{L} + \bm{S}_{Q \bar{Q}}$, where the combined angular momentum $\bm{L} = \bm{L}_{Q \bar{Q}} + \bm{K}$ is the sum of the orbital angular momentum $\bm{L}_{Q \bar{Q}}$ of the HDF and the total angular momentum $\bm{K}$ of the LDF, while $\bm{S}_{Q \bar{Q}} = \bm{S}_Q + \bm{S}_{\bar{Q}}$ is the total heavy-quark spin. 
The eigenvalues relative to the different operators are: $J(J+1)$ for $\bm{J}^2$, $m_J$ for $J_3$, $l(l+1)$ for $\bm{L}^2$, $l_{Q \bar{Q}}(l_{Q \bar{Q}} + 1)$ for $\bm{L}_{Q \bar{Q}}^2$, $s_{Q \bar{Q}} (s_{Q \bar{Q}} + 1)$ for $\bm{S}_{Q \bar{Q}}^2$, and $K(K+1)$ for $\bm{K}^2$.

Quarkonium bound states, specified by $N_Q\! \equiv \!\{ s_{Q \bar{Q}}, l\!=\!l_{Q \bar{Q}}, n, J^{P(C)}, m_J \}$,\footnote{
Since for quarkonium states $K^{PC} = 0^{++}$, the quantum numbers $l$ and $l_{Q \bar{Q}}$ are equal.} 
where $n$ is the principal quantum number, take in the center of mass frame the form 
\begin{equation}\label{eq:quark_state}
\Big| N_Q \Big\rangle = \int d^3\bm{r}  \,
\psi_{\Sigma_g^+}^{(N_Q)} (\bm{r}) \, \bm{\psi}_{\Sigma_g^+}^\dagger(\bm{r},\bm{R}=0)\,|0\rangle.
\end{equation}
The function $\psi_{\Sigma_g^+}^{(N_Q)} (\bm{r})$ is the wavefunction,
\begin{equation}\label{eq:quark_wvf}
\psi_{\Sigma_g^+}^{(N_Q)} (\bm{r}) = \sum_{m_{l}, m_{s_{Q \bar{Q}}}} \mathcal{C}^{J^{P(C)} m_J}_{l m_{l} s_{Q \bar{Q}} m_{s_{Q \bar{Q}}}} \psi_{\Sigma_g^+}^{nl}(r) \; Y_{l}^{m_{l}}(\theta, \phi) \; \chi_{s_{Q \bar{Q}} m_{s_{Q \bar{Q}}}},
\end{equation}
with $\chi_{s_{Q \bar{Q}} m_{s_{Q \bar{Q}}}}$ the spin wavefunctions, $\psi_{\Sigma_g^+}^{nl}(r)$ the radial wavefunctions, $Y_{l}^{m_{l}}(\theta, \phi)$ the spherical harmonics and
$\mathcal{C}^{J^{P(C)} m_J}_{l m_{l} s_{Q \bar{Q}} m_{s_{Q \bar{Q}}}}$ suitable Clebsch--Gordan coefficients.
Similarly, tetraquark bound states, specified by \\
$N_T \equiv \{ s_{Q \bar{Q}}, \kappa, l, n, J^{P(C)}, m_J \}$, have in the center of mass frame the form 
\begin{equation}\label{eq:tetr_state}
\Big|N_T \Big\rangle = \int d^3\bm{r}  \;  \sum_{\lambda}
\psi_{\kappa\lambda}^{(N_T)} (\bm{r}) \,  \bm{\psi}_{\kappa\lambda}^\dagger(\bm{r},\bm{R}=0)|0\rangle\,.
\end{equation}
For $\kappa = 0^{-+}$, the wavefunction reads 
\begin{equation}\label{eq:tetr_wvf1}
\psi_{0^{-+}\, 0}^{(N_T)}\left({\bm r}\right) \equiv  \psi_{\Sigma_u^-}^{(N_T)}\left({\bm r}\right) =
  \sum_{m_{l}, m_{s_{Q \bar{Q}}}} \mathcal{C}^{J^{P(C)} m_J}_{l m_{l} s_{Q \bar{Q}} m_{s_{Q \bar{Q}}}} \psi_{\Sigma_u^-}^{nl}(r) \; v_{l,\,m_l}^0(\theta, \phi) \; \chi_{s_{Q \bar{Q}} m_{s_{Q \bar{Q}}}},
\end{equation}
where $v_{l,\,m_l}^\lambda(\theta, \phi)$ are the angular wavefunctions~\cite{Berwein:2015vca}.  
For $\kappa = 1^{--}$, the wavefunctions $\psi^{(N_T)}_{1^{--}\;\pm1}$ are eigenfunctions of $\bm{K}\cdot\hat{\bm{r}}$ but not of parity.\footnote{
The wavefunction $\psi_{\Sigma_g^{+\prime}}^{(N_T)}$ transforms as the spherical harmonics under parity, whereas  $\psi_{\pm\Pi_g}^{(N_T)}$ transform with the same or with the opposite parity of $\psi_{\Sigma_g^{+\prime}}^{(N_T)}$. 
Since only the $\psi_{+\Pi_g}^{(N_T)}$ tetraquark component can mix with the tetraquark component $\psi_{\Sigma_g^{+ \prime}}^{(N_T)}$ and the quarkonium component $\psi_{\Sigma_g^+}^{(N_Q)}$, from now on we drop the $+$ sign and understand $\Pi_g$ as the positive parity component of $\Pi_g$.} 
The parity eigenfunctions are linear combinations of $\psi^{(N_T)}_{1^{--}\;\pm1}(\bm{r})$: 
\begin{align}
& \psi_{1^{--}\,0}^{(N_T)}\left({\bm r}\right) \equiv \psi_{\Sigma_g^{+\prime}}^{(N_T)}\left({\bm r}\right) =
  \sum_{m_l,\,m_{s_{Q \bar{Q}}}}\,\mathcal{C}^{J^{P(C)} m_J}_{l m_{l} s_{Q \bar{Q}} m_{s_{Q \bar{Q}}}} \,\psi^{n l}_{\Sigma_g^{+\prime}}(r) v_{l,\,m_l}^0(\theta, \phi)\,\bm{\hat{r}}\,\chi_{s_{Q \bar{Q}} m_{s_{Q \bar{Q}}}},
                \label{eq:tetrwf-1}\\
& \psi_{\pm\Pi_g}^{(N_T)}\left({\bm r}\right) =
  \sum_{m_l,\,m_{s_{Q \bar{Q}}}}\frac{\mathcal{C}^{J^{P(C)} m_J}_{l m_{l} s_{Q \bar{Q}} m_{s_{Q \bar{Q}}}}}{\sqrt{2}} \psi^{n l}_{\pm\Pi_g}(r)\,\left(v_{l,\,m_l}^1(\theta, \phi)\,\bm{\hat{r}}_{+}\pm v_{l,\,m_l}^{-1}(\theta, \phi)\,\bm{\hat{r}}_{-}\right)\chi_{s_{Q \bar{Q}} m_{s_{Q \bar{Q}}}}.
 \label{eq:tetrwf-2}
\end{align}

For the leading order BOEFT Lagrangian given in Eq.~\eqref{eq:BOEFTlagrangian}, $l$ and $s_{Q \bar{Q}}$ are good quantum numbers.
The heavy-quark spin decouples from the LDF dynamics, making the BOEFT predictions spin-independent as a consequence of the heavy quark spin symmetry (HQSS).\footnote{
Spin-dependent corrections arise at $\mathcal{O}(1/m_Q)$ for exotics and $\mathcal{O}(1/m_Q^2)$ for quarkonium~\cite{Soto:2020xpm,Oncala:2017hop,Brambilla:2018pyn,Brambilla:2019jfi,Soto:2023lbh}.} 
Hence, all quarkonium or tetraquark states in a given spin multiplet are mass degenerate at leading order.
The $J^{P(C)}$ multiplets for the quarkonium, which have $K^{PC} = 0^{++}$, and for the isospin-0 $(Q \bar{Q} q \bar{q})$ tetraquark, which have $K^{PC} = 1^{--}$ and mix at leading order with the quarkonium ones, are shown in Table~\ref{tab:multiplets}. 

\begin{table}[ht]
\centering
\renewcommand{\arraystretch}{1.6}
\begin{tabular}{|c||c|c||c|c||c|c|}  \hline
  &  $c \bar{c}$ &   & $b \bar{c}$ or $c \bar{b}$ &  & $b \bar{b}$  & \\
$n l$ & $M^{\rm th.}\;(M^{\rm exp.})\; (\mathrm{MeV})$ & $\sqrt{\langle r^2 \rangle}$ (fm)   & $M^{\rm th.}\;(M^{\rm exp.}) \; (\mathrm{MeV})$ & $\sqrt{\langle r^2 \rangle}$ (fm)  & $M^{\rm th.}\;(M^{\rm exp.})\; (\mathrm{MeV})$ &  $\sqrt{\langle r^2 \rangle}$ (fm) \\
\hline\hline
$1S$         & $3127.9$ $(3068.7)$   & $0.4$ & $6337.8$  & $0.3$  & $9445.0 $ $ (9445.0)$  & $0.2$ \\
$2S$         & $3710.0 $ $ (3674.0)$  & $0.7$ & $6890.3$  & $0.6$  & $9988.7 $ $ (10017.3)$  & $0.5$ \\
$3S$         & -  & - & $7278.8$  & $0.9$  & $10331.2$  & $0.7$ \\
$4S$         & -  & - & -  & -  & $10610.0$  & $0.9$ \\\hline
$1P$         & $3538.5 $ $ (3525.3)$  & $0.6$ & $6745.7$  & $0.5$  & $9883.4 $ $ (9899.7)$  & $0.4$ \\
$2P$         & -  & - & $7151.8$  & $0.8$  & $10237.3 $ $ (10260.2)$ & $0.6$ \\
$3P$         & -  & - & -  & -  & $10523.8$ & $0.8$ \\\hline
$1D$         & $3823.1$  & $0.8$ & $7009.8$  & $0.7$  & $10130.6$  & $0.5$ \\
$2D$         & -  & - & -  & -  & $10428.1$  & $0.7$ \\\hline
\end{tabular}
\caption{Spin-averaged spectra obtained from Eq.~\eqref{eq:singleSchr} in the case of charmonium, bottomonium, and $b\bar{c}$ and $c\bar{b}$ systems. 
We include states with orbital angular momenta $l = 0, 1, 2$ ($S$-, $P$-, and $D$-waves) that lie below the corresponding spin-averaged meson-antimeson threshold $M_1\bar{M}_2^{\rm spin \, avg.}$, with $M_{1,2}^{\rm spin \, avg.} = D^{\rm spin \, avg.}, B^{\rm spin \, avg.}$.
We list also the mean square radii of the states.
The dashes refer to quarkonium states above the threshold.
The spin-averaged experimental masses for the different quarkonium multiplets (Table~\ref{tab:multiplets}) are shown in parentheses only if all the states of the multiplet have been observed~\cite{ParticleDataGroup:2024cfk}.}
\label{tab:reference_spectra}
\end{table}

\section{String-breaking effects on the spin-averaged quarkonium spectrum from coupled Schr\"{o}dinger equations: mass shift and state composition}\label{sec:coupled}
After simplifying the angular part as in~\cite{Berwein:2024ztx}, the coupled Schr\"{o}dinger equations for the radial wavefunctions obtained from the leading-order BOEFT Lagrangian~\eqref{eq:BOEFTlagrangian} — previously applied to the $\chi_{c1}(3872)$ description in~\cite{Brambilla:2024imu} — read 
\begin{align}\label{eq:coupledSchr}
&\left[
-\frac{1}{m_Qr^2}\,\partial_rr^2\partial_r+\frac{1}{m_Qr^2}
{\begin{pmatrix}
l\left(l+1\right) & 0 & 0\\[6pt]
0                 & l(l+1)+2        & -2\sqrt{l(l+1)} \\[6pt]
0                 & -2\sqrt{l(l+1)} & l(l+1)
\end{pmatrix}}\right.
\nonumber\\
&\hspace{4.0 cm}\left.
+\begin{pmatrix} V_{\Sigma_{g}^{+}}(r) &  V_{\Sigma_{g}^{+}-\Sigma_{g}^{+\prime}}(r) & 0 \\[6pt]
    V_{\Sigma_{g}^{+}-\Sigma_{g}^{+\prime}}(r) & V_{\Sigma_{g}^{+\prime}}(r) & 0\\[6pt]
      0 & 0 & V_{\Pi_g}(r)\end{pmatrix}
      \right]
      \hspace{-4pt}\begin{pmatrix} \psi_{\Sigma_g^+}^{nl}(r) \\[6pt] \psi_{\Sigma_g^{+\prime}}^{nl}(r) \\[6pt] \psi_{\Pi_g}^{nl}(r)\end{pmatrix}={\mathcal{E}}_{nl} \begin{pmatrix} \psi_{\Sigma_g^+}^{nl}(r) \\[6pt] \psi_{\Sigma_g^{+\prime}}^{nl}(r) \\[6pt] \psi_{\Pi_g}^{nl}(r)\end{pmatrix}.
\end{align}
Because of the HQSS, it is possible to denote the non-degenerate bound states of the different multiplets according to their $nl$ quantum numbers rather than by their $nJ^{P(C)}$ quantum numbers.
Equation~\eqref{eq:coupledSchr} retains the same general form for all values of $l$, with one notable exception: for $l = 0$, the tetraquark sector consists only of the $\Sigma_g^{+'}$ component, as indicated in Table~\ref{tab:multiplets}. 
Therefore, the coupled system reduces to a two-channel equation in this case.
The eigenvalues of the $3\times3$ potential matrix in Eq.~\eqref{eq:coupledSchr} are the \textit{adiabatic potentials}, labeled by the BO quantum numbers $1\Sigma_g^+$, $2\Sigma_g^+$, and $\Pi_g$ and displayed in Fig.~\ref{fig:isospin0}. 
The avoided crossing visible in the figure involves only the $1\Sigma_g^+$ and $2\Sigma_g^+$ adiabatic channels.
We refer to the basis used in Eq.~\eqref{eq:coupledSchr} as \textit{BO mixed basis}. 
It coincides with the basis employed in lattice QCD computations~\cite{Bulava:2019iut, Bulava:2024jpj}.
Two alternative choices, well-known in the literature~\cite{Bruschini:2020voj, Berwein:2024ztx}, are the \textit{BO diabatic basis} and \textit{BO adiabatic basis}.
The first diagonalizes the kinetic matrix and the second one the potential matrix. 
All bases are related by unitary transformations.

In the limit $V_{\Sigma_{g}^{+}-\Sigma_{g}^{+\prime}} \rightarrow 0$, the quarkonium spectrum decouples from the tetraquark one and consists of an infinite series of bound states, whose eigenvalues $E_{nl}$ are determined by the single-channel equation 
\begin{align}\label{eq:singleSchr}
\Biggl( -\frac{1}{m_Qr^2}\,\partial_rr^2\partial_r+\frac{l\left(l+1\right)}{m_Qr^2}
+  V_{\Sigma_{g}^{+}}(r) \Biggr)\, \psi_{\Sigma_g^+}^{nl}(r)
      =E_{nl} \, \psi_{\Sigma_g^+}^{nl}(r).
\end{align}
Table~\ref{tab:reference_spectra} reports the spin-averaged quarkonium spectra from Eq.~\eqref{eq:singleSchr} with the potential given in \eqref{eq:Vcornell} for states below the spin-averaged threshold $M_1 \bar{M}_2^{\rm spin \, avg.}$, where $M_{1,2}^{\rm spin \, avg.} = D^{\rm spin \, avg.}$ or $B^{\rm spin \, avg.}$. 
Results are given for the charmonium, bottomonium, $b\bar{c}$, and $c\bar{b}$ sectors.

For non-vanishing mixing potential, Eq.~\eqref{eq:coupledSchr} predicts a finite number of \textit{predominantly} quarkonium or tetraquark bound states below the $M_1 \bar{M}_2^{\rm spin \, avg.}$ threshold, along with quarkonium resonances and tetraquark scattering states above it. A predominantly quarkonium bound state $nl$ is such that its quarkonium percentage $\displaystyle \% \Sigma_g^{+ \,nl} \equiv \frac{\displaystyle \int dr \, r^2 (\psi_{\Sigma_g^+}^{nl})^2}{\displaystyle \int dr \, r^2 \Big( (\psi_{\Sigma_g^+}^{nl})^2 + (\psi_{\Sigma_g^{+'}}^{nl})^2 + (\psi_{\Pi_g}^{nl})^2 \Big)}  \times 100$ is larger than the corresponding tetraquark one $\% \Sigma_g^{+'\,nl}$ + $\% \Pi_g^{nl}$.
The opposite holds for a predominantly tetraquark bound state.
The open-flavor threshold effect on the mass of an $nl$ quarkonium state to which we refer as \textit{string-breaking correction} is determined from 
\begin{align}\label{eq:str_break_corr}
    \Delta E^{\rm str.br.}_{nl} = \mathcal{E}_{nl} - E_{nl},
\end{align}
i.e. it is the difference between the mass of the predominantly quarkonium state obtained from Eq.~\eqref{eq:coupledSchr}, where the mixing with the tetraquark states is considered, 
and the mass of the quarkonium state solution of Eq.~\eqref{eq:singleSchr}, where the mixing is ignored.
The full set of bound states predicted by Eq.~\eqref{eq:coupledSchr} for $l = 0, 1, 2$ corresponding to spin-averaged $S$-, $P$- and $D$-wave quarkonium states and spin-averaged tetraquark states are reported in Tables~\ref{tab:coupled_spectra_cc}, \ref{tab:coupled_spectra_bb}, and~\ref{tab:coupled_spectra_bc} for respectively the charmonium, bottomonium, and $b \bar{c}$/$c \bar{b}$ sectors together with the associated string-breaking corrections.

\begin{table}[ht]
\centering
\renewcommand{\arraystretch}{1.6}
\begin{tabular}{|c||c|c|c|c|c|c|c|}  \hline
$n l \, (c \bar{c})$ & $M^{\rm th.}(M^{\rm exp.}) \; (\mathrm{MeV})$ & $\sqrt{\langle r^2 \rangle} $ (fm)   & $\%  \; \Sigma_g^+$ & $\% \; \Sigma_g^{+'}$  & $\% \; \Pi_g$ & $E_{\rm bind.}^{D \bar{D}^{\rm spin \, avg.}} \; (\mathrm{MeV})$ & $\Delta E^{\rm str. br.}_{nl}\,(\mathrm{MeV})$ \\
\hline\hline
$1S$         & $3127.6$  $ (3068.7)$  & $0.4$ & $99.9$  & $ 0.1$  &   & $-818.4$ & $-0.4$ \\
$2S$         & $3706.3 $ $ (3674.0)$  & $0.7$ & $99.0$  & $1.0$  &   & $-239.7$ &  $-3.7$ \\ \hline
$1P$         & $3536.9$ $ (3525.3)$  & $0.6$ & $99.7$  & $0.2$  & $0.1$  & $-409.1$ & $-1.6$\\
$\chi_{c1}\left(3872\right)$& $3945.9$  & $4.2$ & $7.6$  & $39.1$  & $53.3$  & $ -0.1 $ &  \\ \hline
$1D$         & $3819.0$  & $0.8$ & $98.6$  & $1.2$  & $0.2$  & $-127.0$ & $-4.1$\\ \hline
\end{tabular}
\caption{Spin-averaged charmonium and tetraquark states lying below the $D \bar{D}^{\rm spin \, avg.}$ threshold obtained from solving Eq.~\eqref{eq:coupledSchr}. 
From the second to the seventh column we show masses, mean square radii, composition, and binding energies with respect to the $D \bar{D}^{\rm spin \, avg.}$ threshold.
The last column shows the string-breaking corrections to the quarkonium states.
The spin-averaged experimental masses for the different quarkonium multiplets 
are shown in parentheses only if all the states of the multiplet have been observed~\cite{ParticleDataGroup:2024cfk}. 
The blank boxes in the sixth column are due to the fact that for $l=0$ only the $\Sigma_g^{+ \prime}$ tetraquark component is present (see Table~\ref{tab:multiplets}).}
\label{tab:coupled_spectra_cc}
\end{table}

\begin{table}[ht]
\centering
\renewcommand{\arraystretch}{1.6}
\begin{tabular}{|c||c|c|c|c|c|c|c|}  \hline
$n l \, (b \bar{b})$ & $M^{\rm th.}(M^{\rm exp.}) \; (\mathrm{MeV})$ & $\sqrt{\langle r^2 \rangle} $ (fm)  & $\%  \; \Sigma_g^+$ & $\% \; \Sigma_g^{+'}$  & $\% \; \Pi_g$ & $E_{\rm bind.}^{B \bar{B}^{\rm spin \, avg.}} \, (\mathrm{MeV})$ & $\Delta E^{\rm str. br.}_{nl}\,(\mathrm{MeV})$ \\
\hline\hline
$1S$         & $9444.9 $ $(9445.0)$  & $0.2$ & $100$  & - &   & $-1182.1$ & $-0.1$ \\
$2S$         & $9987.8$ $(10017.3)$  & $0.5$ & $99.9$  & $0.1$  &   & $-639.4$ & $-0.9$\\
$3S$         & $10327.8$ & $0.7$ & $99.1$  & $0.9$   &   & $-299.5$ & $-3.5$\\
$4S$         & $10593.0$  & $1.0$ & $77.8$  & $22.2$ &   & $-34.3$ & $-17.1$\\ \hline
$1P$         & $9882.9 $ $(9899.7)$  & $0.4$ & $99.9$  & $0.1$  & -  & $-744.1$ & $-0.5$\\
$2P$         & $10235.1 $ $(10260.2)$  & $0.6$ & $99.6$  & $0.4$  & - & $-391.9$ & $-2.2$\\
$3P$         & $10516.1$  & $0.8$ & $95.7$  & $4.2$  & $0.1$ & $-110.9$ & $-7.7$\\
$X_b$        & $10625.9$  & $2.3$ & $1.5$  & $44.3$  & $54.2$ & $-1.1$ & \\\hline
$1D$         & $10124.5$  & $0.5$ & $99.8$  & $0.2$  & -  & $-502.6$ & $-1.1$\\
$2D$         & $10419.0$  & $0.7$ & $98.7$  & $1.3$  & -  & $-208.0$ & $-4.2$\\ \hline
\end{tabular}
\caption{Spin-averaged bottomonium and tetraquark states lying below the $B \bar{B}^{\rm spin \, avg.}$ threshold obtained from solving Eq.~\eqref{eq:coupledSchr}. 
From the second to the seventh column we show masses, mean square radii, composition, and binding energies with respect to the $B \bar{B}^{\rm spin \, avg.}$ threshold. 
The last column shows the string-breaking corrections to the quarkonium states.
The spin-averaged experimental masses for the different quarkonium multiplets 
are shown in parentheses only if all the states of the multiplet have been observed~\cite{ParticleDataGroup:2024cfk}. 
The blank boxes in the sixth column are due to the fact that for $l=0$ only the $\Sigma_g^{+ \prime}$ tetraquark component is present (Table~\ref{tab:multiplets}), while the dashes in the fifth and sixth columns stand for percentages that are smaller than $0.1 \%$.}
\label{tab:coupled_spectra_bb}
\end{table}

\begin{table}[ht]
\centering
\renewcommand{\arraystretch}{1.6}
\begin{tabular}{|c||c|c|c|c|c|c|c|c|}  \hline
$n l \, (b \bar{c}\hspace{0.3em} or \hspace{0.3em} c \bar{b}) $ & $M^{\rm th.} \; (\mathrm{MeV})$ & $\sqrt{\langle r^2 \rangle} $ (fm)   & $\%  \; \Sigma_g^+$ & $\% \; \Sigma_g^{+'}$  & $\% \; \Pi_g$ & $E_{\rm bind.}^{B \bar{D}^{\rm spin \, avg.}} \, (\mathrm{MeV})$ & $\Delta E^{\rm str. br.}_{nl}\,(\mathrm{MeV})$ \\
\hline\hline
$1S$         & $6337.5$  & $0.3$ & $100$  & -  &  & $-949.4$ & $-0.2$ \\
$2S$         & $6888.1$  & $0.6$ & $99.6$  & $0.4$  &   & $-398.9$ & $-2.2$\\
$3S$         & $7262.6$  & $1.0$ & $79.8$  & $20.2$ &   & $-24.4$ & $-16.2$ \\\hline
$1P$         & $6744.7$  & $0.5$ & $99.8$  & $0.1$  & $0.1$  & $-542.3$ & $-1.0$ \\
$2P$         & $7146.0$  & $0.8$ & $97.5$  & $2.3$  & $0.2$ & $-141.0$ & $-5.7$ \\\hline
$1D$         & $7007.4$  & $0.7$ & $99.5$  & $0.5$  & -  & $-279.6$ & $-2.4$\\\hline
\end{tabular}
\caption{Spin-averaged $b \bar{c}$ (or $c \bar{b}$) states lying below the $D \bar{B}^{\rm spin \, avg.}$ (or $B \bar{D}^{\rm spin \, avg.}$) threshold obtained from solving Eq.~\eqref{eq:coupledSchr}. 
From the second to the seventh column we show masses, mean square radii, composition, and binding energies with respect to the $D \bar{B}^{\rm spin \, avg.}$ (or $B \bar{D}^{\rm spin \, avg.}$) threshold. 
The last column shows the string-breaking corrections to the quarkonium states. 
The blank boxes in the sixth column are due to the fact that for $l=0$ only the $\Sigma_g^{+ \prime}$ tetraquark component is present (Table~\ref{tab:multiplets}), while the dashes in the fifth and sixth column stand for percentages that are smaller than $0.1 \%$.}
\label{tab:coupled_spectra_bc}
\end{table}

The spin-averaged $\chi_{c1}(3872)$ bound state is obtained by fine-tuning the parameter $\Lambda_{1^{--}}$ in Eq.~\eqref{eq:QQbar_V} to reproduce the mass of the observed state~\cite{Brambilla:2024imu}, while $X_b$ represents its counterpart in the bottomonium sector.\footnote{
The binding energy of the $X_b$ state differs by about 10~MeV from the value reported in~\cite{Brambilla:2024imu}, due to the different parametrization of the static tetraquark potentials at large distances: 
in this work we employ a one-pion exchange parameterization (see Section~\ref{subsec:parametrization}, 
whereas Ref.~\cite{Brambilla:2024imu} uses a two-pion exchange parameterization. 
For the rest, the composition of the state is found to be nearly identical in the two cases.} 
We emphasize that the observed $\chi_{c1}(3872)$ has $J^{PC} = 1^{++}$, whereas our prediction corresponds to the  spin-averaged state of the $J^{PC} = \{1^{+-}, (0,1,2)^{++} \}$ multiplet lying about $100 \, \mathrm{keV}$ below the $D \bar{D}^{\rm spin \, avg.}$ threshold. A more detailed discussion about the calibration of the $\Lambda_{1^{--}}$ adjoint meson mass and the sensitivity of the spin-averaged $\chi_{c1}(3872)$ properties---namely, its binding energy, radius, and composition---to the value of this parameter, is given in Section~V D of~\cite{Brambilla:2026jqz}.
The state entails about $90 \%$ of tetraquark and $10 \%$ of quarkonium component, this last playing a pivotal role in explaining some of the decay properties of the state.\footnote{
The spin-splitting pattern of the states in the multiplet along with the radiative decays of $\chi_{c1}(3872)$ have been discussed in~\cite{Braaten:2024tbm,Brambilla:2024imu} and are partially discussed as well in Section~\ref{subsec:phenom_impl}.} 
In our description, the state is a different one with respect to the $2P$ charmonium state, which lies a few tens of $\mathrm{MeV}$ above the $D \bar{D}^{\rm spin \, avg.}$ threshold and is a predominantly quarkonium state.
The recent detection of $h_c(4000)$ and $\chi_{c1}(4010)$ with $J^{PC} = 1^{+-}$ and $1^{++}$ at  LHCb~\cite{LHCb:2024vfz} in that mass region appears to support our interpretation. 
A BOEFT calculation for both the states above the $D \bar{D}^{\rm spin \, avg.}$  threshold, accounting also for the mixing with the $c \bar{c} s \bar{s}$ tetraquark states, shall be addressed in future work.

Including the mixing with tetraquark states results in a systematic downward shift in energy for all quarkonium states. 
The shift ranges from about 1~MeV to 15~MeV and grows as the state approaches the spin-averaged meson–antimeson threshold, where the tetraquark admixture becomes more substantial. 
For instance, in states such as the $4S$ bottomonium or the $3S$ $B_c$ the tetraquark component can be as large as 20\%. 
The magnitude of the shift is not sensitive to the heavy-quark flavor itself, but rather to the energy gap between the state and the corresponding spin-averaged threshold.
To explicitly resolve the tetraquark composition of the eigenstates, we also solve Eq.~\eqref{eq:coupledSchr} in the BO diabatic basis. 
This allows us to isolate the $S$-, $P$-, and $D$-wave tetraquark components individually. 
Table~\ref{tab:coupled_diabatic} lists the composition of the states with a tetraquark admixture larger than $10\%$. 
The $S$-wave tetraquark component strongly dominates over the $D$-wave one for both the $\chi_{c1}(3872)$ and the $X_b$ state.

\begin{table}
\centering
\renewcommand{\arraystretch}{1.6}
\begin{tabular}{|c||c|c|c|c|c|}  \hline
  & $M^{\rm th.} \; (\mathrm{MeV})$ & $\% Q \bar{Q}$   & $\% Q \bar{Q} q \bar{q}^{S}$ & $\% Q \bar{Q} q \bar{q}^{P}$ & $\% Q \bar{Q} q \bar{q}^{D}$\\
\hline
$c \bar{c}$         &           &       &        &     &       \\
$\chi_{c1}(3872)$   & $3945.9$  & $7.6$ & $91.4$ & -   & $1.0$ \\\hline
$b \bar{b}$         &           &       &        &     &       \\
$4S$                & $10593.0$ & $77.8$&    -   &$22.2$&  -   \\
$X_b$               & $10625.9$ & $1.5$ & $97.0$ & -   & $1.5$ \\\hline
$b \bar{c}$/$c \bar{b}$         &           &       &        &     &       \\
$3S$                & $7262.6$  & $79.8$&   -    &$20.2$&  -   \\\hline
\end{tabular}
\caption{The composition of the spin-averaged $c \bar{c}$, $b \bar{b}$, $b \bar{c}$/$c \bar{b}$  states with more than $10 \%$ of tetraquark component in the BO diabatic basis. 
The tetraquark components are classified according to the orbital angular momentum $l_{Q \bar{Q}} = 0$, $1$, $2$ in $S$-, $P$-, and $D$- wave components.
The dashes stand for tetraquark components that are absent in the corresponding states.
}
\label{tab:coupled_diabatic}
\end{table}

Our predictions agree with experimental data within $60\,\mathrm{MeV}$, which is a reasonable result considering that the computation is valid at leading order in the non-relativistic expansion, i.e. without including relativistic or spin-dependent corrections that appear at higher order in the effective field theory Lagrangian~\cite{Brambilla:2004jw}.
Based on power-counting arguments, the obtained predictions are valid up to 
corrections suppressed by $\Lambda_{\rm QCD}/m_Q$ and $v^2$, 
with $v^2 \sim 0.3$ for $c \bar{c}$ and $v^2 \sim 0.1$ for $b \bar{b}$ systems~\cite{Bodwin:1994jh}.

To assess the sensitivity of our results to model details, we perform several consistency checks.
First, if we model the tetraquark static potentials $V_{\Sigma_g^{+'}}$ and $V_{\Pi_g}$ as constant lines — corresponding to the asymptotic meson–antimeson mass sum, as has been common in the BO literature~\cite{Bicudo:2019ymo, Bruschini:2020voj, Bicudo:2020qhp, Bruschini:2021cty, Bruschini:2021sjh, Bicudo:2022ihz, TarrusCastella:2022rxb} — the calculated string-breaking corrections change by less than $1\,\mathrm{MeV}$. 
This confirms that these corrections are largely insensitive to the short-distance details of the tetraquark potentials. 
However, a constant potential does not support predominantly tetraquark bound states such as the $\chi_{c1}(3872)$ or $X_b$.
Second, varying the heavy-quark masses in the ranges 1.3 GeV $\lesssim m_c \lesssim 2.0$ GeV and 4.7 GeV $\lesssim  m_b \lesssim 5.3$ GeV changes only weakly the string-breaking corrections, with $\Delta E_{nl}^{\mathrm{\rm str. br.}}$ varying by less than $20 \%$ across the entire mass interval.
In contrast, the string-breaking corrections exhibit strong sensitivity to the mixing coupling~$g$. 
Increasing $g$ by a factor of five enhances $\Delta E_{nl}^{\rm str. br.}$ by  at least an order of magnitude (see Table~\ref{tab:mixing_dep} in Section~\ref{subsec:3P0}).
Unlike purely phenomenological approaches (e.g. the $^{3}P_0$ model), the coupling $g$ in the BO framework is constrained by lattice QCD. 
Its precise determination is therefore essential for a reliable quantification of string-breaking effects, a point that we revisit in Section~\ref{subsec:literature_pheno}.

\section{Tetraquark spin-splitting effects on quarkonium spectrum}\label{sec:thr-spl}
We discuss now the inclusion of $\mathcal{O}(1/m_Q)$ tetraquark spin-dependent potentials in the BOEFT framework and assess their impact on the quarkonium spectrum.  
The explicit derivation of the coupled Schr\"odinger equations for the $J^{PC}=1^{++}$ case is presented in Section~\ref{subsec:spin-splitted_eq}, 
while equations for other $S$-, $P$-, and $D$-wave quarkonium states are collected in Appendix~\ref{app:meson-meson basis}.  
The phenomenological consequences of the inclusion of these higher-order terms in the spectrum are analyzed in Section~\ref{subsec:phenom_impl}, with the resulting spectrum provided in Appendix~\ref{app:tetraquark spin splitting}.

\subsection{The coupled Schr\"odinger equations with threshold spin-splitting corrections }\label{subsec:spin-splitted_eq}
As outlined in Section~\ref{sec:BOEFT}, the leading-order BOEFT Lagrangian in Eq.~\eqref{eq:BOEFTlagrangian} yields the coupled Schr\"odinger Eqs.~\eqref{eq:coupledSchr}, which — due to HQSS — predict the mass patterns of the different quarkonium and tetraquark degenerate multiplets labeled by the $nl$ quantum numbers. 
The degeneracy is lifted once $\mathcal{O}(1/m_Q)$ spin-dependent tetraquark potentials are included.
The $\mathcal{O}(1/m_Q)$ tetraquark potentials contain both a spin-independent part, which shifts all states of a multiplet equally, and a spin-dependent part, which instead shifts them differently and breaks their degeneracy~\cite{Soto:2020xpm}. 
This is different from the quarkonium potential that remains spin-independent at $\mathcal{O}(1/m_Q)$~\cite{Brambilla:2000gk}.
In the present study, we focus on the $\mathcal{O}(1/m_Q)$ spin-dependent mixing between the tetraquark potentials $V_{\Sigma_u^-}$ and $V_{\Sigma_g^{+'}}$, $V_{\Pi_g}$. 
We do not consider mixings involving hybrid states (see Refs.~\cite{Oncala:2017hop,Oncala:2025mqj}) or other exotic configurations, leaving their investigation for future work.

The expressions of the $\mathcal{O}(1/m_Q)$ spin-dependent potentials for exotic hadrons in terms of generalized Wilson loops can be found in~\cite{Soto:2020xpm}.
At present, results for the BO potentials beyond the static limit are available only for quarkonium~\cite{Koma:2006si,Koma:2007jq} and for the lowest hybrid potentials~\cite{Schlosser:2025tca}. 
Spin-splitting effects in hybrid multiplets have been estimated either from direct lattice calculations~\cite{HadronSpectrum:2012gic,Cheung:2017tnt,Ryan:2020iog} or by fitting the non-perturbative parameters entering the spin-dependent BO potentials to those lattice results~\cite{Brambilla:2019jfi,TarrusCastella:2019lyq}.
For tetraquark states, no first-principles determinations of the spin-dependent potentials exist. 
Existing calculations rely on approximate expressions that use the experimentally observed mass differences between the heavy-light mesons ($D$, $D^*$, $B$, $B^*$)\footnote{
In the present work — consistently with the lattice QCD  computations of Refs.~\cite{Bulava:2019iut,Bulava:2024jpj} — the $u$ and $d$ quarks are taken to be mass degenerate (isospin limit). 
Hence, we use isospin-averaged masses: $m_D = (m_{D^+} + m_{D^0})/2$, $m_B = (m_{B^+} + m_{B^0})/2$, etc.} 
and the physical requirement that a tetraquark evolves into a meson--antimeson pair at large distances~\cite{Braaten:2024tbm, Brambilla:2024imu}. 
In this section, following the same approach, the $\mathcal{O}(1/m_Q)$ tetraquark spin-dependent potential is fixed to reproduce the experimentally observed spin splittings of the lowest heavy-light meson--antimeson pairs $M^{(*)} \bar{M}^{(*)}$, where $M^{(*)} = D^{(*)}, B^{(*)}$.
We parameterize the potential as 
\begin{equation}\label{eq:splitting}
    V_{SS} = \delta_Q (\bm{S}_Q \cdot \bm{K}_{\bar{q}} + \bm{S}_{\bar{Q}} \cdot \bm{K}_q),
\end{equation}
where $\bm{S}_Q$ ($\bm{S}_{\bar{Q}}$) and $\bm{K}_{\bar{q}}$ ($\bm{K}_q$) are the heavy quark (antiquark) spin and light antiquark (quark) total angular momenta in the meson $M^{(*)}$ (antimeson $\bar{M}^{(*)}$), respectively. 
The constant $\delta_Q$,  which is of order $1/m_Q$, is fixed to reproduce the observed spin-splitting between $M$ and $M^*$: 
$\delta_c = 141$ $\mathrm{MeV}$ for charmed mesons and $\delta_b = 45$ $\mathrm{MeV}$ for bottom mesons~\cite{ParticleDataGroup:2024cfk}.\footnote{
The procedure to account for threshold spin-splitting corrections in the coupled Schr\"odinger equations for the $b \bar{c}$ and $c \bar{b}$ systems is not discussed here since it is slightly more cumbersome than for the $c\bar c$ and $b\bar b$ cases, although conceptually similar. 
If, for example, we consider the $b \bar{c}$ system, the $\mathcal{O}(1/m_Q)$ tetraquark spin-dependent potential reads: $V_{SS}^{b \bar{c}} = \delta_b \, \bm{S}_b \cdot \bm{K}_{\bar{q}} + \delta_c \, \bm{S}_{\bar c} \cdot \bm{K}_q$, giving rise to four non-degenerate meson-antimeson thresholds $B \bar{D}$, $B^* \bar{D}$, $B \bar{D}^*$, $B^* \bar{D}^*$ with energy shifts $- 3(\delta_b + \delta_c)/4$, $(\delta_b - 3 \delta_c)/4$, $(\delta_c - 3 \delta_b)/4$, $(\delta_b + \delta_c)/4$ to the $B \bar{D}^{\rm spin\, avg}$ threshold, respectively. 
Proceeding then in a similar way as done in this section, it is possible to derive the analogs of Eqs.~\eqref{eq:1++BOEFT} and~\eqref{eq:1++mes-mes} for the $b \bar{c}$ and $c \bar{b}$ systems.}
The potential $V_{SS}$ induces energy shifts of $-3\delta_Q/2$, $-\delta_Q/2$, and $\delta_Q/2$ respectively to the $M \bar{M}$, $M^* \bar{M}$ (or $M \bar{M}^*$), and $M^* \bar{M}^*$ thresholds with respect to the $M \bar{M}^{\rm spin \, avg.}$ threshold. 
Previous studies have already estimated the spin-splitting effects on the tetraquark components of the spin-averaged $\chi_{c1}(3872)$ multiplet using Eq.~\eqref{eq:splitting} for $V_{SS}$~\cite{Braaten:2024tbm,Brambilla:2024imu}.

The spin $s_Q$, $s_{\bar{Q}}$ and angular momentum $K_q$, $K_{\bar{q}}$ quantum numbers can be arranged in several ways and combined with $l_{Q \bar{Q}}$ to define a tetraquark state with a given $J^{PC}$. 
Different arrangements correspond to different bases.
We work primarily with two bases: 
the BO basis (introduced in Section~\ref{subsec:parametrization} to define tetraquark bound states) and the meson-antimeson basis. 
The latter is particularly useful for matching the components of the solutions of the coupled Schr\"odinger equations to the corresponding $M^{(*)}\bar{M}^{(*)}$ thresholds.

\begin{table}[ht]
\small
\scriptsize
\begin{minipage}{0.35\linewidth} 
\scalebox{0.95}{%
\renewcommand{\arraystretch}{1.5} 
\begin{tabular}{|c|c||c|c|c|}
\hline
 &  & $J^{PC}$ & $J^{PC}$ & $J^{PC}$ \\
 &  & $s_{M^{(*)} \bar{M}^{(*)}}=0$ & $s_{M^{(*)} \bar{M}^{(*)}}=1$ & $s_{M^{(*)} \bar{M}^{(*)}}=2$ \\
\hline\hline
$M \bar{M}$     &               & $0^{++}$ & &\\
$M \bar{M}^*_{CP}$   &  $l_{Q \bar{Q}} = 0$ &  &$1^{++}$, $1^{+-}$ &\\ 
$M^* \bar{M}^*$ &               & $0^{++}$& $1^{+-}$ & $2^{++}$\\\hline 
$M \bar{M}$     &               & $1^{--}$ & &\\
$M \bar{M}^*_{CP}$   &  $l_{Q \bar{Q}} = 1$ &  &$(0,1,2)^{--}$, $(0,1,2)^{-+}$ &\\ 
$M^* \bar{M}^*$ &              & $1^{--}$& $(0,1,2)^{-+}$ & $(0,1,2)^{--}$\\\hline 
$M \bar{M}$     &               & $2^{++}$ & &\\
$M \bar{M}^*_{CP}$   &  $l_{Q \bar{Q}} = 2$ &  &$(1,2,3)^{++}$, $(1,2,3)^{+-}$ &\\ 
$M^* \bar{M}^*$ &               & $2^{++}$& $(1,2,3)^{+-}$ & $(0,1,2,3,4)^{++}$\\\hline 
$M \bar{M}$     &              & $3^{--}$ & &\\
$M \bar{M}^*_{CP}$   &  $l_{Q \bar{Q}} = 3$ &  &$(2,3,4)^{--}$, $(2,3,4)^{-+}$ &\\ 
$M^* \bar{M}^*$ &               & $3^{--}$& $(2,3,4)^{-+}$ & $(1,2,3,4,5)^{--}$\\\hline 
$M \bar{M}$     &              & $4^{++}$ & &\\
$M \bar{M}^*_{CP}$   &  $l_{Q \bar{Q}} = 4$ &  &$(3,4,5)^{++}$, $(3,4,5)^{+-}$ &\\ 
$M^* \bar{M}^*$ &              & $4^{++}$& $(3,4,5)^{+-}$ & $(2,3,4,5,6)^{++}$\\\hline 
\end{tabular}
}
\end{minipage}%
\hfill 
\begin{minipage}{0.48\linewidth} 
\scalebox{1.25}{%
\renewcommand{\arraystretch}{1.5} 
\begin{tabular}{|c|c|c||c|c|c|}
\hline
LDF $\#$ & BO $\#$ &  & $J^{PC}$ & $J^{PC}$ & \\
$k^{PC}$ & $\Lambda_\eta^{\sigma}$ & $l$& $s_{Q \bar{Q}} = 0$ & $s_{Q \bar{Q}} = 1$ & $l_{Q \bar{Q}}$\\
\hline\hline
 & $\Sigma_u^-$ & $0$ & $0^{++}$ & $1^{+-}$ & $0$\\
$0^{-+}$ & $\Sigma_u^-$ & $1$ & $1^{--}$ & $(0,1,2)^{-+}$ & $1$ \\ 
 & $\Sigma_u^-$ &  $2$ & $2^{++}$ & $(1,2,3)^{+-}$ & $2$\\         
 & $\Sigma_u^-$ & $3$ & $3^{--}$ &  $(2,3,4)^{-+}$ & $3$\\\hline
 & $\Sigma_g^{+'}, \Pi_g$ & $1$ & $1^{+-}$ & $(0,1,2)^{++}$ & $(0,2)$\\   
 & $\Sigma_g^{+'}$ &  $0$ & $0^{-+}$ & $1^{--}$ & $1$\\ 
 $1^{--}$ & $\Pi_g$ &  $1$ & $1^{-+}$ & $(0,1,2)^{--}$ & $1$\\
 & $\Sigma_g^{+'}, \Pi_g$ & $2$ & $2^{-+}$ & $(1,2,3)^{--}$ & $(1,3)$\\
 & $\Pi_g$ & $2$ & $2^{+-}$ & $(1,2,3)^{++}$ & $2$\\   
 & $\Sigma_g^{+'}, \Pi_g$ & $3$ & $3^{+-}$ & $(2,3,4)^{++}$ & $(2,4)$\\
 & $\Pi_g$ & $3$ & $3^{-+}$ & $(2,3,4)^{--}$ & $3$\\\hline 
\end{tabular}
}
\end{minipage}%
\caption{The $J^{PC}$ multiplets for tetraquark states in the meson-antimeson diabatic basis (left table) and BO diabatic basis (right table), given in terms of their respective quantum numbers (see Eqs.~\eqref{eq:tetr_mesmesbasis} and~\eqref{eq:tetr_BObasis}) for the lowest values of $l_{Q \bar{Q}}$ (meson-antimeson basis) and $l$ (BO basis). 
The quantum numbers $s_Q = {1}/{2}$, $s_{\bar{Q}} = {1}/{2}$, $K_q = {1}/{2}$, $K_{\bar{q}} = {1}/{2}$, $j_M = j_{\bar{M}} = 0$, $j_{M^*} = j_{\bar{M}^*} = 1$ are not displayed since their values are fixed.
The notation $M \bar{M}^*_{CP}$ stands for the two linear combinations $(M \bar{M}^* \pm M^* \bar{M})/\sqrt{2}$, which are $CP$ eigenstates for $Q \bar{Q} = c \bar{c}, b\bar{b}$ with eigenvalues $\pm 1$.}
\label{tab:BOEFTvsmes-mes}
\end{table}

A tetraquark bound state in any BO basis (mixed, adiabatic, adiabatic) can be written as 
\begin{align}\label{eq:tetr_BObasis}
\Big|[s_{Q \bar{Q}}, (K, l_{Q \bar{Q}})\,l]\, J \Big\rangle,
\end{align}
where the $K$, $l_{Q \bar{Q}}$ quantum numbers are combined  within round brackets to define $l$, which, in turn, is summed with $s_{Q \bar{Q}}$  within square brackets to define the $J$ of the state.\footnote{
Since tetraquark and quarkonium states are degenerate in $m_J$, we do not show this quantum number in the rest of the paper.
Moreover, we drop the label relative to the principal quantum number $n$, $P$ and $C$,  as they are not relevant for the present discussion.}
In contrast, a tetraquark bound state in any meson-antimeson basis (mixed, diabatic, adiabatic)\footnote{
Meson-antimeson mixed, adiabatic, and diabatic bases can be defined similarly to the BO basis, as apparent later in this section (see Eq.~\eqref{eq:1++mes-mes}).} 
can be written as
\begin{equation}\label{eq:tetr_mesmesbasis}
\Big|[(j_{M^{(*)}}, j_{\bar{M}^{(*)}})\, s_{M^{(*)} \bar{M}^{(*)}}, l_{Q \bar{Q}}] \, J\Big\rangle,
\end{equation}
where $j_{M^{(*)}}$, $j_{\bar{M}^{(*)}}$, and $s_{M^{(*)} \bar{M}^{(*)}}$ are the quantum numbers associated with the $\bm{J}_{M^{(*)}} = \bm{S}_Q + \bm{K}_{\bar{q}}$, $\bm{J}_{\bar{M}^{(*)}} = \bm{S}_{\bar{Q}} + \bm{K}_q$, and $\bm{S}_{M^{(*)} \bar{M}^{(*)}} = \bm{J}_{M^{(*)}} + \bm{J}_{\bar{M}^{(*)}}$ operators. 
In this case, the meson and antimeson quantum numbers $j_{M^{(*)}}$, $j_{ \bar{M}^{(*)}}$ are first combined into $s_{M^{(*)} \bar{M}^{(*)}}$, and then this last is summed to the heavy quark-antiquark orbital angular momentum $l_{Q \bar{Q}}$ to define the total $J$ of the state. 
The lowest $J^{PC}$ multiplets for BO diabatic and meson-antimeson diabatic states, arranged by increasing energies, are shown in Tables~\ref{tab:BOEFTvsmes-mes}.\footnote{
The determination of parity and charge conjugation for BO or meson-antimeson states is discussed in detail in~\cite{Berwein:2024ztx, Bruschini:2020voj}. 
Note that for both types of states, adding one unit of angular momentum reverses the sign of $P$ and $C$, while it preserves the sign of their product. 
Furthermore, the values of $l_{Q \bar{Q}}$ in the right panel of Table~\ref{tab:BOEFTvsmes-mes} reveal that $S$-wave quarkonium states, which have  $J^{PC} = 0^{-+}, 1^{--}$, mix with $P$-wave tetraquark BO states, 
$P$-wave quarkonium states, which have $J^{PC} = 1^{+-}, (0,1,2)^{++}$,  mix with $S$- or $D$-wave tetraquark BO states, and so on, consistently with the selection rules derived in~\cite{Braaten:2024stn}.}
The angular momentum recoupling relation, which expresses a tetraquark state in the meson--antimeson diabatic basis in terms of states in the BO diabatic basis, involving Wigner $6j$ and $9j$ symbols~\cite{Braaten:2024stn}, is:
\begin{align}\label{eq:rearrangement}
\Big|[(j_{M^{(*)}}, j_{\bar{M}^{(*)}})\, s_{M^{(*)} \bar{M}^{(*)}}, l_{Q \bar{Q}}] \, J\Big\rangle = \sum_{k,s_{Q \bar{Q}},l}
N (-)^{2K_q + l_{Q \bar{Q}} + j} \sqrt{\tilde{j}_{M^{(*)}} \, \tilde{j}_{\bar{M}^{(*)}} \, \tilde{s}_{M^{(*)} \bar{M}^{(*)}}} (-)^{K + s_{Q \bar{Q}}} \sqrt{ \tilde{K} \, \tilde{s}_{Q \bar{Q}} \, \tilde{l}}  \nonumber \\
\left.
\quad \times 
\left\{ \begin{array}{ccc}
s_{Q \bar{Q}} & K & s_{M^{(*)} \bar{M}^{(*)}} \\
l_{Q \bar{Q}} & j & l
\end{array} \right\} \left\{ \begin{array}{ccc}
s_Q & s_{\bar{Q}} & s_{Q \bar{Q}} \\
K_q & K_{\bar{q}} & K \\
j_{M^{(*)}} & j_{\bar{M}^{(*)}} & l
\end{array} \right\} \Big|[s_{Q \bar{Q}}, (K, l_{Q \bar{Q}}) \,l] \, J \Big\rangle \right.,
\end{align}
where $\tilde{a} \equiv 2a + 1$ and $N$ is a normalization constant equal to $1$ for  the $M \bar{M}$, $M^{*} \bar{M}^*$ thresholds and $\sqrt{2}$ otherwise. 

We now detail how the $\mathcal{O}(1/m_Q)$ tetraquark spin-dependent potential $V_{SS}$ modifies the spin-averaged BOEFT coupled equations. 
 We focus on the $J^{PC}=1^{++}$ channel relevant for the $\chi_{c1}(3872)$ state.  
First, we note that there are three tetraquark states with $J^{PC}=1^{++}$ in both the BO adiabatic and meson--antimeson bases (see Table~\ref{tab:BOEFTvsmes-mes}). 
The resulting coupled Schr\"odinger equations for the radial wavefunctions in the BO diabatic basis, including spin-splitting effects, read 
\begin{align}\label{eq:1++BOEFT}
\hspace{+0cm}\left[
-\frac{1}{m_Qr^2}\,\partial_r r^2 \partial_r + \frac{1}{m_Qr^2}
{\begin{pmatrix}
2 & 0 & 0 & 0\\[4pt]
0 & 0 & 0 & 0 \\[4pt]
0 & 0 & 6 & 0 \\[4pt]
0 & 0 & 0 & 6 \\[4pt]
\end{pmatrix}}\right. \hspace{0cm} \left. + \begin{pmatrix} 
V_{\Sigma_{g}^+}(r) & \frac{1}{\sqrt{3}} V_{\Sigma_g^+ - \Sigma_g^{+\prime}}(r) & \sqrt{\frac{2}{3}} V_{\Sigma_g^+ - \Sigma_g^{+\prime}}(r) & 0 \\[6pt]
\frac{1}{\sqrt{3}} V_{\Sigma_g^+ - \Sigma_g^{+\prime}}(r) & \frac{V_{\Sigma_g^{+'}} + 2 V_{\Pi_g}}{3}  & \frac{\sqrt{2} \,(V_{\Sigma_g^{+'}} - V_{\Pi_g})}{3} & 0 \\[6pt]
\sqrt{\frac{2}{3}} V_{\Sigma_g^+ - \Sigma_g^{+\prime}}(r) & \frac{\sqrt{2}\, (V_{\Sigma_g^{+'}} - V_{\Pi_g})}{3} & \frac{2 V_{\Sigma_g^{+'}} + V_{\Pi_g}}{3} & 0 \\[6pt]
0 & 0 & 0 & V_{\Pi_g}(r)
\end{pmatrix} 
\right. \nonumber \\
&\hspace{-9cm}\left. + \begin{pmatrix} 
0 & 0 & 0 & 0 \\[4pt]
0 & -\frac{\delta_Q}{2} & 0 & 0 \\[4pt]
0 & 0 & \frac{\delta_Q}{4} & \frac{\sqrt{3}\delta_Q}{4} \\[4pt]
0 & 0 & \frac{\sqrt{3}\delta_Q}{4} & -\frac{\delta_Q}{4}
\end{pmatrix}  \right]
\begin{pmatrix} \psi_{\Sigma_g^+} \\[4pt] \psi_{\Sigma^{+\prime}_g + 2 \Pi_g} \\[4pt] \psi_{2 \Sigma^{+\prime}_g + \Pi_g} \\[4pt] \psi_{\Pi_g} \end{pmatrix}  
=  \mathcal{E} \begin{pmatrix} \psi_{\Sigma_g^+} \\[4pt] \psi_{\Sigma^{+\prime}_g + 2 \Pi_g} \\[4pt] \psi_{2 \Sigma^{+\prime}_g + \Pi_g} \\[4pt] \psi_{\Pi_g} \end{pmatrix},
\end{align}
where the diagonal entries $0$, $2$, $6$ in the kinetic energy matrix correspond to the orbital angular momentum quantum numbers $l_{Q \bar{Q}} = 0$, $1$, $2$, respectively. 
To avoid overloading the notation, we suppress in the radial wavefunction components the quarkonium $N_Q$ and tetraquark $N_T$ quantum numbers. 
The wavefunction components are named according to the corresponding diagonal entries of the static potential matrix, $\psi_{\Sigma_g^+}$, $\psi_{\Sigma^{+\prime}_g+2 \Pi_g}$, $\psi_{2\Sigma^{+\prime}_g+\Pi_g}$ and $\psi_{\Pi_g}$.
The energy eigenvalues are simply denoted by $\mathcal{E}$.

At leading order, the coupled Schr\"odinger equations for the first three wavefunction components coincide with Eq.~\eqref{eq:coupledSchr} expressed in the BO diabatic basis. 
The fourth component, which decouples at this order, couples once threshold spin-splitting corrections are introduced.
The relations between the BO diabatic states and the meson--antimeson diabatic ones, following from Eq.~\eqref{eq:rearrangement}, are
\begin{align}\label{eq:BOvsmes-mes}
\Big| \Sigma_g^{+'} + 2 \Pi_g \Big\rangle &= - \Big| \frac{M \bar{M}^* + M^* \bar{M}}{\sqrt{2}}, l_{Q \bar{Q}} = 0 \Big\rangle, \\
\Big| 2 \Sigma_g^{+'} + \Pi_g \Big\rangle &= \frac{1}{2} \Big| \frac{M \bar{M}^* + M^* \bar{M}}{\sqrt{2}}, l_{Q \bar{Q}} = 2 \Big\rangle - \frac{\sqrt{3}}{2} \Big| M^* \bar{M}^*, l_{Q \bar{Q}} = 2 \Big\rangle, \\
\Big| \Pi_g \Big\rangle &= -\frac{\sqrt{3}}{2} \Big| \frac{M \bar{M}^* + M^* \bar{M}}{\sqrt{2}}, l_{Q \bar{Q}} = 2 \Big\rangle - \frac{1}{2} \Big| M^* \bar{M}^*, l_{Q \bar{Q}} = 2 \Big\rangle,
\end{align}
where we have defined\footnote{
We recall that tetraquark states in the diabatic meson-antimeson basis approach the corresponding spin-splitted meson-antimeson states — after which they are named — only in the large $r$ limit.}
\begin{align}
    &\Big| \Sigma_g^{+'} + 2 \Pi_g \Big\rangle \equiv \Big|[ s_{Q \bar{Q}} = 1, (K = 1, l_{Q \bar{Q}} = 0)\, l = 1] \, J=1  \Big\rangle,\\
    &\Big| 2 \Sigma_g^{+'} + \Pi_g \Big\rangle \equiv \Big|[s_{Q \bar{Q}} = 1, (K = 1, l_{Q \bar{Q}} = 2)\, l = 1] \, J=1 \Big\rangle,\\
    &\Big| \Pi_g \Big\rangle \equiv \Big|[s_{Q \bar{Q}} = 1, (K = 1, l_{Q \bar{Q}} = 2)\, l = 2]\, J=1 \Big\rangle,\\
    &\Big| \frac{M \bar{M}^* + M^* \bar{M}}{\sqrt{2}}, l_{Q \bar{Q}} = 0 \Big\rangle \equiv \frac{1}{\sqrt{2}} \bigg( \Big|[ (j_{M} = 0, j_{\bar{M}^{*}} = 1)\, s_{M \bar{M}^{*}} = 1, l_{Q \bar{Q}} = 0] \, J=1  \Big\rangle + \Big| j_{M} \leftrightarrow j_{\bar{M}^*} \Big\rangle \biggr),\\
    &\Big| \frac{M \bar{M}^* + M^* \bar{M}}{\sqrt{2}}, l_{Q \bar{Q}} = 2 \Big\rangle \equiv  \frac{1}{\sqrt{2}} \bigg( \Big|[ (j_{M} = 0, j_{\bar{M}^{*}} = 1)\, s_{M \bar{M}^{*}} = 1, l_{Q \bar{Q}} = 2]\, J=1  \Big\rangle + \Big| j_{M} \leftrightarrow j_{\bar{M}^*} \Big\rangle \biggr),\\
    &\Big| M^* \bar{M}^*, l_{Q \bar{Q}} = 2 \Big\rangle \equiv \Big|[(j_{M^*} = 1, j_{\bar{M}^*} = 1)\, s_{M^* \bar{M}^*} = 2, l_{Q \bar{Q}} = 2]\, J=1  \Big\rangle.
\end{align} 
From the expression of $V_{SS}$, diagonal in the meson-antimeson basis, the expression of the tetraquark spin-splitting matrix in the BO diabatic basis can be derived. 

The inclusion of $\mathcal{O}(1/m_Q)$ tetraquark spin-dependent potentials mixes tetraquark states with different heavy-quark spins $s_{Q\bar{Q}}$ or angular momenta $l$, provided they share the same $J^{PC}$ quantum numbers. 
The mixing reflects the breaking of the HQSS, as $s_{Q\bar{Q}}$ and $l$ are no longer good quantum numbers.
While the leading-order spin-averaged Eqs.~\eqref{eq:coupledSchr} have a universal form, 
the tetraquark splitting matrix encoding the $\mathcal{O}(1/m_Q)$ contributions depends on the $J^{PC}$ quantum numbers. 
This dependence also lifts the degeneracy among quarkonium multiplets through their mixing with tetraquark channels.
Furthermore, quarkonium states with different internal quantum numbers but identical $J^{PC}$ quantum numbers — for instance, $\ket{[s_{Q\bar{Q}}=1,l_{Q\bar{Q}}=0] \,J=1}$ and $\ket{[s_{Q\bar{Q}}=1,l_{Q\bar{Q}}=2]\,J=1}$ — mix via their common coupling to tetraquark states. 
This gives rise to BOEFT eigenstates that are predominantly $S$- or $D$-wave, $P$- or $F$-wave, ..., depending on the dominant quarkonium component, as solutions of the same coupled equations 
(see Eqs.~\eqref{eq:1--BOEFT}, ~\eqref{eq:1--mes-mes}, ~\eqref{eq:2++BOEFT}, ~\eqref{eq:2++mes-mes} and Tables ~\ref{tab:JPC=1--}, ~\ref{tab:JPC=2++}), i.e. $S$-$D$, $P$-$F$, ... mixings in the BOEFT.

If we now write the coupled Schr\"odinger equations~\eqref{eq:1++BOEFT} in the meson-antimeson diabatic basis, characterized by a diagonal spin-splitting matrix, they have the form:
\begin{align}\label{eq:1++mes-mes}
&\hspace{0cm}\left[
-\frac{1}{m_Qr^2}\,\partial_rr^2\partial_r+\frac{1}{m_Qr^2}
{\begin{pmatrix}
2 & 0 & 0  & 0  \\[4pt]
0 & 0 & 0 & 0 \\[4pt]
0 & 0 & 6 & 0  \\[4pt]
0 & 0 & 0 & 6  \\
\end{pmatrix}}\right.
+\begin{pmatrix} V_{\Sigma_{g}^+} &  \frac{V_{\Sigma_g^+ - \Sigma_g^{+\prime}}}{\sqrt{3}} &  - \frac{V_{\Sigma_g^+ - \Sigma_g^{+\prime}}}{\sqrt{6}} & \frac{V_{\Sigma_g^+ - \Sigma_g^{+\prime}}}{\sqrt{2}} \\[6pt]
    \frac{V_{\Sigma_g^+ - \Sigma_g^{+\prime}}}{\sqrt{3}} & \frac{V_{\Sigma_g^{+\prime}} + 2 V_{\Pi_g}}{3} & -\frac{V_{\Sigma_g^{+\prime}} - V_{\Pi_g}}{3\sqrt{2}} & \frac{V_{\Sigma_g^{+\prime}} - V_{\Pi_g}}{\sqrt{6}}  \\[6pt]
     - \frac{V_{\Sigma_g^+ - \Sigma_g^{+\prime}}}{\sqrt{6}} & -\frac{V_{\Sigma_g^{+\prime}} - V_{\Pi_g}}{3\sqrt{2}} & \frac{V_{\Sigma_g^{+\prime}} + 5 V_{\Pi_g}}{6} & \frac{- V_{\Sigma_g^{+\prime}} +  V_{\Pi_g}}{2 \sqrt{3}} \\[6pt]
      \frac{V_{\Sigma_g^+ - \Sigma_g^{+\prime}}}{\sqrt{2}} & \frac{V_{\Sigma_g^{+\prime}} - V_{\Pi_g}}{\sqrt{6}} & \frac{- V_{\Sigma_g^{+\prime}} +  V_{\Pi_g}}{2 \sqrt{3}} & \frac{V_{\Sigma_g^{+\prime}} +  V_{\Pi_g}}{2} \\[6pt]
      \end{pmatrix} 
\nonumber\\
&\hspace{+8cm}\left.
+\begin{pmatrix}
 0 & 0 & 0 & 0   \\[4pt]
 0 & -\frac{\delta_Q}{2}& 0 & 0  \\[4pt]
 0 & 0 & -\frac{\delta_Q}{2} & 0  \\[4pt]
 0 & 0 &  0 & \frac{\delta_Q}{2}  \\[4pt]
\end{pmatrix}     
      \right]
      \hspace{-4pt}\begin{pmatrix}  \psi_{Q \bar{Q}} \\[4pt] 
    \psi^S_{M \bar{M}^{*}} \\[4pt]
    \psi^D_{M \bar{M}^{*}} \\[4pt]
    \psi_{M^* \bar{M}^{*}} \\ 
      \end{pmatrix}={\mathcal{E}} 
      \begin{pmatrix} \psi_{Q \bar{Q}} \\[4pt] 
    \psi^S_{M \bar{M}^{*}} \\[4pt] 
    \psi^D_{M \bar{M}^{*}} \\[4pt] 
    \psi_{M^* \bar{M}^{*}} \\ 
      \end{pmatrix}, 
\end{align}
where the superscripts $S$, $D$ refer to the values $l_{Q \bar{Q}} = 0$, $2$ of the wavefunction components and are introduced to distinguish the two $M \bar{M}^{*}$ components entering the coupled equations. 
The diagonal elements of the spin-splitting matrix give the shifts induced by $V_{SS}$ on each tetraquark potential relative to the spin-averaged value. 
The tetraquark wavefunction components are labeled by the corresponding spin-splitted meson-antimeson thresholds, while the remaining component corresponds to quarkonium. 
In this basis, since the tetraquark states are not labeled by $l$ and $s_{Q\bar{Q}}$, mixing occurs among all channels already at the static potential level. 
The quarkonium-tetraquark mixing potential coefficients in this basis, calculated previously in~\cite{Bruschini:2023zkb}, exactly match those obtained via a unitary transformation from the BO diabatic basis.
The off-diagonal mixing terms between different tetraquark components, proportional to $V_{\Sigma_g^\prime} - V_{\Pi_g}$, are small in both bases: 
they are $\mathcal{O}(r^2)$ at short distances and vanish at large distances, where both potentials approach the same spin-averaged threshold. 
In the degenerate limit $V_{\Sigma_g^{+'}} = V_{\Pi_g}$, all such mixings vanish and the diagonal elements in the tetraquark sector become identical. 
If we further set $V_{\Sigma_g^{+ \prime}} = V_{\Pi_g} = m_{M \bar{M}^{\rm spin \, avg.}}$, the potential matrix reduces to a form used in several earlier works~\cite{Bruschini:2020voj, Bruschini:2021cty, Bruschini:2021sjh, Bicudo:2022ihz, Zhang:2025bex}.
A detailed comparison with previous studies is in Section~\ref{sec:literature}. 

The procedure can be straightforwardly generalized to arbitrary values of $J^{PC}$ and the coupled equations relative to the cases $J^{PC} = 0^{-+}$, $1^{+-}$, $2^{-+}$, $0^{++}$, $1^{--}$, $2^{++}$ and $2^{--}$ in the BO and meson-antimeson diabatic bases can be found in Appendix~\ref{app:meson-meson basis}.
The fact that $V_{\Sigma_u^-}$ does not enter in Eqs.~\eqref{eq:1++BOEFT} and~\eqref{eq:1++mes-mes} is a peculiarity of the $J^{PC} = 1^{++}$ and a few others cases, but it does not hold for a general value of $J^{PC}$.
On the contrary, the $\Sigma_u^-$ components of the wavefunction play a prominent role in the description of some states, as discussed in Section~\ref{subsec:phenom_impl}. 
Finally, we note that if the exact expression of the spin-dependent tetraquark potential retains the same form as Eq.~\eqref{eq:splitting}, i.e. $V_{SS}^{\rm exact}(r) = \delta_Q^{\rm exact}(r) (\bm{S}_Q \cdot \bm{K}_{\bar{q}} + \bm{S}_{\bar{Q}} \cdot \bm{K}_q)$, the structure of the coupled Schr\"odinger equations  
\eqref{eq:1++BOEFT}, \eqref{eq:1++mes-mes} and in Appendix~\ref{app:meson-meson basis} 
would remain the same after the replacement $\delta_Q \rightarrow \delta_Q(r)$.

Due to the breaking of the HQSS, identifying the quarkonium states obtained from the coupled Schr\"odinger equations — once $\mathcal{O}(1/m_Q)$ spin-dependent tetraquark potentials are included — requires specifying the $nlJ$ quantum numbers, where $l \equiv l_{Q\bar{Q}}$. 
This is in contrast to the spin-averaged case, where the labels $nl$ suffice.
We define the \textit{threshold spin-splitting corrections} with respect to the spin-averaged quarkonium state labeled by $nl$ as 
\begin{equation}\label{eq:tetrspl}
\Delta E_{nlJ}^{\rm thr. spl.} = \mathcal{E}_{nlJ} - \mathcal{E}_{nl},
\end{equation}
with $\mathcal{E}_{nl}$ indicating the spin-averaged eigenvalue of Eq.~\eqref{eq:coupledSchr}, while $\mathcal{E}_{nlJ}$ gives the eigenvalue once tetraquark spin-splitting effects are included. 
In Appendix~\ref{app:tetraquark spin splitting}, the full set of quarkonium and tetraquark bound states together with the associated spin-splitting corrections predicted by Eqs.~\eqref{eq:0-+BOEFT}, ~\eqref{eq:1+-BOEFT}, ~\eqref{eq:2-+BOEFT}, ~\eqref{eq:0++BOEFT}, ~\eqref{eq:1--BOEFT}, ~\eqref{eq:1++BOEFT}, ~\eqref{eq:2--BOEFT}, ~\eqref{eq:2++BOEFT} for the different $J^{PC}$ numbers and below the lowest meson-antimeson thresold entering the coupled system are reported in Tables~\ref{tab:JPC=0-+}, ~\ref{tab:JPC=1+-}, ~\ref{tab:JPC=2-+}, ~\ref{tab:JPC=0++}, ~\ref{tab:JPC=1--}, ~\ref{tab:JPC=1++}, ~\ref{tab:JPC=2--}, ~\ref{tab:JPC=2++}, where the predicted states are compared with the corresponding experimental ones, if there are.

\begin{figure}[ht]
\center
\includegraphics*[width=11cm,clip=true]{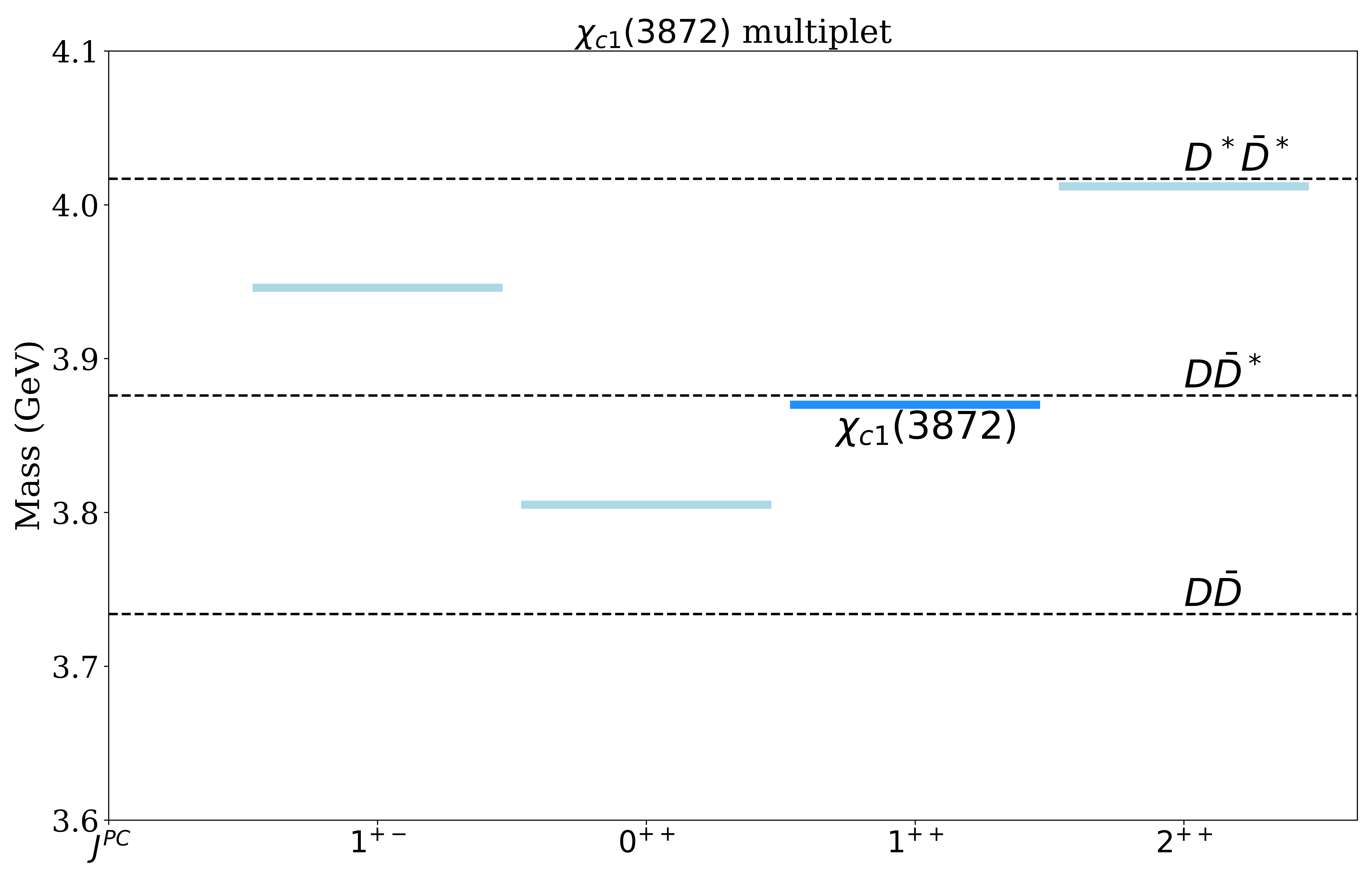}
\caption{The spin structure of the $\chi_{c1}(3872)$ multiplet, once the $\mathcal{O}(1/m_Q)$ tetraquark spin-dependent potential from Eq.~\eqref{eq:splitting} is taken into account. 
The state with $J^{PC} = 1^{++}$, shown as a dark blue line and identified with $\chi_{c1}(3872)$, is obtained by solving Eq.~\eqref{eq:1++BOEFT} after fine-tuning the adjoint meson mass to $\Lambda_{1^{--}}$.
The remaining states in the multiplet (shown as light blue lines) lie above the lowest threshold entering the corresponding coupled equations, 
specifically the $D \bar{D}$ threshold for the states with $J^{PC} = 0^{++}, 2^{++}$ (Eqs.~\eqref{eq:0++BOEFT} and~\eqref{eq:2++BOEFT}) and the $D \bar{D}^*$ threshold for the state with $J^{PC} = 1^{+-}$ (Eq.~\eqref{eq:1+-BOEFT}).
In these cases, 
the spin-splitting corrections are computed from first-order perturbation theory, as discussed in~\cite{Brambilla:2024imu}.}
\label{fig:spin-splitting}
\end{figure}

\begin{figure}[ht]
\center
\includegraphics*[width=11cm,clip=true]{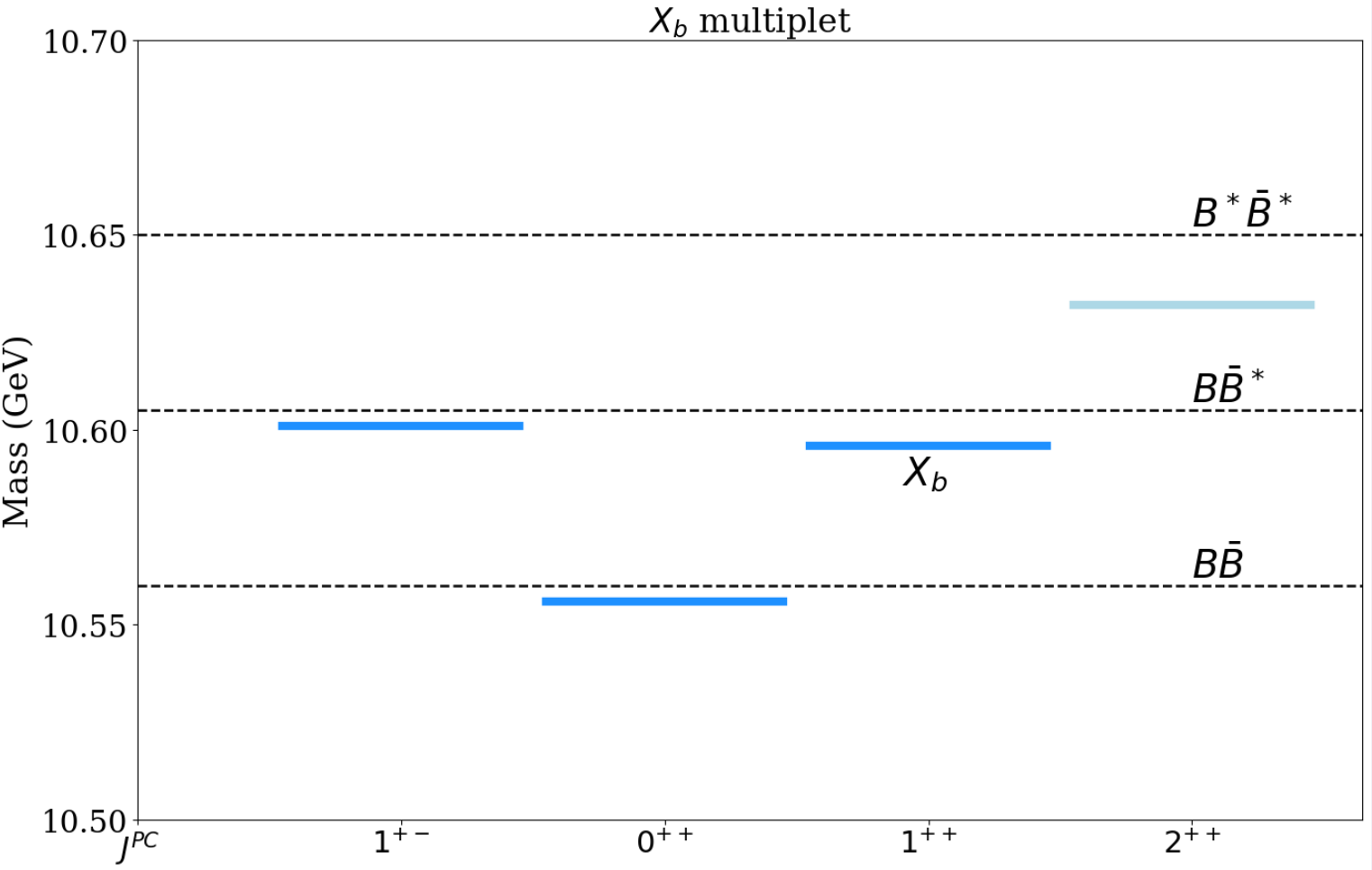}
\caption{The spin structure of the $X_b$ multiplet after including the $\mathcal{O}(1/m_Q)$ tetraquark spin-dependent potential of Eq.~\eqref{eq:splitting}. 
Three states with $J^{PC} = 1^{+-}, 0^{++}$, and $1^{++}$ (the latter denoted $X_b$) appear as dark blue lines and are obtained by solving Eqs.~\eqref{eq:1+-BOEFT}, \eqref{eq:0++BOEFT}, and \eqref{eq:1++BOEFT}, respectively, with the adjoint meson mass $\Lambda_{1^{--}}$ taken from the $\chi_{c1}(3872)$ case. 
The spin-splitting for the fourth state ($J^{PC} = 2^{++}$) is computed from first-order perturbation theory, as described in Ref.~\cite{Brambilla:2024imu}, since it lies above the $B\bar{B}$ threshold entering the coupled equations~\eqref{eq:2++BOEFT}.
It is shown as a light blue line.
}
\label{fig:spin-splitting2}
\end{figure}

\subsection{Spectrum}\label{subsec:phenom_impl} 
To obtain a state with $J^{PC} = 1^{++}$ approximately $100\,\mathrm{keV}$ below the physical $D\bar{D}^*$ threshold from Eq.~\eqref{eq:1++BOEFT} — with properties, i.e. radius and composition, similar to those of the spin-averaged $\chi_{c1}(3872)$ (see Table~\ref{tab:coupled_spectra_cc}) — the adjoint meson mass must be fine-tuned to $\Lambda_{1^{--}} =  0.965$~GeV. 
This requires redefining the static tetraquark potential in Eq.~\eqref{eq:QQbar_V} accordingly. 
The corresponding $0^{-+}$ adjoint meson mass is then fixed via the relation
 $\Lambda_{0^{-+}} - \Lambda_{1^{--}}$
$ = 0.044\,\mathrm{GeV}$, taken from~\cite{Foster:1998wu}.
The new value of $\Lambda_{1^{--}}$
differs by about $50\,\mathrm{MeV}$ from the adjoint meson mass previously used in the case of the spin-averaged $\chi_{c1}(3872)$ below the $D\bar{D}^{\rm spin \, avg.}$ threshold.
Solving the analog equations for $J^{PC} = 0^{++}$, $1^{+-}$, and $2^{++}$ with the adjoint meson mass set to the new value shows that all other states of the multiplet lie above the lowest $D^{(*)}\bar{D}^{(*)}$ threshold included in the coupled system. 
These predictions, shown in Fig.~\ref{fig:spin-splitting}, are consistent with the findings of~\cite{Brambilla:2024imu}, once the spin corrections for the thresholds and tetraquark states are included.\footnote{
In~\cite{Brambilla:2024imu}, the spin-splitting corrections of these tetraquark states are estimated in two  different ways.
The first method assumes an analog spin structure between the lowest tetraquark and the lowest quarkonium hybrid multiplet, given the lack of direct lattice calculations for the former.
Spin splitting corrections are then estimated from the averaged lattice results of~\cite{HadronSpectrum:2012gic,Cheung:2016bym} for the lowest $c \bar{c}$ multiplet and the lattice results of~\cite{Ryan:2020iog} for the lowest $b \bar{b}$ multiplet. 
The second method consists in accounting for the spin-dependent potential $V_{SS}$ of Eq.~\eqref{eq:splitting} at first order in perturbation theory and using the rearrangement formula~\eqref{eq:rearrangement} to express the BO tetraquark states of the $\chi_{c1}(3872)$ multiplet as linear superpositions of the heavy-light meson-antimeson thresholds $D \bar{D}$, $D \bar{D}^*$, $D^* \bar{D}^*$ 
in the charmonium sector and the thresholds $B \bar{B}$, $B \bar{B}^*$, $B^* \bar{B}^*$ in the bottomonium sector.
The spin corrections to the tetraquark states obtained with the two methods are consistent with each other.} 
The $J^{PC} = 2^{++}$ state may be tentatively identified with the $R_2$ structure reported by Belle at $4.014 \,\mathrm{GeV}$~\cite{Belle:2021nuv} or with the $\chi_{c2}(3930)$, this second choice being favored based on the expected structure of the $\chi_{c1}(3872)$ and charmonium $2P$ multiplets.

In the bottomonium sector, computing the spectrum with the newly determined value of the adjoint meson mass $\Lambda_{1^{--}}$
yields the following multiplet structure: the states with $J^{PC} = 1^{+-}$, $1^{++}$, and $0^{++}$ lie below the $B\bar{B}^*$ and $B\bar{B}$ thresholds, respectively, while the $2^{++}$ state lies above the $B\bar{B}$ threshold entering the corresponding coupled equations (see Fig.~\ref{fig:spin-splitting2}).
As in the spin-averaged case, all the $X_b$ multiplet states are predominantly tetraquarks, with quarkonium admixtures from 2\% to 10\%. 
Moreover, both the $1^{+-}$ and $0^{++}$ states of the $X_b$ multiplet exhibit a substantial $\Sigma_u^-$ admixture --- approximately $35\%$ and $20\%$, respectively --- underscoring the prominent role of the $V_{\Sigma_u^-}$ potential and the necessity of its precise determination to address their description. 
The magnitudes of the spin-splitting corrections are consistent with earlier estimates~\cite{Brambilla:2024imu}, though they are approximately $20\,\mathrm{MeV}$ larger. 
Importantly, the predicted mass hierarchy of the multiplet — with $0^{++}$ the lowest state, followed by $1^{++}$ and $1^{+-}$ — 
agrees with that of~\cite{Brambilla:2024imu} (see Tables~\ref{tab:JPC=0++},~\ref{tab:JPC=1++},~\ref{tab:JPC=1+-}).\footnote{
Due to the change in the parametrization of the static tetraquark potentials at large $r$ from two-pion to one-pion exchange, the masses of the various $X_b$ states within the multiplet differ by about 25~MeV from those found in~\cite{Brambilla:2024imu}. 
Despite this shift, the size of the spin-splitting corrections is comparable in magnitude, and the same mass hierarchy among the multiplet states is preserved in the two cases.}

Like the string-breaking corrections discussed in Section~\ref{sec:coupled}, threshold spin-splitting corrections shift the quarkonium masses downward. 
The shifts, which range from 1~MeV to 5~MeV for quarkonium states, are, however, notably smaller than string-breaking effects and increase as the quarkonium states approach the lowest meson-antimeson threshold with which they mix.
This behavior has a clear physical interpretation. 
Spin-splitting corrections affect quarkonium states via their tetraquark admixture. 
Consequently, they are suppressed for predominantly quarkonium states, which have negligible tetraquark components, but become substantial for predominantly tetraquark states, as shown in Figs.~\ref{fig:spin-splitting} and~\ref{fig:spin-splitting2}.
The mechanism can be understood more precisely from the spin-splitting potential matrix in Eq.~\eqref{eq:1++BOEFT} using first-order perturbation theory. 
Treating the  matrix as a perturbation and considering a predominantly quarkonium state for which the tetraquark components approximately vanish, $\psi_{\Sigma_g^{+ \prime} + 2 \Pi_g} \approx \psi_{2 \Sigma_g^{+ \prime} + \Pi_g} \approx \psi_{\Pi_g} \approx 0$,  
the expectation value of the spin-splitting matrix on such a state is (almost) zero.
Conversely, for a predominantly $S$-wave tetraquark state like the $\chi_{c1}(3872)$ — corresponding to the $\psi_{\Sigma_g^{+ \prime} + 2 \Pi_g}$ component with $l_{Q\bar{Q}}=0$, while $\psi_{\Sigma_g^+} \approx \psi_{2 \Sigma_g^{+ \prime} + \Pi_g} \approx \psi_{\Pi_g} \approx 0$ — the first-order correction is $-\delta_Q/2$.
Similarly, using Eqs.~\eqref{eq:0++BOEFT}, \eqref{eq:2++BOEFT}, and \eqref{eq:1+-BOEFT} and looking at predominantly $S$-wave tetraquark states, 
the spin corrections are $-\delta_Q$, $\delta_Q/2$, and zero for the $J^{PC}=0^{++}$, $2^{++}$, and $1^{+-}$ members of the multiplet, respectively. 
These predictions agree exactly with those of Refs.~\cite{Braaten:2024tbm, Brambilla:2024imu}, which were derived from an equivalent first-order perturbative treatment.

We briefly examine the impact of the meson-antimeson spin-splitting terms originating from $V_{SS}$ on the quarkonium spectrum.
The $4S$ bottomonium state with $J^{PC}=0^{-+}$ ($s_{Q\bar{Q}}=0$) lies closer to the $B\bar{B}^*$ threshold than the corresponding spin-averaged state does to $B\bar{B}^{\rm spin \, avg.}$. 
Consequently, its tetraquark admixture increases to about $37\%$ (compared to about $20\%$ in the spin-averaged case), 
highlighting the importance of including tetraquark spin-splitting effects for a realistic description.
Similarly, the $3P$ bottomonium state with $J^{PC}=0^{++}$ ($s_{Q\bar{Q}}=1$) sees its tetraquark component rise to about $11\%$ (compared to about $5\%$ in the spin-averaged case), as it lies only a few tens of $\mathrm{MeV}$ below the $B\bar{B}$ threshold.
The other low-lying quarkonium states are almost unaffected by the introduction of threshold spin-splitting effects, and their quarkonium percentage is nearly identical to the spin-averaged case.

The experimental splittings of about $100\, \mathrm{MeV}$ between $\eta_c(1S)$ and $J/\psi$, $50\, \mathrm{MeV}$ between  $\eta_c(2S)$ and $\psi(2S)$, and $100\, \mathrm{MeV}$ among $h_c(1P)$, $\chi_{c0}(1P)$, $\chi_{c1}(1P)$ and $\chi_{c2}(1P)$, and of about $60\,\mathrm{MeV}$ between $\eta_b(1S)$ and $\Upsilon(1S)$, $25\,\mathrm{MeV}$ between $\eta_b(2S)$ and $\Upsilon(2S)$, $50\,\mathrm{MeV}$ among $h_b(1P)$, $\chi_{b0}(1P)$, $\chi_{b1}(1P)$ and $\chi_{b2}(1P)$, and $40\,\mathrm{MeV}$ among $h_b(2P)$, $\chi_{b0}(2P)$, $\chi_{b1}(2P)$ and $\chi_{b2}(2P)$~\cite{ParticleDataGroup:2024cfk} cannot be accommodated exclusively by the corrections considered here. 
They originate principally from the missing $1/m_Q^2$ quarkonium spin-dependent potentials. 
These are well known (see e.g.~\cite{Brambilla:2004jw}) and have been computed in the lattice gauge theory~\cite{Bali:1996bw,Bali:1997am,Koma:2006fw,Koma:2009ws,Koma:2012bc}. 
Moreover, they have been incorporated in the computation of the hyperfine corrections to the quarkonium spectrum in several phenomenological studies, e.g. in Refs.~\cite{Li:2009ad,Ferretti:2013faa,Ferretti:2013vua,Man:2024mvl,Ni:2025gvx,Hao:2025vmw,Ahmad:2025mue,Zhang:2025bex}.
A consistent inclusion of the spin-dependent quarkonium potentials in the presence of mixing with open flavor thresholds is, however, still to be done, 
as well as it is the derivation and computation of the $1/m_Q^2$ spin-dependent tetraquark potentials.

\section{String-breaking effects on the spin-averaged quarkonium spectrum from a self-energy calculation: comparing the BOEFT framework and the $^3P_0$ model}\label{sec:self-en}
In Section~\ref{subsec:self-en}, we compute string-breaking corrections to spin-averaged quarkonium states from the self-energy of the free quarkonium propagator resulting from the leading-order BOEFT Lagrangian, and we compare the results with the ones obtained from the coupled-channel approach in Section~\ref{sec:coupled}. 
In Section~\ref{subsec:3P0}, we review how the same calculation can be done within the phenomenological $^3P_0$ model, identifying its corresponding BOEFT mixing potential.

\subsection{The BOEFT framework}\label{subsec:self-en}
In the limit $V_{\Sigma_g^+ - \Sigma_g^{+'}} \rightarrow 0$, the quarkonium and tetraquark spectra decouple. 
Quarkonia are then a discrete spectrum of bound states, while tetraquarks are a continuum of scattering states for energies larger than $m_{M_1 \bar{M}_2^{\rm spin\,avg.}}$, where $M_{1,2}^{\rm spin \, avg.} = D^{\rm spin \, avg.}, B^{\rm spin \, avg.}$.\footnote{ 
Equation~\eqref{eq:coupledSchr} in the limit $V_{\Sigma_g^+ - \Sigma_g^{+'}} \rightarrow 0$ gives no tetraquark bound state solution if we adopt the tetraquark potential parametrization~\eqref{eq:QQbar_V} and keep the adjoint meson mass $\Lambda_{1^{--}}$ at the value fixed after~\eqref{eq:QQbar_V}.}
The mixing term in~\eqref{eq:L_mixing} couples the two sectors through their respective $\Sigma_g^+$ and $\Sigma_g^{+\prime}$ wavefunction components. 
In perturbation theory, its effect can be resummed into a self-energy correction to the quarkonium propagator.
The free quarkonium and tetraquark propagators, together with the interaction vertex, follow from the BOEFT Lagrangian~\eqref{eq:BOEFTlagrangian}. 
The pure quarkonium and tetraquark states are given in Eqs.~\eqref{eq:quark_state} and~\eqref{eq:tetr_state}, respectively.
Following the notation of Section~\ref{sec:coupled}, we denote the radial wavefunctions of the pure quarkonium states as $\psi_{\Sigma_g^+}^{nl}(r)$ with eigenvalues $E_{nl}$. 
For the pure tetraquark scattering states, the two radial components are $\psi_{\Sigma_g^{+\prime}}^{E^\prime l}(r)$ and $\psi_{\Pi_g}^{E^\prime l}(r)$, where the continuous energy $E^\prime > m_{M_1 \bar{M}_2^{\rm spin \, avg.}}$ replaces the discrete quantum number $n$  used for bound states. 
They are normalized as 
$\displaystyle \int_0^{+\infty}  dr\, r^2\,   
\Bigl(
\psi^{E^{\prime \prime} l^{\prime\prime}}_{\Sigma_g^{+ \prime}}(r)\,\psi^{E^{\prime} l^{\prime}}_{\Sigma_g^{+ \prime}}(r) + \psi^{E^{\prime \prime} l^{\prime\prime}}_{\Pi_g}(r)\,\psi^{E^{\prime} l^{\prime}}_{\Pi_g}(r) \Bigr)
= 2 \pi \,\delta(E^\prime - E^{\prime\prime}) \delta_{l^{\prime \prime}l^\prime}$.

String-breaking corrections to the mass and width of a spin-averaged quarkonium state may be expressed in terms of the self-energy of the quarkonium propagator evaluated on the quarkonium state labeled by the quantum numbers $n$ and $l$ at the quarkonium energy $E_{nl}$.
This self-energy, $\Sigma^{nn}_{l} (E_{nl})$, written in the BO mixed basis and accounting only for mixing of a given quarkonium state $nl$ with the intermediate $\Sigma_g^{+\prime}$ tetraquark scattering states at $E^\prime > m_{M_1 \bar{M}_2^{\rm spin \, avg.}}$ reads
\begin{equation}
\Sigma^{nn}_{l} (E_{nl}) =  \int_{m_{M_1 \bar{M}_2^{\rm spin \, avg.}}}^{+ \infty} \hspace{-3mm} \frac{dE^{\prime}}{2 \pi} \, 
\frac{a^n_{l}(E^\prime)^2}{E_{nl} - E^\prime  + i \epsilon}, \label{eq:self-en}
\end{equation}
with
\begin{equation}
a^n_{l}(E^{\prime}) = \int_0^{+\infty} dr \, r^2 \, \psi^{*\,nl}_{\Sigma_g^+}(r) \, V_{\Sigma_g^+ - \Sigma_g^{+\prime}}(r) \, \psi_{\Sigma_g^{+\prime}}^{E^{'}l}(r).
\label{eq:self-en2}
\end{equation}
For the overlap integral in Eq.~\eqref{eq:self-en2} to be non-zero 
the orbital angular momentum quantum number $l$ must be the same for both the quarkonium and tetraquark state.  
The real part of the self-energy provides the \textit{string-breaking correction} to the mass of the quarkonium, i.e. the mass shift induced by the mixing with the threshold, embedded in the tetraquark potential.
For quarkonium bound states ($E_{nl} < m_{M_1 \bar{M}_2^{\rm spin \, avg.}}$), the self-energy is purely real. 
For quarkonium resonances above threshold ($E_{nl} > m_{M_1 \bar{M}_2^{\rm spin \,avg.}}$), an imaginary part arises. 
The imaginary part gives the decay rate of the resonance into the spin-averaged meson–antimeson threshold. 
The expressions for the real and imaginary parts of the self-energy are\footnote{
In the case of several \textit{non-mutually interacting} exotics, all interacting at leading order with the quarkonium via its $\Sigma_g^+$ component, like the isospin-0 $Q \bar{Q} q \bar{q}$, $Q \bar{Q} s \bar{s}$ tetraquarks, whose static potentials have been found non-interacting in~\cite{Bulava:2024jpj}, the self-energy has the form 
$$
\Sigma^{nn}_{l\, {\rm tot}} (E_{nl}) = \Sigma^{nn}_{l\, Q \bar{Q} q \bar{q}} (E_{nl}) + \Sigma^{nn}_{l\, Q \bar{Q} s \bar{s}} (E_{nl}) + ..., 
$$
where each term quantifies the shift of the quarkonium state due to the mixing of the quarkonium with a different exotic state.}
\begin{align}
{\rm Re}\,(\Sigma^{nn}_{l} (E_{nl})) & = {\rm P.V.} \int_{m_{M_1 \bar{M}_2^{\rm spin \, avg.}}}^{+ \infty}  \hspace{-3mm} \frac{dE^{'}}{2 \pi} \frac{a^n_{l}(E^{'})^2}{E_{nl} - E^{'}}, \label{eq:realima}\\
-{\rm Im}\, (\Sigma^{nn}_{l} (E_{nl})) &= \frac{\Gamma}{2}( Q \bar{Q}_{nl} \rightarrow M_1 \bar{M_2}^{\rm spin \, avg.}) = \frac{a^n_{l}(E_{nl})^2}{2},
\label{eq:realimb}
\end{align}
where P.V. stands for principal value.
The results for the string-breaking corrections computed from Eq.~\eqref{eq:realima} for $l=0,1,2$ and for all the quarkonium bound states below $M_1 M_2^{\rm spin \, avg.}$ in the charmonium, bottomonium, $b \bar{c}$/$c \bar{b}$  sectors are reported in Table~\ref{tab:selfen}.
The study of the string-breaking corrections for the lowest quarkonium resonances is left for future work.

\begin{table}[ht]
    \centering
    \renewcommand{\arraystretch}{1.3}
   \begin{tabular}{|c||c|c|c|}  \hline
  & $c \bar{c}$  & $b \bar{b}$  & $b \bar{c}$/$c\bar{b}$ \\
$n l$ & $\quad {\rm Re}\,(\Sigma^{nn}_{l}{(E_{nl})})$ (MeV)   & $\quad {\rm Re}\,(\Sigma^{nn}_{l}{(E_{nl})})$ (MeV)  & $\quad {\rm Re}\,(\Sigma^{nn}_{l}{(E_{nl})})$ (MeV)  \\
\hline\hline
$1S$         & $-0.3$  & $-0.1$ & $-0.2$   \\
$2S$         & $-3.4$  & $-0.8$ & $-2.0$   \\
$3S$         &   -         & $-2.8$ & $-19.1$   \\
$4S$         &   -         & $-14.8$ &    -         \\\hline
$1P$         & $-1.9$  & $-0.8$ & $-1.7$   \\
$2P$         &   -         & $-3.2$ & $-8.5$   \\
$3P$         &   -         & $-9.1$ &     -        \\\hline
$1D$         & $-5.1$  & $-1.9$ & $-3.3$   \\
$2D$         &   -        & $-5.9$ &     -        \\\hline
\end{tabular}
\caption{The string-breaking corrections ${\rm Re}\,(\Sigma^{nn}_{l}(E_{nl}))$ obtained from Eq.~\eqref{eq:realima} for the charmonium, bottomonium, and $b\bar{c}$/$c \bar{b}$ spin-averaged states lying below $M_1 \bar{M}_2^{\rm spin\,avg.}$, with $M_{1,2}^{\rm spin \, avg.} = D^{\rm spin \, avg.}$ or $B^{\rm spin \, avg.}$. 
The dashes indicate quarkonium states above the threshold, for which the string-breaking corrections have not been computed. 
The values can be directly compared with the $\Delta E^{\rm str. \, br.}_{nl}$ reported in Tables~\ref{tab:coupled_spectra_cc}, ~\ref{tab:coupled_spectra_bb}, ~\ref{tab:coupled_spectra_bc} that give the values of the string-breaking corrections in the coupled system approach.}
\label{tab:selfen}
\end{table}

\begin{figure}[ht]
\centering 
\includegraphics*[width=11cm,clip=true]{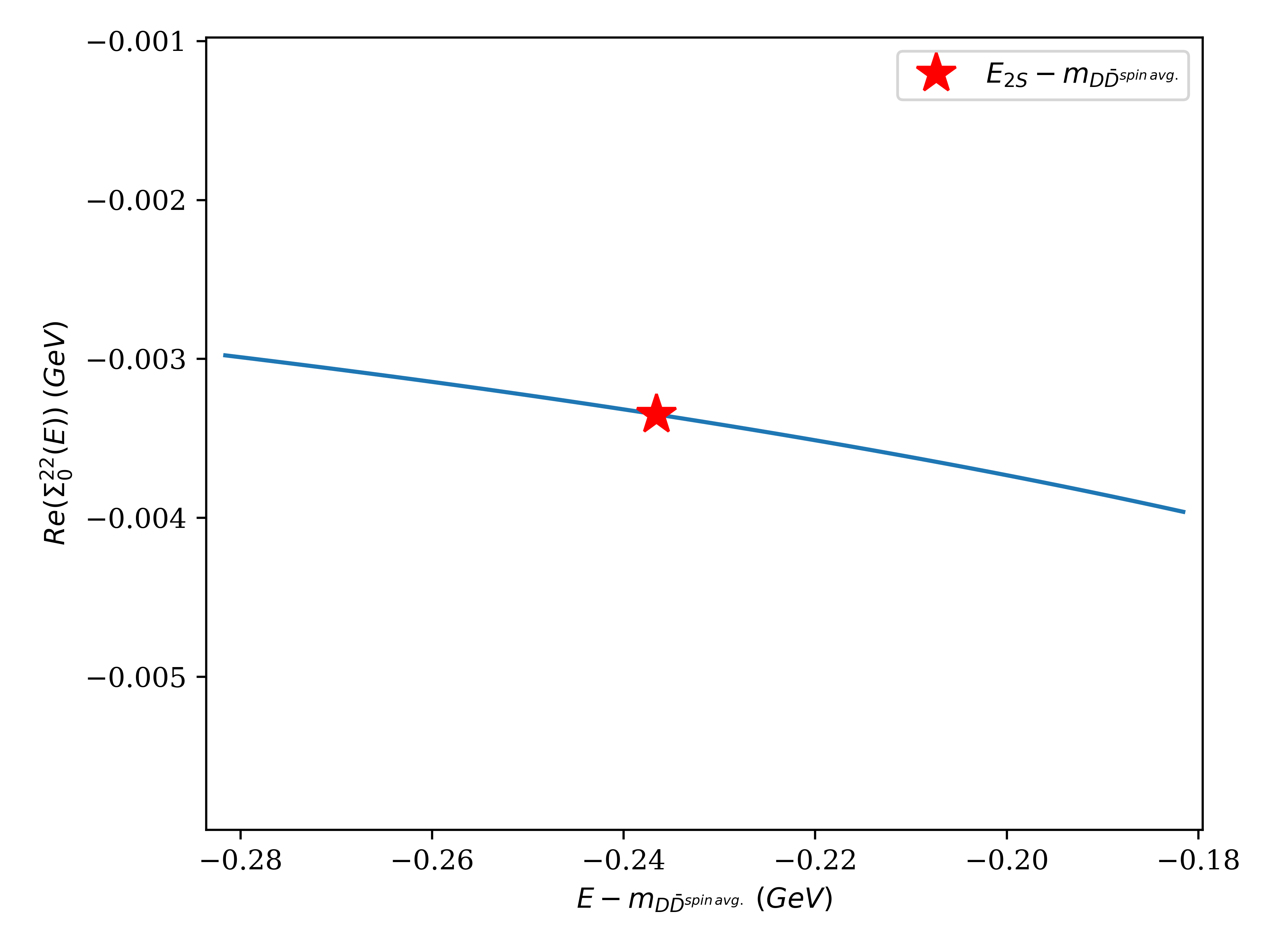}
\caption{The dependence of the self-energy correction ${\rm Re}(\Sigma_0^{22}(E))$ for the $2S$ spin-averaged charmonium state on the energy gap $E - m_{D \bar{D}^{\rm spin \, avg.}}$ over a range of about $100\, \mathrm{MeV}$, centered around $E_{2S} - m_{D \bar{D}^{\rm spin \, avg.}}$, marked with a red star.
}
\label{fig:self-en}
\end{figure}

The results in Table~\ref{tab:selfen} agree well with those from the coupled Schr\"odinger equation approach (Tables~\ref{tab:coupled_spectra_cc}, \ref{tab:coupled_spectra_bb} and \ref{tab:coupled_spectra_bc}), showing the equivalence of the two methods for computing string-breaking corrections.
The real part of the self-energy, $\mathrm{Re}\,(\Sigma_l^{nn}{(E)})$, shows only a weak dependence on the energy gap $E - m_{M_1 \bar{M}_2^{\rm spin \, avg.}}$, varying by at most a few $\mathrm{MeV}$ over an interval of $100\,\mathrm{MeV}$ centered at $E_{nl} - m_{M_1 \bar{M}_2^{\rm spin \,avg.}}$. 
This behavior is illustrated for the spin-averaged $2S$ charmonium state in Fig.~\ref{fig:self-en} and is similar for the other bound states. 
If we compute the same corrections using tetraquark potentials modeled as constant lines fixed at $m_{M_1 \bar{M}_2^{\rm spin \,avg.}}$, we find close agreement with the results obtained with the full potentials. 
This supports the conclusion, already drawn in Section~\ref{sec:coupled}, that string-breaking corrections are largely insensitive to the short-distance details of the tetraquark potentials.
Moreover, mutual interactions between different quarkonium states below threshold due to the mixing with the tetraquark scattering states have been computed from the self-energy and checked to have a negligible effect, altering the string-breaking corrections by less than $1\,\mathrm{MeV}$ in all cases considered.
The self-energy approach used here has been widely adopted in potential models and BOEFT studies~\cite{Eichten:1978tg,Ono:1983rd,Eichten:2004uh,Barnes:2007xu,Li:2009ad,Danilkin:2009hr,Ferretti:2012zz,Ferretti:2013faa,Ferretti:2013vua,Ferretti:2014xqa,Ferretti:2018tco,Bruschini:2021cty,Bruschini:2021sjh,TarrusCastella:2022rxb,Man:2024mvl,Sultan:2025dfe,Hao:2025vmw}. 
A detailed comparison with this body of work is provided in Section~\ref{sec:literature}.
The next section presents a similar self-energy calculation within the phenomenological $^3P_0$ model and compares with the BOEFT result.

\subsection{The $^3P_0$ model and the equivalent BOEFT mixing potential}\label{subsec:3P0}
The first phenomenological model proposed to incorporate threshold effects on quarkonium states — what we also term string-breaking corrections — was the $^3P_0$ model. 
Originally introduced to describe strong quarkonium decays~\cite{LeYaouanc:1972vsx,LeYaouanc:1977fsz}, it was subsequently adopted in several other studies within the same context~\cite{Ono:1980js,Ono:1983rd,Heikkila:1983wd}. 
In the $^3P_0$ model, a phenomenological constant $\gamma$ parametrizes the mixing between quarkonium and threshold states. 
The value of $\gamma$ is typically fitted to reproduce the experimental strong decay widths.
Here we compare the real part of the self-energy at leading order in the BOEFT, Eq.~\eqref{eq:realima}, with the analogous expression in the $^3P_0$ model. 
This allows us to relate the corresponding mixing potentials: in the BOEFT, the mixing is constrained by lattice QCD calculations in the string-breaking region~\cite{Bulava:2024jpj}, while in the $^3P_0$ model, $\gamma$ is fitted to data.

If we compute the $^3P_0$ decay width of a quarkonium state $nlJ$ (with $l \equiv l_{Q\bar{Q}}$) into $M^{(*)}\bar{M}^{(*)}$ from the imaginary part of the self-energy contribution to the quarkonium propagator, like in Eq.~\eqref{eq:realimb}, then the corresponding real part in the chiral and heavy quark limit for the heavy-light meson mass~\cite{Neubert:1996wg} reads\footnote{See Appendix~\ref{app:comparison_3P0_BOEFT} for more details on the derivation of Eq.~\eqref{eq:3P0_real1} from the $^3P_0$ expression of the strong decay width, Eq.~\eqref{eq:3P0_immaginary}.}
\begin{align}\label{eq:3P0_real1}
{\rm Re}\,\Sigma_{^3P_0} (Q \bar{Q}_{nl_{Q \bar{Q}}J}, M^{(*)} \bar{M}^{(*)}) =  
{\rm P.V.} \int_0^{+ \infty} dk \, \frac{ m_Q k^2}{(k^{nJ})^2 - k^2} \, \gamma^2  \, \Big| \Big\langle I_{M^{(*)}}, I_{3\,M^{(*)}}, I_{\bar{M}^{(*)}}, I_{3\,\bar{M}^{(*)}} \Big| I_{Q \bar{Q}}, I_{3\,Q \bar{Q}} \Big\rangle \Big|^2 \nonumber\\
\times \frac{\left[ \begin{array}{ccc}
I_Q & I_{\bar{Q}} & I_{Q \bar{Q}} \\
I_{\bar{q}} & I_q & I_{q \bar{q}} \\
I_{M^{(*)}} & I_{\bar{M}^{(*)}} & I_{M^{(*)} \bar{M}^{(*)}}
\end{array} \right]^2}{8}  \Bigg( \biggl(C_{l_{Q \bar{Q}} + 1, J}^{M^{(*)} \bar{M}^{(*)}}\biggr)^2 \mathcal{I}_{l_{Q \bar{Q}} + 1,nJ}(k r)^{ 2}  + \biggl(C_{l_{Q \bar{Q}} - 1, J}^{M^{(*)} \bar{M}^{(*)}}\biggr)^2  \mathcal{I}_{l_{Q \bar{Q}} - 1,nJ}(kr)^2 \Bigg),
\end{align}
with 
\begin{align}\label{eq:coeff}
C_{l_{Q \bar{Q}} + 1, J}^{M^{(*)} \bar{M}^{(*)}} = \sqrt{\frac{3(l_{Q \bar{Q}} + 1)}{2 l_{Q \bar{Q}} + 3}} \left[ \begin{array}{ccc}
s_Q & s_{\bar{Q}} & s_{Q \bar{Q}} \\
s_{\bar{q}} & s_q & s_{q \bar{q}} \\
s_{M^{(*)}} & s_{\bar{M}^{(*)}} & s_{M^{(*)}\bar{M}^{(*)}}
\end{array} \right]
&\left[ \begin{array}{ccc}
l_{Q \bar{Q}} & s_{Q\bar{Q}} & J \\
l_{q \bar{q}} & s_{q \bar{q}} & K \\
l_{Q \bar{Q}} + 1 & s_{M^{(*)}\bar{M}^{(*)}} & J
\end{array} \right], \\
\label{eq:coeff2}
C_{l_{Q \bar{Q}} - 1, J}^{M^{(*)} \bar{M}^{(*)}} =\sqrt{\frac{3l_{Q \bar{Q}} }{2 l_{Q \bar{Q}} - 1}} \left[ \begin{array}{ccc}
s_Q & s_{\bar{Q}} & s_{Q \bar{Q}} \\
s_{\bar{q}} & s_q & s_{q \bar{q}} \\
s_{M^{(*)}} & s_{\bar{M}^{(*)}} & s_{M^{(*)}\bar{M}^{(*)}}
\end{array} \right]
&\left[ \begin{array}{ccc}
l_{Q \bar{Q}} & s_{Q\bar{Q}} & J \\
l_{q \bar{q}} & s_{q \bar{q}} & K \\
l_{Q \bar{Q}}  - 1 & s_{M^{(*)}\bar{M}^{(*)}} & J
\end{array}\right].
\end{align}
The quantum number $I$, depending on its subscript, indicates the isospin of the heavy (anti)quark, light (anti)quark, quark-antiquark pairs, or (anti)meson, and similarly for the spin $s$ and the orbital angular momentum $l$. 
The overlap integrals are given by 
\begin{align}\label{eq:overlap1}
\mathcal{I}_{l_{Q \bar{Q}} + 1,nJ}(k) &= \int dr r^2 \psi_{Q \bar{Q}}^{nl_{Q \bar{Q}}J}(r) \Biggl( \int dp \, p^3 \psi_{M^{\rm spin \, avg.}}^*(p) \, \psi_{\bar{M}^{\rm spin \, avg.}}^*(p)\, j_1(pr) \Biggr) j_{l_{Q \bar{Q}} + 1}(kr),   \\ 
\label{eq:overlap2}
\mathcal{I}_{l_{Q \bar{Q}} - 1,nJ}(k) &= \int dr r^2 \psi_{Q \bar{Q}}^{nl_{Q \bar{Q}}J}(r) \Biggl( \int dp \, p^3 \psi_{M^{\rm spin \, avg.}}^*(p) \, \psi_{\bar{M}^{\rm spin \, avg.}}^*(p) \, j_1(pr) \Biggr) j_{l_{Q \bar{Q}} - 1}(kr).
\end{align}
The function $\psi_{Q \bar{Q}}^{nl_{Q \bar{Q}}J}$ is the radial quarkonium wavefunction typically obtained from some phenomenological potential model 
that in general may also include $1/m_Q^{2}$ suppressed spin-dependent terms; 
$\psi_{M^{\rm spin \, avg.}}$, $\psi_{\bar{M}^{\rm spin \, avg.}}$ are the meson and antimeson wavefunctions; 
$j_{l_{Q \bar{Q}} \pm 1}$ are spherical Bessel functions.
The momentum is defined as $k^{nJ} \equiv \sqrt{m_Q E_{\mathrm{bind}, nJ}^{M\bar{M}^{\rm spin \, avg.}}}$, where $E_{\mathrm{bind}, nJ}^{M\bar{M}^{\rm spin \,avg.}} \equiv E_{nl_{Q\bar{Q}}J} - m_{M\bar{M}^{\rm spin \, avg.}}$ and all meson–antimeson thresholds are degenerate in the limit considered for  Eq.~\eqref{eq:3P0_real1}.\footnote{
In Eq.~\eqref{eq:3P0_real1}, the different meson-antimeson thresholds are degenerate in mass to $m_{M \bar{M}^{\rm spin \, avg.}}$, but are still defined by different $s_{M^{(*)}}$, $s_{\bar{M}^{(*)}}$, $s_{M^{(*)} \bar{M}^{(*)}}$ quantum numbers, giving rise to separate self-energy contributions differing by a numerical prefactor.}
The isospin rearrangement factor reads~\cite{Ono:1980js}
\begin{align}\label{eq:9j-symbol1}
\left[ \begin{array}{ccc}
I_Q & I_{\bar{Q}} & I_{Q \bar{Q}} \\
I_{\bar{q}} & I_q & I_{q \bar{q}} \\
I_{M^{(*)}} & I_{\bar{M}^{(*)}} & I_{M^{(*)}\bar{M}^{(*)}}
\end{array} \right]  &\equiv \Bigg|\Bigg\langle \left[ (I_Q, I_{\bar{q}})I_{M^{(*)}}, (I_{\bar{Q}}, I_q)I_{\bar{M}^{(*)}}\right] I_{M^{(*)}\bar{M}^{(*)}} \Bigg| \left[ (I_Q, I_{\bar{Q}})I_{Q \bar{Q}}, (I_{q}, I_{\bar{q}})I_{\bar{q \bar{q}}}\right] I_{M^{(*)}\bar{M}^{(*)}} \Bigg\rangle \Bigg|^2  \nonumber\\
& \hspace{-2cm} =\sqrt{(2I_{Q \bar{Q}} + 1)} \sqrt{(2I_{q \bar{q}} + 1)} \sqrt{(2I_{M^{(*)}} + 1)} \sqrt{(2I_{\bar{M}^{(*)}}+1)} \left\{ \begin{array}{ccc}
I_Q & I_{\bar{Q}} & I_{Q \bar{Q}} \\
I_{\bar{q}} & I_q & I_{q \bar{q}} \\
I_{M^{(*)}} & I_{\bar{M}^{(*)}} & I_{M^{(*)}\bar{M}^{(*)}}
\end{array} \right\}, 
\end{align}
with the last factor being a Wigner 9-j symbol, while the other isospin factor in the first line of Eq.~\eqref{eq:3P0_real1} is a Clebsch--Gordan coefficient involving the quarkonium isospin state and the meson-antimeson one. 
The isospin quantum numbers are $I_Q = I_{\bar{Q}} = I_{Q \bar{Q}} = I_{3Q \bar{Q}} = 0$, $I_q = I_{\bar{q}} = 1/2$, $I_M = I_{\bar{M}} = 1/2$, $I_{q \bar{q}} = 0$, $I_{Q \bar{Q}} = I_{M^{(*)} \bar{M}^{(*)}} = 0$.
Similar expressions hold for the spin and angular momentum rearrangement factors.

As in~\cite{Bruschini:2025paj}, we define the form factor
\begin{align}\label{eq:overlap}
\mathcal{F}(r) = \int dp \, p^3 \, \psi_{M^{\rm spin \, avg.}}^*(p) \, \psi^*_{\bar{M}^{\rm spin \, avg.}}(p) \, j_1(pr).
\end{align}
By assuming a Gaussian shape for the heavy-light meson and antimeson wavefunctions, or equivalently by taking the $1S$ wavefunction to be that one of an harmonic oscillator, $\displaystyle \psi_{M^{\rm spin \, avg.}}(p) = \frac{2 L^{\frac{3}{2}}}{\pi^{\frac{1}{4}}} e^{-\frac{p^2 L^2}{2}}$~\cite{LeYaouanc:1972vsx}, with $L$ related to the size of the meson, the previous expression can be simplified into~\cite{Bruschini:2025paj}
\begin{align}\label{eq:overlap_anal}
\mathcal{F}(r)_{\rm anal.} = \frac{r}{2L^2} e^{-\frac{r^2}{4L^2}}.
\end{align}

For a meanigful comparison with the equivalent BOEFT expression, we need to set the tetraquark potentials constant: $V_{\Sigma_g^{+ \prime}}^{\rm flat} = V_{\Pi_g}^{\rm flat} = m_{M \bar{M}^{\rm spin \, avg.}} = m_{M^{\rm spin \, avg.}} + m_{\bar{M}^{\rm spin \, avg.}}$.\footnote{
In all the $^3P_0$ model studies, it is not realized that the tetraquark potentials have repulsive behavior at small $r$ and approach asymptotically the heavy-light meson-antimeson thresholds at large $r$ while conserving the BO quantum numbers. 
Instead, the tetraquark potentials are parametrized by constant lines, whose values are fixed by the sum of the meson and antimeson masses, and the corresponding states are usually referred to as free meson-antimeson or threshold states rather than tetraquark states.} 
In this case, the tetraquark scattering states reduce to $\displaystyle \psi_{\Sigma_g^{+'}}^{El_{Q \bar{Q}}}(r) \equiv \sqrt{\frac{2}{\pi}} j_{l_{Q \bar{Q}}}(kr)$, with $k = \sqrt{m_Q (E - m_{M \bar{M}^{\rm spin \, avg.}})}$.
The real part of the BOEFT self-energy in the diabatic basis, analogous to Eq.~\eqref{eq:realima}, takes the form\footnote{
Besides taking the potential constant, in order to go from Eq.~\eqref{eq:3P0_real1} to Eq.~\eqref{eq:selfen_BOEFT_anal1}, it is necessary to exploit the relation $E = k^2/m_Q + m_{M \bar{M}^{\rm spin \, avg.}}$.
This sets the normalization of the state to be 
$\displaystyle \int dr \, r^2 \,\sqrt{\frac{2}{\pi}} j_{l_{Q \bar{Q}}^\prime \pm 1}(k^\prime r) \sqrt{\frac{2}{\pi}} j_{l_{Q \bar{Q}} \pm 1}(kr)
  = \delta(k^\prime - k)/k^2\, \delta_{l_{Q \bar{Q}}^\prime \pm 1,l_{Q \bar{Q}} \pm 1}$.}
\begin{align}\label{eq:selfen_BOEFT_anal1}
{\rm Re}\,(\Sigma_{l_{Q \bar{Q}}}^{nn}(E_{nl_{Q \bar{Q}}}))^{\rm BOEFT}_{V_{\Sigma_g^{+\prime}} = m_{M \bar{M}^{\rm spin \, avg.}}} = {\rm P.V.} \int_0^{+\infty} dk \, \frac{m_Q k^2 }{m_Q(E_{nl_{Q \bar{Q}}} - m_{M \bar{M}^{\rm spin \, avg.}})-k^2} \, a^{M \bar{M}^{\rm spin \, avg.}}_{nl_{Q \bar{Q}}}(k)^2,
\end{align}
with:
\begin{align}\label{eq:selfen_BOEFT_form_fact1}
a^{M \bar{M}^{\rm spin \, avg.}}_{nl_{Q \bar{Q}}}(k)^2 = \Bigl(C_{l_{Q \bar{Q}} + 1}^{M \bar{M}^{\rm spin \, avg.}}\Bigr)^2  \int dr \, r^2 \, \psi_{\Sigma_g^+}^{*\,nl_{Q \bar{Q}}}(r) \, V_{\Sigma_g^+-\Sigma_g^{+\prime}}(r) \, \sqrt{\frac{2}{\pi}} j_{l_{Q \bar{Q} + 1}}(kr)  \nonumber\\
 + \Bigl(C_{l_{Q \bar{Q}} - 1}^{M \bar{M}^{\rm spin \, avg.}}\Bigr)^2  \int dr \, r^2 \, \psi_{\Sigma_g^+}^{*\,nl_{Q \bar{Q}}}(r) \, V_{\Sigma_g^+-\Sigma_g^{+\prime}}(r) \, \sqrt{\frac{2}{\pi}} j_{l_{Q \bar{Q} - 1}}(kr).
\end{align}
The coefficients $C_{l_{Q\bar{Q}} \pm 1}^{M\bar{M}^{\rm spin \, avg.}}$ correspond to the mixing potential prefactors entering the diabatic potential matrix obtained from Eq.~\eqref{eq:coupledSchr}. 
Equivalently, these coefficients can be extracted from the coupled Schr\"{o}dinger equations derived in Section~\ref{subsec:spin-splitted_eq} and Appendix~\ref{app:meson-meson basis} for specific $J^{PC}$ quarkonium states in the limit $\delta_Q \rightarrow 0$, as discussed below.

\paragraph{$S$-wave quarkonium states.}
We consider here $S$-wave quarkonium states with $J^{PC} = 0^{-+}$ and $1^{--}$, and compare the $^3P_0$ expression in Eq.~\eqref{eq:3P0_real1} with its BOEFT counterpart.
We first identify the relevant quantum numbers. 
By construction the $^3P_0$ quantum numbers are $I_{q \bar{q}} = 0$, $s_{q \bar{q}} = l_{q \bar{q}} = 1 $, $K = 0$.
The isospin quantum numbers being independent of the specific quarkonium or meson-antimeson threshold states have already been fixed above.
Other quantum numbers unambiguously fixed are $s_Q = s_{\bar{Q}} = s_q = s_{\bar{q}} = 1/2$ and $l_{Q \bar{Q}} = 0$. 

Following Table~\ref{tab:BOEFTvsmes-mes}, for the $J^{PC} = 0^{-+}$ quarkonium state with $s_{Q \bar{Q}} = 0$, the non-vanishing $P$-wave threshold contributions come from the states
\begin{itemize}
    \item[{\it (a)}] $\displaystyle \frac{M \bar{M}^* - M^* \bar{M}}{\sqrt{2}}$ with $s_M = 0$, $s_{\bar{M}^*} = 1$, $s_{M \bar{M}^*} = 1$ or ${s_{M^*}} = 1$, $s_{\bar{M}} = 0$, $s_{M \bar{M}^*} = 1$, giving  $C_{1,0}^{(M\bar{M}^* - M^* \bar{M})/\sqrt{2}} = {1}/{\sqrt{2}}$,
    \item[{\it (b)}] $M^* \bar{M}^*$ with ${s_{M^*}} = s_{\bar{M}^*} = 1$, $s_{M^* \bar{M}^*} = 1$, giving $C_{1,0}^{M^*\bar{M}^*,s_{M^* \bar{M}^*} = 1} = {1}/{\sqrt{2}}$.
\end{itemize}
It holds that $\big(C_{1,0}^{(M\bar{M}^* - M^* \bar{M})/\sqrt{2}}\big)^2 + \bigl(C_{1,0}^{M^*\bar{M}^*,s_{M^* \bar{M}^*} = 1}\bigr)^2 \equiv \big(C_{1}^{M \bar{M}^{\rm spin \, avg.}}\bigr)^2 =  1$. 

For the $J^{PC} = 1^{--}$ quarkonium state with $s_{Q \bar{Q}} = 1$, they come from the states
\begin{itemize}
    \item[{\it (a)}] $M \bar{M}$ with $s_M = s_{\bar{M}} = s_{M \bar{M}} = 0$, giving $C_{1,1}^{M\bar{M}} = {1}/{(2 \sqrt{3})}$,
    \item[{\it (b)}] $\displaystyle \frac{M \bar{M}^* + M^* \bar{M}}{\sqrt{2}}$ with $s_M = 0$, $s_{\bar{M}^*} = 1$, $s_{M \bar{M}} = 1$ or $s_{M^*} = 1$, $s_{\bar{M}} = 0$, $s_{M^* \bar{M}} = 1$, giving $C_{1,1}^{(M\bar{M}^* - M^* \bar{M})/\sqrt{2}} = {1}/{\sqrt{3}}$, 
    \item[{\it (c)}] $M^* \bar{M}^*$ with $s_{M^*} = s_{\bar{M}^*} = 1$, $s_{M^* \bar{M}^*} = 0,2$, giving $C_{1,1}^{M^*\bar{M}^*,s_{M^* \bar{M}^*} = 0} = {1}/{6}$, $C_{1,1}^{M^*\bar{M}^*,s_{M^* \bar{M}^*} = 2} = {\sqrt{5}}/{3}$.
\end{itemize}
It holds that $\bigl(C_{1,1}^{M\bar{M}}\bigr)^2 + \bigl(C_{1,1}^{(M\bar{M}^* - M^* \bar{M})/\sqrt{2}}\bigr)^2 + \bigl(C_{1,1}^{M^*\bar{M}^*,s_{M^* \bar{M}^*} = 0}\bigr)^2 + \bigl(C_{1,1}^{M^*\bar{M}^*,s_{M^* \bar{M}^*} = 2}\bigr)^2 \equiv \big(C_{1}^{M \bar{M}^{\rm spin \, avg.}}\bigr)^2 = 1$.
All the previous coefficients computed in the $^3P_0$ model agree up to a phase with the coefficients of the quarkonium-tetraquark mixing potential $V_{\Sigma_g^+ - \Sigma_g^{+\prime}}$ appearing in the potential matrices of Eqs.~\eqref{eq:0-+mes-mes} and~\eqref{eq:1--mes-mes} in the meson-antimeson basis, 
which have been obtained independently within the BOEFT.\footnote{
A difference in phase can be reabsorbed into a redefinition of the corresponding tetraquark state component. \label{footnote:coeff} }

In the limit of Eq.~\eqref{eq:3P0_real1}, the masses of the $M^{(*)} \bar{M}^{(*)}$ thresholds are degenerate. 
This allows to factor out the overlap integrals from the mixing coefficients $C_{l_{Q \bar{Q}}+1, J}^{M^{(*)} \bar{M}^{(*)}}$ and $C_{l_{Q \bar{Q}}-1, J}^{M^{(*)} \bar{M}^{(*)}}$. 
The real part of the self-energy for quarkonium states with $J^{PC} = 0^{-+}, 1^{--}$ in the $^3P_0$ model can then be written as\footnote{
For $S$-wave quarkonium ($l_{Q\bar{Q}}=0$), only the term with $C_{l_{Q\bar{Q}}+1, J}^{M^{(*)} \bar{M}^{(*)}}$ contributes, as the $C_{l_{Q\bar{Q}}-1, J}^{M^{(*)} \bar{M}^{(*)}}$ coefficient vanishes.}
\begin{align}\label{eq:3P0_self_swav1}
&{\rm Re}\,\Sigma_{^3P_0}(Q \bar{Q}_{n00}, \frac{M \bar{M}^* + M^* \bar{M}}{\sqrt{2}} + M^* \bar{M}^*) = {\rm P.V.}\, \int_0^{+\infty} dk \frac{m_Q k^2}{(k^{n0})^2 - k^2} \, \frac{\gamma^2}{16}  \nonumber\\ 
&\qquad\qquad \times \left( \left(C_{1,0}^{(M\bar{M}^* - M^* \bar{M})/\sqrt{2}}\right)^2 + \left(C_{1,0}^{M^*\bar{M}^*,s_{M^* \bar{M}^*} = 1}\right)^2 \right) \, \mathcal{I}_{1,n0}(k)^2,\\ 
\label{eq:3P0_self_swav12}
&{\rm Re}\, \Sigma_{^3P_0}(Q \bar{Q}_{n01}, M\bar{M} + \frac{M \bar{M}^* + M^* \bar{M}}{\sqrt{2}} + M^* \bar{M}^*) = {\rm P.V.}\, \int_0^{+\infty} dk \frac{m_Q k^2}{(k^{n1})^2 - k^2} \, \frac{\gamma^2}{16}  \nonumber\\ 
&\qquad\qquad \times \left( \left(C_{1,1}^{M\bar{M}}\right)^2+  \left(C_{1,1}^{(M\bar{M}^* + M^* \bar{M})/\sqrt{2}}\right)^2 + \left(C_{1,1}^{M^*\bar{M}^*,s_{M^* \bar{M}^*} = 0}\right)^2 + \left(C_{1,1}^{M^*\bar{M}^*,s_{M^* \bar{M}^*} = 2}\right)^2 \right) \, \mathcal{I}_{1,n1}(k)^2,
\end{align}
where in the last line the $s_{M^* \bar{M}^*}$ value is given to distinguish between the $s_{M^* \bar{M}^*} = 0,2$ contributions.

On the other hand, for the BOEFT calculation, the coupled Eqs.~\eqref{eq:0-+BOEFT} and~\eqref{eq:1--BOEFT} relative to $J^{PC} = 0^{-+}$ and $1^{--}$ in the limit $\delta_Q \rightarrow0$ of degenerate thresholds take the block-diagonal form 
\begin{align}\label{eq:0-+blockdiag}
&\left[
-\frac{1}{m_Qr^2}\,\partial_rr^2\partial_r+ \frac{1}{m_Q r^2}{\begin{pmatrix}
0 & 0 & 0\\[4pt]
0                 & 2        & 0 \\[4pt]
0                 & 0 & D_l
\end{pmatrix}} +  
{\begin{pmatrix}
V_{\Sigma_g^+} & V_{\Sigma_g^+ - \Sigma_g^{+\prime}} & 0 \\[6pt]
V_{\Sigma_g^+ - \Sigma_g^{+\prime}}  &  m_{M \bar{M}^{\rm spin \, avg.}}   & 0 \\[6pt]
0                 & 0 & D_V \\[6pt]
\end{pmatrix}}\right]
\begin{pmatrix} \psi_{\Sigma_g^+}^{n0} \\[6pt] \psi_{\Sigma_g^{+ \prime}}^{n1} \\[6pt] \psi_{D}\end{pmatrix}  ={\mathcal{E}} \begin{pmatrix} \psi_{\Sigma_g^+}^{n0} \\[6pt] \psi_{\Sigma_g^{+ \prime}}^{n1} \\[6pt] \psi_{D}\end{pmatrix}, 
\end{align}
where $\psi_{\Sigma_g^+}^{n0}$ and $\psi_{\Sigma_g^{+ \prime}}^{n1}$\footnote{
Unlike for the wavefunction components in Eq.~\eqref{eq:coupledSchr}, here the orbital angular momentum quantum number in the superscript of the wavefunction is $l_{Q \bar{Q}}$, 
reflecting the partial-wave decomposition in the diabatic basis.
\label{footnote:l}}
denote the $S$-wave quarkonium and the $P$-wave tetraquark spin-averaged wavefunctions, respectively, while $D_l$, $D_V$, which are $1 \times 1$ matrices if $J^{PC} = 0^{-+}$ and $6 \times 6$ matrices if $J^{PC} = 1^{--}$,  identify a block of the equation that decouples in the
$\delta_Q\to0$ limit with $\psi_D$ giving the related wavefunction components.\footnote{
The $J^{PC} = 1^{--}$ quantum numbers can come from a predominantly $S$-wave or a predominantly $D$-wave quarkonium state. 
The $D$-wave block is not relevant for the current discussion, and therefore has been included in Eq.~\eqref{eq:0-+blockdiag} into the $D_l$ and $D_V$ parts.}
The upper block of Eq.~\eqref{eq:0-+blockdiag} fixes the value of the mixing potential prefactor $C_{1}^{M \bar{M}^{\rm spin \, avg.}} = 1$ in Eq.~\eqref{eq:selfen_BOEFT_form_fact1}.
The real part of the self-energy for $S$-wave quarkonium states in the BOEFT is then given by 
\begin{align}\label{eq:selfen_BOEFT1}
&\hspace{-1.5cm}{\rm Re}\,(\Sigma_{0}^{nn}(E_{n0}))^{\rm BOEFT}_{V_{\Sigma_g^{+\prime}} = m_{M \bar{M}^{\rm spin \, avg.}}}= {\rm P.V.} \int_0^{+\infty} dk \, \frac{m_Q k^2}{m_Q(E_{n0} - m_{M \bar{M}^{\rm spin \, avg.}}) - k^2}  \nonumber\\
&\hspace{6.5cm}\times \Biggl(\int dr r^2 \psi_{\Sigma_g^+ }^{*n0}(r) V_{\Sigma_g^+ - \Sigma_g^{+ \prime}}(r) \sqrt{\frac{2}{\pi}} j_{1}(k r)\Biggr)^2.
\end{align}

We now compare Eq.~\eqref{eq:selfen_BOEFT1} with Eqs.~\eqref{eq:3P0_self_swav1} and~\eqref{eq:3P0_self_swav12}. 
To enforce the HQSS consistently in the $^3P_0$ model, we set to zero also spin-dependent terms in the quarkonium potential. 
This leads to the following degeneracies $E_{n00} = E_{n01} \equiv E_{n0}$, $\psi_{Q \bar{Q}}^{n00} = \psi_{Q \bar{Q}}^{n01} \equiv \psi_{Q \bar{Q}}^{n0}$, and $k^{n0} =  k^{n1} = \sqrt{m_Q(E_{n0} - m_{M \bar{M}^{\rm spin \, avg.}})}$. 
Hence, matching the real part of the self-energy computed in the BOEFT, i.e. Eq.~\eqref{eq:selfen_BOEFT1}, with the self-energy contributions obtained in the $^3P_0$ model, i.e. Eqs.~\eqref{eq:3P0_self_swav1} and~\eqref{eq:3P0_self_swav12}, in the limit of vanishing spin corrections both in the thresholds and quarkonium potential, we get 
\begin{align}\label{eq:equate}
\int dr \, r^2 \, \psi_{\Sigma_g^+}^{*n0}(r) \, V_{\Sigma_g^+ - \Sigma_g^{+ \prime}}^{^3P_0}(r) \, \sqrt{\frac{2}{\pi}} j_1(k r) \overset{!}{=} \frac{\gamma}{4} \int dr \, r^2 \, \psi_{Q \bar{Q}}^{*\,n0}(r) \, \mathcal{F}(r) \, j_{1}(kr).
\end{align}
From the above equality, after identifying the wavefunctions, $\psi_{\Sigma_g^+}^{n0} = \psi_{Q \bar{Q}}^{n0}$, it follows
\begin{align}\label{eq:equivalence1}
V_{\Sigma_g^+-\Sigma_g^{+ \prime}}^{^3P_0} = \frac{\sqrt{\pi}}{4 \sqrt{2}} \gamma \mathcal{F}(r).
\end{align}
where the superscript $^3P_0$ indicates that this mixing potential can be interpreted as the $^3P_0$ equivalent mixing potential in the BOEFT.
In the case of a Gaussian heavy-light meson wavefunction, it takes the analytical expression
\begin{align}\label{eq:equivalence_analyt1}
V_{\Sigma_g^+-\Sigma_g^{+ \prime}}^{^3P_0, \, {\rm anal.}} = \frac{\sqrt{\pi}}{8 \sqrt{2} L^2} \, \gamma \, r \, e^{-r^2/4L^2},
\end{align}
which is consistent with Eq.~(29) of Ref.~\cite{Bruschini:2025paj}.\footnote{ 
The mixing parameter $\gamma$ introduced in this section, as well as in Appendices~\ref{app:comparison_3P0_BOEFT}  and~\ref{app:comparison_S_P}, quantifies the mixing between a quarkonium state and the isospin-averaged thresholds $M^{(*)} \bar{M}^{(*)}$. In constrast, in Eq.~(1) of~\cite{Bruschini:2025paj}, $\gamma$ quantifies the mixing with the distinct thresholds $M^{(*)\,+} M^{(*)\,-}$ and $M^{(*)\,0} \bar{M}^{*\,(0)}$. 
It follows that our mixing coupling $\gamma$ corresponds to the one introduced in~\cite{Bruschini:2025paj} multiplied by a factor $\sqrt{2}$.
If we now compare Eq.~\eqref{eq:equivalence_analyt1} with Eq.~(29) in~\cite{Bruschini:2025paj}, and account for the factor $\sqrt{2}$ of difference in the definition of the mixing parameter, we find that the two equations are equal.}

\paragraph{$P$-wave quarkonium states and further generalizations.}
We give here the result for $P$-wave quarkonium states; further details can be found in Appendix~\ref{app:comparison_S_P}.
If we sum over the non-vanishing threshold contributions for all the different quarkonium states of the multiplet, $J^{PC} = 1^{+-}, (0,1,2)^{++}$, we find for the real part of the self-energy in the $^3P_0$ model
\begin{align}\label{eq:3P0_self_pwav1}
&{\rm Re}\,\Sigma_{^3P_0}(Q \bar{Q}_{n1J}, M\bar{M} + \frac{M \bar{M}^* + M^* \bar{M}}{\sqrt{2}} + M^* \bar{M}^*) = \nonumber\\
&\hspace{1.5cm} {\rm P.V.} \int_0^{+\infty} dk \, \frac{m_Q k^2}{(k^{nJ})^2 - k^2} \, \frac{\gamma^2}{16} \, \Biggl( \bigl(C_{2}^{M \bar{M}^{\rm spin \, avg.}}\bigr)^2  \mathcal{I}_{2,nJ}(k)^2 + \bigl(C_{0}^{M \bar{M}^{\rm spin \, avg.}}\bigr)^2  \mathcal{I}_{0,nJ}(k)^2\Biggr),
\end{align}
where $C_{2}^{M \bar{M}^{\rm spin \, avg.}} = \sqrt{{2}/{3}}$ and $C_{0}^{M \bar{M}^{\rm spin \, avg.}} = \sqrt{{1}/{3}}$. 
We have left the value of $J$ unspecified, as this form is common to all quarkonium states of the multiplet.

The corresponding real part of the self-energy contribution, computed in the BOEFT in the diabatic basis, reads 
\begin{align}\label{eq:selfen_BOEFT_pwav1}
&{\rm Re}\,(\Sigma_{1}^{nn}(E_{n1}))^{\rm BOEFT}_{V_{\Sigma_g^{+\prime}} = m_{M \bar{M}^{\rm spin \, avg.}}}= \nonumber\\
&\hspace{1.5cm} {\rm P.V.} \int_0^{+\infty} dk \, \frac{m_Q k^2}{m_Q(E_{n1} - m_{M \bar{M}^{\rm spin \, avg.}}) - k^2} \, \Biggl( \bigl(C_{2}^{M \bar{M}^{\rm spin \, avg.}} \bigr)^2 \, \mathcal{I}_{2,n}(k)^2 + \bigl(C_{0}^{M \bar{M}^{\rm spin \, avg.}} \bigr)^2 \, \mathcal{I}_{0,n}(k)^2 \Biggr),
\end{align}
with 
\begin{align}
\mathcal{I}_{2,n} (k) &=   \int dr \, r^2 \, \psi_{\Sigma_g^+}^{*n1}(r) V_{\Sigma_g^+ - \Sigma_g^{+ \prime}}^{^3P_0}(r) \, \sqrt{\frac{2}{\pi}} j_{2}(k r) , \\
 \mathcal{I}_{0,n} (k) &=  \int dr \, r^2 \, \psi_{\Sigma_g^+}^{*n1}(r) V_{\Sigma_g^+ - \Sigma_g^{+ \prime}}^{^3P_0}(r) \, \sqrt{\frac{2}{\pi}} j_{0}(k r),
\end{align}
where $C_{2}^{M \bar{M}^{\rm spin \, avg.}} = \sqrt{2/3}$ and $C_{0}^{M \bar{M}^{\rm spin \, avg.}} = \sqrt{1/3}$ are fixed from the analog of Eq.~\eqref{eq:0-+blockdiag} for $P$-wave quarkonium states reported in Eq.~\eqref{eq:1+-blockdiag} of Appendix~\ref{app:comparison_S_P}.
If we now compare Eqs.~\eqref{eq:3P0_self_pwav1} and ~\eqref{eq:selfen_BOEFT_pwav1}, 
this yields identical results to Eqs.~\eqref{eq:equivalence1} and~\eqref{eq:equivalence_analyt1}.

The same calculation can be generalized to $D$-wave quarkonium states
or to higher partial waves. 
For $D$-wave quarkonia, Eqs.~\eqref{eq:3P0_self_pwav1} and~\eqref{eq:selfen_BOEFT_pwav1} are valid 
after the replacements 
$C_0^{M \bar{M}^{\rm spin\, avg.}} = \sqrt{
{1}/3} \rightarrow C_1^{M \bar{M}^{\rm spin\, avg.}} = \sqrt{2/5}$ and $C_2^{M \bar{M}^{\rm spin\, avg.}} = \sqrt{
{2}/3} \rightarrow C_3^{M \bar{M}^{\rm spin\, avg.}} =\sqrt{3/5}$, and of the corresponding wavefunctions. 
The expressions of the coefficients for generic $l_{Q\bar Q}$ is $\displaystyle C_{l_{Q\bar{Q} + 1 }}^{M \bar{M}^{\rm spin \, avg.}} = \sqrt{\frac{l_{Q \bar{Q}} + 1}{2l_{Q \bar{Q}} + 1}}$ and $\displaystyle C_{l_{Q\bar{Q} - 1 }}^{M \bar{M}^{\rm spin \, avg.}} = \sqrt{\frac{l_{Q \bar{Q}}}{2l_{Q \bar{Q}} + 1}}$.

\section{Comparison with the literature and the experimental data}\label{sec:literature}
We now compare our results with those in the literature, both from a theoretical perspective in Sections~\ref{subsec:literature_pheno} and~\ref{subsec:literature_BOEFT}, 
highlighting the improvements introduced in this study with respect to existing ${^3P_0}$ and BO studies, and from an experimental one in Section~\ref{subsec:spectrum}, interpreting our findings in the light of existing data.

\subsection{Phenomenological studies of string-breaking corrections: the $^3P_0$ and CCCM models}\label{subsec:literature_pheno}
Over the last decades, the $^3P_0$ model has been the most popular approach to calculate string-breaking corrections or strong decay widths of quarkonium states into threshold states.
We now compare the expression of the $V_{\Sigma_g^+-\Sigma_g^{+ \prime}}$ mixing potential adopted in this study with its phenomenologically equivalent version from $^3P_0$ model studies, derived in the Eqs.~\eqref{eq:equivalence1} and~\eqref{eq:equivalence_analyt1}.  

\begin{figure}[ht]
\centering 
\includegraphics*[width=11.0cm,clip=true]{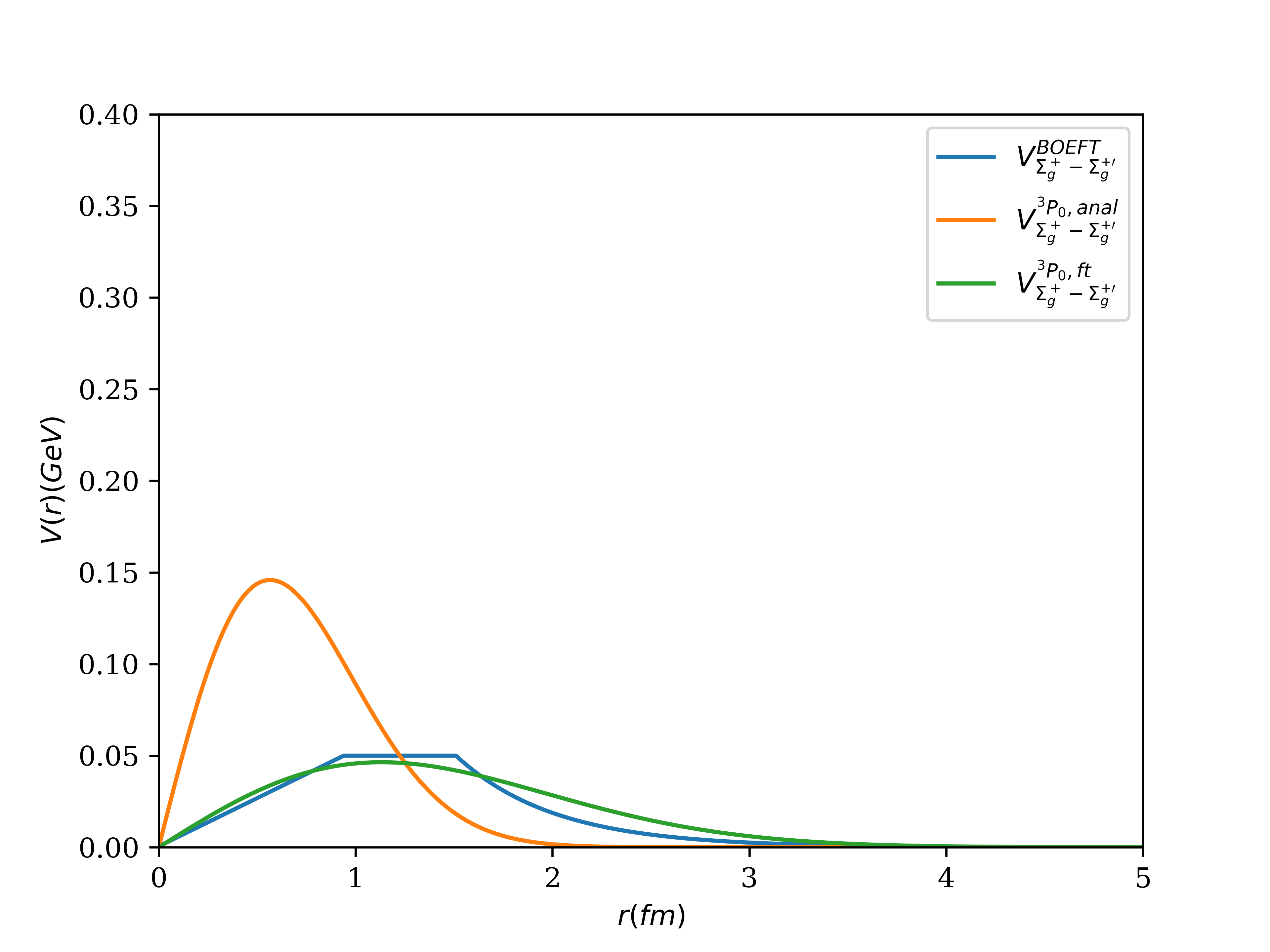}
\caption{The BOEFT static mixing potential of Eq.~\eqref{eq:Vmix}, denoted $V^{\rm BOEFT}_{\Sigma_g^+ - \Sigma_g^{+ \prime}}$ in the legend, compared with the corresponding mixing potential obtained from the static $^3P_0$ model and given in Eq.~\eqref{eq:equivalence_analyt1}.
$V^{^3P_0,\,{\rm anal.}}_{\Sigma_g^+ - \Sigma_g^{+ \prime}}$ is the $^3P_0$ potential for $\gamma = 2.2$, 
which is the value used in~\cite{Ono:1980js}, and $L = 0.4$~fm for the meson size.
$V^{^3P_0,\, {\rm ft.}}_{\Sigma_g^+ - \Sigma_g^{+ \prime}}$ is the potential \eqref{eq:equivalence_analyt1} with the parameters $\gamma = 1.4$ and $L = 0.8$~fm tuned to reproduce at best the BOEFT mixing potential \eqref{eq:Vmix}.
}
\label{fig:comparison}
\end{figure}

As a result of using Gaussian heavy-light meson wavefunctions as in~\cite{LeYaouanc:1972vsx,Bruschini:2025paj}, 
Eq.~\eqref{eq:equivalence_analyt1} has two unknown parameters: the light quark pair-creation strength $\gamma$ and a length scale $L$ related to the meson size.
In the original papers~\cite{LeYaouanc:1972vsx,Ono:1980js} as well as in some recent studies~\cite{Barnes:2005pb,Barnes:2007xu,Ahmad:2025mue,Gao:2025tob,Ni:2025gvx}, $\gamma$ is treated as a flavor-independent parameter for the different quark-antiquark pairs created from the vacuum.
It is treated as a flavor-dependent parameter instead in~\cite{Ferretti:2012zz,Ferretti:2013faa,Ferretti:2013vua,Sultan:2025dfe,Ahmad:2025hcr}. 
In all cases, $\gamma$ (and any flavor-dependent variant) is fitted to reproduce the observed quarkonium decay widths via Eq.~\eqref{eq:3P0_immaginary}.
 
The potential \eqref{eq:equivalence_analyt1} exhibits the same asymptotic behaviors as $V_{\Sigma_g^+-\Sigma_g^{+\prime}}$ in Eq.~\eqref{eq:Vmix}, i.e. linear at short distances ($r \rightarrow 0$) and exponentially suppressed at large distances ($r \rightarrow +\infty$).
We adopt $\gamma = 2.2$ (the value used in~\cite{Ono:1980js}) and $L = 0.4\,\mathrm{fm}$, and compare the resulting potential with $V_{\Sigma_g^+-\Sigma_g^{+\prime}}$.
The two potentials are plotted in Fig.~\ref{fig:comparison}, together with $V^{^3P_0,\, {\rm ft}}_{\Sigma_g^+ - \Sigma_g^{+ \prime}}$, which is the $^3P_0$ potential obtained by fine-tuning the parameters $\gamma$ and $L$ to mimic at best the behavior of the BOEFT mixing potential \eqref{eq:Vmix}.
$V_{\Sigma_g^+-\Sigma_g^{+ \prime}}^{^3P_0, \, {\rm anal.}}$ peaks well before the string-breaking region, and its maximum differs by a factor $3$ from that of $V_{\Sigma_g^+-\Sigma_g^{+\prime}}$, denoted for clarity $V_{\Sigma_g^+-\Sigma_g^{+\prime}}^{\rm BOEFT}$ in Fig.~\ref{fig:comparison}.

\begin{table}[ht]
\centering
\renewcommand{\arraystretch}{1.7} 
\begin{tabular}{|c||c|c|c|c|c|c|}  \hline
$n l (c \bar{c})$ & $g$ & $2\,g$   & $3\,g$ & $4\,g$  & $5\,g$ & $6\,g$ \\
\hline\hline
$\Delta E^{\rm str. br.}_{1s}$~{(MeV)}         & $-0.4$  & $-1.5$ & $-3.4$  & $-6.0$  & $-9.4$  & $-13.7$ \\
$\Delta E^{\rm str. br.}_{2s}$~{(MeV)}         & $-3.7$   & $-14.7$ & $-32.0$  & $-54.4$  & $-81.1$  & $-110.9$ \\ \hline
$n l (b \bar{b})$ & $g$ & $2\,g$   & $3\,g$ & $4\,g$  & $5\,g$ & $6\,g$ \\ \hline
$\Delta E^{\rm str. br.}_{1s}$~{(MeV)}         & $-0.1$  & $-0.4$ & $-0.8$  & $-1.5$  & $-2.3$  & $-3.4$ \\
$\Delta E^{\rm str. br.}_{2s}$~{(MeV)}         & $-0.9$   & $-3.9$ & $-8.8$  & $-15.7$  & $-24.6$  & $-35.4$ \\
$\Delta E^{\rm str. br.}_{3s}$~{(MeV)}         & $-3.5$  & $-13.8$ & $-30.2$  & $-51.7$   & $-77.3$  & $-106.2$ \\
$\Delta E^{\rm str. br.}_{4s}$~{(MeV)}         & $-17.1$  & $-45.0$ & $-77.3$  & $-111.4$ & $-146.5$  & $-182.4$ \\ \hline
\end{tabular}
\caption{We list the values of the string-breaking corrections \eqref{eq:str_break_corr} at increasing values of the mixing coupling $g=50$~MeV for $S$-wave charmonium and bottomonium states. 
A similar dependence on the mixing is also observed for $P$- and $D$-wave states.}
\label{tab:mixing_dep}
\end{table}

As already pointed out in Section~\ref{sec:coupled}, the string-breaking corrections $\Delta E_{nl}^{\rm str. br.}$ (see Eq.~\eqref{eq:str_break_corr}) are extremely sensitive to the value of $g$. 
In Table~\ref{tab:mixing_dep}, we report the corrections to $S$-wave charmonium and bottomonium states  
computed in the BOEFT coupled Schr\"{o}dinger equation approach at increasing values of $g$.

While early implementations of the $^3P_0$ model~\cite{LeYaouanc:1972vsx,Ono:1980js,Ono:1983rd,Heikkila:1983wd} just incorporated spin-independent quarkonium potentials, more recent studies have included $\mathcal{O}(1/m_Q^2)$ spin-dependent terms (spin–spin, spin–orbit, tensor potentials) into the quarkonium potential to reproduce the experimental splittings of the lowest multiplets~\cite{Barnes:2005pb,Pennington:2007xr,Barnes:2007xu,Li:2009ad,Ferretti:2012zz,Ferretti:2013faa,Ferretti:2014xqa,Ferretti:2018tco,Ni:2025gvx,Sultan:2025dfe,Ahmad:2025mue,Gao:2025tob}. 
From an effective field theory perspective a fully consistent treatment at a given order requires including potential corrections up to the same order in both the threshold and mixing potentials. 
 
Another phenomenological model, developed concurrently with the $^3P_0$ model, is the  Cornell coupled-channel model (CCCM)~\cite{Eichten:1978tg,Eichten:1979ms,Eichten:2004uh,Eichten:2005ga} that like the $^3P_0$ model describes open-flavor threshold effects due to the mixing between quarkonium and meson--antimeson channels.
At the time of its formulation, lattice QCD input was scarce, and the LDF dynamics governing the mixing remained poorly constrained.\footnote{
These early studies, as in the $^3P_0$ model case, did not recognize that tetraquark states reduce asymptotically to free meson--antimeson pairs. 
Consequently, they adopted flat constant threshold potentials.} 
A universal Cornell potential was proposed in order to describe both the quarkonium potential and light quark pair creation, the latter inducing mixing between quarkonium and threshold states.
The Cornell parameters were fitted to properties of the lowest charmonium states, and string-breaking corrections were computed via a self-energy approach similar to the one used in Section~\ref{sec:self-en}. 
Following the same line of thought as in Section~\ref{subsec:3P0}, one can compare the real part of the BOEFT self-energy in Eq.~\eqref{eq:realima} with its CCCM counterpart.
Even without going into a detailed quantitative comparison, it is possible to appreciate immediately the different dynamics assumed for the mixing potential $V_{\Sigma_g^+-\Sigma_g^{+\prime}}$, based on a Cornell-like confining potential such as $V_{\Sigma_g^+}$, with respect to the BOEFT or the $^3P_0$ model.

Most of the phenomenological studies mentioned in this section succeed in providing 
a reasonably accurate description of the experimental spectra, at most $100 \, \mathrm{MeV}$ away from the data, 
by appropriately fitting the different model parameters (heavy quark masses, Cornell potential parameters, quark-pair creation constants, and so on). 
The accuracy is similar to the one achieved within the BOEFT formalism at leading order (see  Tables~\ref{tab:coupled_spectra_cc},~\ref{tab:coupled_spectra_bb},~\ref{tab:coupled_spectra_bc} in Section~\ref{sec:coupled} and Tables~\ref{tab:JPC=0-+},~\ref{tab:JPC=1+-},~\ref{tab:JPC=2-+},~\ref{tab:JPC=0++},~\ref{tab:JPC=1++},~\ref{tab:JPC=1--},~\ref{tab:JPC=2++},~\ref{tab:JPC=2--} in Appendix~\ref{app:tetraquark spin splitting}). 
The agreement with spectral data should not hide, however, the fact that the description of quarkonium states as well as their mixing with threshold states varies significantly from one model to another, with string-breaking corrections ranging from a few tens of MeV~\cite{Pennington:2007xr} to several hundreds MeV~\cite{Barnes:2007xu} for the lowest charmonium states, and similarly for bottomonium states. 
This heterogeneity arises primarily from the lack of a clear connection between these models and QCD, leading to very different LDF dynamics — such as in the CCCM and $^3P_0$ model — and the introduction of parameters without a precise field-theoretical definition, which, because of that, can be easily adjusted to the different phenomenological needs.\footnote{
In Ref.~\cite{Li:2009ad} it is shown that similar spectra and string-breaking corrections can be obtained for the lowest states in the charmonium sector using either a $^3P_0$ model or a (single-channel) screened potential model via a tuning of the parameters entering in the two models.} 
In particular, the large values of string-breaking corrections obtained in the majority of $^3P_0$ model studies appear to come from tuning $\gamma$ and therefore effectively $g$ to larger values than the ones indicated by LQCD, see Fig.~\ref{fig:comparison} and Table~\ref{tab:mixing_dep}.
As for the CCCM, the different asymptotic behavior of the Cornell potential in comparison to that of $V_{\Sigma_q^+-\Sigma_g^{+\prime}}$ seems to be at the origin of the typically large string-breaking corrections, between 50~MeV and 180~MeV,  
obtained in that model.
In summary, despite their practical utility for phenomenological studies, string-breaking models lack a first-principles foundation. 
This leads to an underlying dynamics that is often very different from model to model and incompatible with the most recent LQCD findings.
In contrast, the BOEFT approach provides a rigorous and systematic treatment of QCD non-perturbative dynamics 
that is encoded in well-defined gauge-invariant quantities, like generalized Wilson loops, computable (and computed) in LQCD.

Regarding the description of the $\chi_{c1}(3872)$, we identify it as a solution to the coupled equations~\eqref{eq:coupledSchr}. 
It may be understood as originating from the interplay between the short- and long-range behavior of the tetraquark potentials with a residual, approximately $10 \%$, $P$-wave quarkonium component.
This interpretation differs from the one proposed in~\cite{Ferretti:2013faa, Ferretti:2014xqa, Ferretti:2018tco, Gao:2025tob, Li:2009ad, Man:2024mvl,Ahmad:2025hcr}, where the $\chi_{c1}(3872)$ is instead identified with the $\chi_{c1}(2P)$ state, receiving substantial self-energy corrections along with a significant threshold component in its wavefunction due to its proximity to the $D \bar{D}^*$ threshold.
As discussed in Section~\ref{sec:coupled}, our interpretation — that the $\chi_{c1}(3872)$ is distinct from the spin-average $2P$ quarkonium state — is further supported by the recent LHCb observation of the $h_c(4000)$ and $\chi_{c1}(4010)$ states with $J^{PC} = 1^{+-}$ and $ 1^{++}$, respectively~\cite{LHCb:2024vfz}.

\subsection{BO studies of string-breaking corrections}\label{subsec:literature_BOEFT}
We compare here briefly with some previous Born--Oppenheimer-based studies.
They typically do not implement all aspects of the 
Born--Oppenheimer effective field theory employed in this work (e.g. short distance constraints, complete set of mixing potentials, ...) and/or use older sets of lattice data.

In a first set of studies that we consider~\cite{Bruschini:2020voj,Bruschini:2021sjh,Zhang:2025bex}, 
the authors solve coupled Schr\"{o}dinger equations similar to the ones in Section~\ref{subsec:phenom_impl} and Appendix~\ref{app:meson-meson basis} in the meson–antimeson diabatic basis for various $J^{PC}$ values. 
These studies investigate low-lying quarkonium states by accounting for mixing with the lowest $S$-wave thresholds. 
Self-energy corrections are computed via expressions analog to Eq.~\eqref{eq:selfen_BOEFT_anal1} for the real part and similarly for the imaginary part in~\cite{Bruschini:2021cty,Bruschini:2021sjh}.
The thresholds $D_{(s)}\bar{D}_{(s)}$, $D_{(s)}\bar{D}_{(s)}^*$, and $D_{(s)}^*\bar{D}_{(s)}^*$ are included in the charmonium sector, and the thresholds $B_{(s)}\bar{B}_{(s)}$, $B_{(s)}\bar{B}_{(s)}^*$, and $B_{(s)}^*\bar{B}_{(s)}^*$ are included in the bottomonium sector. 
In Refs.~\cite{Bruschini:2020voj,Bruschini:2021sjh,Bruschini:2021cty}, threshold potentials are taken as constants fixed at the physical meson–antimeson masses. 
Hence, the short-range octet repulsion, a proper characteristic of tetraquark potentials, is omitted. 
Moreover, the correct unitary coefficients for quarkonium–threshold mixing (e.g. those multiplying $V_{\Sigma_g^+-\Sigma_g^{+\prime}}$ in Eqs.~\eqref{eq:1++mes-mes}, \eqref{eq:0-+mes-mes}, etc) were only computed later in~\cite{Bruschini:2023zkb}. 
Threshold off-diagonal potential matrix terms are also neglected, whereas in our framework they are proportional to linear combinations of the $\Sigma_g^{+'}$, $\Pi_g$, and $\Sigma_u^-$ tetraquark potentials (see the Schr\"odinger equations~\eqref{eq:1++BOEFT}, ~\eqref{eq:1++mes-mes}, and in Appendix~\ref{app:meson-meson basis}).
Notably, the mixing potential adopted in these studies — though not explicitly constrained by first-principles calculations, but rather assumed to be Gaussian with its maximum fixed to reproduce the $\chi_{c1}(3872)$ mass — is in qualitative agreement with ours, as both are centered on the string-breaking region with comparable strength. 
The results obtained in this way for conventional quarkonium states align with ours: 
both show modest string-breaking corrections of a few MeV for the lowest states and progressively larger threshold admixtures for higher excitations. 
For example, in Ref.~\cite{Bruschini:2021sjh} the $4S$ bottomonium state with $J^{PC}=1^{--}$ gets about $27\%$ threshold components, while the $3P$ bottomonium states with $J^{PC}=(0,1,2)^{++}$ get admixtures between 3\% and 8\%, consistent with our results.
The agreement stems from two main factors: the relative insensitivity of string-breaking corrections to the short-distance parametrization of the tetraquark potentials, and the similarity in magnitude and shape of the adopted mixing potential. 
Crucially, while this simplified treatment suffices for conventional quarkonium states (where tetraquark admixtures are small), a more accurate description of tetraquark static potentials becomes essential for exotic states like the $\chi_{c1}(3872)$.
Regarding Ref.~\cite{Zhang:2025bex}, the inclusion of $\mathcal{O}(1/m_Q^2)$ terms substantially improves the description of the charmonium spectrum, particularly the experimental spin-splittings of the low-lying multiplets. 
However, a fully consistent $1/m_Q$ expansion requires including higher-order terms in both threshold and mixing potentials, which have not yet been computed in lattice QCD.

In a second set of studies that we consider~\cite{Bicudo:2019ymo,Bicudo:2020qhp,Bicudo:2022ihz}, the authors compute bottomonium spectra including string-breaking corrections from the two lowest spin-averaged thresholds, $B\bar{B}^{\rm spin\,avg.}$ and $B_s\bar{B}_s^{\rm spin\,avg.}$, but they neglect threshold spin splittings in the coupled equations.
The quarkonium and mixing static potentials are parametrized using lattice data from~\cite{Bali:2005fu}. 
Threshold potentials are taken as constants fixed at the spin-averaged meson–antimeson masses, thus omitting the short-range octet repulsion characteristic of the tetraquark potentials. 
These studies also disregard the off-diagonal potential matrix terms that are linear combinations of the tetraquark static potentials.
The quarkonium-threshold mixing coefficients given in Eq.~(12) of Ref.~\cite{Bicudo:2022ihz} agree with those derived in Section~\ref{subsec:3P0}.
States below threshold exhibit a decreasing quarkonium component as they approach the threshold: for $\Upsilon(1S)$, $\Upsilon(2S)$ and $\Upsilon(3S)$ it is about 85\%, for $\Upsilon(4S)$ it is about 70\%, and similarly for the $P$- and $D$-wave bottomonium states~\cite{Bicudo:2022ihz}.  
The reason for the threshold effects being larger than in our study, even for the lowest bottomonium states, may be attributed to the different mixing potential employed in these studies with respect to ours, given in Eq.~\eqref{eq:Vmix}. 
The mixing potential in~\cite{Bicudo:2019ymo,Bicudo:2020qhp,Bicudo:2022ihz} is based on the lattice data of Ref.~\cite{Bali:2005fu}.
The lattice data of~\cite{Bali:2005fu} has been obtained from a 2 flavor simulation on a 24$^3\times$ 40 lattice at a fixed lattice spacing of 0.083~fm and with a pion mass of about 654~MeV. 
It shows a mixing potential that peaks around 0.3 fm, well before the string-breaking region identified in~\cite{Bulava:2024jpj}, and is considerably stronger than the mixing potential \eqref{eq:Vmix}.\footnote{
The mixing potential in~\cite{Bali:2005fu} and in the related works \cite{Bicudo:2019ymo,Bicudo:2020qhp,Bicudo:2022ihz,TarrusCastella:2022rxb} is negative, 
while the mixing potential used in this work, Eq.~\eqref{eq:Vmix}, is positive.
This is due to a different convention on the phase of the diabatic quarkonium state: the one adopted here and the one in~\cite{Bali:2005fu} differ by a factor $-1$.
}
There seems to be no obvious justification in QCD for such a behavior, which may be a lattice artifact due to the lattice simulation being far away from the physical point.  
If this is really the case may be proved or disproved only by future LQCD studies.
It is worth remarking that Refs.~\cite{Bicudo:2019ymo,Bicudo:2020qhp,Bicudo:2022ihz} employ the emergent spherical wave method, which enables also the investigation of bottomonium states above threshold, such as the $\Upsilon(10753)$, $\Upsilon(10860)$, and $\Upsilon(11020)$.

Finally, we comment briefly on the study of Ref.~\cite{TarrusCastella:2022rxb}. 
This study uses the BOEFT framework to compute string-breaking corrections induced by the spin-averaged thresholds $M\bar{M}^{\rm spin\,avg.}$ and $M_s\bar{M}_s^{\rm spin\,avg.}$ as self-energy corrections to the quarkonium propagator, both in the charmonium and bottomonium sectors — a calculation similar to the one in Section~\ref{sec:self-en}.
As in the works discussed above, the threshold potentials are modeled by constant lines fixed at the spin-averaged meson–antimeson masses, while the quarkonium and mixing potentials are parametrized using lattice data of~\cite{Bali:2005fu}. 
The resulting string-breaking corrections are significantly larger than ours (even when neglecting the $M_s\bar{M}_s^{\rm spin\,avg.}$ threshold that we do not consider), 
ranging from 70~MeV to 20~MeV for states below threshold.  
Moreover, they are larger for low-lying states than for high-lying states. 
This feature contradicting common physical intuition may be a consequence of using in~\cite{TarrusCastella:2022rxb} the lattice data of~\cite{Bali:2005fu}, about which we commented above.

\subsection{Comparison with the experimental spectrum}\label{subsec:spectrum}
\begin{figure}[ht]
\centering 
\includegraphics*[width=11.0cm,clip=true]{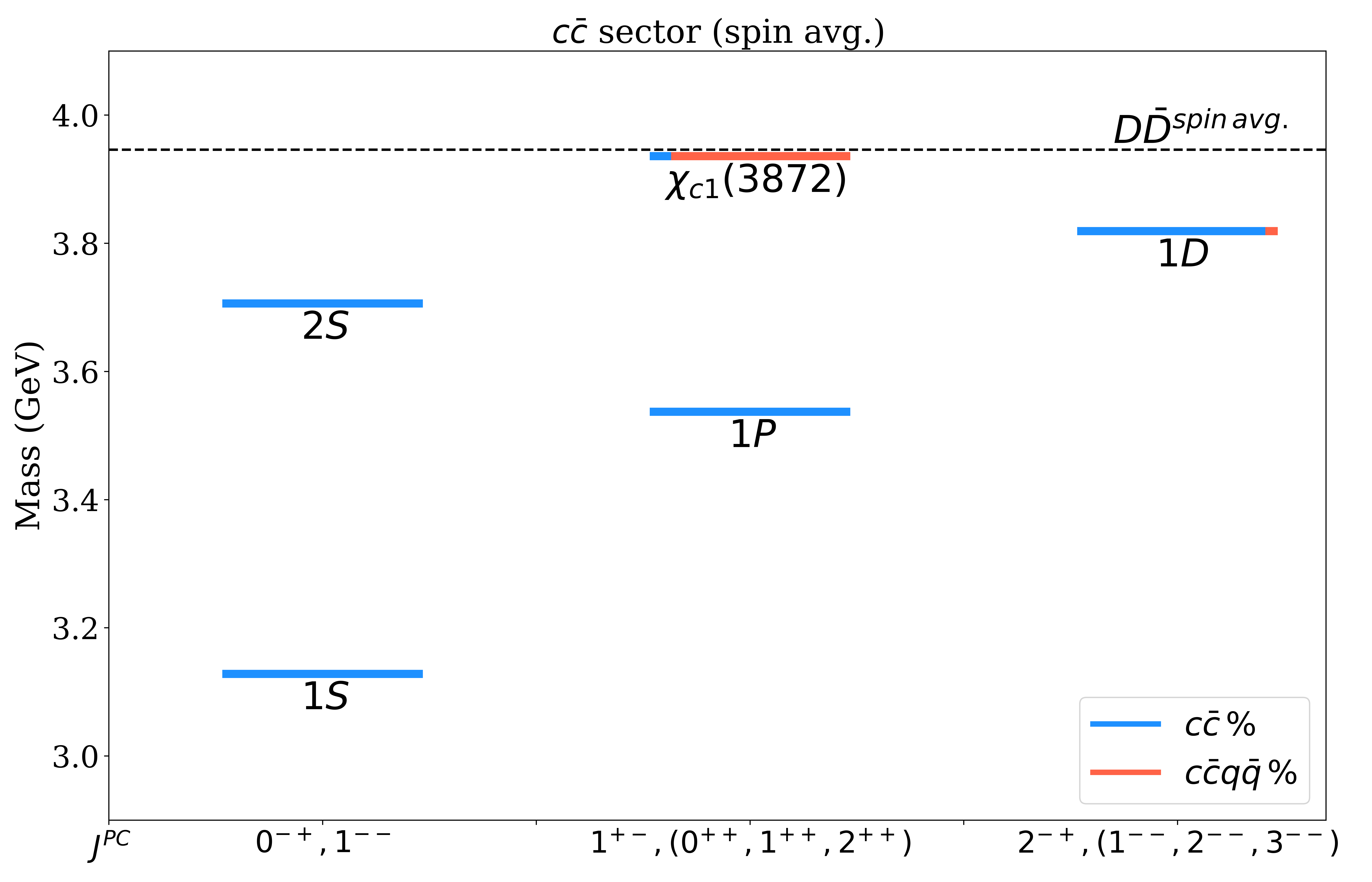}
\caption{Spin-averaged spectrum in the charmonium sector below the $D\bar{D}^{\rm spin\,avg.}$ threshold following from Eq.~\eqref{eq:coupledSchr}. 
States are labeled by their $nl$ quantum numbers.
The horizontal segments representing the states are color-coded according to their composition, with blue lines indicating pure quarkonium states, red lines indicating pure tetraquark states, and partially blue/red lines indicating a mixture of quarkonium and tetraquark components.
}
\label{fig:charm}
\end{figure}

\begin{figure}[ht]
\centering 
\includegraphics*[width=11.0cm,clip=true]{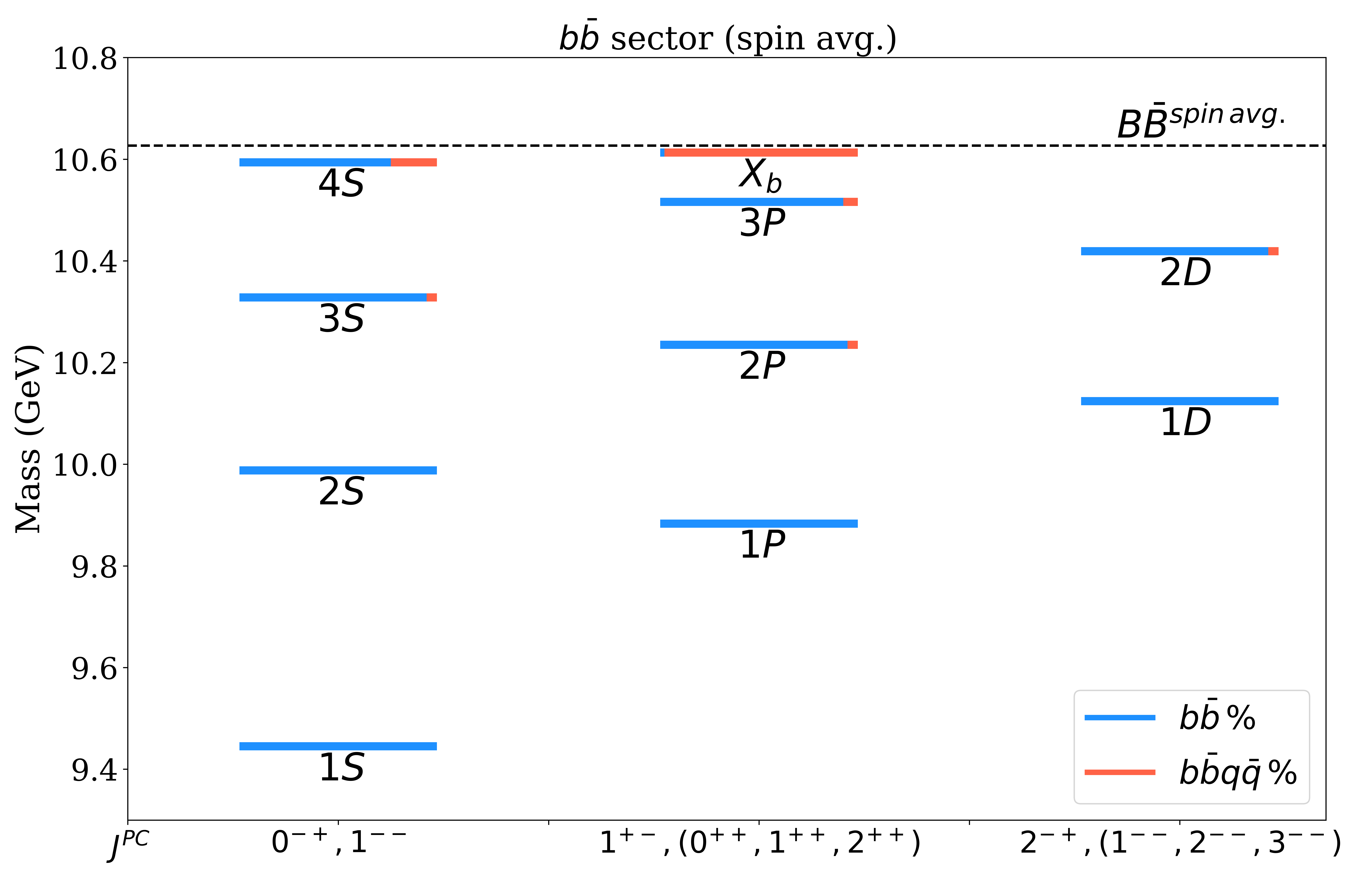}
\caption{Spin-averaged spectrum in the bottomonium sector below the $B\bar{B}^{\rm spin\,avg.}$ threshold following from  Eq.~\eqref{eq:coupledSchr}. 
States are represented as in Fig.~\ref{fig:charm}.
}
\label{fig:bottom}
\end{figure}

\begin{figure}[ht]
\centering 
\includegraphics*[width=11.0cm,clip=true]{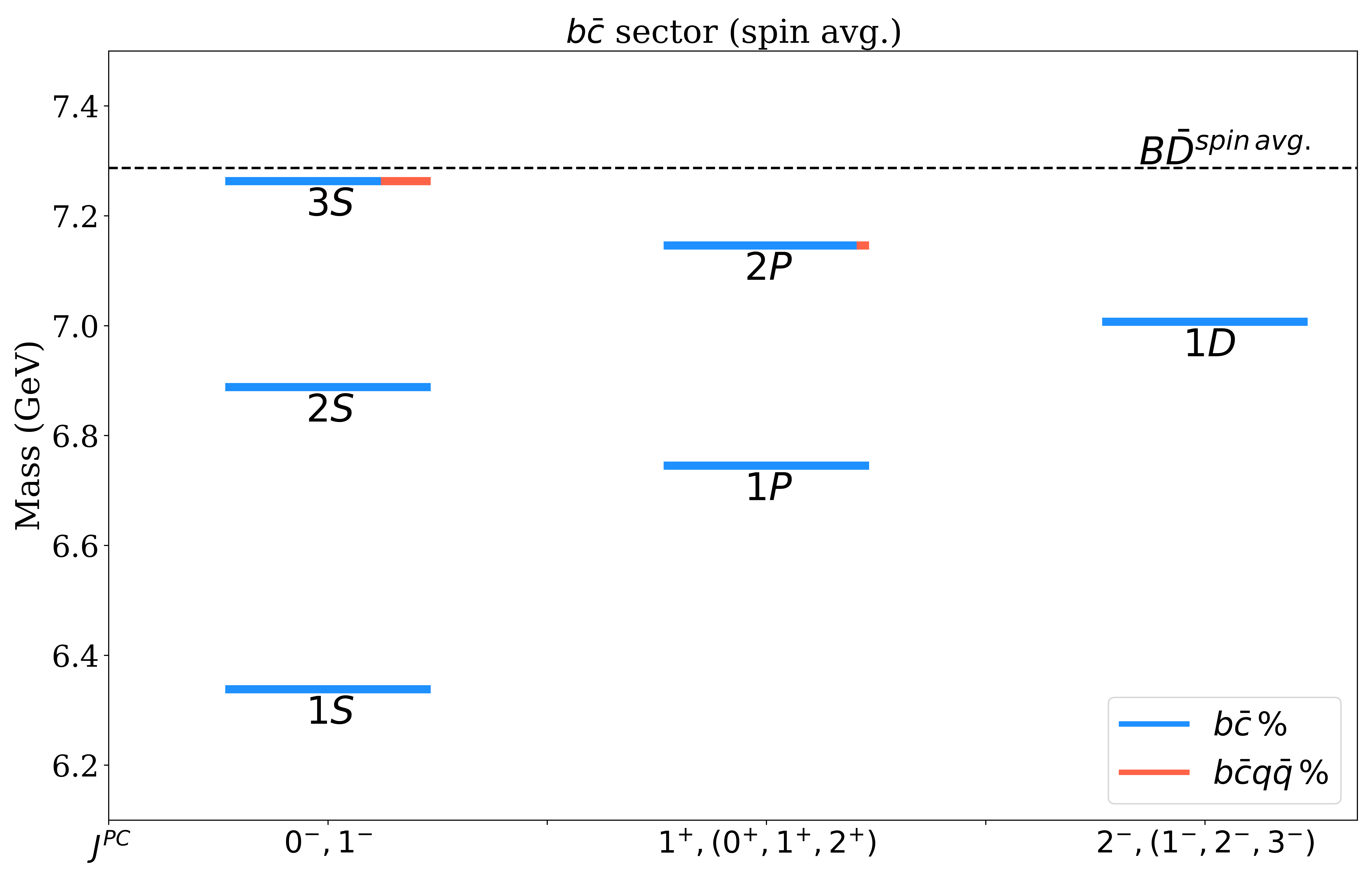}
\caption{Spin-averaged spectrum in the $b \bar{c}$ sector below the $B\bar{D}^{\rm spin\,avg.}$ threshold following from Eq.~\eqref{eq:coupledSchr}. 
States are represented as in Fig.~\ref{fig:charm}.
}
\label{fig:bc}
\end{figure}

The comparison between our calculated spin-averaged quarkonium masses and the corresponding values measured in experiments can be found in the Tables~\ref{tab:coupled_spectra_cc} and~\ref{tab:coupled_spectra_bb} of Section~\ref{sec:coupled}, and,  
after including $\mathcal{O}(1/m_Q)$ tetraquark spin-dependent potentials, in the Tables~\ref{tab:JPC=0-+}--\ref{tab:JPC=2++} of Appendix~\ref{app:tetraquark spin splitting}.  
Given the modest size of the tetraquark spin-splitting corrections, which are in between 1~MeV and 5~MeV, 
in the following, we restrict our discussion to the spin-averaged spectra.
Figures~\ref{fig:charm},~\ref{fig:bottom} and~\ref{fig:bc} show the mass spectra below the corresponding $M_1 \bar{M}_2^{\rm spin \, avg.}$ thresholds
in the charmonium, bottomonium and $b \bar{c}$/$c \bar{b}$  sectors, respectively,  
predicted by solving Eq.~\eqref{eq:coupledSchr} in the different cases.

In the charmonium sector, the two lowest $S$-wave states are identified with the spin-averaged $\eta_c(1S)$ and $J/\psi$ states, and with the spin-averaged $\eta_c(2S)$ and $\psi(2S)$ states.
The lowest $P$-wave state is the spin-average of the $h_c(1P)$, $\chi_{c0}(1P)$, $\chi_{c1}(1P)$ and $\chi_{c2}(1P)$ states.
The next-to-lowest state above the $1P$ state is the spin-averaged $\chi_{c1}(3872)$ multiplet. 
The $\chi_{c1}(3872)$ itself is the $J^{PC} = 1^{++}$ member of the multiplet~\cite{Brambilla:2024imu}, 
while the $J^{PC} = 2^{++}$ partner may be possibly identified with the $R_2$ structure observed by Belle at 4014~MeV~\cite{Belle:2021nuv} or with the $\chi_{c2}(3930)$.
Going further up in the masses, we expect a predominantly quarkonium $2P$ spin-averaged state located a few tens of $\mathrm{MeV}$ above the spin-averaged $D\bar{D}$ threshold. 
This multiplet should contain the two states $h_c(4000)$ and $\chi_{c1}(4010)$ recently observed at LHCb~\cite{LHCb:2024vfz}.
The study of these multiplets, incorporating also the mixing with the $c \bar{c} s \bar{s}$ tetraquark states, is left for future work. 
Finally, the lowest $D$-wave state is identified with the spin-average of the $\psi(3770)$, $\psi_2(3823)$ and $\psi_3(3842)$ states, which provide the complete set of spin triplet states in the $D$-wave multiplet according to~\cite{Brambilla:2019esw, ParticleDataGroup:2024cfk}. 

In the bottomonium sector, the four lowest $S$-wave states are identified with the spin-averaged $\eta_b(1S)$ and $\Upsilon(1S)$ states, the spin-averaged $\eta_b(2S)$ and $\Upsilon(2S)$ states, and so on.
Of particular significance is the $\Upsilon(4S)$ state, positioned approximately 30~MeV below the $B\bar{B}^{\rm spin\,avg.}$ threshold. 
It develops a substantial tetraquark component (up to 35\% for the $J^{PC} = 0^{-+}$ state once the tetraquark spin-splitting has been taken into account, see Table~\ref{tab:JPC=0-+}), whose existence may be supported experimentally by the observation of the HQSS violating decay mode $\Upsilon(4S)\rightarrow h_b(1P) \eta$ ~\cite{ParticleDataGroup:2024cfk}.
The $\Upsilon(10860)$, $\Upsilon(11020)$, and the likely exotic 
$\Upsilon(10753)$ state reported by Belle~\cite{Belle:2019cbt} lie above the spin-averaged threshold and thus fall outside the scope of our current analysis. 
In the $P$-wave sector, we identify the three lowest states with the spin-averaged $h_b(1P)$, $\chi_{b0}(1P)$, $\chi_{b1}(1P)$ and $\chi_{b2}(1P)$ states, the spin-averaged $h_b(2P)$, $\chi_{b0}(2P)$, $\chi_{b1}(2P)$ and $\chi_{b2}(2P)$ states,   and the spin-averaged multiplet containing the states $\chi_{b1}(3P)$ and $\chi_{b2}(3P)$. 
This last multiplet consists also of states with a significant tetraquark component (up to 10\% for the $J^{PC} = 0^{++}$ state once the tetraquark spin-splitting has been taken into account). 
The $X_b$ state emerges as the bottomonium analog to the $\chi_{c1}(3872)$~\cite{Brambilla:2024imu}.
Finally, the $\Upsilon_2(1D)$ state is identified with the $J^{PC} = 2^{--}$ state of the multiplet associated with the lowest spin-averaged $D$-wave state.

In the $b \bar{c}$/$c \bar{b}$ sector, we identify the two lowest $S$-wave states with the spin-averaged multiplets containing the states $B_c^+$ and $B_c(2S)^\pm$. 
Moreover, we predict a spin-averaged $3S$ state lying about 20~MeV below the $B\bar {D}^{\rm spin \, avg.}$ threshold with a substantial tetraquark component of approximately 20\%.
Finally, the recent observation of the two lowest $B_c(1P)^+$ states at respectively 6704.8~MeV and 6752.4~MeV by the LHCb collaboration~\cite{LHCb:2025ubr,LHCb:2025uce} aligns well, within a few tens of MeV, with our prediction for the mass of the spin-averaged $1P$ state.

All our leading-order predictions agree with the experimental findings within 60~MeV. 
Moreover, the discrepancy between the predictions and the data is consistent with the size of the neglected higher-order corrections to the potentials as estimated for quarkonium by several potential models~\cite{Falkensteiner:1984su,Barchielli:1988zp, Eichten:1980mw,Buchmuller:1980su,Falkensteiner:1984su,Eichten:2005ga,Laschka:2012cf,Deng:2016ktl,Ni:2025gvx}. 
Regarding the $S$–$D$ mixing, the BOEFT formalism incorporates part of it once we consider the $\mathcal{O}(1/m_Q)$ tetraquark spin-dependent potential $V_{SS}$ given in Eq.~\eqref{eq:splitting} at second order or higher in perturbation theory (see Eq.~\eqref{eq:1--BOEFT}).
The same is true for $P$-$F$ mixing (see Eq. \eqref{eq:2++BOEFT}). 
Tables~\ref{tab:JPC=1--} and~\ref{tab:JPC=2++} show that in our calculations these mixings have only a minor influence on the masses and compositions of the states.

\section{Conclusions}\label{sec:conclusions}
In this work, we revisit the long-standing problem of quantifying the effects of open-flavor thresholds on the quarkonium spectrum — a subject extensively studied over the past decades through various phenomenological models, most notably the ${}^3P_0$ model~\cite{LeYaouanc:1972vsx,LeYaouanc:1977fsz,Ono:1980js,Heikkila:1983wd,Ono:1983rd,Tornqvist:1984xz,Roberts:1992esl,Barnes:2005pb,Pennington:2007xr,Barnes:2007xu,Li:2009ad,Danilkin:2009hr,Ferretti:2012zz,Ferretti:2013faa,Ferretti:2013vua,Ferretti:2014xqa,Ferretti:2018tco,Man:2024mvl,Blossier:2024dhm,Bruschini:2025paj,Hao:2025vmw,Ni:2025gvx,M:2025gnf,Sultan:2025dfe,Ahmad:2025mue,Gao:2025tob,Man:2025vmm,Ahmad:2025hcr,Farina:2020slb,Zhao:2025kno,Chen:2025pvk,Rui:2025olj}.
We reframe the problem within the Born--Oppenheimer effective field theory.
In the BOEFT, open-flavor threshold effects originate from the mixing between the static quarkonium potential and the lowest static tetraquark potential with the same Born--Oppenheimer quantum numbers.
The two potentials cross at a distance of about 1.2~fm.
The ensuing avoided level crossing mechanism generates a static eigenenergy, or adiabatic potential, that embeds at short distances the attractive Coulomb interaction and shows at large distances a flat constant behaviour.
The latter accounts for the string breaking, i.e. the breaking of quarkonium into the lowest open-flavor threshold allowed by the conservation of the Born--Oppenheimer quantum numbers.
We compute open-flavor threshold effects, first, by taking the mixing at leading order, i.e., without spin effects.
In this situation, the quarkonium potential breaks into a spin-averaged threshold.
Then, we include $1/m_Q$ spin effects.
The inclusion of spin effects splits the threshold into its spin components.

The lowest static tetraquark potentials, $V_{\Sigma_g^{+\prime}}$, $V_{\Pi_g}$ and $V_{\Sigma_u^-}$, exhibit a repulsive color-octet behavior at short distances and approach the corresponding spin-averaged meson–antimeson threshold asymptotically at large distances~\cite{Berwein:2024ztx, Braaten:2024tbm, Brambilla:2024imu}.
The $V_{\Sigma_g^{+\prime}}$ potential mixes with $V_{\Pi_g}$ at small distances, due to the restoration of spherical symmetry, and with the quarkonium potential $V_{\Sigma_g^{+}}$ at a distance of about 1.2~fm through the avoided level crossing mechanism.
The static Born--Oppenheimer potentials are constrained by symmetries and available lattice QCD data in the string-breaking region~\cite{Bulava:2024jpj}. 
They are the only dynamical inputs in the leading-order BOEFT Lagrangian, which at this order is spin-independent.  
The potentials $V_{\Sigma_g^{+\prime}}$ and $V_{\Pi_g}$ depend on a single free parameter, the adjoint meson mass $\Lambda_{1^{--}}$, and the potential $V_{\Sigma_u^-}$ depends on the adjoint meson mass $\Lambda_{0^{-+}}$.
The adjoint meson masses relevant for this study have not been computed yet in lattice QCD in the isospin $I=0$ sector, while a recent lattice study with high statistics for the lowest adjoint mesons in the isospin $I=1$ sector can be found in~\cite{Sharma:2026htv}.
By carefully tuning $\Lambda_{1^{--}}$ and fixing $\Lambda_{0^{-+}}$ on the difference of the two adjoint meson masses computed in~\cite{Foster:1998wu}, we obtain a $\chi_{c1}(3872)$ state that is approximately $100\,\mathrm{keV}$ below the spin-averaged $D\bar{D}$ threshold, and henceforth consistent with experimental observations.  
The state has a large radius of about $4$~fm due to its proximity to the threshold and is composed of roughly $90\%$ tetraquark, with a small quarkonium component that is, however, crucial in explaining its production and decay properties~\cite{Brambilla:2024imu,Lai:2025tpw,Brambilla:2026ujo}.
In the bottomonium sector, we find an analog state, $X_b$, with about $98\%$ tetraquark content, located roughly $1\,\mathrm{MeV}$ below the corresponding threshold \footnote{Detailed discussion on the sensitiveness of the $\chi_{c1}\left(3872\right)$ and bottom counterpart $X_b$ binding energy, radius and composition to the adjoint meson mass $\Lambda_{1^{--}}$ is studied in  Ref.~\cite{Brambilla:2026jqz}.}. 
In our BOEFT framework, the $\chi_{c1}(3872)$ turns out to be a distinct state from the spin-averaged $2P$ charmonium state, 
which lies a few tens of $\mathrm{MeV}$ above the $D\bar{D}^{\rm spin\, avg.}$ threshold and has only a minor tetraquark admixture. 
This interpretation is supported by the recent LHCb observations of the $h_c(4000)$ and $\chi_{c1}(4010)$ states~\cite{LHCb:2024vfz}.
We identify the lowest quarkonium states with the observed states in the charmonium, bottomonium and $b\bar{c}$/$c\bar{b}$ sectors. 
For this last sector, in particular, in addition to the $B_c^+$ and $B_c(2S)^\pm$ states, 
we reproduce correctly the recently observed $B_c(1P)^+$ state~\cite{LHCb:2025ubr,LHCb:2025uce}.  
Overall, our results match the experimental spectrum within $60\,\mathrm{MeV}$, despite omitting higher-order quarkonium and tetraquark spin-dependent and relativistic corrections.

In contrast to our previous work~\cite{Brambilla:2024imu}, which employs a two-pion exchange parametrization for the static tetraquark potentials at large distances, we adopt here a one-pion exchange parameterization. 
This choice is motivated by the fact that the one-pion contribution appears to contribute at large distances, even in the spin-averaged case. 
The change in parametrization induces a shift of about 90~MeV in the adjoint meson mass, while leaving all properties of the $\chi_{c1}(3872)$ state — such as its composition, mass, and radius — nearly identical. 
As a consequence of the shift in the adjoint meson mass, the mass of the $X_b$ state increases by about 10~MeV, 
whereas the quarkonium masses remain largely unaffected in comparison to~\cite{Brambilla:2024imu}.

We quantify open-flavor threshold effects, also called string-breaking corrections, on the quarkonium spectrum.
Threshold effects are induced by the mixing between the quarkonium and tetraquark static potentials. 
We compute them by means of two complementary approaches: 
by solving coupled Schr\"{o}dinger equations involving both quarkonium and tetraquark channels or by computing the self-energy correction to the quarkonium propagator induced by the quarkonium-tetraquark mixing.
In both cases, we find that the mixing shifts the quarkonium energy levels downwards by an amount that is between 1~MeV and 15~MeV, with the effect being negligible for states far from the open-flavor thresholds and increasingly significant for those closer to it. 
Notably, the $4S$ and $3P$ spin-averaged bottomonium states and the $3S$ $b \bar{c}$/$c \bar{b}$ spin-averaged state exhibit sizable string-breaking corrections with a sizable tetraquark admixture of approximately $20 \%$, $5 \%$, $20 \%$, respectively. 
This composition of the states may be instrumental in explaining their decay and production patterns.
The magnitude of the string-breaking corrections shows only weak sensitivity to the short-distance parametrization of the tetraquark potentials.

Furthermore, we derive for the first time within the BOEFT the coupled Schr\"odinger equations with the $\mathcal{O}(1/m_Q)$ tetraquark spin-dependent potential included. 
We quantify the impact of the spin-dependent potential on the quarkonium spectrum by using an approximate potential tuned to reproduce the experimental splittings of the lowest meson-antimeson thresholds.
In this way, the spin symmetry is explicitly broken in the tetraquark sector and indirectly lifted in the quarkonium sector through the quarkonium-tetraquark mixing.
We retune the adjoint meson mass $\Lambda_{1^{--}}$ to obtain a $\chi_{c1}(3872)$ state approximately 100~keV below the physical $D\bar{D}^*$ threshold, 
finding a difference of about $50\,\mathrm{MeV}$ with the value of the adjoint meson mass found in the spin-averaged case.
The spin-splitting pattern predicted by our coupled equations for the $\chi_{c1}(3872)$ multiplet agrees with earlier estimates based on first-order perturbation theory and lattice QCD calculations of the lowest hybrid multiplets~\cite{Brambilla:2024imu}. 
We tentatively associate the $2^{++}$ state of the multiplet with either the $R_2$ structure reported by the Belle experiment~\cite{Belle:2021nuv} or the $\chi_{c2}(3930)$, with the latter assignment being preferred in view of the expected structure of the $\chi_{c1}(3872)$ and $2P$ charmonium multiplets.
For the states in the $X_b$ multiplet, we find threshold spin-splitting corrections roughly 20~MeV larger than those predicted from perturbation theory, although both approaches yield the same mass hierarchy. 
All the $X_b$ multiplet states are predominantly tetraquarks, with quarkonium admixtures from 2\% to 10\%. Moreover, both the $1^{+-}$ and $0^{++}$ states of the $X_b$ multiplet exhibit a substantial $\Sigma_u^-$ admixture, highlighting the necessity of a precise determination of the $V_{\Sigma_u^-}$ potential, which, like $V_{\Pi_g}$ has not been computed yet on the lattice.

Threshold spin-splitting corrections to the lowest quarkonium states are at most 5~MeV, while we observe negligible $S$-$D$ mixing effects for the $J^{PC}=1^{--}$ states. 
Including threshold spin-splittings increases the tetraquark admixture of quarkonium states closest to the spin-averaged threshold. 
For example, the $4S$ bottomonium state with $J^{PC}=0^{-+}$ ($s_{Q\bar{Q}}=0$) gets about $37\%$ tetraquark content (compared to $20\%$ at leading order) and the $3P$ state with $J^{PC}=0^{++}$ ($s_{Q\bar{Q}}=1$) gets about $11 \%$ (compared to $5\%$ at leading order).
Spin-dependent potentials have been computed in lattice QCD in the quarkonium sector, but not in the tetraquark and mixing sectors. 
This prevents at present a fully model independent assessment of the spin-dependent corrections. 

We compare the expression of the real part of the self-energy derived from the leading-order BOEFT Lagrangian with the one from the ${}^3P_0$ model~\cite{Ono:1980js}. 
In the chiral and heavy quark limits, the two self-energy expressions have the same form, allowing the identification of the corresponding ${}^3P_0$ mixing potential.
Assuming Gaussian meson wavefunctions, the so obtained ${}^3P_0$ potential reproduces the asymptotic short- and long-distance behavior of the BOEFT mixing potential adopted here. 
However, its peak lies before the string-breaking region, and its maximum is roughly three times larger than that of the BOEFT potential.
String-breaking corrections reported in the literature typically range from 10~MeV to 100~MeV~\cite{Pennington:2007xr,Barnes:2007xu}, which is one to two orders of magnitude larger than our BOEFT estimates. 
The discrepancy stems mainly from the strong sensitivity of these corrections to the expression of the mixing potential, 
which in the ${}^3P_0$ model is generally stronger and of a different shape than in the BOEFT case. 

Finally, we compare our results with existing studies based on the Born--Oppenheimer approximation. 
Although these studies typically employ some simplifying assumptions — such as modeling the tetraquark potentials as constant lines (thereby neglecting their short-range repulsive behavior) — and/or do not incorporate the most recent lattice QCD data for the mixing potential, they yield a description of the lowest quarkonium states that is quite consistent with ours~\cite{Bruschini:2020voj,Bruschini:2021sjh,Zhang:2025bex,Bicudo:2019ymo,Bicudo:2020qhp,Bicudo:2022ihz,TarrusCastella:2022rxb}.
The magnitude of string-breaking corrections and the resulting tetraquark admixtures in quarkonium states are highly sensitive to the specific form of the mixing potential. 
In cases where differences arise~\cite{Bicudo:2019ymo,Bicudo:2020qhp,Bicudo:2022ihz,TarrusCastella:2022rxb}, the studies in the literature predict larger string-breaking corrections and higher tetraquark percentages, a discrepancy that can be traced back to the different lattice data adopted to constrain the mixing potential.

We believe that a rigorous treatment of open-flavor threshold effects in the quarkonium spectrum requires a framework that systematically implements the non-relativistic expansion and provides an unambiguous field-theoretical definition of the potentials governing the systems. 
In this respect, the BOEFT is well-suited to the task. 
Its input potentials are defined in terms of generalized Wilson loops, which are computable and (partially) computed in lattice QCD. 
The same formalism can be extended to describe other states containing a heavy quark–antiquark pair plus light degrees of freedom, such as hybrids, pentaquarks, etc. 
Also in these cases, further progress will depend crucially on the availability of resources to compute the relevant potentials in lattice QCD.

\section*{Acknowledgements}
We acknowledge the DFG cluster of excellence ORIGINS funded by the Deutsche Forschungsgemeinschaft under Germany’s Excellence Strategy-EXC-2094-390783311. 
N.B. acknowledges the ERC Advanced Grant ERC-2023-ADG-Project EFT-XYZ. The work of A.M at the Physical Research Laboratory, Ahmedabad, is supported by the Department of Space (DOS), Government of India.
T.S. thanks Julian Mayer-Steudte, Roberto Mussa, Merlin Reichard, and Jaume Tarr\`us Castell\`a for helpful discussions.

\appendix

\section{$^3P_0$ model: derivation of Eq.~\eqref{eq:3P0_real1}}\label{app:comparison_3P0_BOEFT}
In this appendix, we provide some additional details on the relation between Eq.~\eqref{eq:3P0_real1} given in Section~\ref{subsec:3P0} and the $^3P_0$ model formula for the strong decay width. 
According to the $^3P_0$ model, the strong decay width of a quarkonium state with definite $J^{PC}$, decaying at rest into an $S$-wave heavy-light meson–antimeson pair, is given by~\cite{Ono:1980js,Bruschini:2025paj}:
\begin{align}\label{eq:3P0_immaginary}
&\Gamma_{^3P_0}(Q \bar{Q}_{nl_{Q \bar{Q}}} \rightarrow M^{(*)} \bar{M}^{(*)}) = 2 \pi \frac{E_{M^{(*)}} E_{\bar{M}^{(*)}}}{E_{nl_{Q \bar{Q}}J}} k^{nJ}_{M^{(*)} \bar{M}^{(*)} } \, \gamma^2   \nonumber\\ 
&\hspace{1cm}\times \Big| \Big\langle I_{M^{(*)}}, I_{3M^{(*)}}, I_{\bar{M}^{(*)}}, I_{3\bar{M}^{(*)}} \Big| I_{Q \bar{Q}}, I_{3Q \bar{Q}} \Big\rangle \Big|^2
 \left[ \begin{array}{ccc}
I_Q & I_{\bar{Q}} & I_{Q \bar{Q}} \\
I_{\bar{q}} & I_q & I_{q \bar{q}} \\
I_{M^{(*)}} & I_{\bar{M}^{(*)}} & I_{M^{(*)} \bar{M}^{(*)}}
\end{array} \right]^2 \left[ \begin{array}{ccc}
s_Q & s_{\bar{Q}} & s_{Q \bar{Q}} \\
s_{\bar{q}} & s_q & s_{q \bar{q}} \\
s_{M^{(*)}} & s_{\bar{M}^{(*)}} & s_{M^{(*)}\bar{M}^{(*)}}
\end{array} \right]^2 \nonumber\\
&
\hspace{1cm}\times \left(
\left[ \begin{array}{ccc}
l_{Q \bar{Q}} & s_{Q\bar{Q}} & J \\
l_{q \bar{q}} & s_{q \bar{q}} & K \\
l_{Q \bar{Q} + 1} & s_{M^{(*)}\bar{M}^{(*)}} & J
\end{array} \right]^2 
\left|  \frac{i^{l_{Q \bar{Q}}}}{\sqrt{8}} \sqrt{\frac{3(l_{Q \bar{Q}} + 1)}{2 l_{Q \bar{Q}} + 3}}   \mathcal{I}_{l_{Q \bar{Q} + 1},nJ}( k^{nJ}_{M^{(*)} \bar{M}^{(*)} } r)\right|^2 \right.\nonumber \\ 
&
\left.\hspace{1.5cm}
+ \left[ \begin{array}{ccc}
l_{Q \bar{Q}} & s_{Q\bar{Q}} & J \\
l_{q \bar{q}} & s_{q \bar{q}} & K \\
l_{Q \bar{Q} - 1} & s_{M^{(*)}\bar{M}^{(*)}} & J
\end{array} \right]^2  \left| \frac{i^{l_{Q \bar{Q}}}}{\sqrt{8}} \sqrt{\frac{3l_{Q \bar{Q}} }{2 l_{Q \bar{Q}} - 1}} \mathcal{I}_{l_{Q \bar{Q} - 1},nJ}( k^{nJ}_{M^{(*)} \bar{M}^{(*)} } r)\right|^2 \right),
\end{align}
where $\mathcal{I}_{l_{Q \bar{Q} - 1},nJ}( k^{nJ}_{M^{(*)} \bar{M}^{(*)} } r)$ and $\mathcal{I}_{l_{Q \bar{Q} + 1,nJ}}(k^{nJ}_{M^{(*)} \bar{M}^{(*)} } r)$ are two overlap integrals defined below. 
The isospin and spin operators satisfy $\bm{I}_Q + \bm{I}_{\bar{Q}} = \bm{I}_{Q \bar{Q}}$, $\bm{I}_q + \bm{I}_{\bar{q}} = \bm{I}_{q \bar{q}}$, $\bm{I}_Q + \bm{I}_{\bar{q}} = \bm{I}_{M^{(*)}}$, $\bm{I}_{\bar{Q}} + \bm{I}_q = \bm{I}_{\bar{M}^{(*)}}$, $\bm{I}_{Q \bar{Q}} + \bm{I}_{q \bar{q}} = \bm{I}_{M^{(*)}} + \bm{I}_{\bar{M}^{(*)}}$\footnote{
In Eqs.~\eqref{eq:tetr_wvf1}, \eqref{eq:tetrwf-1} and \eqref{eq:tetrwf-2}, which define the quarkonium and tetraquark wavefunctions, the isospin part of the wavefunctions is not explicitly written since there we are just dealing with isospin-0 tetraquark states.}
and $\bm{S}_Q + \bm{S}_{\bar{Q}} = \bm{S}_{Q \bar{Q}}$, $\bm{S}_q + \bm{S}_{\bar{q}} = \bm{S}_{q \bar{q}}$, $\bm{S}_Q + \bm{S}_{\bar{q}} = \bm{S}_{M^{(*)}}$, $\bm{S}_{\bar{Q}} + \bm{S}_q = \bm{S}_{\bar{M}^{(*)}}$, $\bm{S}_{Q \bar{Q}} + \bm{S}_{q \bar{q}} = \bm{S}_{M^{(*)}\bar{M}^{(*)}} = \bm{S}_{M^{(*)}} + \bm{S}_{\bar{M}^{(*)}}$. 
The LDF operators satisfy $\bm{L}_{q \bar{q}} + \bm{S}_{q \bar{q}} = \bm{K}$.
Since we take the $u$ and $d$ masses as degenerate, and compute the decay into the sum of the $M^{(*)\,0} \bar{M}^{(*)\,0}$ and $M^{(*)\,+}M^{(*)\,-}$ thresholds, which we call the isospin-averaged thresholds $M^{(*)} \bar{M}^{(*)}$, the parameter $\gamma$ quantifies in Eq.~\eqref{eq:3P0_immaginary} the mixing between a quarkonium state and the isospin-averaged threshold. 
The difference between $\gamma$ referring to a specific isospin threshold and the isospin-averaged threshold is a factor $\sqrt{2}$.
The relative momentum between the meson-antimeson pair is given by $k^{nJ}_{M^{(*)} \bar{M}^{(*)}} = \sqrt{2 \mu_{M^{(*)}\bar{M}^{(*)}} (E_{n l_{Q \bar{Q}}J} - m_{M^{(*)}} - m_{\bar{M}^{(*)}}})$ with $\mu_{M^{(*)}\bar{M}^{(*)}}$ the meson-antimeson reduced mass, up to higher order relativistic corrections.  
The meson mass can be expressed in the heavy-quark expansion as $m_{M^{(*)}} = m_Q + \Lambda_{\bar{q}} + \mathcal{O}(1/m_Q)$ with $\Lambda_{\bar{q}}$ being the binding energy and light quark mass contribution in the static limit. 
Spin splittings and heavy-quark kinetic terms appear at $\mathcal{O}(1/m_Q)$~\cite{Isgur:1991wq}. 
The quarkonium mass can be written as
$E_{nl_{Q \bar{Q}}J} = m_{M^{(*)}} + m_{\bar{M}^{(*)}} + E_{{\rm bind.},\,nl_{Q \bar{Q}}J}^{M^{(*)} \bar{M}^{(*)}}$, with $E_{{\rm bind.},\,nl_{Q \bar{Q}}J}^{M^{(*)} \bar{M}^{(*)}}$ indicating its binding energy with respect to the $M^{(*)} \bar{M}^{(*)}$ threshold.
The two overlap integrals are~\cite{Ono:1980js,Bruschini:2025paj}:
\begin{align}\label{eq:overlap3}
\mathcal{I}_{l_{Q \bar{Q}} + 1,nJ}(k^{nJ}_{M^{(*)} M^{(*)} }) =  \int dr \, r^2 \, \int& dp \, p^2\,  \psi_{M^{(*)}}^*(p) \, \psi_{\bar{M}^{(*)}}^*(p) \, \psi_{Q \bar{Q}}^{nl_{Q \bar{Q}}J}(r) 
\Big[p \, j_1(pr) \,  j_{l_{Q \bar{Q}} + 1}(h_Qk^{nJ}_{M^{(*)} \bar{M}^{(*)} }r) \nonumber \\
&+ h_q \, k^{nJ}_{M^{(*)} \bar{M}^{(*)} } \, j_0(pr) \, j_{l_{Q \bar{Q}}}(h_Q k^{nJ}_{M^{(*)} \bar{M}^{(*)} }r) \Big],  \\ 
\label{eq:overlap4}
\mathcal{I}_{l_{Q \bar{Q}} - 1,nJ}(k^{nJ}_{M^{(*)} M^{(*)} }) =  \int dr \, r^2 \, \int& dp \, p^2 \,  \psi_{M^{(*)}}^*(p) \, \psi_{\bar{M}^{(*)}}^*(p) \, \psi_{Q \bar{Q}}^{nl_{Q\bar
Q}J}(r) 
\Big[p \, j_1(pr) \, j_{l_{Q \bar{Q}} - 1}(h_Qk^{nJ}_{M^{(*)} \bar{M}^{(*)} }r) \nonumber\\
&- h_q \, k^{nJ}_{M^{(*)} \bar{M}^{(*)} } \, j_0(pr) \, j_{l_{Q \bar{Q}}}(h_Q k^{nJ}_{M^{(*)} \bar{M}^{(*)} }r) \Big].
\end{align}
Here $h_q = m_q/(m_Q + m_q)$ and $h_Q = m_Q/(m_Q + m_q)$, where $m_q$ ($m_{\bar{q}}$) is the mass of the light quark (antiquark) in the heavy-light meson (antimeson) $M$ ($\bar{M}$). 
The function $\psi_{Q \bar{Q}}^{nl_{Q \bar{Q}}J}$ is the radial quarkonium wavefunction, while $j_{l_{Q \bar{Q}} \pm 1}$ are the free radial meson-antimeson wavefunctions expressed in terms of spherical Bessel functions.

To go from Eq.~\eqref{eq:3P0_immaginary} to 
Eq.~\eqref{eq:3P0_real1}, we take the chiral limit ($m_q \rightarrow 0$) and retain the leading order term in the heavy-quark expansion, such that $E_{M^{(*)}} E_{\bar{M}^{(*)}}/E_{nl_{Q \bar{Q}}J} \rightarrow m_Q/2$, $h_Q \rightarrow 1$, $h_q \rightarrow 0$, $k^{nJ}_{M^{(*)} \bar{M}^{(*)}} \rightarrow k^{nJ} = \sqrt{m_Q E_{\mathrm{bind},\,nJ}^{M\bar{M}^{\rm spin\,avg.}}}$, 
where $E_{{\rm bind.}, nJ}^{M \bar{M}^{\rm spin \, avg.}} = E_{nl_{Q \bar{Q}}J} - m_{M^{\rm spin \, avg.}} - m_{\bar{M}^{\rm spin \, avg.}}$ is the binding energy relative to the spin-averaged threshold and $m_M, m_{M^*} \rightarrow m_{M^{\mathrm{\rm spin\,avg.}}}$. 
Interpreting Eq.~\eqref{eq:3P0_immaginary} as the imaginary part of the self-energy contribution to the quarkonium propagator, similarly to Eq.~\eqref{eq:realimb}, the corresponding real part is 
\begin{align}\label{eq:3P0_real}
& {\rm Re}\, \Sigma_{^3P_0}(Q \bar{Q}_{nl_{Q \bar{Q}}J}, M^{(*)} \bar{M}^{(*)}) = {\rm P.V.} \int_0^{+ \infty} dk \, \frac{ m_Q k^2}{(k^{nJ})^2 - k^2} \, \gamma^2 \,  \Big| \Big\langle I_M, I_{3 \, M}, I_{\bar{M}}, I_{3 \,\bar{M}} \Big| I_{Q \bar{Q}}, I_{3 \,Q \bar{Q}} \Big\rangle \Big|^2 \nonumber\\
&\hspace{1cm} \times \frac{\left[ \begin{array}{ccc}
I_Q & I_{\bar{Q}} & I_{Q \bar{Q}} \\
I_{\bar{q}} & I_q & I_{q \bar{q}} \\
I_{M^{(*)}} & I_{\bar{M}^{(*)}} & I_{M^{(*)} \bar{M}^{(*)}}
\end{array} \right]^2}{8}  \Bigg( \biggl(C_{l_{Q \bar{Q}} + 1, J}^{M^{(*)} \bar{M}^{(*)}}\biggr)^2 \mathcal{I}_{l_{Q \bar{Q} + 1},nJ}(k r)^{ 2}  + \biggl(C_{l_{Q \bar{Q}} - 1, J}^{M^{(*)} \bar{M}^{(*)}}\biggr)^2  \mathcal{I}_{l_{Q \bar{Q} - 1},nJ}(kr)^2 \Bigg),
\end{align}
with 
\begin{align}\label{eq:coeff2}
C_{l_{Q \bar{Q} + 1}, J}^{M^{(*)} \bar{M}^{(*)}} = \sqrt{\frac{3(l_{Q \bar{Q}} + 1)}{2 l_{Q \bar{Q}} + 3}} \left[ \begin{array}{ccc}
s_Q & s_{\bar{Q}} & s_{Q \bar{Q}} \\
s_{\bar{q}} & s_q & s_{q \bar{q}} \\
s_{M^{(*)}} & s_{\bar{M}^{(*)}} & s_{M^{(*)}\bar{M}^{(*)}}
\end{array} \right]
&\left[ \begin{array}{ccc}
l_{Q \bar{Q}} & s_{Q\bar{Q}} & J \\
l_{q \bar{q}} & s_{q \bar{q}} & K \\
l_{Q \bar{Q} + 1} & s_{M^{(*)}\bar{M}^{(*)}} & J
\end{array} \right],\\
\label{eq:coeff3}
C_{l_{Q \bar{Q} - 1}, J}^{M^{(*)} \bar{M}^{(*)}} =\sqrt{\frac{3l_{Q \bar{Q}} }{2 l_{Q \bar{Q}} - 1}} \left[ \begin{array}{ccc}
s_Q & s_{\bar{Q}} & s_{Q \bar{Q}} \\
s_{\bar{q}} & s_q & s_{q \bar{q}} \\
s_{M^{(*)}} & s_{\bar{M}^{(*)}} & s_{M^{(*)}\bar{M}^{(*)}}
\end{array} \right]
&\left[ \begin{array}{ccc}
l_{Q \bar{Q}} & s_{Q\bar{Q}} & J \\
l_{q \bar{q}} & s_{q \bar{q}} & K \\
l_{Q \bar{Q} - 1} & s_{M^{(*)}\bar{M}^{(*)}} & J
\end{array} \right], 
\end{align}
and the overlap integrals in the chiral and heavy quark limit given by 
\begin{align}\label{eq:overlap3}
\mathcal{I}_{l_{Q \bar{Q}} + 1,nJ}(k) &= \int dr \, r^2 \, \psi_{\Sigma_g^+}^{nl_{Q \bar{Q}}J}(r) \Biggl( \int dp \, p^3 \, \psi_{M^{\rm spin \, avg.}}^*(p) \, \psi_{\bar{M}^{\rm spin \, avg.}}(p)^* j_1(pr) \Biggr) j_{l_{Q \bar{Q}} + 1}(kr),   \\ 
\label{eq:overlap4}
\mathcal{I}_{l_{Q \bar{Q}} - 1,nJ}(k) &= \int dr \, r^2 \, \psi_{\Sigma_g^+}^{nl_{Q \bar{Q}}J}(r) \Biggl( \int dp \, p^3 \, \psi_{M^{\rm spin \, avg.}}^*(p) \, \psi_{\bar{M}^{\rm spin \, avg.}}^*(p) j_1(pr) \Biggr) j_{l_{Q \bar{Q}} - 1}(kr).   
\end{align}
Equation~\eqref{eq:3P0_real} coincides with Eq.~\eqref{eq:3P0_real1} of  Section~\ref{subsec:3P0}. 

We conclude by computing the isospin factors entering Eqs.~\eqref{eq:3P0_immaginary} and \eqref{eq:3P0_real}, which are the same for the various quarkonium and threshold states analyzed in the main text.
The isospin quantum numbers are $I_Q = I_{\bar{Q}} = I_{Q \bar{Q}} = I_{3Q \bar{Q}} = 0$, $I_q = I_{\bar{q}} = 1/2$, $I_M = I_{\bar{M}} = 1/2$. 
Moreover, in the $^3P_0$ model, $I_{q \bar{q}} = 0$, and from isospin conservation 
$I_{M^{(*)} \bar{M}^{(*)}} = 0$. 
Since $I_{3 \,M} = - I_{3 \, \bar{M}} = \pm 1/2$, we get 
\begin{align}\label{eq:isospin_factor}
\left[ \begin{array}{ccc}
I_Q & I_{\bar{Q}} & I_{Q \bar{Q}} \\
I_{\bar{q}} & I_q & I_{q \bar{q}} \\
I_M & I_{\bar{M}} & I_{M^{(*)} \bar{M}^{(*)}}
\end{array} \right]^2 = 1, \quad\quad\quad \Big| \Big\langle I_M, I_{3 \,M}, I_{\bar{M}}, I_{3 \,\bar{M}} \Big| I_{Q \bar{Q}}, I_{3 \,Q \bar{Q}} \Big\rangle \Big|^2 = \frac{1}{2},
\end{align}
as in~\cite{Bruschini:2025paj}.

\section{$^3P_0$ model vs BOEFT framework: $P$-wave quarkonium case 
}\label{app:comparison_S_P}
In this Appendix, we sum over all the non-vanishing threshold contributions to the real part of the self-energy for the $P$-wave quarkonium multiplet states having $J^{PC} = 1^{+-}$, $(0,1,2)^{++}$. 
This allows us to fix the coefficient $C_2^{M \bar{M}^{\rm spin \, avg.}}$,  relative to the mixing with the $D$-wave threshold, and the coefficient $C_0^{M \bar{M}^{\rm spin \, avg.}}$, relative to the mixing with the $S$-wave threshold, entering Eq.~\eqref{eq:3P0_self_pwav1} in the $^3P_0$ model.\footnote{
According to Eqs.~\eqref{eq:coeff2} and~\eqref{eq:coeff3} the coefficients are labeled by $l_{Q\bar Q}\pm 1$, where $l_{Q\bar Q}$ is the orbital angular momentum of the heavy quark-antiquark pair in the quarkonium state.
For $0^{-+}$ and $1^{--}$ quarkonium states, $l_{Q\bar Q}=0$ and the angular momentum label of the coefficients can only be 1, i.e. the quarkonium couples only with a $P$-wave threshold, 
while for $1^{+-}$ and $(0,1,2)^{++}$ quarkonium states,  $l_{Q\bar Q}=1$ and the angular momentum label of the coefficients can assume the values 0 or 2, i.e. the quarkonium couples with $S$- and $D$-wave thresholds.
}
The real part of the self-energy projected on the different quarkonium states in the chiral and heavy quark limits reads
\begin{align}
&{\rm Re}\, \Sigma_{^3P_0}(Q \bar{Q}_{n11^{+-}}, \frac{M \bar{M}^* + M^* \bar{M}}{\sqrt{2}} + M^* \bar{M}^*) = {\rm P.V.} \int_0^{+\infty} dk \, \frac{m_Q k^2}{(k^{n1})^2 - k^2} \, \frac{\gamma^2}{16} \nonumber\\ 
&\hspace{1.6cm} \times \left[ \Biggl( \left(C_{2,1}^{(M\bar{M}^* - M^* \bar{M})/\sqrt{2}}\right)^2 + \left(C_{2,1}^{M^*\bar{M}^*,s_{M^* \bar{M}^*} = 1}\right)^2 \Biggr) \mathcal{I}_{2,n1}(kr)^2  \right.  \nonumber\\ 
&\hspace{1.8cm} \left. + \Biggl( \left(C_{0,1}^{(M\bar{M}^* - M^* \bar{M})/\sqrt{2}}\right)^2 + \left(C_{0,1}^{M^*\bar{M}^*,s_{M^* \bar{M}^*} = 1}\right)^2 \Biggr) \mathcal{I}_{0,n1}(kr)^2  \right], 
\label{eq:3P0_self_pwav1+-}\\ 
&{\rm Re}\, \Sigma_{^3P_0}(Q \bar{Q}_{n10}, M \bar{M} + M^* \bar{M}^*) = {\rm P.V.} \int_0^{+\infty} dk \, \frac{m_Q k^2}{(k^{n0})^2 - k^2} \, \frac{\gamma^2}{16}  
\nonumber\\ 
&\hspace{1.6cm} \times \left[ \left(C_{2,0}^{M^*\bar{M}^*,s_{M^* \bar{M}^*} = 2}\right)^2  \mathcal{I}_{2,n0}(kr)^2  
+ \Biggl( \left(C_{0,0}^{M\bar{M}}\right)^2 + \left(C_{0,0}^{M^*\bar{M}^*,s_{M^* \bar{M}^*} = 0}\right)^2 \Biggr) \mathcal{I}_{0,n0}(kr)^2  \right], 
\label{eq:3P0_self_pwav0++}\\ 
&{\rm Re}\, \Sigma_{^3P_0}(Q \bar{Q}_{n11^{++}}, \frac{M \bar{M}^* + M^* \bar{M}}{\sqrt{2}} + M^* \bar{M}^*) = {\rm P.V.} \int_0^{+\infty} dk \, \frac{m_Q k^2}{(k^{n1})^2 - k^2} \, \frac{\gamma^2}{16} \nonumber\\
&\hspace{1.6cm} \times \left[ \Biggl(\left(C_{2,1}^{(M\bar{M}^* + M^* \bar{M})/\sqrt{2}}\right)^2+ \left(C_{2,1}^{M^*\bar{M}^*,s_{M^* \bar{M}^*} = 2}\right)^2\Biggr)  \mathcal{I}_{2,n1}(kr)^2  + \left(C_{0,1}^{(M\bar{M}^* + M^* \bar{M})/\sqrt{2}}\right)^2  \mathcal{I}_{0,n1}(kr)^2 \right], \label{eq:3P0_self_pwav1++}\\ 
&{\rm Re}\, \Sigma_{^3P_0}(Q \bar{Q}_{n12}, M\bar{M} + \frac{M \bar{M}^* + M^* \bar{M}}{\sqrt{2}} + M^* \bar{M}^*) = {\rm P.V.} \int_0^{+\infty} dk \, \frac{m_Q k^2}{(k^{n2})^2 - k^2} \, \frac{\gamma^2}{16} \nonumber\\
&\hspace{1.6cm} \times \Bigg[ \Biggl(\left(C_{2,2}^{M\bar{M}}\right)^2+ \left(C_{2,2}^{(M\bar{M}^* + M^* \bar{M})/\sqrt{2}}\right)^2+ \left(C_{2,2}^{M^*\bar{M}^*,s_{M^* \bar{M}^*} = 0}\right)^2 + \left(C_{2,2}^{M^*\bar{M}^*,s_{M^* \bar{M}^*} = 2}\right)^2\Biggr)  \mathcal{I}_{2,n2}(kr)^2  
\nonumber\\ 
&\hspace{1.8cm} +\left(C_{0,2}^{M^*\bar{M}^*,s_{M^* \bar{M}^*} = 2}\right)^2  \mathcal{I}_{0,n2}(kr)^2\Bigg], \label{eq:3P0_self_pwav2++}
\end{align}
where in Eqs.~\eqref{eq:3P0_self_pwav1+-} and \eqref{eq:3P0_self_pwav1++} the PC quantum numbers have been specified to distinguish between $J^{PC} = 1^{+-}$ and $1^{++}$ states. 
The non-vanishing coefficients, whose general expressions are given in Eqs. \eqref{eq:coeff2} and \eqref{eq:coeff3}, take the values
\begin{itemize}
\item[{\it (a)}] $C_{2,1}^{(M\bar{M}^* - M^* \bar{M})/\sqrt{2}} = 1/\sqrt{3}$, 
    $C_{2,1}^{M^*\bar{M}^*,s_{M^* \bar{M}^*} = 1} = 1/\sqrt{3}$, 
    $C_{0,1}^{(M\bar{M}^* - M^* \bar{M})/\sqrt{2}} = 1/\sqrt{6} $, 
    $C_{0,1}^{M^*\bar{M}^*,s_{M^* \bar{M}^*} = 1} = 1/\sqrt{6}$,
\item[{\it (b)}] $C_{2,0}^{M^*\bar{M}^*,s_{M^* \bar{M}^*} = 2}$ $=$ $\sqrt{2/3}$, 
    $C_{0,0}^{M\bar{M}} = 1/2$, 
    $C_{0,0}^{M^*\bar{M}^*,s_{M^* \bar{M}^*} = 0} = 1/(2 \sqrt{3})$,
\item[{\it (c)}] $C_{2,1}^{(M\bar{M}^* + M^* \bar{M})/\sqrt{2}}=1/\sqrt{6}$, 
    $C_{2,1}^{M^*\bar{M}^*,s_{M^* \bar{M}^*} = 2}=1/\sqrt{2}$, 
    $C_{0,1}^{(M\bar{M}^* + M^* \bar{M})/\sqrt{2}} = 1/\sqrt{3}$,
\item[{\it (d)}] $C_{2,2}^{M\bar{M}} = 1/\sqrt{10}$, 
    $C_{2,2}^{(M\bar{M}^* + M^* \bar{M})/\sqrt{2}} = \sqrt{3/10}$, $C_{2,2}^{M^*\bar{M}^*,s_{M^* \bar{M}^*} = 0}$ $ = 1/\sqrt{30}$, 
    $C_{2,2}^{M^*\bar{M}^*,s_{M^* \bar{M}^*} = 2} = \sqrt{7/30}$, \\
    $C_{0,2}^{M^*\bar{M}^*,s_{M^* \bar{M}^*} = 2} = 1/\sqrt{3}$.
\end{itemize}
All the coefficients reported above, derived in the $^3P_0$ model, agree up to a phase  with the prefactors of the mixing potential $V_{\Sigma_g^+ - \Sigma_g^{+\prime}}$ in the static potential matrices written in the meson-antimeson basis for each $J^{PC}$, cf. Eqs.~\eqref{eq:1+-mes-mes}, \eqref{eq:0++mes-mes}, \eqref{eq:1++mes-mes}, and \eqref{eq:2++mes-mes}.\footnote{
See footnote~\ref{footnote:coeff}.}
Moreover, it holds that 
\begin{align}
&\big(C_{2,1}^{(M\bar{M}^* - M^* \bar{M})/\sqrt{2}} \big)^2 + \big(C_{2,1}^{M^*\bar{M}^*,s_{M^* \bar{M}^*} = 1} \big)^2 = 2/3, \quad 
\big(C_{0,1}^{(M\bar{M}^* - M^* \bar{M})/\sqrt{2}} \big)^2 + \big( C_{0,1}^{M^*\bar{M}^*,s_{M^* \bar{M}^*} = 1}\big)^2 = 1/3, \label{eq:sumCi}\\
&\bigl(C_{2,0}^{M^*\bar{M}^*,s_{M^* \bar{M}^*} = 2}\bigr)^2 = 2/3, \quad 
\bigl(C_{0,0}^{M\bar{M}}\bigr)^2 + \bigl(C_{0,0}^{M^*\bar{M}^*,s_{M^* \bar{M}^*} = 0}\bigr)^2 = 1/3, \\
& \bigl(C_{2,1}^{(M\bar{M}^* + M^* \bar{M})/\sqrt{2}}\bigr)^2 + \bigl(C_{2,1}^{M^*\bar{M}^*,s_{M^* \bar{M}^*} = 2} \bigr)^2 = 2/3, \quad  
\bigl(C_{0,1}^{(M\bar{M}^* + M^* \bar{M})/\sqrt{2}} \bigr)^2=1/3, \\ 
&\bigl( C_{2,2}^{M\bar{M}} \bigr)^2 + \bigl(C_{2,2}^{(M\bar{M}^* + M^* \bar{M})/\sqrt{2}} \bigr)^2 + \bigl( C_{2,2}^{M^*\bar{M}^*,s_{M^* \bar{M}^*} = 0}\bigr)^2 + \bigl( C_{2,2}^{M^*\bar{M}^*,s_{M^* \bar{M}^*} = 2} \bigr)^2 = 2/3, \quad
(C_{0,2}^{M^*\bar{M}^*, s_{M^* \bar{M}^*} = 2})^2 = 1/3. \label{eq:sumCf}
\end{align}
The fact that the sums of the squares of the non-vanishing coefficients is equal for all the quarkonium states is a consequence of these states becoming degenerate in the considered limit, hence they mix with the threshold in the same way.
Replacing Eqs.~\eqref{eq:sumCi}-\eqref{eq:sumCf} into Eqs.~\eqref{eq:3P0_self_pwav1+-}-\eqref{eq:3P0_self_pwav2++} and comparing with Eq.~\eqref{eq:3P0_self_pwav1} sets $C_{2}^{M \bar{M}^{\rm spin \, avg.}} = \sqrt{{2}/{3}}$ and $C_{0}^{M \bar{M}^{\rm spin \, avg.}} = \sqrt{{1}/{3}}$.

For what concerns the BOEFT calculation, we proceed as for the $S$-wave quarkonium case discussed in Section~\ref{subsec:3P0}. 
In order to fix the mixing potential prefactors entering  Eq.~\eqref{eq:selfen_BOEFT_pwav1}, we take the limit $\delta_Q \to 0$ of Eqs.~\eqref{eq:1+-BOEFT}, \eqref{eq:0++BOEFT}, \eqref{eq:1++BOEFT} and \eqref{eq:2++BOEFT}, so that all four coupled equations acquire the common block-diagonal form 
\begin{align}\label{eq:1+-blockdiag}
&\hspace{-0.3 cm}\left[
-\frac{1}{m_Qr^2}\,\partial_rr^2\partial_r+ \frac{1}{m_Q r^2}
{\begin{pmatrix}
2 & 0 & 0 & 0\\[4pt]
0                 & 0 & 0 & 0\\[4pt]
0                 & 0 & 6 & 0\\[4pt]
0                 & 0 & 0 & D_l
\end{pmatrix}} + 
{\begin{pmatrix}
 V_{\Sigma_g^+} & \sqrt{\frac{1}{3}} V_{\Sigma_g^+ - \Sigma_g^{+\prime}} & \sqrt{\frac{2}{3}} V_{\Sigma_g^+ - \Sigma_g^{+\prime}} & 0 \\[6pt]
\sqrt{\frac{1}{3}}V_{\Sigma_g^+ - \Sigma_g^{+\prime}}  &  m_{M \bar{M}^{\rm spin \, avg.}} & 0 & 0 \\[6pt]
\sqrt{\frac{2}{3}} V_{\Sigma_g^+ - \Sigma_g^{+\prime}}  & 0 &  m_{M \bar{M}^{\rm spin \, avg.}}  & 0\\[6pt]
0                 & 0 & 0 & D_V \\[6pt]
\end{pmatrix}}
\right] \nonumber\\
&\hspace{9cm}
\times
 \begin{pmatrix} \psi_{\Sigma_g^+}^{n1} \\[6pt] \psi_{\Sigma_g^{+ \prime} + 2\Pi_g}^{n0} \\[6pt] \psi_{2\Sigma_g^{+ \prime} + \Pi_g}^{n2} \\[6pt] \psi_D \end{pmatrix}={\mathcal{E}} \begin{pmatrix} \psi_{\Sigma_g^+}^{n1} \\[6pt] \psi_{\Sigma_g^{+ \prime} + 2\Pi_g}^{n0} \\[6pt] \psi_{2\Sigma_g^{+ \prime} + \Pi_g}^{n2} \\[6pt] \psi_D \end{pmatrix},
\end{align}
where $\psi_{\Sigma_g^+}^{n1}$, $\psi_{\Sigma_g^{+ \prime} + 2\Pi_g}^{n0}$ and $\psi_{2\Sigma_g^{+ \prime} + \Pi_g}^{n2}$\footnote{
See footnote~\ref{footnote:l}.}
denote the $P$-wave quarkonium and the $S$- and $D$-wave tetraquark spin-averaged wavefunctions, respectively, while $D_l$, $D_V$ identify a separate block of the equation that decouples in the $\delta_Q\to0$ limit with $\psi_D$ the related wavefunction components.
The values of $0$, $2$, $6$ along the main diagonal of the centrifugal barrier matrix correspond to the orbital angular momentum values $l_{Q \bar{Q}} = 0$, $1$, $2$, respectively.
Equation~\eqref{eq:1+-blockdiag} fixes the mixing coefficients of the real part of the BOEFT self-energy reported in Eq.~\eqref{eq:selfen_BOEFT_pwav1} to $C_{2}^{M \bar{M}^{\rm spin \, avg.}} = \sqrt{{2}/{3}}$, $C_{0}^{M \bar{M}^{\rm spin \, avg.}} = \sqrt{{1}/{3}}$.

\section{Coupled Schr\"{o}dinger equations with spin-splitting potential at $\mathcal{O}(1/m_Q)$ in the diabatic and meson-antimeson bases}\label{app:meson-meson basis}
Section~\ref{sec:thr-spl} details how to include the $\mathcal{O}(1/m_Q)$ tetraquark spin-dependent potential $V_{SS}$ given in \eqref{eq:splitting} in the BOEFT coupled Schr\"{o}dinger equations for the case $J^{PC}=1^{++}$, giving expressions in both the BO diabatic and meson–antimeson bases (Eqs.~\eqref{eq:1++BOEFT} and \eqref{eq:1++mes-mes}, respectively). 
In this appendix, we derive analog coupled Schr\"{o}dinger equations for the cases $J^{PC} = 0^{-+}$, $1^{+-}$, $2^{-+}$, $0^{++}$, $1^{--}$, $2^{++}$ and $2^{--}$.
The resulting expressions are presented in both the BO diabatic and meson–antimeson bases and involve $S$-, $P$-, $D$- and $F$-wave quarkonium wavefunction components.

Quarkonium components couple directly to tetraquark states in the meson–antimeson basis even in the static limit because $s_{M^{(*)} \bar{M}^{(*)}}$ and $l_{Q \bar{Q}}$ (in the meson-antimeson  basis the states are labeled $\ket{[(j_{M^{(*)}}, j_{\bar{M}^{(*)}})\, s_{M^{(*)} \bar{M}^{(*)}}, l_{Q \bar{Q}}] J}$) are not good quantum numbers, unlike $s_{Q \bar{Q}}$ and $l$ in the BO diabatic basis (in this basis the states are labeled $\ket{[s_{Q \bar{Q}}, (K, l_{Q \bar{Q}})\,l], J}$). 
Diagonal shifts in the spin-splitting matrix written in the meson-antimeson basis, $0$, $-3\delta_Q/2$, $-\delta_Q/2$ and $\delta_Q/2$, identify the quarkonium ($Q \bar{Q}$), $M \bar{M}$, $(M^* \bar{M} \pm M \bar{M}^*)/{\sqrt{2}}$, and $M^* \bar{M}^*$ threshold components, respectively ($M = D, B$).
The mixing coefficients for different $J^{PC}$ in the meson–antimeson basis were calculated in~\cite{Bruschini:2023zkb} and agree with those derived here in the BOEFT in the meson-antimeson basis.

In this appendix, for ease of notation, we use $V_{\Sigma_g^+} \equiv V_{\Sigma_g}$, $V_{\Pi_g}$, $V_{\Sigma_g^{+ \prime}} \equiv V_{\Sigma_g^\prime}$, $V_{\Sigma_u^-} \equiv V_{\Sigma_u}$, $V_{\Sigma_g^+ - \Sigma_g^{+ \prime}} \equiv V_{\Sigma_g - \Sigma_g^\prime}$. 
We adopt a similar notation for the wavefunctions, where we furthermore suppress all the quarkonium, $N_Q$, and tetraquark, $N_T$, quantum numbers.
The energy eigenvalues are simply denoted as $\mathcal{E}$.
In the BO basis, the different wavefunctions are named after the diagonal entries of the static potential matrix, while in the meson-antimeson basis, they are named after the spin-splitted meson-antimeson thresholds or quarkonium component they refer to.\footnote{
The shorthand notation $\psi_{M \bar{M}^*}$ indicates the radial part of the CP eigenstates $(M \bar{M}^* \pm M^* \bar{M})/{\sqrt{2}}$.}
The superscripts $S$, $P$, $D$, ..., when appearing, indicate the $l_{Q \bar{Q}} = 0$, $1$, $2$, ... values of the heavy quark-antiquark orbital angular momentum and are introduced to differentiate two (or more) quarkonium/tetraquark wavefunction components entering the coupled equations. 

In all equations of this section, only the $J^{PC}$ quantum numbers are displayed. 
Values of $0$, $2$, $6$, ... along the main diagonal of the centrifugal barrier matrix correspond to the values $l_{Q \bar{Q}} = 0$, $1$, $2$, ...~.  
The remaining quantum numbers can be read off from Tables~\ref{tab:multiplets} and~\ref{tab:BOEFTvsmes-mes} for quarkonium and tetraquark states, respectively.
When multiple wavefunction components share the same $l_{Q \bar{Q}}$ value — as in Eq.~\eqref{eq:1--mes-mes} or in Eq.~\eqref{eq:2++mes-mes} — the $s_{M^{(*)} \bar{M}^{(*)}}$ quantum number is indicated as a superscript to distinguish between distinct tetraquark states.

\subsection{$J^{PC} = 0^{-+}$}

BO diabatic basis:

\begin{align}\label{eq:0-+BOEFT}
&\left[
-\frac{1}{m_Qr^2}\,\partial_rr^2\partial_r+\frac{1}{m_Qr^2}
{\begin{pmatrix}
0 & 0 & 0\\[4pt]
0                 & 2        & 0 \\[4pt]
0                 & 0 & 2
\end{pmatrix}}\right.
\left.
+\begin{pmatrix} V_{\Sigma_g} &  V_{\Sigma_g - \Sigma_g^\prime} & 0 \\[4pt]
    V_{\Sigma_g - \Sigma_g^\prime} & V_{\Sigma_{g}^{\prime}} & 0\\[4pt]
      0 & 0 & V_{\Sigma_u}\end{pmatrix} 
+ \begin{pmatrix} 0 &  0 & 0 \\[4pt]
    0 & 0 & \frac{\delta_Q}{2}\\[4pt]
      0 & \frac{\delta_Q}{2} & 0\end{pmatrix}      
      \right]
      \hspace{-4pt}\begin{pmatrix} \psi_{\Sigma_g} \\[4pt] \psi_{\Sigma_g^{\prime}} \\[4pt] \psi_{\Sigma_u }\end{pmatrix}={\mathcal{E}} \begin{pmatrix} \psi_{\Sigma_g} \\[4pt] \psi_{\Sigma_g^{\prime}} \\[4pt] \psi_{\Sigma_u}\end{pmatrix}, 
\end{align}

meson-antimeson diabatic basis:

\begin{align}\label{eq:0-+mes-mes}
&\left[
-\frac{1}{m_Qr^2}\,\partial_rr^2\partial_r+\frac{1}{m_Qr^2}
{\begin{pmatrix}
0 & 0 & 0\\[4pt]
0                 & 2       & 0 \\[4pt]
0                 & 0 & 2
\end{pmatrix}}\right.
\left. +\begin{pmatrix} V_{\Sigma_{g}} &  \frac{V_{\Sigma_g - \Sigma_g^\prime}}{\sqrt{2}} & \frac{V_{\Sigma_g - \Sigma_g^\prime}}{\sqrt{2}} \\[4pt]
    \frac{V_{\Sigma_g - \Sigma_g^\prime}}{\sqrt{2}} & \frac{V_{\Sigma_{g}^{\prime}}+V_{\Sigma_u}}{2} & \frac{V_{\Sigma_{g}^{\prime}}-V_{\Sigma_u}}{2}\\[4pt]
      \frac{V_{\Sigma_g - \Sigma_g^\prime}}{\sqrt{2}} & \frac{V_{\Sigma_{g}^{\prime}}-V_{\Sigma_u}}{2} & \frac{V_{\Sigma_{g}^{\prime}}+V_{\Sigma_u}}{2}\end{pmatrix} 
+ \begin{pmatrix} 0 &  0 & 0 \\[4pt]
    0 & -\frac{\delta_Q}{2} & 0\\[4pt]
      0 & 0 & \frac{\delta_Q}{2}\end{pmatrix}      
      \right]
      \hspace{-4pt}\begin{pmatrix} \psi_{Q \bar{Q}} \\[4pt] \psi_{M \bar{M}^*} \\[4pt] \psi_{M^* \bar{M}^*}\end{pmatrix}={\mathcal{E}} \begin{pmatrix} \psi_{Q \bar{Q}} \\[4pt] \psi_{M \bar{M}^*} \\[4pt] \psi_{M^* \bar{M}^*}\end{pmatrix}.
\end{align}

\subsection{$J^{PC} = 1^{+-}$}

BO diabatic basis:

\begin{align}\label{eq:1+-BOEFT}
&\hspace{-0.5 cm}\left[
-\frac{1}{m_Qr^2}\,\partial_rr^2\partial_r+\frac{1}{m_Qr^2}
{\begin{pmatrix}
2 & 0 & 0  & 0  & 0   \\[4pt]
0                 & 0        & 0 & 0 & 0\\[4pt]
0                 & 0 & 6 & 0  & 0 \\[4pt]
0                 & 0 & 0 & 0  & 0 \\[4pt]
0                 & 0 & 0 & 0  & 6 \\
\end{pmatrix}}\right.
+\begin{pmatrix} V_{\Sigma_{g}} &  \frac{V_{\Sigma_g - \Sigma_g^\prime}}{\sqrt{3}} &  \frac{\sqrt{2} \, V_{\Sigma_g - \Sigma_g^\prime}}{\sqrt{3}} & 0 & 0 \\[4pt]
    \frac{V_{\Sigma_g - \Sigma_g^\prime}}{\sqrt{3}} & \frac{V_{\Sigma_g^\prime} + 2 V_{\Pi_g}}{3} & \frac{\sqrt{2} \,(V_{\Sigma_g^\prime} - V_{\Pi_g})}{3}& 0 & 0 \\[4pt]
      \frac{\sqrt{2} \, V_{\Sigma_g - \Sigma_g^\prime}}{\sqrt{3}} & \frac{\sqrt{2} \,(V_{\Sigma_g^\prime} - V_{\Pi_g})}{3} & \frac{2 V_{\Sigma_g^\prime} + V_{\Pi_g}}{3}& 0 &0 \\[4pt]
      0 & 0 & 0& V_{\Sigma_u} &0 \\[4pt]
      0 & 0 & 0& 0 &V_{\Sigma_u} \\
      \end{pmatrix} 
      \nonumber\\
&\hspace{+8.0 cm}\left.+
\begin{pmatrix}
 0 & 0 & 0 & 0 & 0 \\[4pt]
 0 & 0 & 0 & \frac{\delta_Q }{2} & 0 \\[4pt]
 0 & 0 & 0 & 0 & \frac{\delta_Q }{2} \\[4pt]
 0 & \frac{\delta_Q }{2} & 0 & 0 & 0 \\[4pt]
 0 & 0 & \frac{\delta_Q }{2} & 0 & 0 \\
\end{pmatrix}     
      \right]
      \hspace{-4pt}\begin{pmatrix} \psi_{\Sigma_g} \\[5pt] \psi_{\Sigma_g^{\prime}+2\Pi_g} \\[5pt] \psi_{2\Sigma_g^{\prime}+\Pi_g} \\[5pt] \psi^S_{\Sigma_u} \\[5pt] \psi^D_{\Sigma_u}
      \end{pmatrix}={\mathcal{E}} \begin{pmatrix} \psi_{\Sigma_g} \\[5pt] \psi_{\Sigma_g^{\prime}+2\Pi_g} \\[5pt] \psi_{2\Sigma_g^{\prime}+\Pi_g} \\[5pt] \psi^S_{\Sigma_u} \\[5pt] \psi^D_{\Sigma_u} \end{pmatrix}, 
\end{align}

meson-antimeson diabatic basis: 

\begin{align}\label{eq:1+-mes-mes}
&\hspace{0 cm}\left[
-\frac{1}{m_Qr^2}\,\partial_rr^2\partial_r+\frac{1}{m_Qr^2}
{\begin{pmatrix}
2 & 0 & 0  & 0  & 0   \\[4pt]
0                 & 0        & 0 & 0 & 0\\[4pt]
0                 & 0 & 6 & 0  & 0 \\[4pt]
0                 & 0 & 0 & 0  & 0 \\[4pt]
0                 & 0 & 0 & 0  & 6 \\
\end{pmatrix}}\right.
 \nonumber\\
&\hspace{+1.5 cm}\left.
+\begin{pmatrix} 
V_{\Sigma_{g}} &  \frac{V_{\Sigma_g - \Sigma_g^\prime}}{\sqrt{6}} &  \frac{V_{\Sigma_g - \Sigma_g^\prime}}{\sqrt{3}} & \frac{V_{\Sigma_g - \Sigma_g^\prime}}{\sqrt{6}} & \frac{V_{\Sigma_g - \Sigma_g^\prime}}{\sqrt{3}} \\[4pt]
\frac{V_{\Sigma_g - \Sigma_g^\prime}}{\sqrt{6}} & \frac{V_{\Sigma_g^\prime} + 2 V_{\Pi_g} + 3 V_{\Sigma_u}}{6} & \frac{V_{\Sigma_g^\prime} - V_{\Pi_g}}{3 \sqrt{2}} & \frac{V_{\Sigma_g^\prime} + 2 V_{\Pi_g} - 3 V_{\Sigma_u}}{6} & \frac{V_{\Sigma_g^\prime} - V_{\Pi_g}}{3 \sqrt{2}} \\[4pt]
\frac{V_{\Sigma_g - \Sigma_g^\prime}}{\sqrt{3}} & \frac{V_{\Sigma_g^\prime} - V_{\Pi_g}}{3 \sqrt{2}} & 
\frac{2 V_{\Sigma_g^\prime} + V_{\Pi_g} + 3 V_{\Sigma_u}}{6}& 
\frac{V_{\Sigma_g^\prime} - V_{\Pi_g}}{3 \sqrt{2}} &
\frac{2 V_{ \Sigma_g^\prime} + V_{\Pi_g} - 3 V_{\Sigma_u}}{6} \\[4pt]
\frac{V_{\Sigma_g - \Sigma_g^\prime}}{\sqrt{6}} & \frac{V_{\Sigma_g^\prime} + 2 V_{\Pi_g} - 3 V_{\Sigma_u}}{6} & 
\frac{V_{\Sigma_g^\prime} - V_{\Pi_g}}{3 \sqrt{2}} &
\frac{V_{\Sigma_g^\prime} + 2 V_{\Pi_g} + 3 V_{\Sigma_u}}{6} &
\frac{V_{\Sigma_g^\prime} - V_{\Pi_g}}{3 \sqrt{2}} \\[4pt]
\frac{V_{\Sigma_g - \Sigma_g^\prime}}{\sqrt{3}} & 
\frac{V_{\Sigma_g^\prime} - V_{\Pi_g}}{3 \sqrt{2}} & 
\frac{2 V_{ \Sigma_g^\prime} + V_{\Pi_g} - 3 V_{\Sigma_u}}{6}&
\frac{V_{\Sigma_g^\prime} - V_{\Pi_g}}{3 \sqrt{2}} &
\frac{2 V_{\Sigma_g^\prime} + V_{\Pi_g} + 3 V_{\Sigma_u}}{6}\\
\end{pmatrix}\right.
\nonumber\\
&\hspace{7.0 cm}\left.
+ \begin{pmatrix} 
0 &  0 & 0  & 0 & 0\\[4pt]
0 & -\frac{\delta_Q}{2} & 0& 0 & 0\\[4pt]
0 & 0 & -\frac{\delta_Q}{2}& 0 & 0 \\[4pt]
0 & 0 & 0& \frac{\delta_Q}{2} & 0 \\[4pt]
0 & 0 & 0& 0 & \frac{\delta_Q}{2} \\
\end{pmatrix}      
\right]
\begin{pmatrix} 
\psi_{Q \bar{Q}} \\[5pt]
\psi^S_{M \bar{M}^{*}} \\[5pt]
\psi^D_{M \bar{M}^{*}} \\[5pt] 
\psi^S_{M^* \bar{M}^{*}} \\[5pt] 
\psi^{D}_{M ^* \bar{M}^{*}}
\end{pmatrix}
={\mathcal{E}} 
\begin{pmatrix} 
\psi_{Q \bar{Q}} \\[5pt]
\psi^S_{M \bar{M}^{*}} \\[5pt]
\psi^D_{M \bar{M}^{*}} \\[5pt] 
\psi^S_{M^* \bar{M}^{*}} \\[5pt] 
\psi^{D}_{M ^* \bar{M}^{*}}
\end{pmatrix}.
\end{align}

\subsection{$J^{PC} = 2^{-+}$}

BO diabatic basis:

\begin{align}\label{eq:2-+BOEFT}
&\hspace{-1 cm}\left[
-\frac{1}{m_Qr^2}\,\partial_rr^2\partial_r+\frac{1}{m_Qr^2}
{\begin{pmatrix}
6 & 0 & 0  & 0  & 0   \\
0                 & 2        & 0 & 0 & 0 \\[4pt]
0                 & 0 & 12 & 0  & 0 \\[4pt]
0                 & 0 & 0 & 2  & 0 \\[4pt]
0                 & 0 & 0 & 0  & 12 \\[4pt]
\end{pmatrix}}\right.
+\begin{pmatrix} V_{\Sigma_{g}} &  \sqrt{\frac{2}{5}} V_{\Sigma_g - \Sigma_g^\prime} &  \sqrt{\frac{3}{5}} V_{\Sigma_g - \Sigma_g^\prime} & 0 & 0 \\[4pt]
    \sqrt{\frac{2}{5}} V_{\Sigma_g - \Sigma_g^\prime} & \frac{2 V_{\Sigma_g^\prime} + 3 V_{\Pi_g}}{5} & \frac{\sqrt{6}\, (V_{\Sigma_g^\prime} - V_{\Pi_g})}{5} & 0 & 0 \\[4pt]
      \sqrt{\frac{3}{5}} V_{\Sigma_g - \Sigma_g^\prime} & \frac{\sqrt{6} \, (V_{\Sigma_g^\prime} - V_{\Pi_g})}{5}  & \frac{3 V_{\Sigma_g^\prime} + 2 V_{\Pi_g}}{5}& 0 &0 \\[4pt]
      0 & 0 & 0& V_{\Sigma_u} &0 \\[4pt]
      0 & 0 & 0& 0 &V_{\Sigma_u} \\
      \end{pmatrix} 
\nonumber\\
&\hspace{7cm}\left.
+\begin{pmatrix}
 0 & 0 & 0 & 0 & 0 \\[4pt]
 0 & 0 & 0 & \frac{\delta_Q }{2} & 0 \\[4pt]
 0 & 0 & 0 & 0 & \frac{\delta_Q }{2} \\[4pt]
 0 & \frac{\delta_Q }{2} & 0 & 0 & 0 \\[4pt]
 0 & 0 & \frac{\delta_Q }{2} & 0 & 0 \\[4pt]
\end{pmatrix}      
      \right]
      \hspace{-4pt}\begin{pmatrix} \psi_{\Sigma_g} \\[5pt] \psi_{2\Sigma_g^{\prime}+3\Pi_g} \\[5pt] \psi_{3\Sigma_g^{\prime}+2\Pi_g} \\[5pt] \psi^P_{\Sigma_u} \\[5pt] \psi^F_{\Sigma_u}
      \end{pmatrix}={\mathcal{E}} \begin{pmatrix} \psi_{\Sigma_g} \\[5pt] \psi_{2\Sigma_g^{\prime}+3\Pi_g} \\[5pt] \psi_{3\Sigma_g^{\prime}+2\Pi_g} \\[5pt] \psi^P_{\Sigma_u} \\[5pt] \psi^F_{\Sigma_u} \end{pmatrix}, 
\end{align}

meson-antimeson diabatic basis:

\begin{align}\label{eq:2-+mes-mes}
&\hspace{-0.5 cm}\left[
-\frac{1}{m_Qr^2}\,\partial_rr^2\partial_r+\frac{1}{m_Qr^2}
{\begin{pmatrix}
6 & 0 & 0  & 0  & 0   \\[4pt]
0                 & 2        & 0 & 0 & 0\\[4pt]
0                 & 0 & 12 & 0  & 0 \\[4pt]
0                 & 0 & 0 & 2  & 0 \\[4pt]
0                 & 0 & 0 & 0  & 12 \\
\end{pmatrix}}\right.
\nonumber\\
&\hspace{+1.5 cm}\left.
+\begin{pmatrix} 
V_{\Sigma_{g}} &  
\frac{1}{\sqrt{5}} V_{\Sigma_g - \Sigma_g^\prime} & 
\sqrt{\frac{3}{10}} V_{\Sigma_g - \Sigma_g^\prime} &
\frac{1}{\sqrt{5}} V_{\Sigma_g - \Sigma_g^\prime} & 
\sqrt{\frac{3}{10}} V_{\Sigma_g - \Sigma_g^\prime} \\[6pt]
\frac{1}{\sqrt{5}} V_{\Sigma_g - \Sigma_g^\prime} & 
\frac{2 V_{\Sigma_g^\prime} + 3 V_{\Pi_g} + 5 V_{\Sigma_u}}{10} & 
 \sqrt{\frac{3}{2}} \frac{V_{\Sigma_g^\prime} - V_{\Pi_g} }{5} & \frac{2V_{ \Sigma_g^\prime} + 3 V_{\Pi_g} - 5 V_{\Sigma_u}}{10} & 
 \sqrt{\frac{3}{2}} \frac{V_{\Sigma_g^\prime} - V_{\Pi_g}}{5}  \\[6pt]
\sqrt{\frac{3}{10}} V_{\Sigma_g - \Sigma_g^\prime} & 
\sqrt{\frac{3}{2}} \frac{V_{\Sigma_g^\prime} - V_{\Pi_g}}{5} & 
\frac{3 V_{\Sigma_g^\prime} + 2 V_{\Pi_g} + 5 V_{\Sigma_u}}{10}& 
\sqrt{\frac{3}{2}} \frac{V_{\Sigma_g^\prime} - V_{\Pi_g}}{5} &
\frac{3 V_{\Sigma_g^\prime} + 2 V_{\Pi_g} - 5 V_{\Sigma_u}}{10} \\[6pt]
\frac{1}{\sqrt{5}} V_{\Sigma_g - \Sigma_g^\prime} & 
\frac{2 V_{\Sigma_g^\prime} + 3 V_{\Pi_g} - 5 V_{\Sigma_u}}{10} & 
\sqrt{\frac{3}{2}} \frac{V_{\Sigma_g^\prime} - V_{\Pi_g}}{5}&
\frac{2 V_{\Sigma_g^\prime} + 3 V_{\Pi_g} + 5 V_{\Sigma_u}}{10} &
\sqrt{\frac{3}{2}} \frac{V_{\Sigma_g^\prime} - V_{\Pi_g}}{5} \\[6pt]
\sqrt{\frac{3}{10}} V_{\Sigma_g - \Sigma_g^\prime} & 
\sqrt{\frac{3}{2}} \frac{V_{\Sigma_g^\prime} - V_{\Pi_g}}{5} & 
\frac{3 V_{\Sigma_g^\prime} + 2 V_{\Pi_g} - 5 V_{\Sigma_u}}{10}& 
\sqrt{\frac{3}{2}} \frac{V_{\Sigma_g^\prime} - V_{\Pi_g}}{5} &
\frac{3 V_{\Sigma_g^\prime} + 2 V_{\Pi_g} + 5 V_{\Sigma_u}}{10} \\
\end{pmatrix}\right.
\nonumber\\
&\hspace{7cm}\left.
+ \begin{pmatrix} 
0 &  0 & 0  & 0 & 0\\[4pt]
0 & -\frac{\delta_Q}{2} & 0& 0 & 0\\[4pt]
0 & 0 & -\frac{\delta_Q}{2}& 0 & 0 \\[4pt]
0 & 0 & 0& \frac{\delta_Q}{2} & 0 \\[4pt]
0 & 0 & 0& 0 & \frac{\delta_Q}{2} \\[4pt]
\end{pmatrix}      
\right]
\begin{pmatrix} 
\psi_{Q \bar{Q}} \\[5pt] 
\psi^P_{M \bar{M}^{*}} \\[5pt] 
\psi^F_{M \bar{M}^{*}} \\[5pt] 
\psi^P_{M^* \bar{M}^{*}} \\[5pt] 
\psi^F_{M ^* \bar{M}^{*}}
\end{pmatrix}
={\mathcal{E}} 
\begin{pmatrix} 
\psi_{Q \bar{Q}} \\[5pt] 
\psi^P_{M \bar{M}^{*}} \\[5pt] 
\psi^F_{M \bar{M}^{*}} \\[5pt] 
\psi^P_{M^* \bar{M}^{*}} \\[5pt] 
\psi^F_{M ^* \bar{M}^{*}}
\end{pmatrix}.
\end{align}

\subsection{$J^{PC} = 0^{++}$}

BO diabatic basis:

\begin{align}\label{eq:0++BOEFT}
&\hspace{-1.0 cm}\left[
-\frac{1}{m_Qr^2}\,\partial_rr^2\partial_r+\frac{1}{m_Qr^2}
{\begin{pmatrix}
2 & 0 & 0  & 0   \\[4pt]
0                 & 0        & 0 & 0 \\[4pt]
0                 & 0 & 6 & 0  \\[4pt]
0                 & 0 & 0 & 0  \\[4pt]
\end{pmatrix}}\right.
+\begin{pmatrix} V_{\Sigma_{g}} &  \frac{V_{\Sigma_g - \Sigma_g^\prime}}{\sqrt{3}} &  \sqrt{\frac{2}{3}} V_{\Sigma_g - \Sigma_g^\prime} & 0 \\[4pt]
    \frac{V_{\Sigma_g - \Sigma_g^\prime}}{\sqrt{3}} & \frac{V_{\Sigma_g^\prime} + 2 V_{\Pi_g}}{3} & \frac{\sqrt{2}\,(V_{\Sigma_g^\prime} - V_{\Pi_g})}{3} & 0  \\[4pt]
     \sqrt{\frac{2}{3}} V_{\Sigma_g - \Sigma_g^\prime} & \frac{\sqrt{2}\,(V_{\Sigma_g^\prime} - V_{\Pi_g})}{3}  & \frac{2 V_{\Sigma_g^\prime} + V_{\Pi_g}}{3}& 0 \\[4pt]
      0 & 0 & 0& V_{\Sigma_u} \\[4pt]
      \end{pmatrix} +
    \nonumber\\
&\hspace{7cm}\left.
\begin{pmatrix}
 0 & 0 & 0 & 0   \\[4pt]
 0 & -\delta_Q & 0 & -\frac{\sqrt{3} \, \delta_Q}{2}  \\[4pt]
 0 & 0 & \frac{\delta_Q}{2} & 0  \\[4pt]
 0 & -\frac{\sqrt{3} \, \delta_Q}{2} & 0 & 0  \\[4pt]
\end{pmatrix}     
      \right]
      \hspace{-4pt}\begin{pmatrix} \psi_{\Sigma_g} \\[6pt] \psi_{\Sigma_g^{\prime}+2\Pi_g} \\[6pt]
      \psi_{2\Sigma_g^{\prime}+\Pi_g} \\[6pt]
      \psi_{\Sigma_u}
      \end{pmatrix}={\mathcal{E}} 
      \begin{pmatrix} \psi_{\Sigma_g} \\[6pt] 
      \psi_{\Sigma_g^{\prime}+2\Pi_g} \\[6pt]
      \psi_{2\Sigma_g^{\prime}+\Pi_g} \\[6pt]
      \psi_{\Sigma_u}
      \end{pmatrix}, 
\end{align}

meson-antimeson diabatic basis:

\begin{align}\label{eq:0++mes-mes}
&\hspace{-0.8 cm}\left[
-\frac{1}{m_Qr^2}\,\partial_rr^2\partial_r+\frac{1}{m_Qr^2}
{\begin{pmatrix}
2 & 0 & 0  & 0   \\[4pt]
0                 & 0  & 0 & 0  \\[4pt]
0                 & 0 & 6  & 0  \\[4pt]
0                 & 0 & 0  & 0  \\
\end{pmatrix}}\right.
+\begin{pmatrix} V_{\Sigma_{g}} &  \frac{V_{\Sigma_g - \Sigma_g^\prime}}{2} &  \sqrt{\frac{2}{3}} V_{\Sigma_g - \Sigma_g^\prime} & - \frac{V_{\Sigma_g - \Sigma_g^\prime}}{2 \sqrt{3}} \\[4pt]
   \frac{V_{\Sigma_g - \Sigma_g^\prime}}{2} & \frac{V_{\Sigma_g^\prime} + 2 V_{\Pi_g} + V_{\Sigma_u}}{4} & \frac{V_{\Sigma_g^\prime} - V_{\Pi_g}}{\sqrt{6}} & \frac{ - V_{\Sigma_g^\prime} - 2 V_{\Pi_g} + 3 V_{\Sigma_u}}{4 \sqrt{3}}  \\[4pt]
    \sqrt{\frac{2}{3}} V_{\Sigma_g - \Sigma_g^\prime} & \frac{V_{\Sigma_g^\prime} - V_{\Pi_g}}{\sqrt{6}} & \frac{2 V_{\Sigma_g^\prime} + V_{\Pi_g}}{3}& -\frac{ V_{\Sigma_g^\prime} - V_{\Pi_g}}{3 \sqrt{2}} \\[4pt]
      - \frac{V_{\Sigma_g - \Sigma_g^\prime}}{2 \sqrt{3}} & \frac{ - V_{\Sigma_g^\prime} - 2 V_{\Pi_g} + 3 V_{\Sigma_u}}{4 \sqrt{3}}  &-\frac{ V_{\Sigma_g^\prime} - V_{\Pi_g}}{3 \sqrt{2}}& \frac{ V_{\Sigma_g^\prime} + 2 V_{\Pi_g} + 9 V_{\Sigma_u}}{12}  \\
      \end{pmatrix} 
\nonumber\\
&\hspace{8.0 cm}\left.
+
\setlength{\arraycolsep}{1pt}
\begin{pmatrix}
 0 & 0 & 0 & 0   \\[4pt]
 0 & -\frac{3\delta_Q}{2} & 0 & 0  \\[4pt]
 0 & 0 & \frac{\delta_Q}{2} & 0  \\[4pt]
 0 & 0 & 0 & \frac{\delta_Q}{2}  \\
\end{pmatrix}    
      \right]
      \hspace{-4pt}\begin{pmatrix} 
      \psi_{Q \bar{Q}} \\[6pt] 
    \psi_{M \bar{M}} \\[6pt] 
    \psi^D_{M^* \bar{M}^{*}} \\[6pt] 
    \psi^S_{M^* \bar{M}^{*}} \\ 
      \end{pmatrix}={\mathcal{E}} 
      \begin{pmatrix} \psi_{Q \bar{Q}} \\[6pt] 
    \psi_{M \bar{M}} \\[6pt] 
    \psi^D_{M^* \bar{M}^{*}} \\[6pt] 
    \psi^S_{M^* \bar{M}^{*}} \\ 
      \end{pmatrix}.
\end{align}

\subsection{$J^{PC} = 1^{--}$}

BO diabatic basis:

\begin{align}\label{eq:1--BOEFT}
&\hspace{+-0.5cm}\left[
-\frac{1}{m_Qr^2}\,\partial_rr^2\partial_r+
\frac{1}{m_Qr^2}
{\begin{pmatrix}
0 & 0 & 0  & 0  & 0 & 0  & 0  \\[4pt]
0                 & 6       & 0 & 0 & 0 & 0  & 0\\[4pt]
0                 & 0 & 2 & 0  & 0 & 0  & 0\\[4pt]
0                 & 0 & 0 & 2  & 0 & 0  & 0\\[4pt]
0                 & 0 & 0 & 0  & 2 & 0  & 0\\[4pt]
0                 & 0 & 0 & 0  & 0 & 12  & 0\\[4pt]
0                 & 0 & 0 & 0  & 0 & 0  &2\\
\end{pmatrix}}\right.
\nonumber\\
&\hspace{+1.5 cm}\left.
+\begin{pmatrix} 
V_{\Sigma_{g}} & 0 &  V_{\Sigma_g - \Sigma_g^\prime} & 0 & 0  & 0 & 0\\[4pt]
0 & V_{\Sigma_{g}} & 0 & 0 & \sqrt{\frac{2}{5}} V_{\Sigma_g - \Sigma_g^\prime} & \sqrt{\frac{3}{5}} V_{\Sigma_g - \Sigma_g^\prime} & 0\\[4pt]
V_{\Sigma_g - \Sigma_g^\prime} & 0 &  V_{\Sigma_g^\prime} & 0 &0  & 0 & 0\\[4pt]
0 & 0 & 0& V_{\Pi_g} &0  & 0 & 0\\[4pt]
0 & \sqrt{\frac{2}{5}} V_{\Sigma_g - \Sigma_g^\prime} & 0& 0 &\frac{2 V_{\Sigma_g^\prime} + 3 V_{\Pi_g}}{5}  & \frac{ \sqrt{6} \, (V_{\Sigma_g^\prime} -  V_{\Pi_g})}{5} & 0\\[4pt]
0 & \sqrt{\frac{3}{5}} V_{\Sigma_g - \Sigma_g^\prime} & 0& 0 &\frac{ \sqrt{6} \, (V_{\Sigma_g^\prime} -  V_{\Pi_g}) }{5}  & \frac{3 V_{\Sigma_g^\prime} + 2 V_{\Pi_g}}{5} & 0\\[4pt]
0 & 0 & 0& 0 &0  & 0 & V_{\Sigma_u}\\[4pt]
\end{pmatrix}\right.
\nonumber\\
&\hspace{+3.5 cm}\left. +
\begin{pmatrix}
0 & 0 & 0 & 0 & 0 & 0 & 0\\[4pt]
0 & 0 & 0 & 0 & 0 & 0 & 0\\[4pt]
0 & 0 & 0 & \frac{\delta_Q }{\sqrt{3}} &0 & 0 & -\frac{\delta_Q }{\sqrt{12}}\\[4pt]
0 & 0 & \frac{\delta_Q }{\sqrt{3}} & -\frac{\delta_Q}{4} & \frac{\sqrt{5} \,\delta_Q}{4 \sqrt{3}} & 0 & \frac{\delta_Q }{2}\\[4pt]
0 & 0 & 0 & \frac{\sqrt{5} \,\delta_Q}{4 \sqrt{3}} & -\frac{3 \, \delta_Q}{4} & 0 & - \frac{\sqrt{5} \,\delta_Q }{\sqrt{12}}\\[4pt]
0 & 0 & 0 & 0 & 0 & \frac{\delta_Q }{2} & 0\\[4pt]
0 & 0 & -\frac{\delta_Q }{\sqrt{12}} & \frac{\delta_Q }{2} & - \frac{ \sqrt{5} \, \delta_Q}{\sqrt{12}} & 0 & 0\\[4pt]
\end{pmatrix}      
\right]
\begin{pmatrix} 
\psi^S_{\Sigma_g} \\[6pt] 
\psi^D_{\Sigma_g} \\[6pt] 
\psi_{\Sigma_g^{\prime}} \\[6pt] 
\psi_{\Pi_g} \\[6pt] 
\psi_{2\Sigma_g^{\prime}+3\Pi_g} \\[6pt] 
\psi_{3\Sigma_g^{\prime}+2\Pi_g} \\[6pt] 
\psi_{\Sigma_u} \\
\end{pmatrix}
={\mathcal{E}} 
\begin{pmatrix} 
\psi^S_{\Sigma_g} \\[6pt] 
\psi^D_{\Sigma_g} \\[6pt] 
\psi_{\Sigma_g^{\prime}} \\[6pt] 
\psi_{\Pi_g} \\[6pt] 
\psi_{2\Sigma_g^{\prime}+3\Pi_g} \\[6pt] 
\psi_{3\Sigma_g^{\prime}+2\Pi_g} \\[6pt] 
\psi_{\Sigma_u} \\
\end{pmatrix}, 
\end{align}

meson-antimeson diabatic basis:

\begin{align}\label{eq:1--mes-mes}
&\hspace{+0 cm}\left[
-\frac{1}{m_Qr^2}\,\partial_rr^2\partial_r+
\frac{1}{m_Qr^2}
{\begin{pmatrix}
0 & 0 & 0  & 0  & 0 & 0  & 0  \\[4pt]
0                 & 6       & 0 & 0 & 0 & 0  & 0\\[4pt]
0                 & 0 & 2 & 0  & 0 & 0  & 0\\[4pt]
0                 & 0 & 0 & 2  & 0 & 0  & 0\\[4pt]
0                 & 0 & 0 & 0  & 2 & 0  & 0\\[4pt]
0                 & 0 & 0 & 0  & 0 &  12  & 0\\[4pt]
0                 & 0 & 0 & 0  & 0 & 0  & 2\\
\end{pmatrix}}\right.
\nonumber\\
&\hspace{+0.0 cm}\left.
+\begin{pmatrix} 
V_{\Sigma_{g}} & 0 &  
\frac{1}{2 \sqrt{3}} V_{\Sigma_g - \Sigma_g^\prime} & 
-\frac{1}{\sqrt{3}} V_{\Sigma_g - \Sigma_g^\prime} & 
-\frac{1}{6} V_{\Sigma_g - \Sigma_g^\prime}  & 
0 & 
\frac{\sqrt{5}}{3} V_{\Sigma_g - \Sigma_g^\prime}\\[6pt]
0 & V_{\Sigma_{g}} & 
\frac{1}{\sqrt{6}} V_{\Sigma_g - \Sigma_g^\prime} &
\frac{1}{\sqrt{6}} V_{\Sigma_g - \Sigma_g^\prime} &
-\frac{1}{3 \sqrt{2}} V_{\Sigma_g - \Sigma_g^\prime} & \sqrt{\frac{3}{5}} V_{\Sigma_g - \Sigma_g^\prime} & 
\frac{1}{3 \sqrt{10}} V_{\Sigma_g - \Sigma_g^\prime}\\[6pt]
\frac{1}{2 \sqrt{3}} V_{\Sigma_g - \Sigma_g^\prime} &
\frac{1}{\sqrt{6}} V_{\Sigma_g - \Sigma_g^\prime} &
\frac{V_{\Sigma_g^\prime} + 2 V_{\Pi_g} + V_{\Sigma_u}}{4} & 
0 &
-\frac{V_{\Sigma_g^\prime} + 2 V_{\Pi_g} -3 V_{\Sigma_u}}{4 \sqrt{3}} &
\frac{V_{\Sigma_g^\prime} - V_{\Pi_g}}{\sqrt{10}} & 
\frac{V_{\Sigma_g^\prime} - V_{\Pi_g}}{\sqrt{15}}\\[6pt]
-\frac{1}{\sqrt{3}} V_{\Sigma_g - \Sigma_g^\prime} &
\frac{1}{ \sqrt{6}} V_{\Sigma_g - \Sigma_g^\prime} & 
0& 
\frac{V_{\Sigma_g^\prime} + V_{\Pi_g}}{2} &
0  & 
\frac{V_{\Sigma_g^\prime} - V_{\Pi_g}}{\sqrt{10}} & 
\sqrt{\frac{3}{5}} \frac{-V_{\Sigma_g^\prime} + V_{\Pi_g}}{2}\\[6pt]
-\frac{1}{6} V_{\Sigma_g - \Sigma_g^\prime} &
-\frac{1}{3 \sqrt{2}}V_{\Sigma_g - \Sigma_g^\prime} &
-\frac{V_{\Sigma_g^\prime} + 2 V_{\Pi_g} -3 V_{\Sigma_u}}{4 \sqrt{3}}&
0 &
\frac{V_{\Sigma_g^\prime} + 2 V_{\Pi_g} + 9 V_{\Sigma_u}}{12} &
\frac{- V_{\Sigma_g^\prime} + V_{\Pi_g}}{\sqrt{30}}  & 
\frac{- V_{\Sigma_g^\prime} + V_{\Pi_g}}{3 \sqrt{5}}\\[6pt]
0 &
\sqrt{\frac{3}{5}} V_{\Sigma_g - \Sigma_g^\prime} &
\frac{V_{\Sigma_g^\prime} - V_{\Pi_g}}{\sqrt{10}}&
\frac{V_{\Sigma_g^\prime} - V_{\Pi_g}}{\sqrt{10}} &
\frac{- V_{\Sigma_g^\prime} + V_{\Pi_g}}{\sqrt{30}}  &
\frac{3 V_{\Sigma_g^\prime} + 2 V_{\Pi_g}}{5} &
\frac{V_{\Sigma_g^\prime} - V_{\Pi_g}}{5 \sqrt{6}}\\[6pt]
\frac{\sqrt{5}}{3} V_{\Sigma_g - \Sigma_g^\prime} & 
\frac{1}{3 \sqrt{10}} V_{\Sigma_g - \Sigma_g^\prime} & 
\frac{V_{\Sigma_g^\prime} - V_{\Pi_g}}{\sqrt{15}}&
\sqrt{\frac{3}{5}} \, \frac{-V_{\Sigma_g^\prime} + V_{\Pi_g}}{2} &
\frac{- V_{\Sigma_g^\prime} + V_{\Pi_g}}{3 \sqrt{5}}  &
\frac{V_{\Sigma_g^\prime} - V_{\Pi_g}}{5 \sqrt{6}} & 
\frac{17 V_{\Sigma_g^\prime} + 13 V_{\Pi_g}}{30}\\
\end{pmatrix}\right.
\nonumber\\
&\hspace{5.0 cm}\left. +
\begin{pmatrix}
0 & 0 & 0 & 0 & 0 & 0 & 0\\[4pt]
0 & 0 & 0 & 0 & 0 & 0 & 0\\[4pt]
0 & 0 & -\frac{3 \delta_Q}{2}   & 0 &0 & 0 & 0\\[4pt]
0 & 0 & 0 & -\frac{\delta_Q}{2} & 0 & 0 & 0\\[4pt]
0 & 0 & 0 & 0 & \frac{\delta_Q}{2}  & 0 & 0\\[4pt]
0 & 0 & 0 & 0 & 0 & \frac{\delta_Q}{2} & 0\\[4pt]
0 & 0 & 0 & 0 & 0 & 0 & \frac{\delta_Q}{2} \\[4pt]
\end{pmatrix}      
\right]
\begin{pmatrix} 
\psi^S_{Q \bar{Q}} \\[6pt] 
\psi^D_{Q \bar{Q}} \\[6pt] 
\psi_{M \bar{M}} \\[6pt] 
\psi_{M \bar{M}^{*}} \\[6pt] 
\psi^{P, s_{M^* \bar{M}^*} = 0}_{M ^* \bar{M}^{*}} \\[6pt]
\psi^{F, s_{M^* \bar{M}^*} = 2}_{M^* \bar{M}^{*}} \\[6pt] 
\psi^{P, s_{M^* \bar{M}^*} = 2}_{M ^* \bar{M}^{*}}
\end{pmatrix}
={\mathcal{E}} 
\begin{pmatrix} 
\psi^S_{Q \bar{Q}} \\[6pt] 
\psi^D_{Q \bar{Q}} \\[6pt] 
\psi_{M \bar{M}} \\[6pt] 
\psi_{M \bar{M}^{*}} \\[6pt] 
\psi^{P, s_{M^* \bar{M}^*} = 0}_{M ^* \bar{M}^{*}} \\[6pt]
\psi^{F, s_{M^* \bar{M}^*} = 2}_{M^* \bar{M}^{*}} \\[6pt] 
\psi^{P, s_{M^* \bar{M}^*} = 2}_{M ^* \bar{M}^{*}}
\end{pmatrix}.
\end{align}

\subsection{$J^{PC} = 2^{++}$ }

BO diabatic basis:

\begin{align}\label{eq:2++BOEFT}
&\hspace{ 0 cm}\left[
-\frac{1}{m_Qr^2}\,\partial_rr^2\partial_r+
\frac{1}{m_Qr^2}
{\begin{pmatrix}
2 & 0 & 0  & 0  & 0 & 0  & 0 & 0 \\[4pt]
0                 & 12      & 0 & 0  & 0 & 0  & 0 & 0\\[4pt]
0              & 0 & 0 & 0  & 0 & 0  & 0 & 0\\[4pt]
0              & 0 & 0 & 6  & 0 & 0  & 0 & 0\\[4pt]
0              & 0 & 0 & 0  & 6 & 0  & 0 & 0\\[4pt]
0              & 0 & 0 & 0  & 0 & 6  & 0 & 0\\[4pt]
0              & 0 & 0 & 0  & 0 & 0  & 20 & 0\\[4pt]
0              & 0 & 0 & 0  & 0 & 0  & 0 & 6\\
\end{pmatrix}}\right.
\nonumber\\
&\hspace{ +1 cm}\left.
+\begin{pmatrix} 
V_{\Sigma_{g}} & 0 &  \sqrt{\frac{1}{3}} V_{\Sigma_g - \Sigma_g^\prime} & \sqrt{\frac{2}{3}} V_{\Sigma_g - \Sigma_g^\prime} & 0  & 0 & 0 & 0\\[4pt]
0 & V_{\Sigma_{g}} & 0 & 0 & 0 & \sqrt{\frac{3}{7}} V_{\Sigma_g - \Sigma_g^\prime} & \frac{2}{\sqrt{7}} V_{\Sigma_g - \Sigma_g^\prime} & 0\\[4pt]
\sqrt{\frac{1}{3}} V_{\Sigma_g - \Sigma_g^\prime}  & 0 &  \frac{V_{\Sigma_g^\prime} + 2 V_{\Pi_g}}{3} & \frac{ \sqrt{2} \, (V_{\Sigma_g^\prime} -  V_{\Pi_g})}{3} &0  & 0 & 0 & 0\\[4pt]
\sqrt{\frac{2}{3}} V_{\Sigma_g - \Sigma_g^\prime}  & 0 & \frac{ \sqrt{2} \,(V_{\Sigma_g^\prime} -  V_{\Pi_g})}{3} & \frac{2 V_{\Sigma_g^\prime} + V_{\Pi_g}}{3}&0  & 0 & 0 & 0\\[4pt]
0 & 0 & 0& 0 & V_{\Pi_g}  & 0 & 0 & 0\\[4pt]
0 & \sqrt{\frac{3}{7}} V_{\Sigma_g - \Sigma_g^\prime} & 0& 0 &0  & \frac{3 V_{\Sigma_g^\prime} + 4 V_{\Pi_g}}{7} & \frac{ 2 \sqrt{3} \,(V_{\Sigma_g^\prime} -  V_{\Pi_g})}{7}  & 0\\[4pt]
0 & \frac{2}{\sqrt{7}} V_{\Sigma_g - \Sigma_g^\prime} & 0& 0 &0  & \frac{ 2 \sqrt{3} \,(V_{\Sigma_g^\prime} -  V_{\Pi_g}) }{7} & \frac{4 V_{\Sigma_g^\prime} + 3 V_{\Pi_g}}{7} & 0 \\[4pt]
0 & 0 & 0& 0 &0  & 0 & 0 & V_{\Sigma_u} \\
\end{pmatrix}\right.
\nonumber\\
&\hspace{3cm}\left. +
\begin{pmatrix}
0 & 0 & 0 & 0 & 0 & 0 & 0 & 0\\[4pt]
0 & 0 & 0 & 0 & 0 & 0 & 0 & 0\\[4pt]
0 & 0 & \frac{\delta_Q }{2} & 0 &0 & 0 & 0 & 0\\[4pt]
0 & 0 & 0 & -\frac{\delta_Q}{4} & \frac{3 \sqrt{3} \, \delta_Q}{4 \sqrt{5}} & 0 & 0 & - \frac{ \sqrt{3} \delta_Q}{2 \sqrt{5}}\\[4pt]
0 & 0 & 0 & \frac{3 \sqrt{3} \, \delta_Q}{4 \sqrt{5}} & -\frac{\delta_Q}{12} & \frac{ \sqrt{7} \delta_Q}{3 \sqrt{5}} & 0 & \frac{\delta_Q }{2}\\[4pt]
0 & 0 & 0 & 0 & \frac{ \sqrt{7} \delta_Q}{3 \sqrt{5}} & -\frac{2 \delta_Q}{3} & 0 &  -\frac{ \sqrt{7}\delta_Q}{2 \sqrt{5}}\\[4pt]
0 & 0 & 0 & 0 &0 & 0 & \frac{\delta_Q }{2} & 0\\[4pt]
0 & 0 & 0 & -\frac{\sqrt{3} \delta_Q}{2 \sqrt{5}} & \frac{\delta_Q }{2} & - \frac{\sqrt{7} \delta_Q}{2 \sqrt{5}} & 0 & 0\\[4pt]
\end{pmatrix}      
\right]
\begin{pmatrix} 
\psi^P_{\Sigma_g} \\[6pt] 
\psi^F_{\Sigma_g} \\[6pt]  
\psi_{\Sigma_g^{\prime}+2\Pi_g} \\[6pt]  
\psi_{2\Sigma_g^{\prime}+\Pi_g} \\[6pt]  
\psi_{\Pi_g} \\[6pt] 
\psi_{3\Sigma_g^{\prime}+4\Pi_g} \\[6pt]  
\psi_{4\Sigma_g^{\prime}+3\Pi_g} \\[6pt]  
\psi_{\Sigma_u} \\
\end{pmatrix}
={\mathcal{E}} 
\begin{pmatrix} 
\psi^P_{\Sigma_g} \\[6pt]  
\psi^F_{\Sigma_g} \\[6pt]  
\psi_{\Sigma_g^{\prime}+2\Pi_g} \\[6pt]  
\psi_{2\Sigma_g^{\prime}+\Pi_g} \\[6pt]  
\psi_{\Pi_g} \\[6pt]  
\psi_{3\Sigma_g^{\prime}+4\Pi_g} \\[6pt]  
\psi_{4\Sigma_g^{\prime}+3\Pi_g} \\[6pt]  
\psi_{\Sigma_u} \\
\end{pmatrix},
\end{align}

meson-antimeson diabatic basis:
\begin{align}\label{eq:2++mes-mes}
&\hspace{+0 cm}\left[
-\frac{1}{m_Qr^2}\,\partial_rr^2\partial_r+
\frac{1}{m_Qr^2}
{\begin{pmatrix}
2 & 0 & 0  & 0  & 0 & 0  & 0 & 0 \\[4pt]
0                 & 12       & 0 & 0 & 0 & 0  & 0 & 0\\[4pt]
0              & 0 & 0 & 0  & 0 & 0  & 0 & 0\\[4pt]
0              & 0 & 0 & 6  & 0 & 0  & 0 & 0\\[4pt]
0              & 0 & 0 & 0  &6 & 0  & 0 & 0 \\[4pt]
0              & 0 & 0 & 0  & 0 & 6  & 0 & 0\\[4pt]
0              & 0 & 0 & 0  & 0 & 0  & 20 & 0\\[4pt]
0              & 0 & 0 & 0  & 0 & 0  & 0 & 6 \\[4pt]
\end{pmatrix}}\right.
\nonumber\\
&\hspace{-0.2 cm}\left.
\setlength{\arraycolsep}{0.3pt}
+\begin{pmatrix} 
V_{\Sigma_{g}} & 0 &  \sqrt{\frac{1}{3}} V_{\Sigma_g - \Sigma_g^\prime} & \sqrt{\frac{1}{10}} V_{\Sigma_g - \Sigma_g^\prime} & -\sqrt{\frac{3}{10}} V_{\Sigma_g - \Sigma_g^\prime}  & -\sqrt{\frac{1}{30}} V_{\Sigma_g - \Sigma_g^\prime} & 0 & \sqrt{\frac{7}{30}} V_{\Sigma_g - \Sigma_g^\prime}\\[6pt]
0 & V_{\Sigma_{g}} & 0 & \frac{1}{2} \sqrt{\frac{3}{5}} V_{\Sigma_g - \Sigma_g^\prime} & \sqrt{\frac{1}{5}} V_{\Sigma_g - \Sigma_g^\prime} & -\frac{1}{2 \sqrt{5}} V_{\Sigma_g - \Sigma_g^\prime} & \frac{2}{\sqrt{7}} V_{\Sigma_g - \Sigma_g^\prime} & \frac{1}{\sqrt{35}} V_{\Sigma_g - \Sigma_g^\prime}\\[6pt]
\sqrt{\frac{1}{3}} V_{\Sigma_g - \Sigma_g^\prime}  & 0 &  \frac{V_{\Sigma_g^\prime} + 2 V_{\Pi_g}}{3} &
\frac{V_{\Sigma_g^\prime} - V_{\Pi_g}}{\sqrt{30}} &
\frac{- V_{\Sigma_g^\prime} +  V_{\Pi_g}}{\sqrt{10}}  & 
\frac{- V_{\Sigma_g^\prime} +  V_{\Pi_g}}{3 \sqrt{10}}  & 
0  & 
\sqrt{\frac{7}{10}} \frac{V_{\Sigma_g^\prime}-V_{\Pi_g}}{3}\\[6pt]
\sqrt{\frac{1}{10}} V_{\Sigma_g - \Sigma_g^\prime}  &
\frac{1}{2} \sqrt{\frac{3}{5}} V_{\Sigma_g - \Sigma_g^\prime} &
\frac{V_{\Sigma_g^\prime} - V_{\Pi_g}}{\sqrt{30}}&
\frac{V_{\Sigma_g^\prime} + 2 V_{\Pi_g} + V_{\Sigma_u^-}}{4}&
0  & 
-\frac{V_{\Sigma_g^\prime} + 2 V_{\Pi_g} - 3 V_{\Sigma_u^-}}{4 \sqrt{3}} & 
\frac{\sqrt{3}(V_{\Sigma_g^\prime} - V_{\Pi_g})}{\sqrt{35}} & 
\frac{V_{\Sigma_g^\prime} - V_{\Pi_g}}{\sqrt{21}} \\[6pt]
-\sqrt{\frac{3}{10}} V_{\Sigma_g - \Sigma_g^\prime} &
\sqrt{\frac{1}{5}} V_{\Sigma_g - \Sigma_g^\prime} & 
\frac{- V_{\Sigma_g^\prime} +  V_{\Pi_g}}{\sqrt{10}} & 
0& 
\frac{ V_{\Sigma_g^\prime} +  V_{\Pi_g}}{2}  & 
0 & 
\frac{2(- V_{\Sigma_g^\prime} +  V_{\Pi_g})}{\sqrt{35}} &
\frac{- V_{\Sigma_g^\prime} +  V_{\Pi_g}}{2 \sqrt{7}}\\[6pt]
-\sqrt{\frac{1}{30}} V_{\Sigma_g - \Sigma_g^\prime} & -\frac{1}{2 \sqrt{5}} V_{\Sigma_g - \Sigma_g^\prime} &
\frac{- V_{\Sigma_g^\prime} +  V_{\Pi_g}}{3 \sqrt{10}}&
-\frac{V_{\Sigma_g^\prime} + 2 V_{\Pi_g} - 3 V_{\Sigma_u^-}}{4 \sqrt{3}} &
0  & 
\frac{V_{\Sigma_g^\prime} + 2 V_{\Pi_g} + 9 V_{\Sigma_u^-}}{12} &
\frac{- V_{\Sigma_g^\prime} +  V_{\Pi_g}}{\sqrt{35}} &
\frac{- V_{\Sigma_g^\prime} +  V_{\Pi_g}}{3 \sqrt{7}}\\[6pt]
0 & \frac{2}{\sqrt{7}} V_{\Sigma_g - \Sigma_g^\prime} &
0&
\frac{\sqrt{3}(V_{\Sigma_g^\prime} - V_{\Pi_g})}{\sqrt{35}} &
\frac{2( V_{\Sigma_g^\prime} -  V_{\Pi_g})}{\sqrt{35}} &
\frac{- V_{\Sigma_g^\prime} +  V_{\Pi_g}}{\sqrt{35}} &
\frac{4 V_{\Sigma_g^\prime} + 3 V_{\Pi_g}}{7} & 
\frac{ 2 (V_{\Sigma_g^\prime} -  V_{\Pi_g})}{7 \sqrt{5}}\\[6pt]
\sqrt{\frac{7}{30}} V_{\Sigma_g - \Sigma_g^\prime} & \frac{1}{\sqrt{35}} V_{\Sigma_g - \Sigma_g^\prime} &
\sqrt{\frac{7}{10}} \frac{V_{\Sigma_g^\prime}-V_{\Pi_g}}{3}&
\frac{V_{\Sigma_g^\prime} - V_{\Pi_g}}{\sqrt{21}} &
\frac{- V_{\Sigma_g^\prime} +  V_{\Pi_g}}{2 \sqrt{7}} &
\frac{- V_{\Sigma_g^\prime} +  V_{\Pi_g}}{3 \sqrt{7}} &
\frac{ 2 (V_{\Sigma_g^\prime} -  V_{\Pi_g})}{7 \sqrt{5}} &
\frac{11 V_{\Sigma_g^\prime} + 31 V_{\Pi_g}}{42}\\
\end{pmatrix}\right.
\nonumber\\
&\hspace{4.0 cm}\left. +
\begin{pmatrix}
0 & 0 & 0 & 0 & 0 & 0 & 0 & 0\\[4pt]
0 & 0 & 0 & 0 & 0 & 0 & 0 & 0\\[4pt]
0 & 0 & \frac{\delta_Q }{2} & 0 &0 & 0 & 0 & 0\\[4pt]
0 & 0 & 0 & -\frac{3 \delta_Q}{2} & 0 & 0 & 0 & 0\\[4pt]
0 & 0 & 0 & 0 & -\frac{\delta_Q}{2} & 0 & 0 & 0\\[4pt]
0 & 0 & 0 & 0 & 0 & \frac{\delta_Q}{2} & 0 &  0\\[4pt]
0 & 0 & 0 & 0 &0 & 0 & \frac{\delta_Q }{2} & 0\\[4pt]
0 & 0 & 0 & 0 & 0 & 0 & 0 & \frac{\delta_Q }{2}\\[4pt]
\end{pmatrix}      
\right]
\begin{pmatrix} 
   \psi^P_{Q \bar{Q}} \\[4pt] 
    \psi^F_{Q \bar{Q}} \\[4pt] 
    \psi^{S, s_{M \bar{M}} = 2}_{M^* \bar{M}^{*}} \\[4pt] 
    \psi_{M \bar{M}} \\[4pt] 
    \psi_{M \bar{M}^{*}} \\[4pt] 
    \psi^{D, s_{M^* \bar{M}^*} = 0}_{M^* \bar{M}^{*}} \\[4pt]
    \psi^{G, s_{M^* \bar{M}^*} = 2}_{M^* \bar{M}^{*}} \\[4pt]
    \psi^{D, s_{M^* \bar{M}^*} = 2}_{M^* \bar{M}^{*}} \\
\end{pmatrix}
={\mathcal{E}} 
\begin{pmatrix} 
   \psi^P_{Q \bar{Q}} \\[4pt] 
    \psi^F_{Q \bar{Q}} \\[4pt] 
    \psi^{S, s_{M^* \bar{M}^*} = 2}_{M^* \bar{M}^{*}} \\[4pt] 
    \psi_{M \bar{M}} \\[4pt] 
    \psi_{M \bar{M}^{*}} \\[4pt] 
    \psi^{D, s_{M^* \bar{M}^*} = 0}_{M^* \bar{M}^{*}} \\[4pt]
    \psi^{G, s_{M^* \bar{M}^*} = 2}_{M^* \bar{M}^{*}} \\[4pt]
    \psi^{D, s_{M^* \bar{M}^*} = 2}_{M^* \bar{M}^{*}} \\
\end{pmatrix}.
\end{align}

\subsection{$J^{PC} = 2^{--}$}

BO diabatic basis:

\begin{align}\label{eq:2--BOEFT}
&\hspace{-0.5 cm}\left[
-\frac{1}{m_Qr^2}\,\partial_rr^2\partial_r+\frac{1}{m_Qr^2}
{\begin{pmatrix}
6 & 0 & 0  & 0  & 0   \\[4pt]
0                 & 2        & 0 & 0 & 0\\[4pt]
0                 & 0 & 2 & 0  & 0 \\[4pt]
0                 & 0 & 0 & 12  & 0 \\[4pt]
0                 & 0 & 0 & 0  & 12 \\
\end{pmatrix}}\right.
\left.
+\begin{pmatrix} V_{\Sigma_{g}} &  0 &  \sqrt{\frac{2}{5}} V_{\Sigma_g - \Sigma_g^\prime} & \sqrt{\frac{3}{5}} V_{\Sigma_g - \Sigma_g^\prime} & 0 \\
    0 & V_{\Pi_g} & 0 & 0 & 0 \\[4pt]
     \sqrt{\frac{2}{5}} V_{\Sigma_g - \Sigma_g^\prime} & 0 & \frac{ 2 V_{\Sigma_g^\prime} + 3 V_{\Pi_g}}{5} & \frac{\sqrt{6} \, ( V_{\Sigma_g^\prime} - V_{\Pi_g})}{5}  & 0 \\[4pt]
      \sqrt{\frac{3}{5}} V_{\Sigma_g - \Sigma_g^\prime} & 0 & \frac{\sqrt{6} \, ( V_{\Sigma_g^\prime} - V_{\Pi_g})}{5}  & \frac{ 3 V_{\Sigma_g^\prime} + 2 V_{\Pi_g}}{5} & 0 \\[4pt]
      0 & 0 & 0 & 0 & V_{\Pi_g} \\[4pt]
      \end{pmatrix} \right. \nonumber \\
&\hspace{5.0 cm}\left.+
\begin{pmatrix}
 0 & 0 & 0 & 0 & 0 \\[4pt]
 0 & \frac{\delta_Q }{4} & \frac{ \sqrt{3} \delta_Q}{4} & 0 & 0 \\[4pt]
 0 & \frac{ \sqrt{3} \delta_Q}{4} & -\frac{\delta_Q }{4} & 0 & 0 \\[4pt]
 0 & 0 & 0 & \frac{\delta_Q}{6} & \frac{\sqrt{2} \delta_Q}{3} \\[4pt]
 0 & 0 & 0 & \frac{\sqrt{2} \delta_Q}{3} & -\frac{\delta_Q}{6} \\[4pt]
\end{pmatrix}      
      \right]
      \hspace{-4pt}\begin{pmatrix} \psi_{\Sigma_g} \\[6pt] \psi^P_{\Pi_g} \\[6pt] \psi_{2\Sigma_g^{\prime}+3\Pi_g} \\[6pt] \psi_{3\Sigma_g^{\prime}+2\Pi_g} \\[6pt] \psi^F_{\Pi_g}
      \end{pmatrix}={\mathcal{E}} 
      \begin{pmatrix} \psi_{\Sigma_g} \\[6pt] \psi^P_{\Pi_g} \\[6pt] \psi_{2\Sigma_g^{\prime}+3\Pi_g} \\[6pt] \psi_{3\Sigma_g^{\prime}+2\Pi_g} \\[6pt] \psi^F_{\Pi_g} \end{pmatrix}, 
\end{align}

meson-antimeson diabatic basis:

\begin{align}\label{eq:2--mes-mes}
&\hspace{- 0 cm}\left[
-\frac{1}{m_Qr^2}\,\partial_rr^2\partial_r+\frac{1}{m_Qr^2}
{\begin{pmatrix}
6 & 0 & 0  & 0  & 0   \\[4pt]
0                 & 2        & 0 & 0 & 0\\[4pt]
0                 & 0 & 2 & 0  & 0 \\[4pt]
0                 & 0 & 0 & 12  & 0 \\[4pt]
0                 & 0 & 0 & 0  & 12 \\[4pt]
\end{pmatrix}}\right.
\nonumber\\
&\hspace{2.0 cm}\left.
+\begin{pmatrix} V_{\Sigma_{g}} &  \sqrt{\frac{3}{10}} V_{\Sigma_g - \Sigma_g^\prime} &  \sqrt{\frac{1}{10}} V_{\Sigma_g - \Sigma_g^\prime} & -\sqrt{\frac{1}{5}} V_{\Sigma_g - \Sigma_g^\prime} & \sqrt{\frac{2}{5}} V_{\Sigma_g - \Sigma_g^\prime} \\[6pt]
    \sqrt{\frac{3}{10}} V_{\Sigma_g - \Sigma_g^\prime} & \frac{3 V_{\Sigma_g^\prime} + 7 V_{\Pi_g}}{10} & \frac{\sqrt{3} \, (V_{\Sigma_g^\prime} - V_{\Pi_g})}{10} & \sqrt{\frac{3}{2}} \frac{-V_{\Sigma_g^\prime} + V_{\Pi_g}}{5} & \frac{\sqrt{3} \, (V_{\Sigma_g^\prime} - V_{\Pi_g})}{5}  \\[6pt]
      \sqrt{\frac{1}{10}} V_{\Sigma_g - \Sigma_g^\prime} & \frac{\sqrt{3} \, (V_{\Sigma_g^\prime} - V_{\Pi_g})}{10} & \frac{ V_{\Sigma_g^\prime} + 9 V_{\Pi_g}}{10} & \frac{- V_{\Sigma_g^\prime} + V_{\Pi_g}}{5 \sqrt{2}} & \frac{ V_{\Sigma_g^\prime} - V_{\Pi_g}}{5} \\[6pt]
      -\sqrt{\frac{1}{5}} V_{\Sigma_g - \Sigma_g^\prime} & \sqrt{\frac{3}{2}}  \frac{-V_{\Sigma_g^\prime} + V_{\Pi_g}}{5} & \frac{- V_{\Sigma_g^\prime} + V_{\Pi_g}}{5 \sqrt{2}} & \frac{ V_{\Sigma_g^\prime} + 4 V_{\Pi_g}}{5} & \frac{\sqrt{2} \, (-V_{\Sigma_g^\prime} + V_{\Pi_g} )}{5}  \\[6pt]
      \sqrt{\frac{2}{5}} V_{\Sigma_g - \Sigma_g^\prime} & \frac{\sqrt{3}\,(V_{\Sigma_g^\prime} - V_{\Pi_g})}{5} & \frac{ V_{\Sigma_g^\prime} - V_{\Pi_g}}{5}& \frac{\sqrt{2} \, (-V_{\Sigma_g^\prime} + V_{\Pi_g} )}{5}  & \frac{2 V_{\Sigma_g^\prime} + 3 V_{\Pi_g}}{5} \\
      \end{pmatrix} \right. \nonumber \\
&\hspace{7cm}\left.+
\begin{pmatrix}
 0 & 0 & 0 & 0 & 0 \\[4pt]
 0 & -\frac{\delta_Q }{2} & 0 & 0 & 0 \\[4pt]
 0 & 0 & \frac{\delta_Q }{2} & 0 & 0 \\[4pt]
 0 & 0 & 0 & -\frac{\delta_Q }{2} & 0 \\[4pt]
 0 & 0 & 0 & 0 & \frac{\delta_Q }{2} \\[4pt]
\end{pmatrix}      
      \right]
      \hspace{-4pt}\begin{pmatrix} \psi_{Q \bar{Q}} \\[6pt] 
    \psi^P_{M \bar{M}^{*}} \\[6pt] 
    \psi^P_{M^* \bar{M}^{*}} \\[6pt] 
    \psi^F_{M \bar{M}^{*}} \\[6pt] 
    \psi^F_{M^* \bar{M}^{*}} \\
      \end{pmatrix}={\mathcal{E}} 
      \begin{pmatrix} \psi_{Q \bar{Q}} \\[6pt] 
    \psi^P_{M \bar{M}^{*}} \\[6pt] 
    \psi^P_{M^* \bar{M}^{*}} \\[6pt] 
    \psi^F_{M \bar{M}^{*}} \\[6pt] 
    \psi^F_{M^* \bar{M}^{*}} \\ \end{pmatrix}.
\end{align}

\section{$J^{PC}$  quarkonium states at $\mathcal{O}(1/m_Q)$: summary tables}\label{app:tetraquark spin splitting}
In Tables~\ref{tab:JPC=0-+}, ~\ref{tab:JPC=1+-}, ~\ref{tab:JPC=2-+}, ~\ref{tab:JPC=0++}, ~\ref{tab:JPC=1++}, ~\ref{tab:JPC=1--}, ~\ref{tab:JPC=2++} and  ~\ref{tab:JPC=2--}
we show the complete set of predominantly quarkonium and tetraquark bound states predicted by Eqs.~\eqref{eq:0-+BOEFT}, ~\eqref{eq:1+-BOEFT}, ~\eqref{eq:2-+BOEFT}, ~\eqref{eq:0++BOEFT}, ~\eqref{eq:1++BOEFT}, ~\eqref{eq:1--BOEFT}, ~\eqref{eq:2++BOEFT},  and ~\eqref{eq:2--BOEFT} below the lowest $M^{(*)} \bar{M}^{(*)}$ threshold (with $M^{(*)} = D^{(*)}, B^{(*)}$). 
Each table lists the mass, mean square radius, composition, binding energy relative to the spin-averaged meson–antimeson threshold $M \bar{M}^{\rm spin\, avg.}$, and threshold-splitting correction.
We consider states both in the charmonium and bottomonium sectors.
For consistency with Appendix~\ref{app:meson-meson basis}, the same notation for the wavefunction components in the BO diabatic basis is adopted here to denote the  percentages relative to these components.
Quarkonium bound states are labeled by the $nl$ quantum numbers of the dominant spin-averaged quarkonium component, whereas tetraquark bound states follow the naming convention of the  corresponding spin-averaged state. 
The $J^{PC}$ and $s_{Q\bar{Q}}$ quantum numbers appear in the tables' captions. 
In Table~\ref{tab:JPC=mes-mes}, we give in the meson–antimeson basis the composition of all bound states whose tetraquark component exceeds $10\%$ after including the spin-dependent potential $V_{SS}$.

\begin{table}
\centering
\renewcommand{\arraystretch}{1.6}
\begin{tabular}{|c||c|c|c|c|c|c|c|c|}  \hline
$nl$ & $M^{\rm th.}(M^{\rm exp.}) $ & $\sqrt{\langle r^2 \rangle} $   & $\%{\Sigma_g}$ & $\%{\Sigma_g^{'}}$  & $\%{\Sigma_u}$ & $E_{\rm bind.}^{M \bar{M}^{\rm spin\,avg.}}$ & $\Delta E^{\rm thr.\,spl.} $ \\
 & $(\mathrm{MeV})$ & $(\rm fm)$   &  &   &  & $(\mathrm{MeV})$ & $(\mathrm{MeV})$ \\
\hline\hline
$c \bar{c} $ &  &  & &   &  &  & \\
\hline
$1S$         & $3127.6$ $(2984.1)$  & $0.4$ & $ 99.9$  & $ 0.1$  &  & $-818.4$ & - \\
$2S$         & $3706.0$ $(3637.8)$  & $0.7$ & $98.8$  & $1.1$ & $0.1$  & $-240.0$ &  $-0.3$\\\hline
$b \bar{b} $ &  &    & &   &  &  & \\
\hline
$1S$         & $9444.9$ $(9398.7)$  & $0.2$ & $100$  &  &  & $-1182.1$ & - \\
$2S$         & $9987.8$ $(9999.0)$  & $0.5$ & $99.9$  & $0.1$ &   & $-639.2$ & - \\
$3S$         & $10327.5$  & $0.7$ & $99.0$  & $1.0$  &  & $-299.5$ & $-0.1$ \\
$4S$         & $10588.4$  & $1.0$ & $62.5$  & $32.8$ & $4.7$  & $-38.6$ & $-4.6$ \\\hline
\end{tabular}
\caption{The table lists the predominantly quarkonium bound states with $J^{PC} = 0^{-+}$ and $s_{Q \bar{Q}} = 0$ obtained from Eq.~\eqref{eq:0-+BOEFT} and lying below the physical $M \bar{M}^*$ threshold with $M = D, B$ in the charmonium and bottomonium sectors. 
The table includes the computed masses, mean square radii, compositions, binding energies with respect to the $M \bar{M}^{\rm spin \, avg.}$ threshold, and threshold spin-splitting corrections. 
The $nl$ quantum numbers used to label the quarkonium bound states are the same used to label the corresponding spin-averaged quarkonium states.  
The different components of the states are named after the diagonal entries of the static potential matrix in the BO diabatic basis. 
The dashes indicate threshold spin-splitting corrections lower than $0.1 \, \mathrm{MeV}$, while the blank boxes indicate tetraquark components whose percentages are inferior to $0.1 \%$.
If known, the experimental masses from~\cite{ParticleDataGroup:2024cfk} are shown in parentheses.
}
\label{tab:JPC=0-+}
\end{table}

\begin{table}[h!]
\centering
\renewcommand{\arraystretch}{1.6}
\begin{tabular}{|c||c|c|c|c|c|c|c|c|c|c|}  \hline
$nl $ & $M^{\rm th.}(M^{\rm exp.})(\mathrm{MeV})$ & $\sqrt{\langle r^2 \rangle}$   & $\%{\Sigma_g}$ & $\%{\Sigma_g^{'} + 2 \Pi_g}$ & $\%{2\Sigma_g^{'} + \Pi_g}$ & $\%{\Sigma_u^S} $ & $\%{\Sigma_u^D} $ & $E_{\rm bind.}^{M \bar{M}^{\rm spin \, avg.}}$ & $\Delta E^{\rm thr.\,spl.}$ \\
 & $(\mathrm{MeV})$ & $(\rm fm)$   &  &  &  &  &  & $(\mathrm{MeV})$ & $(\mathrm{MeV})$ \\
\hline\hline
$c \bar{c} $ &  &  & &   &  &  & &  & \\
\hline
$1P$         & $3536.8$ $(3525.4)$  & $0.6$ & $99.7$  & $0.2$  & $0.1$ &  &  & $-409.2$ & $-0.1$ \\\hline
$b \bar{b} $ &  &    & &   & &  & &  & -\\
\hline
$1P$         & $9882.9$ $(9899.3)$  & $0.4$ & $ 99.9$  &  & $0.1$ &  & & $-744.1$ & - \\
$2P$         & $10235.0$ $(10259.8)$  & $0.6$ & $99.5$  & $0.3$ & $0.2$ &  &  & $-392.0$ & $-0.1$ \\
$3P$         & $10515.5$  & $0.8$ & $94.7$  & $2.6$ & $2.7$ &  & & $-111.7$ & $-0.6$ \\
$X_b$         & $10601.3$  & $0.6$ & $1.8$  & $60.6$ & $0.7$ & $36.8$ & $0.1$ & $-25.9$ & $-24.6$ \\\hline
\end{tabular}
\caption{The table lists the predominantly quarkonium and tetraquark bound states with $J^{PC} = 1^{+-}$ and $s_{Q \bar{Q}} = 0$ obtained from Eq.~\eqref{eq:1+-BOEFT} and lying below the physical $M \bar{M}^*$ threshold with $M = D, B$ in the charmonium and bottomonium sectors. 
For the tetraquark bound state we use the same name as for the spin-averaged state.
All the rest is as in Table~\ref{tab:JPC=0-+}.
}
\label{tab:JPC=1+-}
\end{table}

\begin{table}
\centering
\renewcommand{\arraystretch}{1.6}
\begin{tabular}{|c||c|c|c|c|c|c|c|c|c|c|}  \hline
$nl $ & $M^{\rm th.}(M^{\rm exp.}) $ & $\sqrt{\langle r^2 \rangle}$   & $\%{\Sigma_g}$ & $\%{2 \Sigma_g^{'} + 3 \Pi_g}$ & $\%{3 \Sigma_g^{'} + 2 \Pi_g}$ & $\%{\Sigma_u^P} $ & $\%{\Sigma_u^F} $ & $E_{\rm bind.}^{M \bar{M}^{\rm spin\,avg.}} $ & $\Delta E^{\rm thr.\,spl.}$ \\
 & $(\mathrm{MeV})$ & $(\rm fm)$   &  &  &  &  &  & $(\mathrm{MeV})$ & $(\mathrm{MeV})$ \\
\hline\hline
$c \bar{c} $ &  &  & &   &  &  & &  & \\
\hline
$1D$         & $3818.5$  & $0.8$ & $98.0$  & $1.4$  & $0.4$ & $0.2$ & & $-127.5$ & $-0.5$ \\\hline
$b \bar{b} $ &  &    & &   & &  & &  &  \\
\hline
$1D$         & $10124.5$  & $0.5$ & $99.8$  & $0.1$  & $0.1$ &  & & $-502.6$ & - \\
$2D$         & $10418.8$  & $0.7$ & $98.6$  & $0.8$ & $0.6$ &  &  & $-208.2$ & $-0.2$ \\\hline
\end{tabular}
\caption{
The table lists the predominantly quarkonium bound states with $J^{PC} = 2^{-+}$ and $s_{Q \bar{Q}} = 0$ obtained from Eq.~\eqref{eq:2-+BOEFT} and lying below the physical $M \bar{M}^*$ threshold with $M = D, B$ in the charmonium and bottomonium sectors. 
All the rest is as in Table~\ref{tab:JPC=0-+}.
}
\label{tab:JPC=2-+}
\end{table}

\begin{table}
\centering
\renewcommand{\arraystretch}{1.6}
\begin{tabular}{|c||c|c|c|c|c|c|c|c|c|}  \hline
$nl $ & $M^{\rm th.}(M^{\rm exp.}) \; (\mathrm{MeV})$ & $\sqrt{\langle r^2 \rangle}$   & $\%{\Sigma_g}$ & $\%{\Sigma_g^{'} + 2 \Pi_g}$ & $\%{2 \Sigma_g^{'} + \Pi_g}$ & $\%{\Sigma_u} $  & $E_{\rm bind.}^{M \bar{M}^{\rm spin\,avg.}}$ & $\Delta E^{\rm thr.\,spl.}$ \\
 & $(\mathrm{MeV})$ & $(\rm fm)$   &  &  &  &   & $(\mathrm{MeV})$ & $(\mathrm{MeV})$ \\
\hline\hline
$c \bar{c} $ &  &  & &   &  &  & &  \\
\hline
$1P$         & $3536.5$ $(3414.7)$  & $0.6$ & $99.4$  & $0.4$  & $0.2$ &  & $-409.5$ & $-0.4$ \\\hline
$b \bar{b} $ &  &    & &   & &  & &  \\
\hline
$1P$         & $9882.9$ $(9859.4)$  & $0.4$ & $99.9$  & $ 0.1$  &  & & $-744.1$ & - \\
$2P$         & $10234.9$ $(10232.5)$  & $0.6$ & $99.4$  & $0.3$ & $0.3$ &   & $-392.0$ & $-0.1$ \\
$3P$         & $10513.7$  & $0.8$ & $88.9$  & $8.3$ & $2.0$ & $0.8$  & $-113.3$ & $-2.8$ \\
$X_b$         & $10555.9$  & $0.6$ & $7.1$  & $73.1$ & $0.1$ & $19.7$  & $-71.3$ & $-82.9$ \\\hline
\end{tabular}
\caption{
The table lists the predominantly quarkonium and tetraquark bound states with $J^{PC} = 0^{++}$ and $s_{Q \bar{Q}} = 1$ obtained from Eq.~\eqref{eq:0++BOEFT} and lying below the physical $M \bar{M}$ threshold with $M = D, B$ in the charmonium and bottomonium sectors. 
All the rest is as in Table~\ref{tab:JPC=1+-}.
}
\label{tab:JPC=0++}
\end{table}

\begin{table}
\centering
\renewcommand{\arraystretch}{1.6}
\begin{tabular}{|c||c|c|c|c|c|c|c|c|c|}  \hline
$nl $ & $M^{\rm th.}(M^{\rm exp.})$ & $\sqrt{\langle r^2 \rangle}$   & $\%{\Sigma_g}$ & $\%{\Sigma_g^{'} + 2 \Pi_g}$ & $\%{2 \Sigma_g^{'} + \Pi_g}$ & $\%{\Pi_g} $  & $E_{\rm bind.}^{M \bar{M}^{\rm spin\,avg.}}$ & $\Delta E^{\rm thr.\,spl.}$ \\
 & $(\mathrm{MeV})$ & $(\rm fm)$   &  &  &  &   & $(\mathrm{MeV})$ & $(\mathrm{MeV})$ \\
\hline\hline
$c \bar{c} $ &  &  & &   &  &  & &  \\
\hline
$1P$         & $3536.7$ $(3510.7)$  & $0.6$ & $99.7$  & $0.2$  & $0.1$ &  & $-409.2$ & $-0.2$ \\
$\chi_{c1}(3872)$         & $3875.4$ $(3871.6)$  & $4.3$ & $5.2$  & $94.3$  & $0.5$ &  & $-70.6$ & $-70.5$ \\\hline
$b \bar{b} $ &  &    & &   & &  & &  \\
\hline
$1P$         & $9882.9$ $(9892.8)$  & $0.4$ & $99.9$  &   & $0.1$ & & $-744.1$ & - \\
$2P$         & $10235.0$ $(10255.5)$  & $0.6$ & $99.5$  & $0.3$ & $0.2$ &   & $-392.0$ & - \\
$3P$         & $10515.2$ $(10513.4)$  & $0.8$ & $94.1$  & $3.4$ & $2.4$ &   & $-111.4$ & $-0.9$ \\
$X_b$         & $10595.9$  & $0.6$ & $3.2$  & $96.2$ & $0.5$ & $0.1$ & $-31.0$ & $-29.9$ \\\hline
\end{tabular}
\caption{
The table lists the predominantly quarkonium and tetraquark bound states with $J^{PC} = 1^{++}$ and $s_{Q \bar{Q}} = 1$ obtained from Eq.~\eqref{eq:1++BOEFT} and lying below the physical $M \bar{M}^*$ threshold with $M = D, B$ in the charmonium and bottomonium sectors. 
All the rest is as in Table~\ref{tab:JPC=1+-}.
}
\label{tab:JPC=1++}
\end{table}

\begin{table}
\centering
\scalebox{1.0}{%
\renewcommand{\arraystretch}{1.6}
\begin{tabular}{|c||c|c|c|c|c|c|c|c|c|c|c|c|}  \hline
$nl $ & $M^{\rm th.} (M^{\rm exp.})$ & $\sqrt{\langle r^2 \rangle}$   & $\%{\Sigma_g^{S}}$ & $\%{\Sigma_g^{D}}$ & $\%{\Sigma_g^{'}}$ & $\%{\Pi_g} $& $\%{2 \Sigma_g^{'} + 3\Pi_g}$ & $\%{3 \Sigma_g^{'} + 2\Pi_g}$  & $\%{\Sigma_u^-} $ &$E_{\rm bind.}^{M \bar{M}^{\rm spin\,avg.}}$ & $\Delta E^{\rm thr.\,spl.}$\\
 & $(\mathrm{MeV})$ & $(\rm fm)$   &  &  &  &  &  &  &  &$(\mathrm{MeV})$ & $(\mathrm{MeV})$ \\
\hline\hline
$c \bar{c} $ &  &  & &   &  &  & & &  & & \\
\hline
$1S$         & $3127.6$ $(3096.9)$  & $0.4$ & $100$  &  & &  & & &  & $-818.4$ & - \\
$2S$         & $3705.6$ $(3686.1)$  & $0.7$ & $98.2$  &  & $1.4$ & $0.1$  & $0.1$ & & $0.2$  & $-240.3$ & $-0.6$ \\\hline
$b \bar{b} $ &  &    & & &  & &  & &  & & - \\
\hline
$1S$         & $9444.9$ $(9460.4)$  & $0.2$ & $100$  &   & &  & & & & $-1182.1$ & - \\
$2S$         & $9987.7$ $(10023.4)$  & $0.5$ & $99.8$  & & $0.2$ & &  & &  & $-639.3$ & - \\
$1D$         & $10124.4$  & $0.5$ &   & $99.8$ & & & $0.1$ & $0.1$ &  & $-502.6$ & $-0.1$ \\
$3S$         & $10327.5$ $(10355.1)$  & $0.7$ & $99.0$  & & $1.0$ & &  & &  & $-299.5$ & $-0.1$ \\
$2D$         & $10418.6$  & $0.7$ &  & $98.4$ & & & $0.6$ & $1.0$&  & $-208.4$ & $-0.3$ \\\hline
\end{tabular}}
\caption{The table lists the predominantly quarkonium bound states with $J^{PC} = 1^{--}$ and $s_{Q \bar{Q}} = 1$ obtained from Eq.~\eqref{eq:1--BOEFT} and lying below the physical $M \bar{M}$ threshold with $M = D, B$ in the charmonium and bottomonium sectors. 
In order to differentiate between the percentages relative to the $S$- and $D$-wave quarkonium components, we add the superscripts $S$ and $D$ to the corresponding entries. 
All the rest is as in Table~\ref{tab:JPC=0-+}.
}
\label{tab:JPC=1--}
\end{table}

\begin{table}
\centering
\scalebox{0.9}{%
\renewcommand{\arraystretch}{1.6}
\begin{tabular}{|c||c|c|c|c|c|c|c|c|c|c|c|c|c|}  \hline
$nl $ & $M^{\rm th.}(M^{\rm exp.})$ & $\sqrt{\langle r^2 \rangle}$   & $\%{\Sigma_g^{P}}$ & $\%{\Sigma_g^{F}}$ & $\%{\Sigma_g^{\prime} + 2\Pi_g}$ & $\%{2 \Sigma_g^\prime + \Pi_g}$  & $\%{\Pi_g} $ & $\%{3 \Sigma_g^\prime + 4 \Pi_g}$ & $\%{4 \Sigma_g^\prime + 3 \Pi_g}$  & $\%{\Sigma_u}$ &$E_{\rm bind.}^{M \bar{M}^{\rm spin\,avg.}}$ & $\Delta E^{\rm thr.\,spl.}$ \\
 & $(\mathrm{MeV})$ & $(\rm fm)$ &  &  &  &  &  &  &  &  &$(\mathrm{MeV})$ & $(\mathrm{MeV})$ \\
\hline\hline
$c \bar{c} $ &  &  & &   &  &  &  & & &  & &\\
\hline
$1P$         & $3536.9$ $(3556.2)$  & $0.6$ & $99.7$  &  & $0.1$ & $0.2$ & &  & & & $-409.1$ & - \\\hline
$b \bar{b} $ &  &    & & &  & &  & &  &  & & \\
\hline
$1P$         & $9882.9$ $(9912.2)$  & $0.4$ & $99.9$  &  &  & $0.1$ &  & & & & $-744.1$ & - \\
$2P$         & $10235.0$ $(10268.7)$  & $0.6$ & $99.5$  &  & $0.2$ & $0.3$ & &  & &  & $-392.0$ & -\\
$1F$         & $10314.7$  & $0.6$ &   & $99.5$ &  & &  & $0.3$ & $0.2$ & & $-312.3$ & - \\
$3P$         & $10515.6$ $(10524.0)$  & $0.8$ & $95.0$  &  & $1.5$ & $3.0$ & $0.4$ &  & & $0.1$ & $-111.4$ & $-0.2$ \\\hline
\end{tabular}}
\caption{The table lists the predominantly quarkonium bound states with $J^{PC} = 2^{++}$ and $s_{Q \bar{Q}} = 1$ obtained from Eq.~\eqref{eq:2++BOEFT} and lying below the physical $M \bar{M}$ threshold with $M = D, B$ in the charmonium and bottomonium sectors. 
In order to differentiate between the percentages relative to the $P$- and $F$-wave quarkonium components, we add the superscripts $P$ and $F$ to the corresponding entries. 
All the rest is as in Table~\ref{tab:JPC=0-+}.
}
\label{tab:JPC=2++}
\end{table}

\begin{table}
\centering
\renewcommand{\arraystretch}{1.6}
\begin{tabular}{|c||c|c|c|c|c|c|c|c|c|c|}  \hline
$nl $ & $M^{\rm th.}(M^{\rm exp.}) \; (\mathrm{MeV})$ & $\sqrt{\langle r^2 \rangle}$   & $\%{\Sigma_g}$ &  $\%{\Pi_g^P} $& $\%{2 \Sigma_g^{'} + 3 \Pi_g}$ & $\%{3 \Sigma_g^{'} + 2\Pi_g}$  & $\%{\Pi_g^F} $ &$E_{\rm bind.}^{M \bar{M}^{\rm spin\,avg.}}$ & $\Delta E^{\rm thr.\,spl.}$ \\
 & $(\mathrm{MeV})$ & $(\rm fm)$   &  &  &  &  &  & $(\mathrm{MeV})$ & $(\mathrm{MeV})$ \\
\hline\hline
$c \bar{c} $ &  &  & &   &  &  &  & & \\
\hline
$1D$         & $3818.2$ $(3823.5)$  & $0.8$ & $97.6$  & $0.1$  & $1.9$ & $0.4$ &  & $-127.8$ & $-0.8$ \\\hline
$b \bar{b} $ &  &    & & &  & &  & &   \\
\hline
$1D$         & $10124.4$ $(10163.7)$  & $0.5$ & $99.8$  &    & $0.1$ & $0.1$ & & $-502.6$ & $-0.1$ \\
$2D$         & $10418.7$  & $0.7$ & $98.5$  & & $0.9$ & $0.6$ &   & $-208.3$ & $-0.2$ \\\hline
\end{tabular}
\caption{The table lists the predominantly quarkonium bound states with $J^{PC} = 2^{--}$ and $s_{Q \bar{Q}} = 1$ obtained from Eq.~\eqref{eq:2--BOEFT} and lying below the physical $M \bar{M}^*$ threshold with $M = D, B$ in the charmonium and bottomonium sectors. 
In order to differentiate between the percentages relative to the $P$- and $F$-wave tetraquark  components, we add the superscripts $P$ and $F$ to the corresponding entries. 
All the rest is as in Table~\ref{tab:JPC=0-+}.
}
\label{tab:JPC=2--}
\end{table}

\begin{table}
\centering
\scalebox{1.0}{%
\renewcommand{\arraystretch}{1.6}
\begin{tabular}{|c||c|c|c|c|c|c|c|c|c|}  \hline
  & $M^{\rm th.} \; (\mathrm{MeV})$ & $\% Q \bar{Q}$   & $\% M \bar{M}^{S}$ & $\% M \bar{M}^{*S}$ & $\% M \bar{M}^{*P}$ & $\% M \bar{M}^{*D}$ & $\% M^* \bar{M}^{*S}$ & $\% M^* \bar{M}^{*P}$ & $\% M^* \bar{M}^{*D}$\\
\hline\hline
$J^{PC} = 0^{-+}$         &           &       &        &     &       &   &   &  &  \\
$4S\,(b \bar{b})$   & $10588.4$  & $62.5$ & - & -   & $6.7$ & -  & -  &  $30.8$ & - \\\hline
$J^{PC} = 1^{+-}$         &           &       &        &     &       &   &   &  &  \\
$X_b\,(b \bar{b})$   & $10601.3$  & $1.8$ & - & $2.0$   & - & $0.2$ & $95.3$  & $0.7$  & - \\\hline
$J^{PC} = 0^{++}$         &           &       &        &     &       &   &   &  &  \\
$3P\,(b \bar{b})$   & $10513.7$  & $88.9$ & $8.6$ & -   & - & -  & $2.0$  & - & $0.5$\\
$X_b\,(b \bar{b})$   & $10555.9$  & $7.0$ & $92.4$ & -   & - & -  & $0.1$  & - & $0.5$\\\hline
$J^{PC} = 1^{++}$         &           &       &        &     &       &   &   &  &  \\
$\chi_{c1}(3872)\,(c \bar{c})$   & $3875.4$  & $5.2$ & - & $94.6$  & - &   & -  & - & $1.2$\\
$X_b\,(b \bar{b})$   & $10595.9$  & $3.2$ & - & $96.1$  & - & $0.1$  & -  & - & $0.6$\\\hline
\end{tabular}}
\caption{The table lists in the meson-antimeson basis the composition of the quarkonium and tetraquark states below the corresponding thresholds with more than $10 \%$ of tetraquark component. 
The $M^{(*)} \bar{M}^{(*)}$ components are classified according to the value of the heavy quark-antiquark pair orbital angular momentum $l_{Q \bar{Q}} = 0$, $1$, $2$ in $S$-, $P$-, and $D$- wave components, which are identified by the superscripts $S$, $P$ and $D$, respectively.} 
\label{tab:JPC=mes-mes}
\end{table}

\newpage

\bibliography{bibliography}

@article{E598:1974sol,
    author = "Aubert, J. J. and others",
    collaboration = "E598",
    title = "{Experimental Observation of a Heavy Particle $J$}",
    reportNumber = "COO-3069-271",
    doi = "10.1103/PhysRevLett.33.1404",
    journal = "Phys. Rev. Lett.",
    volume = "33",
    pages = "1404--1406",
    year = "1974"
}

@article{Aoki:2025jvi,
    author = "Aoki, Sinya and Doi, Takumi and Lyu, Yan",
    title = "{Left-hand cut and the HAL QCD method}",
    eprint = "2501.16804",
    archivePrefix = "arXiv",
    primaryClass = "hep-lat",
    reportNumber = "YITP-25-10, RIKEN-iTHEMS-Report-25",
    doi = "10.22323/1.466.0089",
    journal = "PoS",
    volume = "LATTICE2024",
    pages = "089",
    year = "2025"
}

@article{Alberti:2016dru,
    author = {Alberti, Maurizio and Bali, Gunnar S. and Collins, Sara and Knechtli, Francesco and Moir, Graham and S\"oldner, Wolfgang},
    title = "{Hadroquarkonium from lattice QCD}",
    eprint = "1608.06537",
    archivePrefix = "arXiv",
    primaryClass = "hep-lat",
    reportNumber = "WUB-16-03, DAMTP-2016-50",
    doi = "10.1103/PhysRevD.95.074501",
    journal = "Phys. Rev. D",
    volume = "95",
    number = "7",
    pages = "074501",
    year = "2017"
}

@article{SLAC-SP-017:1974ind,
    author = "Augustin, J. E. and others",
    collaboration = "SLAC-SP-017",
    title = "{Discovery of a Narrow Resonance in $e^+ e^-$ Annihilation}",
    reportNumber = "SLAC-PUB-1504, LBL-3391",
    doi = "10.1103/PhysRevLett.33.1406",
    journal = "Phys. Rev. Lett.",
    volume = "33",
    pages = "1406--1408",
    year = "1974"
}

@article{Bacci:1974za,
    author = "Bacci, C. and others",
    title = "{Preliminary Result of Frascati (ADONE) on the Nature of a New 3.1-GeV Particle Produced in $e^+$ $e^-$ Annihilation}",
    reportNumber = "LNF-74/61-P",
    doi = "10.1103/PhysRevLett.33.1408",
    journal = "Phys. Rev. Lett.",
    volume = "33",
    pages = "1408",
    year = "1974",
    note = "[Erratum: Phys.Rev.Lett. 33, 1649 (1974)]"
}

@article{Abrams:1974yy,
    author = "Abrams, G. S. and others",
    title = "{The Discovery of a Second Narrow Resonance in $e^+$ $e^-$ Annihilation}",
    reportNumber = "SLAC-PUB-1510, LBL-3605",
    doi = "10.1103/PhysRevLett.33.1453",
    journal = "Phys. Rev. Lett.",
    volume = "33",
    pages = "1453--1455",
    year = "1974"
}

@article{E288:1977efs,
    author = "Innes, Walter R. and others",
    collaboration = "E288",
    title = "{Observation of Structure in the $\Upsilon$ Region}",
    reportNumber = "FERMILAB-PUB-77-086-E",
    doi = "10.1103/PhysRevLett.39.1240",
    journal = "Phys. Rev. Lett.",
    volume = "39",
    pages = "1240--1242",
    year = "1977",
    note = "[Erratum: Phys.Rev.Lett. 39, 1640 (1977)]"
}

@article{CLEO:1980qvy,
    author = "Andrews, D. and others",
    collaboration = "CLEO",
    title = "{Observation of Three Upsilon States}",
    reportNumber = "CLNS 80/445",
    doi = "10.1103/PhysRevLett.44.1108",
    journal = "Phys. Rev. Lett.",
    volume = "44",
    pages = "1108",
    year = "1980"
}

@article{PhysRevLett.44.1111,
  title = {Observation of $\ensuremath{\Upsilon}$, ${\ensuremath{\Upsilon}}^{\ensuremath{'}}$, and ${\ensuremath{\Upsilon}}^{\ensuremath{''}}$ at the Cornell Electron Storage Ring},
  author = {B\"ohringer, T. and Costantini, F. and Dobbins, J. and Franzini, P. and Han, K. and Herb, S. W. and Kaplan, D. M. and Lederman, L. M. and Mageras, G. and Peterson, D. and Rice, E. and Yoh, J. K. and Finocchiaro, G. and Lee-Franzini, J. and Giannini, G. and Schamberger, R. D. and Sivertz, M. and Spencer, L. J. and Tuts, P. M.},
  journal = {Phys. Rev. Lett.},
  volume = {44},
  issue = {17},
  pages = {1111--1114},
  numpages = {0},
  year = {1980},
  month = {Apr},
  publisher = {American Physical Society},
  doi = {10.1103/PhysRevLett.44.1111},
  url = {https://link.aps.org/doi/10.1103/PhysRevLett.44.1111}
}

@article{Rapidis:1977cv,
    author = "Rapidis, Petros A. and others",
    title = "{Observation of a Resonance in $e^+ e^-$ Annihilation Just Above Charm Threshold}",
    reportNumber = "SLAC-PUB-1959, LBL-6484",
    doi = "10.1103/PhysRevLett.39.526",
    journal = "Phys. Rev. Lett.",
    volume = "39",
    pages = "526",
    year = "1977",
    note = "[Erratum: Phys.Rev.Lett. 39, 974 (1977)]"
}

@article{Abrams:1979cx,
    author = "Abrams, G. S. and others",
    title = "{Measurement of the Parameters of the $\psi^{\prime\prime}$(3770) Resonance}",
    reportNumber = "SLAC-PUB-2411, LBL-9938",
    doi = "10.1103/PhysRevD.21.2716",
    journal = "Phys. Rev. D",
    volume = "21",
    pages = "2716",
    year = "1980"
}

@article{CLEO:1980tem,
    author = "Andrews, D. and others",
    collaboration = "CLEO",
    title = "{Observation of a Fourth Upsilon State in $e^+ e^-$ Annihilations}",
    reportNumber = "CLNS 80/452",
    doi = "10.1103/PhysRevLett.45.219",
    journal = "Phys. Rev. Lett.",
    volume = "45",
    pages = "219",
    year = "1980"
}

@article{PhysRevLett.34.43,
  title = {Heavy Quarks and ${e}^{+}{e}^{\ensuremath{-}}$ Annihilation},
  author = {Appelquist, Thomas and Politzer, H. David},
  journal = {Phys. Rev. Lett.},
  volume = {34},
  issue = {1},
  pages = {43--45},
  numpages = {0},
  year = {1975},
  month = {Jan},
  publisher = {American Physical Society},
  doi = {10.1103/PhysRevLett.34.43},
  url = {https://link.aps.org/doi/10.1103/PhysRevLett.34.43}
}

@article{PhysRevLett.34.369,
  title = {Spectrum of Charmed Quark-Antiquark Bound States},
  author = {Eichten, E. and Gottfried, K. and Kinoshita, T. and Kogut, J. and Lane, K. D. and Yan, T. -M.},
  journal = {Phys. Rev. Lett.},
  volume = {34},
  issue = {6},
  pages = {369--372},
  numpages = {0},
  year = {1975},
  month = {Feb},
  publisher = {American Physical Society},
  doi = {10.1103/PhysRevLett.34.369},
  url = {https://link.aps.org/doi/10.1103/PhysRevLett.34.369}
}

@article{Bhanot:1978mj,
    author = "Bhanot, Gyan and Rudaz, Serge",
    title = "{A New Potential for Quarkonium}",
    reportNumber = "CLNS-393",
    doi = "10.1016/0370-2693(78)90362-3",
    journal = "Phys. Lett. B",
    volume = "78",
    pages = "119--124",
    year = "1978"
}

@article{Richardson:1978bt,
    author = "Richardson, John L.",
    title = "{The Heavy Quark Potential and the Upsilon, J/$\psi$ systems}",
    reportNumber = "SLAC-PUB-2229",
    doi = "10.1016/0370-2693(79)90753-6",
    journal = "Phys. Lett. B",
    volume = "82",
    pages = "272--274",
    year = "1979"
}

@article{PhysRevLett.45.103,
  title = {Regge Slope and the $\ensuremath{\Lambda}$ Parameter in Quantum Chromodynamics: An Empirical Approach via Quarkonia},
  author = {Buchm\"uller, W. and Grunberg, G. and Tye, S. -H. H.},
  journal = {Phys. Rev. Lett.},
  volume = {45},
  issue = {2},
  pages = {103--106},
  year = {1980},
  month = {Jul},
  publisher = {American Physical Society},
  doi = {10.1103/PhysRevLett.45.103},
  url = {https://link.aps.org/doi/10.1103/PhysRevLett.45.103}
}

@article{Martin:1980jx,
    author = "Martin, Andre",
    title = "{A fit of Upsilon and Charmonium Spectra}",
    reportNumber = "CERN-TH-2843",
    doi = "10.1016/0370-2693(80)90527-4",
    journal = "Phys. Lett. B",
    volume = "93",
    pages = "338--342",
    year = "1980"
}

@article{Buchmuller:1980su,
    author = "Buchm{\"u}ller, W. and Tye, S. H. H.",
    title = "{Quarkonia and Quantum Chromodynamics}",
    reportNumber = "FERMILAB-PUB-80-094-T",
    doi = "10.1103/PhysRevD.24.132",
    journal = "Phys. Rev. D",
    volume = "24",
    pages = "132",
    year = "1981"
}

@article{Quigg,
    author = "Quigg, Chris",
    title = "{Bound states of heavy quarks and antiquarks}",
    doi = "",
    journal = "9th international symposium on lepton and photon interactions at high energy. , Batavia, IL, USA., 23 - 29 Aug 1979",
    volume = "",
    pages = "",
    year = ""
}

@article{Micu:1968mk,
    author = "Micu, L.",
    title = "{Decay rates of meson resonances in a quark model}",
    doi = "10.1016/0550-3213(69)90039-X",
    journal = "Nucl. Phys. B",
    volume = "10",
    pages = "521--526",
    year = "1969"
}

@article{Carlitz:1970xb,
    author = "Carlitz, Robert D. and Kislinger, M.",
    title = "{Regge amplitude arising from $\mathrm{su}(6)_W$ vertices}",
    doi = "10.1103/PhysRevD.2.336",
    journal = "Phys. Rev. D",
    volume = "2",
    pages = "336--342",
    year = "1970"
}

@article{LeYaouanc:1972vsx,
    author = "Le Yaouanc, A. and Oliver, L. and Pene, O. and Raynal, J. C.",
    title = "{Naive quark pair creation model of strong interaction vertices}",
    doi = "10.1103/PhysRevD.8.2223",
    journal = "Phys. Rev. D",
    volume = "8",
    pages = "2223--2234",
    year = "1973"
}

@article{LeYaouanc:1977fsz,
    author = "Le Yaouanc, A. and Oliver, L. and Pene, O. and Raynal, J. -C.",
    title = "{Strong Decays of $\psi^{\prime \prime}(4.028)$ as a Radial Excitation of Charmonium}",
    reportNumber = "LPTHE 77/25",
    doi = "10.1016/0370-2693(77)90250-7",
    journal = "Phys. Lett. B",
    volume = "71",
    pages = "397--399",
    year = "1977"
}

@article{Eichten:1978tg,
    author = "Eichten, E. and Gottfried, K. and Kinoshita, T. and Lane, K. D. and Yan, Tung-Mow",
    title = "{Charmonium: The Model}",
    reportNumber = "CLNS-375",
    doi = "10.1103/PhysRevD.17.3090",
    journal = "Phys. Rev. D",
    volume = "17",
    pages = "3090",
    year = "1978",
    note = "[Erratum: Phys.Rev.D 21, 313 (1980)]"
}

@article{Eichten:1979ms,
    author = "Eichten, E. and Gottfried, K. and Kinoshita, T. and Lane, K. D. and Yan, Tung-Mow",
    title = "{Charmonium: Comparison with Experiment}",
    reportNumber = "CLNS-425",
    doi = "10.1103/PhysRevD.21.203",
    journal = "Phys. Rev. D",
    volume = "21",
    pages = "203",
    year = "1980"
}

@article{Ono:1980js,
    author = "Ono, Seiji",
    title = "{Strong Decay Widths of Bottomonium States Above Threshold}",
    reportNumber = "PITHA-80/02",
    doi = "10.1103/PhysRevD.23.1118",
    journal = "Phys. Rev. D",
    volume = "23",
    pages = "1118",
    year = "1981"
}

@article{Heikkila:1983wd,
    author = "Heikkila, K. and Ono, Seiji and Tornqvist, N. A.",
    title = "{Heavy $c \bar{c}$ and $b \bar{b}$ quarkonium states and unitarity effects}",
    reportNumber = "HU-TFT-83-4",
    doi = "10.1103/PhysRevD.29.2136",
    journal = "Phys. Rev. D",
    volume = "29",
    pages = "110",
    year = "1984",
    note = "[Erratum: Phys.Rev.D 29, 2136 (1984)]"
}

@article{Ono:1983rd,
    author = "Ono, S. and Tornqvist, N. A.",
    title = "{Continuum Mixing and Coupled Channel Effects in $c \bar{c}$ and $b \bar{b}$ Quarkonium}",
    reportNumber = "CERN-TH-3729",
    doi = "10.1007/BF01558041",
    journal = "Z. Phys. C",
    volume = "23",
    pages = "59",
    year = "1984"
}

@article{Tornqvist:1984xz,
    author = "Tornqvist, Nils A.",
    title = "{Quarkonium and Quark Loops}",
    reportNumber = "HU-TFT-84-47",
    journal = "Acta Phys. Polon. B",
    volume = "16",
    pages = "503",
    year = "1985",
    note = "[Erratum: Acta Phys.Polon.B 16, 683 (1985)]"
}

@article{Roberts:1992esl,
    author = "Roberts, W. and Silvestre-Brac, B.",
    title = "{General method of calculation of any hadronic decay in the $^3P_0$ model}",
    doi = "10.1007/bf01641821",
    journal = "Few Body Syst.",
    volume = "11",
    number = "4",
    pages = "171--193",
    year = "1992"
}

@article{Gell-Mann:1964ewy,
    author = "Gell-Mann, Murray",
    title = "{A Schematic Model of Baryons and Mesons}",
    doi = "10.1016/S0031-9163(64)92001-3",
    journal = "Phys. Lett.",
    volume = "8",
    pages = "214--215",
    year = "1964"
}

@article{Jaffe:1975fd,
    author = "Jaffe, R. L. and Johnson, K.",
    title = "{Unconventional States of Confined Quarks and Gluons}",
    reportNumber = "MIT-CTP-508",
    doi = "10.1016/0370-2693(76)90423-8",
    journal = "Phys. Lett. B",
    volume = "60",
    pages = "201--204",
    year = "1976"
}

@article{Jaffe:1976ig,
    author = "Jaffe, Robert L.",
    title = "{Multi-Quark Hadrons. 1. The Phenomenology of $Q^2 \bar{Q}^2$ Mesons}",
    reportNumber = "SLAC-PUB-1772",
    doi = "10.1103/PhysRevD.15.267",
    journal = "Phys. Rev. D",
    volume = "15",
    pages = "267",
    year = "1977"
}

@article{Jaffe:1976ih,
    author = "Jaffe, Robert L.",
    title = "{Multi-Quark Hadrons. 2. Methods}",
    reportNumber = "SLAC-PUB-1773",
    doi = "10.1103/PhysRevD.15.281",
    journal = "Phys. Rev. D",
    volume = "15",
    pages = "281",
    year = "1977"
}

@article{Belle:2003nnu,
    author = "Choi, S. K. and others",
    collaboration = "Belle",
    title = "{Observation of a narrow charmonium-like state in exclusive $B^\pm \to K^\pm \pi^+ \pi^- J/\psi$ decays}",
    eprint = "hep-ex/0309032",
    archivePrefix = "arXiv",
    doi = "10.1103/PhysRevLett.91.262001",
    journal = "Phys. Rev. Lett.",
    volume = "91",
    pages = "262001",
    year = "2003"
}

@article{BESIII:2013ouc,
    author = "Ablikim, M. and others",
    collaboration = "BESIII",
    title = "{Observation of a Charged Charmoniumlike Structure $Z_c (4020)$ and Search for the $Z_c (3900)$ in $e^+e^- \to \pi^+\pi^-h_c$}",
    eprint = "1309.1896",
    archivePrefix = "arXiv",
    primaryClass = "hep-ex",
    doi = "10.1103/PhysRevLett.111.242001",
    journal = "Phys. Rev. Lett.",
    volume = "111",
    number = "24",
    pages = "242001",
    year = "2013"
}

@article{Belle:2013yex,
    author = "Liu, Z. Q. and others",
    collaboration = "Belle",
    title = "{Study of $e^+e^- \rightarrow \pi^+ \pi^- J/\psi$ and Observation of a Charged Charmoniumlike State at Belle}",
    eprint = "1304.0121",
    archivePrefix = "arXiv",
    primaryClass = "hep-ex",
    reportNumber = "BELLE-PREPRINT-2013-6, KEK-PREPRINT-2013-2",
    doi = "10.1103/PhysRevLett.110.252002",
    journal = "Phys. Rev. Lett.",
    volume = "110",
    pages = "252002",
    year = "2013",
    note = "[Erratum: Phys.Rev.Lett. 111, 019901 (2013)]"
}

@article{Belle:2011aa,
    author = "Bondar, A. and others",
    collaboration = "Belle",
    title = "{Observation of two charged bottomonium-like resonances in $\Upsilon(5S)$ decays}",
    eprint = "1110.2251",
    archivePrefix = "arXiv",
    primaryClass = "hep-ex",
    doi = "10.1103/PhysRevLett.108.122001",
    journal = "Phys. Rev. Lett.",
    volume = "108",
    pages = "122001",
    year = "2012"
}

@article{Belle:2013urd,
    author = "Krokovny, P. and others",
    collaboration = "Belle",
    title = "{First observation of the $Z_b^0$(10610) in a Dalitz analysis of $\Upsilon$(10860) $\to \Upsilon$(nS)$\pi^0 \pi^0$}",
    eprint = "1308.2646",
    archivePrefix = "arXiv",
    primaryClass = "hep-ex",
    doi = "10.1103/PhysRevD.88.052016",
    journal = "Phys. Rev. D",
    volume = "88",
    number = "5",
    pages = "052016",
    year = "2013"
}

@article{BESIII:2015cld,
    author = "Ablikim, M. and others",
    collaboration = "BESIII",
    title = "{Observation of $Z_c(3900)^{0}$ in $e^+e^-\to\pi^0\pi^0 J/\psi$}",
    eprint = "1506.06018",
    archivePrefix = "arXiv",
    primaryClass = "hep-ex",
    doi = "10.1103/PhysRevLett.115.112003",
    journal = "Phys. Rev. Lett.",
    volume = "115",
    number = "11",
    pages = "112003",
    year = "2015"
}

@article{Belle:2008qeq,
    author = "Mizuk, R. and others",
    collaboration = "Belle",
    title = "{Observation of two resonance-like structures in the $\pi^+$ $\chi_{c1}$ mass distribution in exclusive $\bar{B}^0 \rightarrow K^- \pi^+ \chi_{c1}$ decays}",
    eprint = "0806.4098",
    archivePrefix = "arXiv",
    primaryClass = "hep-ex",
    reportNumber = "BELLE-CONF-0848",
    doi = "10.1103/PhysRevD.78.072004",
    journal = "Phys. Rev. D",
    volume = "78",
    pages = "072004",
    year = "2008"
}

@article{Belle:2014nuw,
    author = "Chilikin, K. and others",
    collaboration = "Belle",
    title = "{Observation of a new charged charmoniumlike state in $\bar{B}^0 \rightarrow J/\psi K^-\pi^+$ decays}",
    eprint = "1408.6457",
    archivePrefix = "arXiv",
    primaryClass = "hep-ex",
    reportNumber = "BELLE-PREPRINT-2014-12, KEK-PREPRINT-2014-19",
    doi = "10.1103/PhysRevD.90.112009",
    journal = "Phys. Rev. D",
    volume = "90",
    number = "11",
    pages = "112009",
    year = "2014"
}

@article{BESIII:2020qkh,
    author = "Ablikim, Medina and others",
    collaboration = "BESIII",
    title = "{Observation of a Near-Threshold Structure in the $K^+$ Recoil-Mass Spectra in $e^+e^- \rightarrow K^+(D_s^-D^{*0}+D_s^{*-}D^0$)}",
    eprint = "2011.07855",
    archivePrefix = "arXiv",
    primaryClass = "hep-ex",
    doi = "10.1103/PhysRevLett.126.102001",
    journal = "Phys. Rev. Lett.",
    volume = "126",
    number = "10",
    pages = "102001",
    year = "2021"
}

@article{LHCb:2021uow,
    author = "Aaij, Roel and others",
    collaboration = "LHCb",
    title = "{Observation of New Resonances Decaying to $J/\psi K^+$ and $J/\psi \phi$}",
    eprint = "2103.01803",
    archivePrefix = "arXiv",
    primaryClass = "hep-ex",
    reportNumber = "LHCb-PAPER-2020-044, CERN-EP-2021-025",
    doi = "10.1103/PhysRevLett.127.082001",
    journal = "Phys. Rev. Lett.",
    volume = "127",
    number = "8",
    pages = "082001",
    year = "2021"
}

@article{PhysRevLett.131.131901,
  title = "{Evidence of a $J/\psi K_{S}^{0}$ Structure in ${B}^{0}\rightarrow J/\psi \phi {K}_{S}^{0}$ Decays}",
  author = {Aaij, R. and others},
  journal = {Phys. Rev. Lett.},
  volume = {131},
  issue = {13},
  pages = {131901},
  numpages = {11},
  year = {2023},
  month = {Sep},
  publisher = {American Physical Society},
  doi = {10.1103/PhysRevLett.131.131901},
  url = {https://link.aps.org/doi/10.1103/PhysRevLett.131.131901}
}

@article{LHCb:2021vvq,
    author = "Aaij, Roel and others",
    collaboration = "LHCb",
    title = "{Observation of an exotic narrow doubly charmed tetraquark}",
    eprint = "2109.01038",
    archivePrefix = "arXiv",
    primaryClass = "hep-ex",
    reportNumber = "CERN-EP-2021-165, LHCb-PAPER-2021-031",
    doi = "10.1038/s41567-022-01614-y",
    journal = "Nature Phys.",
    volume = "18",
    number = "7",
    pages = "751--754",
    year = "2022"
}

@article{LHCb:2015yax,
    author = "Aaij, Roel and others",
    collaboration = "LHCb",
    title = "{Observation of $J/\psi p$ Resonances Consistent with Pentaquark States in $\Lambda_b^0 \to J/\psi K^- p$ Decays}",
    eprint = "1507.03414",
    archivePrefix = "arXiv",
    primaryClass = "hep-ex",
    reportNumber = "CERN-PH-EP-2015-153, LHCB-PAPER-2015-029",
    doi = "10.1103/PhysRevLett.115.072001",
    journal = "Phys. Rev. Lett.",
    volume = "115",
    pages = "072001",
    year = "2015"
}

@article{LHCb:2019kea,
    author = "Aaij, Roel and others",
    collaboration = "LHCb",
    title = "{Observation of a narrow pentaquark state, $P_c(4312)^+$, and of two-peak structure of the $P_c(4450)^+$}",
    eprint = "1904.03947",
    archivePrefix = "arXiv",
    primaryClass = "hep-ex",
    reportNumber = "LHCb-PAPER-2019-014 CERN-EP-2019-058",
    doi = "10.1103/PhysRevLett.122.222001",
    journal = "Phys. Rev. Lett.",
    volume = "122",
    number = "22",
    pages = "222001",
    year = "2019"
}

@article{LHCb:2024vfz,
    author = "Aaij, Roel and others",
    collaboration = "LHCb",
    title = "{Observation of New Charmonium or Charmoniumlike States in $B^+ \rightarrow D^{*\pm}D^{\mp}K^+$ Decays}",
    eprint = "2406.03156",
    archivePrefix = "arXiv",
    primaryClass = "hep-ex",
    reportNumber = "LHCb-PAPER-2023-047, CERN-EP-2024-096",
    doi = "10.1103/PhysRevLett.133.131902",
    journal = "Phys. Rev. Lett.",
    volume = "133",
    number = "13",
    pages = "131902",
    year = "2024"
}

@article{Belle:2021nuv,
    author = "Wang, X. L. and others",
    collaboration = "Belle",
    title = "{Study of {\ensuremath{\gamma}}{\ensuremath{\gamma}}{\textrightarrow}{\ensuremath{\gamma}}{\ensuremath{\psi}}(2S) at Belle}",
    eprint = "2105.06605",
    archivePrefix = "arXiv",
    primaryClass = "hep-ex",
    doi = "10.1103/PhysRevD.105.112011",
    journal = "Phys. Rev. D",
    volume = "105",
    number = "11",
    pages = "112011",
    year = "2022"
}

@article{Guo:2017jvc,
    author = "Guo, Feng-Kun and Hanhart, Christoph and Mei\ss{}ner, Ulf-G. and Wang, Qian and Zhao, Qiang and Zou, Bing-Song",
    title = "{Hadronic molecules}",
    eprint = "1705.00141",
    archivePrefix = "arXiv",
    primaryClass = "hep-ph",
    doi = "10.1103/RevModPhys.90.015004",
    journal = "Rev. Mod. Phys.",
    volume = "90",
    number = "1",
    pages = "015004",
    year = "2018",
    note = "[Erratum: Rev.Mod.Phys. 94, 029901 (2022)]"
}

@article{Ali:2017jda,
    author = {Ali, Ahmed and Lange, Jens S\"oren and Stone, Sheldon},
    title = "{Exotics: Heavy Pentaquarks and Tetraquarks}",
    eprint = "1706.00610",
    archivePrefix = "arXiv",
    primaryClass = "hep-ph",
    reportNumber = "DESY-17-071",
    doi = "10.1016/j.ppnp.2017.08.003",
    journal = "Prog. Part. Nucl. Phys.",
    volume = "97",
    pages = "123--198",
    year = "2017"
}

@article{Olsen:2017bmm,
    author = "Olsen, Stephen Lars and Skwarnicki, Tomasz and Zieminska, Daria",
    title = "{Nonstandard heavy mesons and baryons: Experimental evidence}",
    eprint = "1708.04012",
    archivePrefix = "arXiv",
    primaryClass = "hep-ph",
    doi = "10.1103/RevModPhys.90.015003",
    journal = "Rev. Mod. Phys.",
    volume = "90",
    number = "1",
    pages = "015003",
    year = "2018"
}

@book{Ali:2019roi,
    author = "Ali, Ahmed and Maiani, Luciano and Polosa, Antonio D.",
    title = "{Multiquark Hadrons}",
    doi = "10.1017/9781316761465",
    isbn = "978-1-316-76146-5, 978-1-107-17158-9, 978-1-316-77419-9",
    publisher = "Cambridge University Press",
    month = "6",
    year = "2019"
}

@article{Brambilla:2019esw,
    author = "Brambilla, Nora and Eidelman, Simon and Hanhart, Christoph and Nefediev, Alexey and Shen, Cheng-Ping and Thomas, Christopher E. and Vairo, Antonio and Yuan, Chang-Zheng",
    title = "{The $XYZ$ states: experimental and theoretical status and perspectives}",
    eprint = "1907.07583",
    archivePrefix = "arXiv",
    primaryClass = "hep-ex",
    reportNumber = "TUM-EFT 125/19",
    doi = "10.1016/j.physrep.2020.05.001",
    journal = "Phys. Rept.",
    volume = "873",
    pages = "1--154",
    year = "2020"
}

@article{Liu:2019zoy,
    author = "Liu, Yan-Rui and Chen, Hua-Xing and Chen, Wei and Liu, Xiang and Zhu, Shi-Lin",
    title = "{Pentaquark and Tetraquark states}",
    eprint = "1903.11976",
    archivePrefix = "arXiv",
    primaryClass = "hep-ph",
    doi = "10.1016/j.ppnp.2019.04.003",
    journal = "Prog. Part. Nucl. Phys.",
    volume = "107",
    pages = "237--320",
    year = "2019"
}

@article{Chen:2022asf,
    author = "Chen, Hua-Xing and Chen, Wei and Liu, Xiang and Liu, Yan-Rui and Zhu, Shi-Lin",
    title = "{An updated review of the new hadron states}",
    eprint = "2204.02649",
    archivePrefix = "arXiv",
    primaryClass = "hep-ph",
    doi = "10.1088/1361-6633/aca3b6",
    journal = "Rept. Prog. Phys.",
    volume = "86",
    number = "2",
    pages = "026201",
    year = "2023"
}

@article{Eichten:2004uh,
    author = "Eichten, Estia J. and Lane, Kenneth and Quigg, Chris",
    title = "{Charmonium levels near threshold and the narrow state $X(3872) \to \pi^{+}\pi^{-}J/\psi$}",
    eprint = "hep-ph/0401210",
    archivePrefix = "arXiv",
    reportNumber = "FERMILAB-PUB-04-001-T, BUHEP-04-01",
    doi = "10.1103/PhysRevD.69.094019",
    journal = "Phys. Rev. D",
    volume = "69",
    pages = "094019",
    year = "2004"
}

@article{Barnes:2005pb,
    author = "Barnes, T. and Godfrey, S. and Swanson, E. S.",
    title = "{Higher charmonia}",
    eprint = "hep-ph/0505002",
    archivePrefix = "arXiv",
    doi = "10.1103/PhysRevD.72.054026",
    journal = "Phys. Rev. D",
    volume = "72",
    pages = "054026",
    year = "2005"
}

@article{Eichten:2005ga,
    author = "Eichten, Estia J. and Lane, Kenneth and Quigg, Chris",
    title = "{New states above charm threshold}",
    eprint = "hep-ph/0511179",
    archivePrefix = "arXiv",
    reportNumber = "FERMILAB-PUB-05-380-T, BUHEP-05-18",
    doi = "10.1103/PhysRevD.73.014014",
    journal = "Phys. Rev. D",
    volume = "73",
    pages = "014014",
    year = "2006",
    note = "[Erratum: Phys.Rev.D 73, 079903 (2006)]"
}

@article{Pennington:2007xr,
    author = "Pennington, M. R. and Wilson, D. J.",
    title = "{Decay channels and charmonium mass-shifts}",
    eprint = "0704.3384",
    archivePrefix = "arXiv",
    primaryClass = "hep-ph",
    reportNumber = "DCPT-07-28, IPPP-07-14",
    doi = "10.1103/PhysRevD.76.077502",
    journal = "Phys. Rev. D",
    volume = "76",
    pages = "077502",
    year = "2007"
}

@article{Barnes:2007xu,
    author = "Barnes, T. and Swanson, E. S.",
    title = "{Hadron loops: General theorems and application to charmonium}",
    eprint = "0711.2080",
    archivePrefix = "arXiv",
    primaryClass = "hep-ph",
    doi = "10.1103/PhysRevC.77.055206",
    journal = "Phys. Rev. C",
    volume = "77",
    pages = "055206",
    year = "2008"
}

@article{Li:2009ad,
    author = "Li, Bai-Qing and Meng, Ce and Chao, Kuang-Ta",
    title = "{Coupled-Channel and Screening Effects in Charmonium Spectrum}",
    eprint = "0904.4068",
    archivePrefix = "arXiv",
    primaryClass = "hep-ph",
    doi = "10.1103/PhysRevD.80.014012",
    journal = "Phys. Rev. D",
    volume = "80",
    pages = "014012",
    year = "2009"
}

@article{Danilkin:2009hr,
    author = "Danilkin, I. V. and Simonov, Yu. A.",
    title = "{Channel coupling in heavy quarkonia: Energy levels, mixing, widths and new states}",
    eprint = "0907.1088",
    archivePrefix = "arXiv",
    primaryClass = "hep-ph",
    doi = "10.1103/PhysRevD.81.074027",
    journal = "Phys. Rev. D",
    volume = "81",
    pages = "074027",
    year = "2010"
}

@article{Ferretti:2012zz,
    author = "Ferretti, J. and Galata, G. and Santopinto, E. and Vassallo, A.",
    title = "{Bottomonium self-energies due to the coupling to the meson-meson continuum}",
    doi = "10.1103/PhysRevC.86.015204",
    journal = "Phys. Rev. C",
    volume = "86",
    pages = "015204",
    year = "2012"
}

@article{Ferretti:2013faa,
    author = "Ferretti, J. and Galat\`a, G. and Santopinto, E.",
    title = "{Interpretation of the X(3872) as a charmonium state plus an extra component due to the coupling to the meson-meson continuum}",
    eprint = "1302.6857",
    archivePrefix = "arXiv",
    primaryClass = "hep-ph",
    doi = "10.1103/PhysRevC.88.015207",
    journal = "Phys. Rev. C",
    volume = "88",
    number = "1",
    pages = "015207",
    year = "2013"
}

@article{Ferretti:2013vua,
    author = "Ferretti, J. and Santopinto, E.",
    title = "{Higher mass bottomonia}",
    eprint = "1306.2874",
    archivePrefix = "arXiv",
    primaryClass = "hep-ph",
    doi = "10.1103/PhysRevD.90.094022",
    journal = "Phys. Rev. D",
    volume = "90",
    number = "9",
    pages = "094022",
    year = "2014"
}

@article{Ferretti:2014xqa,
    author = "Ferretti, J. and Galat\`a, G. and Santopinto, E.",
    title = "{Quark structure of the $X(3872)$ and $\chi_b(3P)$ resonances}",
    eprint = "1401.4431",
    archivePrefix = "arXiv",
    primaryClass = "nucl-th",
    doi = "10.1103/PhysRevD.90.054010",
    journal = "Phys. Rev. D",
    volume = "90",
    number = "5",
    pages = "054010",
    year = "2014"
}

@article{Ferretti:2018tco,
    author = "Ferretti, J. and Santopinto, E.",
    title = "{Threshold corrections of $\chi_c (2P)$ and $\chi_b (3P)$ states and J /$\psi \rho$ and J /$\psi \omega$ transitions of the $X (3872)$ in a coupled-channel model}",
    eprint = "1806.02489",
    archivePrefix = "arXiv",
    primaryClass = "hep-ph",
    doi = "10.1016/j.physletb.2018.12.052",
    journal = "Phys. Lett. B",
    volume = "789",
    pages = "550--555",
    year = "2019"
}

@article{Man:2024mvl,
    author = "Man, Zi-Long and Shu, Cheng-Rui and Liu, Yan-Rui and Chen, Hong",
    title = "{Charmonium states in a coupled-channel model}",
    eprint = "2402.02765",
    archivePrefix = "arXiv",
    primaryClass = "hep-ph",
    doi = "10.1140/epjc/s10052-024-13132-7",
    journal = "Eur. Phys. J. C",
    volume = "84",
    number = "8",
    pages = "810",
    year = "2024"
}

@article{Blossier:2024dhm,
    author = "Blossier, Ben{\^o}{\i}t and Heitger, Jochen and Neuendorf, Jan and San Jos{\'e}, Teseo",
    title = "{Hadronic decay of vector charmonium from the lattice}",
    eprint = "2411.10123",
    archivePrefix = "arXiv",
    primaryClass = "hep-lat",
    reportNumber = "MS-TP-24-36",
    doi = "10.1007/JHEP04(2025)139",
    journal = "JHEP",
    volume = "04",
    pages = "139",
    year = "2025"
}

@article{Bruschini:2025paj,
    author = "Bruschini, R. and Gonz{\'a}lez, P. and Tarutina, T.",
    title = "{$^3P_0$ model revisited}",
    doi = "10.1103/PhysRevD.111.074042",
    journal = "Phys. Rev. D",
    volume = "111",
    number = "7",
    pages = "074042",
    year = "2025"
}

@article{Hao:2025vmw,
    author = "Hao, Wei and Wang, Guan-Ying and Wang, En and Sultan, M. Atif",
    title = "{Coupled channel effects for the bottom-strange mesons}",
    eprint = "2501.04298",
    archivePrefix = "arXiv",
    primaryClass = "hep-ph",
    doi = "10.1140/epjc/s10052-025-15029-5",
    journal = "Eur. Phys. J. C",
    volume = "85",
    number = "11",
    pages = "1332",
    year = "2025"
}

@article{Ni:2025gvx,
    author = "Ni, Ru-Hui and Deng, Qian and Wu, Jia-Jun and Zhong, Xian-Hui",
    title = "{Bottomonia in an unquenched quark model}",
    eprint = "2501.15110",
    archivePrefix = "arXiv",
    primaryClass = "hep-ph",
    doi = "10.1103/6x3t-x15s",
    journal = "Phys. Rev. D",
    volume = "111",
    number = "11",
    pages = "114027",
    year = "2025"
}

@article{M:2025gnf,
    author = "M, Sreelakshmi and Ranjan, Akhilesh",
    title = "{Mass spectroscopy of charmonium using a screened potential}",
    eprint = "2503.07393",
    archivePrefix = "arXiv",
    primaryClass = "hep-ph",
    doi = "10.1007/s10773-025-05924-8",
    journal = "Int. J. Theor. Phys.",
    volume = "64",
    pages = "58",
    year = "2025"
}

@article{Sultan:2025dfe,
    author = "Sultan, M. Atif and Hao, Wei and Swanson, E. S. and Chang, Lei",
    title = "{Bottomonium meson spectrum with quenched and unquenched quark models}",
    eprint = "2503.10178",
    archivePrefix = "arXiv",
    primaryClass = "hep-ph",
    doi = "10.1140/epja/s10050-025-01607-4",
    journal = "Eur. Phys. J. A",
    volume = "61",
    number = "6",
    pages = "137",
    year = "2025"
}

@article{Ahmad:2025mue,
    author = "Ahmad, Zaki and Asghar, Ishrat and Akram, Faisal and Masud, Bilal",
    title = "{Strong decays of charmonia}",
    eprint = "2504.07605",
    archivePrefix = "arXiv",
    primaryClass = "hep-ph",
    doi = "10.1103/PhysRevD.111.034007",
    journal = "Phys. Rev. D",
    volume = "111",
    number = "3",
    pages = "034007",
    year = "2025"
}

@article{Gao:2025tob,
    author = "Gao, Xiu-Li and Cui, Jun-Xi and Zhou, Yu-Hui and Zhou, Zhi-Yong",
    title = "{Updated analysis of charmonium states in a relativized quark potential model}",
    eprint = "2504.14575",
    archivePrefix = "arXiv",
    primaryClass = "hep-ph",
    journal = "",
    month = "4",
    year = "2025"
}

@article{Man:2025vmm,
    author = "Man, Zi-Long and Luo, Si-Qiang and Liu, Xiang",
    title = "{Is the $3S$-$2D$ mixing strong for the charmonia $\psi(4040)$ and $\psi(4160)$?}",
    eprint = "2507.18536",
    archivePrefix = "arXiv",
    primaryClass = "hep-ph",
    journal = "",
    month = "7",
    year = "2025"
}

@article{Ahmad:2025hcr,
    author = "Ahmad, Zaki and Asghar, Ishrat and Masud, Bilal and Sultan, M. Atif",
    title = "{Charmonium spectrum and its decay properties}",
    eprint = "2508.17841",
    archivePrefix = "arXiv",
    primaryClass = "hep-ph",
    doi = "10.1140/epja/s10050-025-01671-w",
    journal = "Eur. Phys. J. A",
    volume = "61",
    number = "9",
    pages = "200",
    year = "2025"
}

@article{Rui:2025olj,
    author = "Rui, Yi-Yi and Wang, Tianhong and Wang, Zhi-Hui and Feng, Tai-Fu and Wang, Guo-Li",
    title = "{Relativistic Bethe-Salpeter study of OZI-rule allowed strong decays: determination of total width of $h_{c}(2P)$ and $^{3}P_{0}$ model parameter}",
    eprint = "2512.12531",
    journal = "",
    archivePrefix = "arXiv",
    primaryClass = "hep-ph",
    month = "12",
    year = "2025"
}

@article{Kokoski:1985is,
    author = "Kokoski, Richard and Isgur, Nathan",
    title = "{Meson Decays by Flux Tube Breaking}",
    reportNumber = "UTPT-85-05",
    doi = "10.1103/PhysRevD.35.907",
    journal = "Phys. Rev. D",
    volume = "35",
    pages = "907",
    year = "1987"
}

@article{Godfrey:1986wj,
    author = "Godfrey, Stephen and Kokoski, Richard",
    title = "{The Properties of P-Wave Mesons with One Heavy Quark}",
    reportNumber = "TRI-PP-86-51, GIPP-90-6A",
    doi = "10.1103/PhysRevD.43.1679",
    journal = "Phys. Rev. D",
    volume = "43",
    pages = "1679--1687",
    year = "1991"
}

@article{Ackleh:1996yt,
    author = "Ackleh, E. S. and Barnes, Ted and Swanson, E. S.",
    title = "{On the mechanism of open flavor strong decays}",
    eprint = "hep-ph/9604355",
    archivePrefix = "arXiv",
    reportNumber = "ORNL-CTP-96-03",
    doi = "10.1103/PhysRevD.54.6811",
    journal = "Phys. Rev. D",
    volume = "54",
    pages = "6811--6829",
    year = "1996"
}

@article{Segovia:2012cd,
    author = "Segovia, J. and Entem, D. R. and Fern{\'a}ndez, F.",
    title = "{Scaling of the $^3P_0$ Strength in Heavy Meson Strong Decays}",
    eprint = "1205.2215",
    archivePrefix = "arXiv",
    primaryClass = "hep-ph",
    doi = "10.1016/j.physletb.2012.08.005",
    journal = "Phys. Lett. B",
    volume = "715",
    pages = "322--327",
    year = "2012"
}

@article{Kumano:1988ga,
    author = "Kumano, S. and Pandharipande, V. R.",
    title = "{Decay of Mesons in Flux Tube Quark Model}",
    doi = "10.1103/PhysRevD.38.146",
    journal = "Phys. Rev. D",
    volume = "38",
    pages = "146--151",
    year = "1988"
}

@article{Barnes:2002mu,
    author = "Barnes, T. and Black, N. and Page, P. R.",
    title = "{Strong decays of strange quarkonia}",
    eprint = "nucl-th/0208072",
    archivePrefix = "arXiv",
    doi = "10.1103/PhysRevD.68.054014",
    journal = "Phys. Rev. D",
    volume = "68",
    pages = "054014",
    year = "2003"
}

@article{daSilva:2008rp,
    author = "da Silva, D. T. and da Silva, M. L. L. and de Quadros, J. N. and Hadjimichef, D.",
    title = "{Meson decay in a corrected $^3P_0$ model}",
    eprint = "0810.0293",
    archivePrefix = "arXiv",
    primaryClass = "hep-ph",
    doi = "10.1103/PhysRevD.78.076004",
    journal = "Phys. Rev. D",
    volume = "78",
    pages = "076004",
    year = "2008"
}

@article{Farina:2020slb,
    author = "Farina, Christian and Garcia Tecocoatzi, Hugo and Giachino, Alessandro and Santopinto, Elena and Swanson, Eric S.",
    title = "{Heavy hybrid decays in a constituent gluon model}",
    eprint = "2005.10850",
    archivePrefix = "arXiv",
    primaryClass = "hep-ph",
    doi = "10.1103/PhysRevD.102.014023",
    journal = "Phys. Rev. D",
    volume = "102",
    number = "1",
    pages = "014023",
    year = "2020"
}

@article{Zhao:2025kno,
    author = "Zhao, Zheng and Kaewsnod, Attaphon and Xu, Kai and Tagsinsit, Nattapat and Liu, Xuyang and Limphirat, Ayut and Yan, Yupeng",
    title = "{Study of $1^{--}$ P-wave charmoniumlike and bottomoniumlike tetraquark spectroscopy}",
    eprint = "2503.00552",
    archivePrefix = "arXiv",
    primaryClass = "hep-ph",
    journal = "",
    month = "3",
    year = "2025"
}

@article{Chen:2025pvk,
    author = "Chen, Bing and Liu, Xiang",
    title = "{Investigating hybrid mesons with $0^{+-}$ and $2^{+-}$ exotic quantum numbers}",
    eprint = "2503.06116",
    archivePrefix = "arXiv",
    primaryClass = "hep-ph",
    doi = "10.1140/epjc/s10052-025-14509-y",
    journal = "Eur. Phys. J. C",
    volume = "85",
    number = "7",
    pages = "788",
    year = "2025"
}

@article{Cheung:2017tnt,
    author = "Cheung, Gavin K. C. and Thomas, Christopher E. and Dudek, Jozef J. and Edwards, Robert G.",
    collaboration = "Hadron Spectrum",
    title = "{Tetraquark operators in lattice QCD and exotic flavour states in the charm sector}",
    eprint = "1709.01417",
    archivePrefix = "arXiv",
    primaryClass = "hep-lat",
    reportNumber = "DAMTP-2017-33, JLAB-THY-17-2541",
    doi = "10.1007/JHEP11(2017)033",
    journal = "JHEP",
    volume = "11",
    pages = "033",
    year = "2017"
}

@article{Ryan:2020iog,
    author = "Ryan, Sin\'ead M. and Wilson, David J.",
    collaboration = "Hadron Spectrum",
    title = "{Excited and exotic bottomonium spectroscopy from lattice QCD}",
    eprint = "2008.02656",
    archivePrefix = "arXiv",
    primaryClass = "hep-lat",
    doi = "10.1007/JHEP02(2021)214",
    journal = "JHEP",
    volume = "02",
    pages = "214",
    year = "2021"
}

@article{Brambilla:1999xf,
    author = "Brambilla, Nora and Pineda, Antonio and Soto, Joan and Vairo, Antonio",
    title = "{Potential NRQCD: An Effective theory for heavy quarkonium}",
    eprint = "hep-ph/9907240",
    archivePrefix = "arXiv",
    reportNumber = "CERN-TH-99-199, HEPHY-PUB-716-99, UB-ECM-PF-99-06, UWTHPH-1999-34, UB-ECM-PF-99-13",
    doi = "10.1016/S0550-3213(99)00693-8",
    journal = "Nucl. Phys. B",
    volume = "566",
    pages = "275",
    year = "2000"
}

@article{Brambilla:2000gk,
    author = "Brambilla, Nora and Pineda, Antonio and Soto, Joan and Vairo, Antonio",
    title = "{The QCD potential at $O(1/m)$}",
    eprint = "hep-ph/0002250",
    archivePrefix = "arXiv",
    reportNumber = "CERN-TH-2000-053, UB-ECM-PF-00-03, UWTHPH-1999-47",
    doi = "10.1103/PhysRevD.63.014023",
    journal = "Phys. Rev. D",
    volume = "63",
    pages = "014023",
    year = "2001"
}

@article{Brambilla:2004jw,
    author = "Brambilla, Nora and Pineda, Antonio and Soto, Joan and Vairo, Antonio",
    title = "{Effective Field Theories for Heavy Quarkonium}",
    eprint = "hep-ph/0410047",
    archivePrefix = "arXiv",
    reportNumber = "IFUM-805-FT, UB-ECM-PF-04-24",
    doi = "10.1103/RevModPhys.77.1423",
    journal = "Rev. Mod. Phys.",
    volume = "77",
    pages = "1423",
    year = "2005"
}

@article{Berwein:2015vca,
    author = "Berwein, Matthias and Brambilla, Nora and Tarr\'us Castell\`a, Jaume and Vairo, Antonio",
    title = "{Quarkonium Hybrids with Nonrelativistic Effective Field Theories}",
    eprint = "1510.04299",
    archivePrefix = "arXiv",
    primaryClass = "hep-ph",
    reportNumber = "TUM-EFT-45-14",
    doi = "10.1103/PhysRevD.92.114019",
    journal = "Phys. Rev. D",
    volume = "92",
    number = "11",
    pages = "114019",
    year = "2015"
}

@article{Oncala:2017hop,
    author = "Oncala, Rub{\'e}n and Soto, Joan",
    title = "{Heavy Quarkonium Hybrids: Spectrum, Decay and Mixing}",
    eprint = "1702.03900",
    archivePrefix = "arXiv",
    primaryClass = "hep-ph",
    reportNumber = "ICCUB-17-004, NIKHF-2017-005",
    doi = "10.1103/PhysRevD.96.014004",
    journal = "Phys. Rev. D",
    volume = "96",
    number = "1",
    pages = "014004",
    year = "2017"
}

@article{Oncala:2025mqj,
    author = "Oncala, Rub{\'e}n and Soto, Joan",
    title = "{Hybrid to Quarkonia transitions}",
    eprint = "2511.14016",
    archivePrefix = "arXiv",
    primaryClass = "hep-ph",
    journal = "",
    month = "11",
    year = "2025"
}

@article{Brambilla:2018pyn,
    author = "Brambilla, Nora and Lai, Wai Kin and Segovia, Jorge and Tarr\'us Castell\`a, Jaume and Vairo, Antonio",
    title = "{Spin structure of heavy-quark hybrids}",
    eprint = "1805.07713",
    archivePrefix = "arXiv",
    primaryClass = "hep-ph",
    reportNumber = "TUM-EFT-95-17, TUM-EFT 95/17",
    doi = "10.1103/PhysRevD.99.014017",
    journal = "Phys. Rev. D",
    volume = "99",
    number = "1",
    pages = "014017",
    year = "2019",
    note = "[Erratum: Phys.Rev.D 101, 099902 (2020)]"
}

@article{Brambilla:2019jfi,
    author = "Brambilla, Nora and Lai, Wai Kin and Segovia, Jorge and Tarr\'us Castell\`a, Jaume",
    title = "{QCD spin effects in the heavy hybrid potentials and spectra}",
    eprint = "1908.11699",
    archivePrefix = "arXiv",
    primaryClass = "hep-ph",
    reportNumber = "TUM-EFT 129/19",
    doi = "10.1103/PhysRevD.101.054040",
    journal = "Phys. Rev. D",
    volume = "101",
    number = "5",
    pages = "054040",
    year = "2020"
}

@article{Soto:2023lbh,
    author = "Soto, Joan and Valls, Sandra Tom\`as",
    title = "{Hyperfine splittings of heavy quarkonium hybrids}",
    eprint = "2302.01765",
    archivePrefix = "arXiv",
    primaryClass = "hep-ph",
    doi = "10.1103/PhysRevD.108.014025",
    journal = "Phys. Rev. D",
    volume = "108",
    number = "1",
    pages = "014025",
    year = "2023"
}

@article{Brambilla:2017uyf,
    author = "Brambilla, Nora and Krein, Gast\~ao and Tarr\'us Castell\`a, Jaume and Vairo, Antonio",
    title = "{Born-Oppenheimer approximation in an effective field theory language}",
    eprint = "1707.09647",
    archivePrefix = "arXiv",
    primaryClass = "hep-ph",
    reportNumber = "TUM-EFT-69-15",
    doi = "10.1103/PhysRevD.97.016016",
    journal = "Phys. Rev. D",
    volume = "97",
    number = "1",
    pages = "016016",
    year = "2018"
}

@article{Soto:2020xpm,
    author = "Soto, Joan and Tarr\'us Castell\`a, Jaume",
    title = "{Nonrelativistic effective field theory for heavy exotic hadrons}",
    eprint = "2005.00552",
    archivePrefix = "arXiv",
    primaryClass = "hep-ph",
    doi = "10.1103/PhysRevD.102.014012",
    journal = "Phys. Rev. D",
    volume = "102",
    number = "1",
    pages = "014012",
    year = "2020",
    note = "[Erratum: Phys.Rev.D 110, 099901 (2024)]"
}

@article{Berwein:2024ztx,
    author = "Berwein, Matthias and Brambilla, Nora and Mohapatra, Abhishek and Vairo, Antonio",
    title = "{Hybrids, tetraquarks, pentaquarks, doubly heavy baryons, and quarkonia in Born-Oppenheimer effective theory}",
    eprint = "2408.04719",
    archivePrefix = "arXiv",
    primaryClass = "hep-ph",
    reportNumber = "TUM-EFT 185/23",
    doi = "10.1103/PhysRevD.110.094040",
    journal = "Phys. Rev. D",
    volume = "110",
    number = "9",
    pages = "094040",
    year = "2024"
}

@article{Braaten:2024tbm,
    author = "Braaten, Eric and Bruschini, Roberto",
    title = "{Exotic hidden-heavy hadrons and where to find them}",
    eprint = "2409.08002",
    archivePrefix = "arXiv",
    primaryClass = "hep-ph",
    doi = "10.1016/j.physletb.2025.139386",
    journal = "Phys. Lett. B",
    volume = "863",
    pages = "139386",
    year = "2025"
}

@article{Brambilla:2024imu,
    author = "Brambilla, Nora and Mohapatra, Abhishek and Scirpa, Tommaso and Vairo, Antonio",
    title = "{Nature of $\chi_{c1}(3872)$ and $T_{cc}^+(3875)$}",
    eprint = "2411.14306",
    archivePrefix = "arXiv",
    primaryClass = "hep-ph",
    reportNumber = "TUM-EFT 193/24",
    doi = "10.1103/pdy7-hvg7",
    journal = "Phys. Rev. Lett.",
    volume = "135",
    number = "13",
    pages = "131902",
    year = "2025"
}

@article{Brambilla:2025xma,
    author = "Brambilla, Nora and Mohapatra, Abhishek and Vairo, Antonio",
    title = "{Unravelling Pentaquarks with Born-Oppenheimer effective theory}",
    eprint = "2508.13050",
    journal = "",
    archivePrefix = "arXiv",
    primaryClass = "hep-ph",
    reportNumber = "TUM-EFT 198/25",
    month = "8",
    year = "2025"
}

@article{Bicudo:2019ymo,
    author = "Bicudo, Pedro and Cardoso, Marco and Cardoso, Nuno and Wagner, Marc",
    title = "{Bottomonium resonances with $I = 0$ from lattice QCD correlation functions with static and light quarks}",
    eprint = "1910.04827",
    archivePrefix = "arXiv",
    primaryClass = "hep-lat",
    doi = "10.1103/PhysRevD.101.034503",
    journal = "Phys. Rev. D",
    volume = "101",
    number = "3",
    pages = "034503",
    year = "2020"
}

@article{Bruschini:2020voj,
    author = "Bruschini, R. and Gonz\'alez, P.",
    title = "{Diabatic description of charmoniumlike mesons}",
    eprint = "2007.07693",
    archivePrefix = "arXiv",
    primaryClass = "hep-ph",
    doi = "10.1103/PhysRevD.102.074002",
    journal = "Phys. Rev. D",
    volume = "102",
    number = "7",
    pages = "074002",
    year = "2020"
}

@article{Bicudo:2020qhp,
    author = "Bicudo, Pedro and Cardoso, Nuno and Mueller, Lasse and Wagner, Marc",
    title = "{Computation of the quarkonium and meson-meson composition of the $\Upsilon(nS)$  states and of the new $\Upsilon(10753)$ Belle resonance from lattice QCD static potentials}",
    eprint = "2008.05605",
    archivePrefix = "arXiv",
    primaryClass = "hep-lat",
    doi = "10.1103/PhysRevD.103.074507",
    journal = "Phys. Rev. D",
    volume = "103",
    number = "7",
    pages = "074507",
    year = "2021"
}

@article{Bruschini:2021cty,
    author = "Bruschini, R. and Gonz\'alez, P.",
    title = "{Diabatic description of charmoniumlike mesons II: mass corrections and strong decay widths}",
    eprint = "2101.04636",
    archivePrefix = "arXiv",
    primaryClass = "hep-ph",
    doi = "10.1103/PhysRevD.103.074009",
    journal = "Phys. Rev. D",
    volume = "103",
    number = "7",
    pages = "074009",
    year = "2021"
}

@article{Bruschini:2021sjh,
    author = "Bruschini, R. and Gonz\'alez, P.",
    title = "{Diabatic description of bottomoniumlike mesons}",
    eprint = "2105.04401",
    archivePrefix = "arXiv",
    primaryClass = "hep-ph",
    doi = "10.1103/PhysRevD.103.114016",
    journal = "Phys. Rev. D",
    volume = "103",
    number = "11",
    pages = "114016",
    year = "2021"
}

@article{TarrusCastella:2022rxb,
    author = "Tarr\'us Castell\`a, Jaume",
    title = "{Heavy meson thresholds in Born-Oppenheimer effective field theory}",
    eprint = "2207.09365",
    archivePrefix = "arXiv",
    primaryClass = "hep-ph",
    doi = "10.1103/PhysRevD.106.094020",
    journal = "Phys. Rev. D",
    volume = "106",
    number = "9",
    pages = "094020",
    year = "2022"
}

@article{Bicudo:2022ihz,
    author = "Bicudo, Pedro and Cardoso, Nuno and Mueller, Lasse and Wagner, Marc",
    title = "{Study of I=0 bottomonium bound states and resonances in S, P, D, and F waves with lattice QCD static-static-light-light potentials}",
    eprint = "2205.11475",
    archivePrefix = "arXiv",
    primaryClass = "hep-lat",
    doi = "10.1103/PhysRevD.107.094515",
    journal = "Phys. Rev. D",
    volume = "107",
    number = "9",
    pages = "094515",
    year = "2023"
}

@article{Zhang:2025bex,
    author = "Zhang, Zi-Zhao and Li, Rong and Liu, Bo-Chao",
    title = "{Study of charmonium(-like) mesons under a diabatic approach}",
    eprint = "2504.12149",
    archivePrefix = "arXiv",
    primaryClass = "hep-ph",
    doi = "10.1088/1674-1137/ad79d6",
    journal = "Chin. Phys. C",
    volume = "49",
    number = "1",
    pages = "013102",
    year = "2025"
}

@article{TarrusCastella:2024zps,
    author = "Tarr\'us Castell\`a, Jaume",
    title = "{Effect of continuum states on the double-heavy hadron spectra}",
    eprint = "2401.13393",
    archivePrefix = "arXiv",
    primaryClass = "hep-ph",
    doi = "10.1007/JHEP06(2024)107",
    journal = "JHEP",
    volume = "06",
    pages = "107",
    year = "2024"
}

@article{Bodwin:1994jh,
    author = "Bodwin, Geoffrey T. and Braaten, Eric and Lepage, G. Peter",
    title = "{Rigorous QCD analysis of inclusive annihilation and production of heavy quarkonium}",
    eprint = "hep-ph/9407339",
    archivePrefix = "arXiv",
    reportNumber = "ANL-HEP-PR-94-24, FERMILAB-PUB-94-073-T, NUHEP-TH-94-5",
    doi = "10.1103/PhysRevD.55.5853",
    journal = "Phys. Rev. D",
    volume = "51",
    pages = "1125--1171",
    year = "1995",
    note = "[Erratum: Phys.Rev.D 55, 5853 (1997)]"
}

@article{Prelovsek:2019ywc,
    author = "Prelovsek, S. and Bahtiyar, H. and Petkovic, J.",
    title = "{Zb tetraquark channel from lattice QCD and Born-Oppenheimer approximation}",
    eprint = "1912.02656",
    archivePrefix = "arXiv",
    primaryClass = "hep-lat",
    doi = "10.1016/j.physletb.2020.135467",
    journal = "Phys. Lett. B",
    volume = "805",
    pages = "135467",
    year = "2020"
}

@article{Bali:2005fu,
    author = "Bali, Gunnar S. and Neff, Hartmut and Duessel, Thomas and Lippert, Thomas and Schilling, Klaus",
    collaboration = "SESAM",
    title = "{Observation of string breaking in QCD}",
    eprint = "hep-lat/0505012",
    archivePrefix = "arXiv",
    doi = "10.1103/PhysRevD.71.114513",
    journal = "Phys. Rev. D",
    volume = "71",
    pages = "114513",
    year = "2005"
}

@article{Bulava:2019iut,
    author = {Bulava, John and H\"orz, Ben and Knechtli, Francesco and Koch, Vanessa and Moir, Graham and Morningstar, Colin and Peardon, Mike},
    title = "{String breaking by light and strange quarks in QCD}",
    eprint = "1902.04006",
    archivePrefix = "arXiv",
    primaryClass = "hep-lat",
    reportNumber = "WUB/19-00, CP3-Origins-2019-002 DNRF90, MITP/19-010",
    doi = "10.1016/j.physletb.2019.05.018",
    journal = "Phys. Lett. B",
    volume = "793",
    pages = "493--498",
    year = "2019"
}

@article{Bulava:2024jpj,
    author = "Bulava, John and Knechtli, Francesco and Koch, Vanessa and Morningstar, Colin and Peardon, Michael",
    title = "{The quark-mass dependence of the potential energy between static colour sources in the QCD vacuum with light and strange quarks}",
    eprint = "2403.00754",
    archivePrefix = "arXiv",
    primaryClass = "hep-lat",
    reportNumber = "WUB/24-00",
    doi = "10.1016/j.physletb.2024.138754",
    journal = "Phys. Lett. B",
    volume = "854",
    pages = "138754",
    year = "2024"
}

@article{Foster:1998wu,
    author = "Foster, M. and Michael, Christopher",
    collaboration = "UKQCD",
    title = "{Hadrons with a heavy color adjoint particle}",
    eprint = "hep-lat/9811010",
    archivePrefix = "arXiv",
    reportNumber = "LTH-444",
    doi = "10.1103/PhysRevD.59.094509",
    journal = "Phys. Rev. D",
    volume = "59",
    pages = "094509",
    year = "1999"
}

@article{Alasiri:2024nue,
    author = "Alasiri, Fareed and Braaten, Eric and Mohapatra, Abhishek",
    title = "{Born-Oppenheimer potentials for SU(3) gauge theory}",
    eprint = "2406.05123",
    archivePrefix = "arXiv",
    primaryClass = "hep-ph",
    reportNumber = "TUM-EFT 188/24",
    doi = "10.1103/PhysRevD.110.054029",
    journal = "Phys. Rev. D",
    volume = "110",
    number = "5",
    pages = "054029",
    year = "2024"
}

@article{Lyu:2023xro,
    author = "Lyu, Yan and Aoki, Sinya and Doi, Takumi and Hatsuda, Tetsuo and Ikeda, Yoichi and Meng, Jie",
    title = "{Doubly Charmed Tetraquark $T_{cc}^+$ from Lattice QCD near Physical Point}",
    eprint = "2302.04505",
    archivePrefix = "arXiv",
    primaryClass = "hep-lat",
    reportNumber = "RIKEN-iTHEMS-Report-23, YITP-23-14",
    doi = "10.1103/PhysRevLett.131.161901",
    journal = "Phys. Rev. Lett.",
    volume = "131",
    number = "16",
    pages = "161901",
    year = "2023"
}

@article{Neubert:1996wg,
    author = "Neubert, Matthias",
    editor = "Zichichi, A.",
    title = "{Heavy quark effective theory}",
    eprint = "hep-ph/9610266",
    archivePrefix = "arXiv",
    reportNumber = "CERN-TH-96-281",
    journal = "Subnucl. Ser.",
    volume = "34",
    pages = "98--165",
    year = "1997"
}

@article{ParticleDataGroup:2024cfk,
    author = "Navas, S. and others",
    collaboration = "Particle Data Group",
    title = "{Review of particle physics}",
    doi = "10.1103/PhysRevD.110.030001",
    journal = "Phys. Rev. D",
    volume = "110",
    number = "3",
    pages = "030001",
    year = "2024"
}

@article{Eichten:1980mw,
    author = "Eichten, E. and Feinberg, F.",
    title = "{Spin Dependent Forces in QCD}",
    reportNumber = "HUTP-80-A053",
    doi = "10.1103/PhysRevD.23.2724",
    journal = "Phys. Rev. D",
    volume = "23",
    pages = "2724",
    year = "1981"
}

@article{Falkensteiner:1984su,
    author = "Falkensteiner, P. and Grosse, H. and Sch{\"o}berl, F.",
    title = "{Relativistic corrections to quarkonium spectra}",
    reportNumber = "UWThPh-1984-20",
    doi = "10.1016/0370-2693(84)91637-X",
    journal = "Phys. Lett. B",
    volume = "148",
    pages = "194--198",
    year = "1984"
}

@article{Barchielli:1988zp,
    author = "Barchielli, A. and Brambilla, N. and Prosperi, G. M.",
    title = "{Relativistic Corrections to the Quark-Antiquark Potential and the Quarkonium Spectrum}",
    reportNumber = "IFUM-352/FT",
    doi = "10.1007/BF02902620",
    journal = "Nuovo Cim. A",
    volume = "103",
    pages = "59",
    year = "1990"
}

@article{Laschka:2012cf,
    author = "Laschka, Alexander and Kaiser, Norbert and Weise, Wolfram",
    title = "{Charmonium Potentials: Matching Perturbative and Lattice QCD}",
    eprint = "1205.3390",
    archivePrefix = "arXiv",
    primaryClass = "hep-ph",
    doi = "10.1016/j.physletb.2012.07.049",
    journal = "Phys. Lett. B",
    volume = "715",
    pages = "190--193",
    year = "2012"
}

@article{Deng:2016ktl,
    author = "Deng, Wei-Jun and Liu, Hui and Gui, Long-Cheng and Zhong, Xian-Hui",
    title = "{Spectrum and electromagnetic transitions of bottomonium}",
    eprint = "1607.04696",
    archivePrefix = "arXiv",
    primaryClass = "hep-ph",
    doi = "10.1103/PhysRevD.95.074002",
    journal = "Phys. Rev. D",
    volume = "95",
    number = "7",
    pages = "074002",
    year = "2017"
}

@article{Koma:2007jq,
    author = "Koma, Yoshiaki and Koma, Miho and Wittig, Hartmut",
    editor = "Bali, Gunnar and Braun, Vladimir and Gattringer, Christof and Gockeler, Meinulf and Schafer, Andreas and Weisz, Peter and Wettig, Tilo",
    title = "{Relativistic corrections to the static potential at $O(1/m)$ and $O(1/m^2)$}",
    eprint = "0711.2322",
    archivePrefix = "arXiv",
    primaryClass = "hep-lat",
    reportNumber = "MKPH-T-07-15",
    doi = "10.22323/1.042.0111",
    journal = "PoS",
    volume = "LATTICE2007",
    pages = "111",
    year = "2007"
}

@article{Schlosser:2025tca,
    author = "Schlosser, Carolin and Wagner, Marc",
    title = "{Hybrid spin-dependent and hybrid-quarkonium mixing potentials at order $1/m_Q$ from SU(3) lattice gauge theory}",
    eprint = "2501.08844",
    archivePrefix = "arXiv",
    primaryClass = "hep-lat",
    doi = "10.1103/PhysRevD.111.074504",
    journal = "Phys. Rev. D",
    volume = "111",
    number = "7",
    pages = "074504",
    year = "2025"
}

@article{HadronSpectrum:2012gic,
    author = "Liu, Liuming and Moir, Graham and Peardon, Michael and Ryan, Sinead M. and Thomas, Christopher E. and Vilaseca, Pol and Dudek, Jozef J. and Edwards, Robert G. and Joo, Balint and Richards, David G.",
    collaboration = "Hadron Spectrum",
    title = "{Excited and exotic charmonium spectroscopy from lattice QCD}",
    eprint = "1204.5425",
    archivePrefix = "arXiv",
    primaryClass = "hep-ph",
    reportNumber = "TCDMATH-12-04, JLAB-THY-12-1510",
    doi = "10.1007/JHEP07(2012)126",
    journal = "JHEP",
    volume = "07",
    pages = "126",
    year = "2012"
}

@article{Cheung:2016bym,
    author = "Cheung, Gavin K. C. and O'Hara, Cian and Moir, Graham and Peardon, Michael and Ryan, Sin\'ead M. and Thomas, Christopher E. and Tims, David",
    collaboration = "Hadron Spectrum",
    title = "{Excited and exotic charmonium, $D_s$ and $D$ meson spectra for two light quark masses from lattice QCD}",
    eprint = "1610.01073",
    archivePrefix = "arXiv",
    primaryClass = "hep-lat",
    reportNumber = "DAMTP-2016-63",
    doi = "10.1007/JHEP12(2016)089",
    journal = "JHEP",
    volume = "12",
    pages = "089",
    year = "2016"
}

@article{TarrusCastella:2019lyq,
    author = "Tarr\'us Castell\`a, Jaume",
    editor = "Meye, Curtis and Schumacher, Reinhard A.",
    title = "{Spin structure of heavy-quark hybrids}",
    eprint = "1908.05179",
    archivePrefix = "arXiv",
    primaryClass = "hep-ph",
    doi = "10.1063/5.0008570",
    journal = "AIP Conf. Proc.",
    volume = "2249",
    number = "1",
    pages = "020008",
    year = "2020"
}

@article{Braaten:2024stn,
    author = "Braaten, E. and Bruschini, R.",
    title = "{Model-independent predictions for decays of hidden-heavy hadrons into pairs of heavy hadrons}",
    eprint = "2403.12868",
    archivePrefix = "arXiv",
    primaryClass = "hep-ph",
    doi = "10.1103/PhysRevD.109.094051",
    journal = "Phys. Rev. D",
    volume = "109",
    number = "9",
    pages = "094051",
    year = "2024"
}

@article{Bruschini:2023zkb,
    author = "Bruschini, R.",
    title = "{Heavy-quark spin symmetry breaking in the Born-Oppenheimer approximation}",
    eprint = "2303.17533",
    archivePrefix = "arXiv",
    primaryClass = "hep-ph",
    doi = "10.1007/JHEP08(2023)219",
    journal = "JHEP",
    volume = "08",
    pages = "219",
    year = "2023"
}

@article{Belle:2019cbt,
    author = "Mizuk, R. and others",
    collaboration = "Belle",
    title = "{Observation of a new structure near 10.75 GeV}",
    eprint = "1905.05521",
    archivePrefix = "arXiv",
    primaryClass = "hep-ex",
    reportNumber = "BELLE-CONF-1903",
    doi = "10.1007/JHEP10(2019)220",
    journal = "JHEP",
    volume = "10",
    pages = "220",
    year = "2019"
}

@article{LHCb:2025ubr,
    author = "Aaij, Roel and others",
    collaboration = "LHCb",
    title = "{Study of $B_{c}(1P)^{+}$ states in the $B_{c}^{+} \gamma$ mass spectrum}",
    eprint = "2507.02142",
    archivePrefix = "arXiv",
    primaryClass = "hep-ex",
    reportNumber = "CERN-EP-2025-122, LHCb-PAPER-2025-015",
    journal = "",
    month = "7",
    year = "2025"
}

@article{LHCb:2025uce,
    author = "Aaij, Roel and others",
    collaboration = "LHCb",
    title = "{Observation of orbitally excited $B_{c}^{+}$ states}",
    eprint = "2507.02149",
    archivePrefix = "arXiv",
    primaryClass = "hep-ex",
    reportNumber = "LHCb-PAPER-2025-014, CERN-EP-2025-123",
    journal = "",
    month = "7",
    year = "2025"
}

@article{Koma:2006si,
    author = "Koma, Yoshiaki and Koma, Miho and Wittig, Hartmut",
    title = "{Nonperturbative determination of the QCD potential at $O(1/m)$}",
    eprint = "hep-lat/0607009",
    archivePrefix = "arXiv",
    reportNumber = "DESY-06-062, RCNP-TH-06004, MKPH-T-06-12",
    doi = "10.1103/PhysRevLett.97.122003",
    journal = "Phys. Rev. Lett.",
    volume = "97",
    pages = "122003",
    year = "2006"
}

@inproceedings{Bali:1996bw,
    author = "Bali, Gunnar S. and Schilling, Klaus and Wachter, Armin",
    title = "{Spin and velocity dependent corrections to the interquark potential and quarkonia spectra from lattice QCD}",
    booktitle = "{2nd International Conference on Quark Confinement and the Hadron Spectrum}",
    eprint = "hep-ph/9611226",
    archivePrefix = "arXiv",
    reportNumber = "HLRZ-69-96, SHEP-96-26, WUB-96-39",
    pages = "277--280",
    month = "6",
    year = "1996"
}

@article{Bali:1997am,
    author = "Bali, Gunnar S. and Schilling, Klaus and Wachter, Armin",
    title = "{Complete  $\mathcal{O}(v^2)$ corrections to the static interquark potential from SU(3) gauge theory}",
    eprint = "hep-lat/9703019",
    archivePrefix = "arXiv",
    reportNumber = "SHEP-96-28, HLRZ-03-97, WUB-97-11",
    doi = "10.1103/PhysRevD.56.2566",
    journal = "Phys. Rev. D",
    volume = "56",
    pages = "2566--2589",
    year = "1997"
}

@article{Koma:2006fw,
    author = "Koma, Yoshiaki and Koma, Miho",
    title = "{Spin-dependent potentials from lattice QCD}",
    eprint = "hep-lat/0609078",
    archivePrefix = "arXiv",
    reportNumber = "DESY-06-061, MKPH-T-06-11, RCNP-TH-06003",
    doi = "10.1016/j.nuclphysb.2007.01.033",
    journal = "Nucl. Phys. B",
    volume = "769",
    pages = "79--107",
    year = "2007"
}

@article{Koma:2009ws,
    author = "Koma, Yoshiaki and Koma, Miho",
    editor = "Liu, Chuan and Zhu, Yu",
    title = "{Scaling study of the relativistic corrections to the static potential}",
    eprint = "0911.3204",
    archivePrefix = "arXiv",
    primaryClass = "hep-lat",
    doi = "10.22323/1.091.0122",
    journal = "PoS",
    volume = "LAT2009",
    pages = "122",
    year = "2009"
}

@article{Koma:2012bc,
    author = "Koma, Yoshiaki and Koma, Miho",
    editor = "Leinweber, Derek and Kamleh, Waseem and Mahbub, Selim and Matevosyan, Hrayr and Thomas, Anthony and Williams, Anthony G. and Young, Ross and Zanotti, James",
    title = "{Heavy quarkonium spectroscopy in pNRQCD with lattice QCD input}",
    eprint = "1211.6795",
    archivePrefix = "arXiv",
    primaryClass = "hep-lat",
    doi = "10.22323/1.164.0140",
    journal = "PoS",
    volume = "LATTICE2012",
    pages = "140",
    year = "2012"
}

@article{Mateu:2018zym,
    author = "Mateu, Vicent and Ortega, Pablo G. and Entem, David R. and Fern{\'a}ndez, Francisco",
    title = {{Calibrating the Na{\"\i}ve Cornell Model with NRQCD}},
    eprint = "1811.01982",
    archivePrefix = "arXiv",
    primaryClass = "hep-ph",
    reportNumber = "IFT-UAM/CSIC-17-110, IFT-UAM/CSIC-18-109",
    doi = "10.1140/epjc/s10052-019-6808-2",
    journal = "Eur. Phys. J. C",
    volume = "79",
    number = "4",
    pages = "323",
    year = "2019"
}

@article{Brambilla:2026ujo,
    author = "Brambilla, Nora and Butenschoen, Mathias and Hibler, Simon and Mohapatra, Abhishek and Vairo, Antonio and Wang, Xiangpeng",
    title = "{Inclusive hadroproduction of {\ensuremath{\chi}}$_{c1}$(3872), X$_{b}$ and pentaquarks}",
    eprint = "2602.14916",
    archivePrefix = "arXiv",
    primaryClass = "hep-ph",
    reportNumber = "TUM-EFT 203/26",
    doi = "10.1007/JHEP07(2026)260",
    journal = "JHEP",
    volume = "07",
    pages = "260",
    year = "2026"
}

@article{Lai:2025tpw,
    author = "Lai, Wai Kin and Chung, Hee Sok",
    title = "{Hadroproduction data support tetraquark hypothesis for $\chi_{c1(3872)}$}",
    eprint = "2505.06910",
    archivePrefix = "arXiv",
    primaryClass = "hep-ph",
    doi = "10.1103/lkff-d2ph",
    journal = "Phys. Rev. D",
    volume = "112",
    number = "5",
    pages = "054005",
    year = "2025"
}

@article{Isgur:1991wq,
    author = "Isgur, Nathan and Wise, Mark B.",
    title = "{Spectroscopy with heavy quark symmetry}",
    reportNumber = "CEBAF-TH-91-01, WM-91-101, CALT-68-1704",
    doi = "10.1103/PhysRevLett.66.1130",
    journal = "Phys. Rev. Lett.",
    volume = "66",
    pages = "1130--1133",
    year = "1991"
}

@article{Sharma:2026htv,
    author = "Sharma, Sipaz and Urrea-Ni{\~n}o, Juan Andr{\'e}s and Brambilla, Nora and Knechtli, Francesco and Peardon, Michael J.",
    title = "{Imprint of the adjoint meson spectrum in the decay patterns of hidden-bottom tetraquarks}",
    eprint = "2603.29581",
    archivePrefix = "arXiv",
    primaryClass = "hep-lat",
    reportNumber = "TUM-EFT 210/26",
    doi = "10.22323/1.518.0086",
    journal = "PoS",
    volume = "LATTICE2025",
    pages = "086",
    year = "2026"
}

@article{Brambilla:2026jqz,
    author = "Brambilla, Nora and Bruschini, Roberto and Mohapatra, Abhishek and Peng, Fang-Zheng and Scirpa, Tommaso",
    title = "{Quarkoniumlike states above open-flavor thresholds in Born-Oppenheimer EFT}",
    eprint = "2608.04105",
    journal = "",
    archivePrefix = "arXiv",
    primaryClass = "hep-ph",
    month = "8",
    year = "2026"
}
\end{document}